\documentclass[fleqn, final, openany]{unmeethesis}
\usepackage{zj}

\makeatletter
\def\CatchFBT@Fin@l#1[#2]{%
   \begingroup
      \makeatletter #2%
      \scantokens\expandafter{%
         \expandafter\CatchFBT@tok\expandafter{\the\CatchFBT@tok}}%
      \CatchFBT@IsAToken{#1}
         {\global#1\expandafter{\the\CatchFBT@tok}}
         {\xdef#1{\the\CatchFBT@tok}}%
      \ifx\CatchFBT@tok#1\else\global\CatchFBT@tok{}\fi
   \endgroup
}
\makeatother

\begin{document}
\frontmatter


\title{Particle Correlations in Bose-Einstein Condensates}

\author{Zhang Jiang}

\degreesubject{Ph.D., Physics}

\degree{Doctor of Philosophy \\ Physics}

\documenttype{Dissertation}

\previousdegrees{B.S., M.E., University of Science and Technology of China, 2004 \\ M.S., Physics, University of Science and Technology of China, 2008}

\date{July 2014}

\maketitle

\makecopyright

\begin{dedication}
  To my parents. It was you who ignited my curiosity, continued encouraging me, and noticed every progress I had made.\\ 
  \vspace{1em}
  To my beloved wife. Your tenderness and kindheartedness have made me fearless. 
\end{dedication}

\begingroup 
\let\clearpage\relax
\begin{acknowledgments}
\noindent I am extremely lucky to have received much support from my family, friends, and mentors.  Without these people this dissertation would never have been possible.  I am indebted to my dissertation advisor Carl Caves, from whom you can find friendship, knowledge, wisdom, and encouragement.  Carl is one of those physicists who think not only about research, but also pay a lot of attention to advancing their students.  I am owing to Xiao-Nan Wang and Jiong-Ming Zhu, who first showed me how to think like a physicist; Qing Chen, who introduced me to the field of quantum information; Ivan Deutsch, who taught me useful things such as quantum optics and warmly encouraged my progress; Andrew Landahl, for his excellent advice on how to make a good talk; Akimasa Miyake, who cares for me as if I were his own student; and Marco Piani at the Institute for Quantum Computing, with whom I had a very pleasant and fruitful collaboration.

Many thanks to Alex Tacla, from whom I learned how to balance life and research; Shashank Pandey, who made the most serious thinking of our quantum amplifier papers; Matthias Lang, who kept me from a couple of computer disasters; Josh Combes, who was very generous to give me advice on academia and to teach me how to climb like an expert; Vaibhav Madhok, who made life in CQuIC full of surprises; Jonas Anderson, who pointed out a lot of interesting things to me; Chris Cesare, from whom I learned what topological quantum computing is; Ben Baragiola, who decorated our office like a botanic garden; Robert Cook, a playful but serious physicist; Leigh Norris, who saved my life by teaching me how not to make ugly Mathematica plots; Carlos Riofr\'{i}o, whose warmth and care made CQuIC a better place; Krittika Goya, who taught me what a Feshbach resonance is; Robin Blume-Kohout, who helped me to revise my presentation slides; and the rest of current or former CQuIC group guys---Charlie Baldwin, Sergio Boixo, Adrian Chapman, Ninnat Dangniam, Bryan Eastin, Andrew Ferdinand, Chris Ferrie, Steve Flammia, Jonathan Gross, Bob Keating, Jacob Miller, Seth Merkel, Xiaodong Qi, Akash Rakholia, Iris Reichenbach, Anil Shaji, Ezad Shojaee, Munik Shrestha, and Rolando Somma.

I also thank Vicky Bird for booking the flights and hotels and for filling my empty stomach; Alisa Gibson for all the paperwork, even before I came to UNM, that she did so smoothly; and Linda Melville for much helpful advice about living and studying in the \hbox{US}.
\end{acknowledgments} 
\endgroup

\begin{abstract}

\noindent 
The impact of interparticle correlations on the behavior of Bose-Einstein Condensates (BECs) is discussed using two approaches.  In the first approach, the wavefunction of a BEC is encoded in the $N$-particle sector of an extended ``catalytic state''.  Going to a time-dependent interaction picture, we can organize the effective Hamiltonian by powers of ${N}^{-\half}$.   Requiring the terms of order ${N}^{\half}$ to vanish, we get the Gross-Pitaevskii Equation.  Going to the next order, $N^0$, we obtain the number-conserving Bogoliubov approximation.  Our approach allows one to stay in the Schr\"{o}dinger picture and to apply many techniques from quantum optics.  Moreover, it is easier to track different orders in the Hamiltonian and to generalize to the multi-component case.  In the second approach, I consider a state of $N =l\times n$ bosons that is derived by symmetrizing the $n$-fold tensor product of an arbitrary $l$-boson state.  Particularly, we are interested in the pure state case for $l=2$, which we call the Pair-Correlated State (PCS).  I show that PCS reproduces the number-conserving Bogoliubov approximation; moreover, it also works in the strong interaction regime where the Bogoliubov approximation fails.  For the two-site Bose-Hubbard model, I find numerically that the error (measured by trace distance of the two-particle RDMs) of PCS is less than two percent over the entire parameter space, thus making PCS a bridge between the superfluid and Mott insulating phases.  Amazingly, the error of PCS does not increase, in the time-dependent case, as the system evolves for longer times.  I derive both time-dependent and -independent equations for the ground state and the time evolution of the PCS ansatz.  The time complexity of simulating PCS does not depend on $N$ and is linear in the number of orbitals in use.  Compared to other methods, e.g, the Jastrow wavefunction, the Gutzwiller wavefunction, and the multi-configurational time-dependent Hartree method, our approach does not require quantum Monte Carlo nor demanding computational power.

\end{abstract} 

\setlength{\headsep}{1.5em} 

\tableofcontents 

\addcontentsline{toc}{chapter}{{\bf Contents}}

\listoffigures


\newcommand{\symb}[2]{\makebox[6em][l]{#1} #2}

\chapter{List of Symbols}

\symb{$\delta_{jk}$}{Kronecker delta function}\\
\symb{$\delta(x_1, x_2)$}{Dirac delta function}\\
\symb{$\dt A$}{Time derivative of $A$}\\
\symb{$A^\dagger$}{Hermitian conjugate of $A$}\\
\symb{$A^*$}{Complex conjugate of $A$}\\
\symb{$A^\transp$}{Transpose of $A$}\\
\symb{$\det(A)$}{Determinant of $A$}\\
\symb{$\tr(A)$}{Trace of $A$}\\
\symb{$\av{A}$}{Expectation value of $A$}\\
\symb{$\commut{A}{B}$}{Commutator of $A$ and $B$}\\
\symb{$\anticommut{A}{B}$}{Anticommutator of $A$ and $B$}\\
\symb{$\Colon\,\quad\,\Colon$}{Normal ordering of modal operators}\\
\symb{$\re(z)$}{Real part of $z$}\\
\symb{$\im(z)$}{Imaginary part of $z$}\\
\symb{$\uppsi$, $\upphi$}{Upright Greek fonts for field operators}\\
\symb{$\sH$}{\AMS\hspace{.6pt} mathcal fonts for many-body operators}\\
\symb{$\mathsp{H}$}{Slanted sans serif fonts for matrices of symplectic structure}\\
\symb{$\dsH$}{Double stroke fonts for subsets of the Hilbert space}\\
\symb{$\identity$, $\nullmatrix$}{The identity matrix and null matrix}\\
\symb{$\xbf$}{Vectors of spatial degrees of freedom}\\
\symb{$\vec{x}$}{Vectors of other degrees of freedom}\\
\symb{$\a$, $\a^\dagger$ ($\b$, $\b^\dagger$)}{Creation and annihilation operators for bosonic modes}\\
\symb{$\alpha$, $\beta$}{Amplitude of coherent states}\\
\symb{$\psinot$}{Condensate wavefunction}\\
\symb{$\ket{\vac}$}{Vacuum state}\\
\symb{$i$}{Unit imaginary number}\\
\symb{$\mathrm{c.c.}$}{Complex conjugate}\\
\symb{$\mathrm{H.c.}$}{Hermitian conjugate}\\
\symb{$\braket{\psi}{\uppsi}$}{Annihilation operator of the state $\psi(\xbf)$, i.e., $\int\uppsi(\xbf)\,\psi^*(\xbf)\,\dif \xbf$}\\
\symb{$\braket{\uppsi}{\psi}$}{Creation operator of the state $\psi(\xbf)$, i.e., $\int\uppsi^\dagger(\xbf)\,\psi(\xbf)\,\dif \xbf$}\\
\symb{$\ket{\varPsi}$}{Slanted capital Greek letters for many-body quantum states}\\

\newcommand{\acronym}[2]{\makebox[6em][l]{#1} #2}

\chapter{List of Acronyms}

\acronym{1RDM}{single-particle Reduced Density Matrix}\\
\acronym{2RDM}{two-particle Reduced Density Matrix}\\
\acronym{BdG}{Bogoliubov-de Gennes}\\
\acronym{BEC}{Bose-Einstein Condensate}\\
\acronym{BF}{Bargmann-Fock}\\
\acronym{BPCS}{Bosonic Particle-Correlated State}\\
\acronym{DFS}{Double-Fock State}\\
\acronym{ECS}{Extended Catalytic State}\\
\acronym{GP}{Gross-Pitaevskii}\\
\acronym{GPE}{Gross-Pitaevskii Equation}\\
\acronym{GPS}{Gross-Pitaevskii State}\\
\acronym{MCTDHB}{Multiconfigurational Time-Dependent Hartree Method for Bosons}\\
\acronym{ODLRO}{Off-Diagonal Long Range Order}\\
\acronym{PCS}{Pair-Correlated State}\\
\acronym{$q$RDM}{$q$-particle Reduced Density Matrix}\\
\acronym{RDM}{Reduced Density Matrix}\\
\acronym{SLD}{Symmetric Logarithmic Derivative}\\
\acronym{SQUID}{Superconducting Quantum Interference Device}\\
\acronym{TFS}{Twin-Fock State}

\mainmatter

\renewcommand{\lb}[1]{\label{introduction:#1}}
\renewcommand{\rf}[1]{\ref{introduction:#1}}

\chapter{Introduction}
\label{ch:introduction}

\begin{quote}
Today, if you have a demanding job for light, you use an optical laser. In the future, if there is a demanding job for atoms, you may be able to use an atom laser.\\[4pt]
--  Wolfgang Ketterle\ai{Ketterle, Wolfgang}
\end{quote}

\noindent When Charles H. Townes, Nikolay Basov, Aleksandr Prokhorov, and Theodore Maiman were working on masers and lasers in the 1950s and 60s, they probably did not anticipate the extremely wide range of applications of their work.  Lasers are ubiquitous nowadays: from supermarket scanners to remote sensing, from CD players to holographic technologies.  But what makes lasers so useful besides their extremely high powers?  Compared to previous light sources, lasers are both monochromatic and coherent; these properties enable one to take advantage of---much more efficiently---the interference effects of light.  Interference is the key to manipulating waves just like Newton's laws of motion are crucial to controlling the movements of particles.  Lasers allow one to control the waveforms of light.

Perhaps one of the most surprising discoveries of quantum mechanics is the wave-particle duality\si{Wave-particle duality}, or equivalently, that any quantum particle can interfere with itself.  This ``weird'' idea of de Broglie\ai{de Broglie, Louis} turned out to be quite useful beyond its purely theoretical interest; e.g., the diffraction patterns of scattered neutrons can provide information about the positions and movements of the nuclei in a sample.  ``Atomic physics after about 1975 has been interested in controlling the atom's external quantum state \ldots\ .  The ultimate payoff of external state manipulation is atom interferometry which involves making and reading out superpositions of the external quantum state,'' said David E. Pritchard~\cite{pritchard_bose-einstein_2002}\ai{Pritchard, David E.}\si{Atom interferometer}.  Atom interferometers~\cite{kasevich_coherence_2002, cronin_optics_2009}, since their first demonstration~\cite{carnal_youngs_1991}, have had an impact on many fields of science and engineering, for example, acceleration and rotation sensors, gravitational-wave detectors, and various other force and field sensors.  Despite their relatively short history, atomic sensors can already compete with the best classical or laser-based sensors~\cite{gustavson_precision_2000, kitching_atomic_2011}.  A typical principle of operation is as follows: A thermal cesium atomic beam crosses three laser interaction regions where two-photon stimulated Raman transitions between cesium ground states transfer momentum to atoms and divide, deflect, and recombine the atomic wavepackets;  rotation induces a phase shift between the two possible trajectories and causes a change in the detected number of atoms with a particular internal state.

The advantages of atom interferometers include short de Broglie wavelength, narrow frequency response, and long interrogation times.  The trapping of atoms also allows the possibility of steadily splitting the wavefunction in coordinate space, for example the double-well configuration in Fig.~\ref{fig:double_well_interfer}; this property, which is unprecedented with light interferometers\cmc{Couldn't tell what this was supposed to mean?  Splitting in momentum leads to splitting in position, so I am guessing you mean that we can do the splitting in position directly, as in raising a barrier that splits a BEC in two.}\zj{What I mean is: The wavefunction of atom interferometers can be \emph{steadily} splitted in coordinate space.}, is essential to experimental studies of spatially varying fields\cmc{Couldn't tell what you were trying to get across here.  Do you mean that the evanescent fields vary strongly in space and are studies by coupling them to atoms and other small objects?  Not clear what you are saying}\zj{If someone wants to measure the gradient of a field, he must split the wavefunction to two spatially different positions and let the wavefunction evolve under the local field. The local field created by the small object on the atoms, can be measured by an atom interferometer. Anyway, I deleted the evanescent coupling part.} such as gravitational forces.

\section{Bose-Einstein Condensates}

As in the case of light, one needs a unified army of atoms, all marching in step---i.e., an atom laser---to enhance the performance of a matter-wave interferometer, instead of just an ensemble of uncorrelated atoms.  An atom laser\si{Atom laser} can be thought as a monochromatic and coherent matter-wave beam~\cite{bloch_atom_1999}; monochromatic means that most of the atoms in the beam occupy the same quantum state, and coherent means that the phases of atom beams are well defined and can be correlated.  Lasers are made by stimulated emission, which is not an option for atoms; so how can we make an atom laser?  Remember that a laser, where photons share the same quantum state, is a state of light with extremely low entropy; similarly, the entropy of an atom laser must be very low.  One way to get low-entropy states is to cool the atoms, and the temperature at which an atom beam becomes an atom laser is called the critical temperature $T_{\mathrm{\ssC}}$\si{Critical temperature}.  It turns out that for classical (distinguishable) particles, the critical temperature $T_{\mathrm{\ssC}}$ goes to zero when the number of particles $N$ is very large (the thermodynamic limit)\cmc{I don't see what the number of particles has to do with it.  Are you trying to say that for distinguishable particles there is no gap, i.e., the energy levels are dense near the ground state, so there is no condensation to the ground state till the temperature goes to zero?}\zj{This is not completely true.  The energy levels are the same for classical and bosonic particles (assume no interaction).  The degeneracy of excited states for bosons, however, is much smaller than that of classical particles (permutations do not change the bosonic states).  Thus, a large portion of classical particles are excited at low temperature. It turns out that one needs to go to zero temperature for a macroscopic occupation of the ground state when $N$ goes to infinity.}.  Microscopic particles are not classical particles; instead they are indistinguishable particles obeying either Bose-Einstein statistics\si{Bose-Einstein statistics} (bosons) or Fermi-Dirac statistics (fermions)\si{Fermi-Dirac statistics}.  Bosons tend to occupy the same quantum state, while fermions can only occupy different quantum states.  In 1925, based on the earlier work of Satyendra Nath Bose\ai{Bose, Satyendra Nath}, Albert Einstein\ai{Einstein, Albert} pointed out that a large fraction of particles in a boson gas condense to the lowest quantum state at a finite critical temperature $T_{\mathrm{\ssC}}$, which depends only on the mass of the particles and the density of particles at the thermodynamic limit.  This phenomenon, called a Bose-Einstein Condensate (BEC), is a consequence of the Bose-Einstein statistics rather than attractive interactions between the particles; BECs happen when the de Broglie waves of atoms overlap, and this gives an estimate of the critical temperature
\begin{equation}
 T_{\mathrm{\ssC}} \sim \frac{\hbar^2}{m k_\mathrm{\ssB}}\bigg(\frac{N}{V}\bigg)^{2/3}\,,
\end{equation}
where $\hbar$ is the reduced Planck constant, $k_\mathrm{\ssB}$ is the Boltzmann constant, $m$ is the mass per particle, $N$ is the number of particles, and $V$ is the volume of the gas.

To get the ultimate resolution of an atom interferometer, we need a single-mode, coherent source that is also very bright, i.e., a Bose-Einstein condensate.  Yet BECs are so fragile that they have not been observed in nature.  Erwin Schr\"{o}dinger\ai{Schr\"{o}dinger, Erwin} wrote~\cite{schrodinger_statistical_1952}, ``The densities are so high and the temperatures so low \ldots\ the van der Waals corrections are bound to coalesce with the possible effects of degeneration \ldots\ .''  What Schr\"{o}dinger meant was that the atoms will condense into a solid or a liquid before they become a BEC, but what he forgot to consider was a dilute gas in a metastable phase; the challenge is to find a window of the critical temperature $T_{\mathrm{\ssC}}$ accessible by cryogenics, while making the lifetime of the metastable state long enough.  It turned out that the feasible critical temperature $T_{\mathrm{\ssC}}$ is around or below one microkelvin; e.g., the first BEC realized at the NIST-JILA lab by Eric Cornell\ai{Cornell, Eric} and Carl Wieman\ai{Wieman, Carl}~\cite{cornell_nobel_2002} had a temperature of \unit{170}\nano\kelvin.
\begin{figure}[ht] 
   \centering
   \def\svgwidth{0.85\textwidth} 
   \input{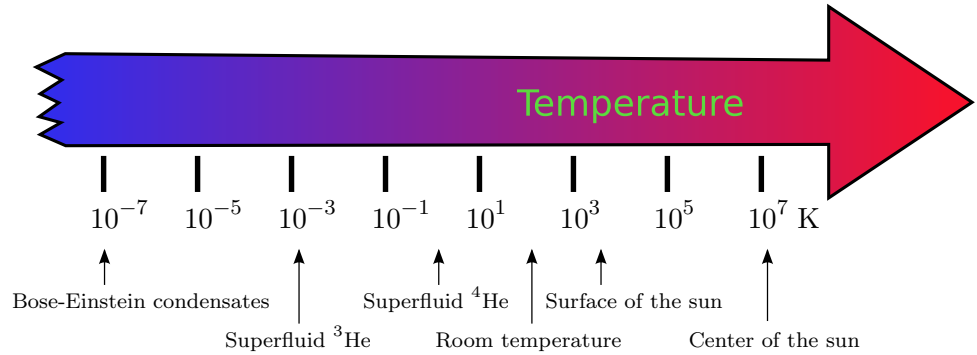}
   \caption[Temperature Gradient from a BEC to the Center of the Sun]{Temperature gradient from a BEC to the center of the sun.}
   \label{fig:temperature_gradient}
\end{figure}

Conventional cryogenics such as the dilution refrigerators provide cooling to the temperature of superfluid $^3\mathrm{He}$, which is about \unit{2}\milli\kelvin, but this temperature is still about $10,000$ times hotter than a BEC.  In addition, it takes a long time to reach these very low temperature using conventional cryogenics, not to mention the associated high maintenance costs.  To chill the atoms more efficiently, physicists invented laser cooling,\si{Laser cooling} which soon became a central topic of modern cryogenics.  But how can lasers, emitting high-energy photons, be used for cooling?  The answer again lies in the low entropy of lasers; even though they emit photons whose energy is about the same as those emitted by the Sun, the laser photons are actually very cold and thus able to extract entropy from the atoms as the photons heat up.  The most common method of laser cooling, Doppler cooling\si{Doppler cooling}, works via the recoil forces exerted by the light on the atoms; no matter which way an atom drifts, it always runs into a laser beam that slows it down.  But Doppler cooling has its limit due to the natural linewidth of the atoms in use, which for Rubidium 85 is around \unit{150}\micro\kelvin.  Other more sophisticated laser cooling techniques, such as Sisyphus cooling and polarization gradient cooling, are able to approach the much lower recoil temperature; the recoil temperature, which is usually on the order of \unit{1}\micro\kelvin, equals the recoil energy deposited in a single atom initially at rest by the spontaneous emission of a single photon.  The final step that takes laser cooled atoms to BECs is evaporative cooling, which works by allowing the atoms with highest energy to escape from the trap where the atoms are stored.  There are many excellent reviews of the theory and experiment of dilute-gas atomic BECs~\cite{taubes_hot_1994, dalfovo_theory_1999, townsend_bose-einstein_1997, ketterle_nobel_2002, cornell_nobel_2002}.

\section{Interferometers with BECs}

Physics with coherent matter waves is an emerging research field~\cite{molmer_quantum_2003, bongs_physics_2004}.  The first observation of interference between two BECs was made in Ketterle's group~\cite{andrews_observation_1997}, owing to the high density of atoms in their Ioffe-Pritchard trap.  Anderson \emph{et al.}~\cite{anderson_macroscopic_1998} observed interference of BECs tunneling from an array of traps.  Stenger \emph{et al.}~\cite{stenger_bragg_1999} developed Bragg spectroscopy to measure properties of a condensate for high-resolution velocimetry.  Shin \emph{et al.}~\cite{shin_atom_2004} succeeded in making a BEC based interferometer by continuously deforming between single- and double-well trapping potentials; the device was then miniaturized by using an atom chip~\cite{shin_interference_2005,schumm_matter-wave_2005}.\footnote{See~\cite{hinds_two-wire_2001} for a theoretical proposal to use atom chips as a substitute for magneto-optical traps: ``A versatile miniature de Broglie waveguide is formed by two parallel current-carrying wires in the presence of a uniform bias field \ldots\ it offers a remarkable range of possibilities for atom manipulation \ldots\ include controlled and coherent splitting of the wave function as well as cooling, trapping, and guiding.''\cmc{The grammar in this quote isn't quite right, which suggests that you didn't quote it correctly.}\zj{You are pretty right.}}  Hadzibabic \emph{et al.}~\cite{hadzibabic_interference_2004} observed matter-wave interference between 30 BECs with uncorrelated phases.  Recently, a matter-wave interferometer that uses optical ionization gratings has also been demonstrated~\cite{haslinger_universal_2013}; this does not require any particular internal level structure and is thus universally applicable to different kinds of particles.  In addition, the first controllable atomic circuit that functions analogously to a Superconducting Quantum Interference Device (SQUID) has been implemented~\cite{wright_driving_2013}.

Before diving into atom interferometers, we review the five essential steps of a general interferometer: (i)~prepare the initial state; (ii)~split the initial state into a coherent superposition of two states; (iii)~apply interactions that affect the two states differently, for example, due to their different spatial locations; (iv)~recombine the evolved states coherently; and (v)~measure the shift of the interference fringes.  In the following, I briefly review two different configurations of atom interferometers based on the above prescription; the atoms are not required to be condensed into a BEC for the first configuration, while BECs are required for the second configuration.

\begin{figure}[ht] 
   \centering
   \def\svgwidth{0.85\textwidth} 
   \input{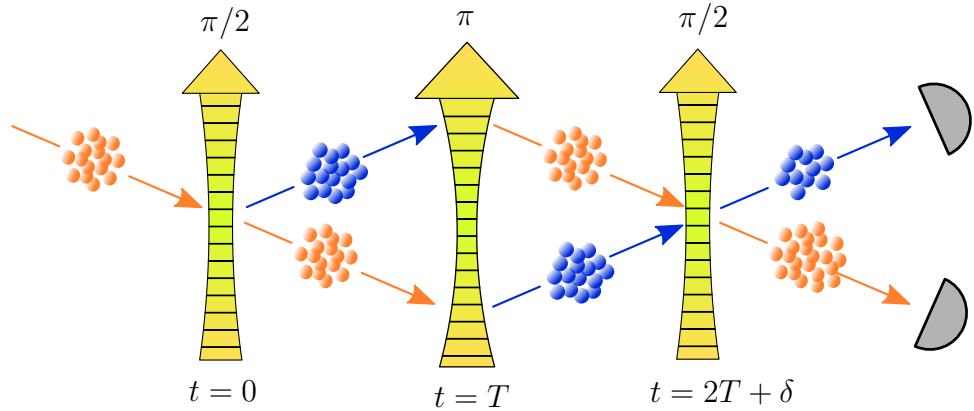}
   \caption[A Typical Atom Interferometer]{A typical atom interferometer: The first $\pi/2$ pulse transfers half of the atoms coherently from the ground state $\ket{g}$ into the other hyperfine state $\ket{e}$. After flying apart for time $T$, the atom states are mirrored by a $\pi$ pulse.  The atoms overlap again at time $t=2T$, and a second $\pi/2$ pulse recombines the two clouds.  The number of atoms left in the state $\ket{g}$ depends on the phase difference accumulated between the two paths. Furthermore, if the atoms are cold enough to form a BEC and the time delay $\delta$ is nonzero, one should be able to observe interference fringes for both of the two clouds at the detectors.}
   \label{fig:atom_interfer}
\end{figure}

In the first configuration, moving clouds of ultracold atoms are subjected to three laser beams as in Fig.~\ref{fig:atom_interfer}.
The lasers are tuned to induce Raman transitions between the two hyperfine ground states of atoms.  The first $\pi/2$ pulse transfers half of the atoms coherently from the ground state $\ket{g}$ into the other hyperfine state $\ket{e}$; the atoms excited to $\ket{e}$ receive a recoil force and separate from those still left in the ground state $\ket{g}$.  After evolving for a time $T$, the atom states are mirrored by a $\pi$ pulse, and the recoil force swaps the velocity of the atoms in the two hyperfine states.  The atoms overlap again at time $t=2T$, and a second $\pi/2$ pulse recombines the two clouds.  The number of atoms left in the state $\ket{g}$ depends on the phase difference accumulated between the two paths.  Furthermore, if the atoms are cold enough to form a BEC and the time delay $\delta$ is nonzero, one should observe interference fringes for both of the two clouds at the detectors.

\begin{figure}[ht] 
   \centering
   \def\svgwidth{0.85\textwidth}
   \input{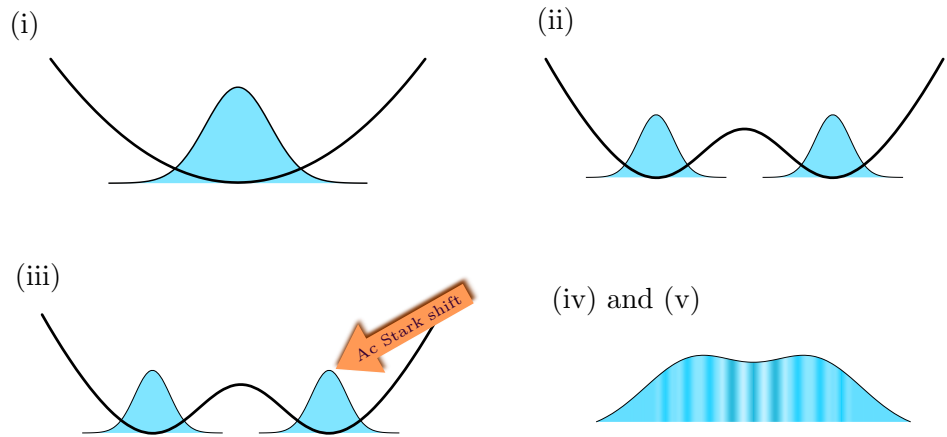}
   \caption[A Double-Well Atom Interferometer]{A double-well atom interferometer~\cite{shin_atom_2004}: (i)~cool the atoms trapped in a single-well potential to form a BEC; (ii)~split the condensate by slowly deforming the single-well potential to a double-well potential; (iii)~apply ac Stark shift potentials to either of the two separated condensates; (iv)~turn off the double-well trapping potential, and let the condensates ballistically expand, overlap, and interfere; and (v)~take an absorption image.}
   \label{fig:double_well_interfer}
\end{figure}

In the other configuration~\cite{shin_atom_2004}, where the atoms are put in a double-well trapping potential, a BEC is considered necessary for the fringe patterns to be visible.  The five steps to perform this interferometer are listed in Fig.~\ref{fig:double_well_interfer}: (i)~cool the atoms trapped in a single-well potential to form a BEC; (ii)~split the condensate by slowly deforming the single-well potential to a double-well potential; (iii)~apply ac Stark shift potentials to either of the two separated condensates; (iv)~turn off the double-well trapping potential, thus letting the condensates ballistically expand, overlap, and interfere; and (v)~take an absorption image.
In this configuration, the atoms are confined in a trapping potential until the measurement is made, while the atoms are free to fly in the first configuration.  This confinement of atoms allows long interrogation time, and the relative phase of two condensates can be measured using a small sample of the atoms.  In addition, confined atom interferometers, especially those using atom chips, can be small and portable.  On the other hand, confined atom interferometers usually operate with high density to achieve large signals, from which several disadvantages follow: (i)~Three-body interactions can cause the atomic gas to form a liquid or solid; (ii)~strong two-body interactions can cause large depletion of the condensate, even at zero temperature; (iii)~the matter-wave dynamics becomes nonlinear; (iv)~interactions suppress the Josephson oscillation and cause phase diffusion; (v)~the potential wells have to be controlled very accurately in stiffness and depth to prevent additional frequency shifts; (vi)~the collective excitations of the condensate have to be controlled carefully, because, for example, sound or shape oscillations may arise if the potential changes too suddenly.

\section{Nonclassical Aspects of BECs}

To describe the interference of BECs, perhaps the most obvious choice is to modify the existing theory for laser light fields.  There are, however, two intrinsic difference between lasers and BECs: (i)~Lasers are superpositions of different number states of photons, while the number of particles in a BEC is usually fixed; (ii)~lasers are composed of noninteracting photons, while particles in a BEC generally interact with one another.  In this section, I discuss how these two differences are handled in the literature.

Similar to a laser light, a BEC is usually described by a coherent state, i.e., an eigenstate of the annihilation operator for a particular state of the single-atom Hilbert space.  A well defined amplitude $\norm{\alpha}$ and phase angle $\theta=\arg(\alpha)$ is associated with this coherent state.  When particle loss is negligible, however, the real condensate is much closer to a number state than to a coherent state.  A number state is distributed evenly across all phase-plane directions, and no definite phase can be attributed to it.  The fictitious phase to the coherent state breaks the rotational symmetry and introduces Goldstone bosons into the Bogoliubov approximation to the condensate~\cite{lewenstein_quantum_1996}; these should be treated as a defect of the mathematical description and not as a physical property of the atomic system.  The Goldstone mode causes the condensate state to deviate linearly in time from a single condensate in a coherent
state (i.e., this is a secular deviation, not an oscillation).  This problem is particularly
pesky when the condensate is in a trapping potential, where the Goldstone mode is
a mixture of the condensate mode and modes orthogonal to it and thus cannot be
removed easily.  The solution to getting rid of the unphysical Goldstone mode is to
adhere to the fact that the condensate has a fixed number of particles, i.e., by using
a Bogoliubov approximation where particle number is conserved~\cite{girardeau_theory_1959, gardiner_particle-number-conserving_1997, castin_low-temperature_1998}.

Due to the interparticle interactions, a BEC based interferometer is very different from a laser based interferometer.  A nonlinear effect known as macroscopic quantum self-trapping in bosonic Josephson junctions has been discussed and observed in~\cite{smerzi_quantum_1997, smerzi_macroscopic_2000, cataliotti_josephson_2001, albiez_direct_2005, ananikian_gross-pitaevskii_2006}.  The collapse and revival of interference patterns of matter waves were observed in optic lattices~\cite{greiner_collapse_2002, schachenmayer_atomic_2011}.  It is worth mentioning that sometimes the nonlinear effect is useful to achieving high sensitivity.  One example is that of fast moving solitons, created in the recombination stage, which can enhance the sensitivity of phase measurements~\cite{negretti_enhanced_2004}.  Nonlinear effects in BECs can be useful for parameter estimation problems~\cite{choi_bose-einstein_2008, boixo_quantum-limited_2009, tacla_nonlinear_2010}.  In particular, they can be used to produce so-called spin squeezed states\si{Spin squeezed state}, which are useful for overcoming the shot-noise limit\si{Standard quantum limit}~\cite{caves_quantum-mechanical_1981, bouyer_heisenberg-limited_1997, orzel_squeezed_2001, dunningham_interferometry_2002, giovannetti_quantum-enhanced_2004, esteve_squeezing_2008, pezze_entanglement_2009, gross_nonlinear_2010, riedel_atom-chip-based_2010, gross_spin_2012}.

To quantify the impact of interparticle interactions on a BEC based interferometer, many authors have used the two-mode model\si{Two-mode model}.  Milburn \emph{et al.}~\cite{milburn_quantum_1997} showed that the mean-field solution is modulated by a quantum collapse and revival sequence.  In~\cite{cirac_quantum_1998, steel_quantum_1998, sorensen_many-particle_2001}, the authors argued that it is feasible to prepare, control, and detect a Schr\"odinger cat state with a two-component BEC.  Spekkens and Sipe~\cite{spekkens_spatial_1999} considered the transition from a single condensate to a fragmented condensate as the central barrier in a double-well trapping potential is raised.  Menotti \emph{et al.}~\cite{menotti_dynamic_2001} explicitly considered the spatial dependence of the mode functions.  Mahmud \emph{et al.}~\cite{mahmud_quantum_2005} analyzed the two-mode model in the quantum phase-space picture.  Several authors~\cite{huang_optimized_2008, lapert_optimal_2012} discussed how to optimally create quantum superpositions and squeezing using the bosonic Josephson Hamiltonian.\footnote{The ground and excited states of the bosonic Josephson Hamiltonian can be solved exactly by using the algebraic Bethe ansatz~\cite{links_algebraic_2003, zhou_exact_2003}.}  More realistically, the influences of the noncondensed modes on the relative phase of the two condensed modes were considered in~\cite{villain_dephasing_1999, sinatra_coherence_2009, gillet_tunneling_2014}.

\section{Beyond Mean Field}

Interesting phenomena such as quantum phase transitions from a superfluid to a Mott insulator in a Bose gas~\cite{jaksch_cold_1998, greiner_quantum_2002} cannot be interpreted within the mean-field approach.  Also, a proper analysis of the atom interferometers discussed above requires one to go beyond mean-field theory.  The most common way to do so is by perturbing the mean field with collective excitations, i.e., the Bogoliubov approximation.  When interparticle interactions are strong or the mean field is unstable, the Bogoliubov approximation fails.  In this section, I introduce some nonperturbative methods that go beyond the mean field theory.

One powerful idea to deal with strong interactions is to construct a many-particle wavefunction from a two-particle state.  In 1950s, Jastrow introduced the wavefunction which bears his name;\ai{Jastrow, Robert}\si{Jastrow wavefunction} the $N$-particle state is a product of two-particle states of all $N(N-1)/2$ pairs,
\begin{align}
 \varPsi(\xbf_1,\xbf_2,\ldots,\xbf_N) \sim \prod_{\substack{j,k=1\\j< k}}^N f(\xbf_j-\xbf_k)\;.
\end{align}
Many famous wavefunctions are of Jastrow type, e.g., the Laughlin wavefunction~\cite{laughlin_anomalous_1983} and the Gutzwiller wavefunction~\cite{gutzwiller_effect_1963, rokhsar_gutzwiller_1991}.\si{Gutzwiller wavefunction}  The Jastrow wavefunction has found wide application in strongly interacting systems: It is used in variational quantum Monte Carlo as a trial wavefunction~\cite{umrigar_optimized_1988}; it is used to show that the one-particle reduced density matrix of $^4\mathrm{He}$ is an extensive quantity, thus verifying that BEC underlies superfluidity~\cite{reatto_bose-einstein_1969}; and it is used to investigate the effect of interatomic correlations and the accuracy of the Gross-Pitaevskii equation~\cite{fabrocini_beyond_1999, dubois_bose-einstein_2001, cowell_cold_2002}.  The validity of the Jastrow wavefunction is discussed in~\cite{kane_general_1991}, where the authors constructed a class of interacting-boson Hamiltonians whose exact ground-state wavefunctions are of Jastrow form.

Another ansatz that many authors have adapted to discuss fragmentation of BECs is the Double-Fock State (DFS), or sometimes Twin-Fock State (TFS); most generally, one can use a many-Fock state that takes the form
\begin{align}
 \ket{\varPsi} = \bigg(\prod_{j=1}^\rank \frac{1}{\sqrt{N_j!}}\big(\a^\dagger_j\big)^{N_j}\bigg)\ketb{\vac}\;,
\end{align}
where $\rank$ is the number of fragments.  Using this ansatz, Streltsov \emph{et al.}~\cite{streltsov_ground-state_2004} argued that fragmentation of the ground state of a BEC only happens when the total number of particles is finite; Mueller \emph{et al.}~\cite{mueller_fragmentation_2006} showed that as degeneracies multiply, so do the varieties of fragmentation;  and Alon \emph{et al.}~\cite{alon_time-dependent_2007} generalized the Gross-Pitaevskii equation to include multiple orbitals. Other authors treated fragmented BECs by evolving the single-particle reduced density matrix with the Bogoliubov back-reaction approximation~\cite{vardi_bose-einstein_2001, tikhonenkov_quantum_2007}.

One interesting numerical method to simulate the dynamics of $N$ interacting bosons is the Multiconfigurational Time-Dependent Hartree Method for Bosons (MCTDHB)~\cite{masiello_multiconfigurational_2005, alon_multiconfigurational_2008}; instead of fixing the single-particle orbitals, MCTDHB uses time-dependent orbitals, which reduces the number of orbitals required to reach a certain precision.  However, it still has to simulate the full quantum dynamics in a $D$-dimensional Hilbert space, with $D=\binom{N+\rank-1}{ \rank-1}\simeq N^{\rank-1}$, where $\rank$ is the number of orbitals.

\section{Outline of This Dissertation}

Chapter~\chref{ch:basics_bec} is an introduction to the basics of BECs.  The Gross-Pitaevskii (GP) equation is derived using an approach that ``projects'' the evolved state back to the product manifold; I show that while the error introduced by the ``projection'' is large for the entire wavefunction, it is quite small for the few-particle Reduced Density Matrices (RDMs).  This explains why the GP equation works so well in practice, because only the few-particle RDMs---not the whole wavefunction---can be measured in a lab.  Also, the relative phase of two BECs is discussed, with a brief explanation of the MIT experiment~\cite{andrews_observation_1997}.

In Chap.~\chref{ch:n_conserving}, I introduce an alternative way to derive the number-conserving Bogoliubov approximation, where the many-body wave function of a BEC is ``encoded" in the $N$-particle sector of an \emph{extended catalytic state}, which is a coherent state for the condensate mode and an arbitrary state for the noncondensed modes.  By going to a time-dependent interaction picture, the coherent state is displaced to the vacuum, where all the field operators are small compared to ${N}^{1/2}$.  The resulting Hamiltonian can then be organized by powers of ${N}^{-1/2}$.  Requiring the terms of order ${N}^{1/2}$ to vanish, we get the Gross Pitaevskii equation for the condensate wave function.  Going to the next order, $N^0$, we are able to derive equations equivalent to those found by Castin and Dum~\cite{castin_low-temperature_1998} for a number-conserving Bogoliubov approximation.  In contrast to other approaches, the extended-catalytic-state approach allows us to calculate the state evolution in the Schr\"{o}dinger picture instead of the Heisenberg picture.  In addition, many techniques from quantum optics, such as quasi-probability distributions, can be used directly in our approach.  Moreover, it is much easier to track different orders in the Hamiltonian with our approach, which allows one to go to approximations beyond second order.  Last but not the least, the number-conserving Bogoliubov approximations for multicomponent cases, which are useful for BEC based interferometers, become much easier with our approach.

In Chap.~\chref{ch:pcs_1}, I consider a state of $N =l\times n$ bosons that is derived by symmetrizing the $n$-fold tensor product of an arbitrary $l$-boson state.  The rationale behind this approach comes from the BBGKY hierarchy: The errors to the many-particle RDMs only weakly affect the few-particle RDMs.  Particularly, we are interested in the pure state case for $l=2$, which we call the Pair-Correlated State (PCS).  I show that the PCS approach reproduces the number-conserving Bogoliubov approximation when depletion is low; moreover, PCS allows one to go to the strong interaction regime where the Bogoliubov approximation fails.  The normalization factor of the PCS is calculated in the large $N$ limit; the few-particle reduced density matrices can then be derived by taking derivatives of the normalization factor with respect to the PCS parameters.  For the two-mode case, these matrix elements are related to the PCS parameters by modified Bessel functions.  In the basis that the single-particle RDM is diagonalized, the two-particle RDMs of PCS has large corrections corresponding to annihilating two particles in one mode and then creating two particles in the other mode; although such correlations are crucial for strongly correlated bosonic systems, they are always zero in the double-Fock ansatz.

In Chap.~\chref{ch:pcs_2}, the two-site Bose-Hubbard model is used to benchmark the PCS ansatz.  I find that the error (measured by the trace distance between the two-particle RDMs) of PCS to the exact solution is less than two percent over the entire parameter space (the error of DFS is roughly 10 times larger).  Thus PCS serves as a bridge between the superfluid and the Mott insulating phases.  More interestingly, for the time-dependent case, numerical simulations suggest that the error of PCS does not become larger as time increases.  I derive both time-dependent and -independent equations for the ground state and the time evolution of the PCS ansatz, along with a condition for fragmentation.  The time complexity of simulating PCS does not depend on $N$ and is linear in the Schmidt rank of the two-particle state used to construct PCS.  Compared to other methods, e.g, the Jastrow wavefunction, the Gutzwiller wavefunction, and the multi-configurational time-dependent Hartree method, our approach does not require quantum Monte Carlo nor a particularly demanding computational power.

\section{Other Work}

In addition to the work on BECs reported in this dissertation, I also took advantage of the broad research interests in the Center for Quantum Information and Control (CQuIC) at the University of New Mexico to work on a variety of other research topics.  In the following, I describe briefly some published work that I participated in during my PhD study.  Typically, my role in these other projects arose from my realizing that I could make a contribution, based on my expertise in quantum measurement and information theory, to problems that were discussed in the group meeting of my supervisor, Professor Caves.

A quantum linear amplifier makes a small signal larger, so that it can be perceived by instruments incapable of resolving the original signal, while sacrificing as little as possible in signal-to-noise ratio.  Quantum mechanics limits how well this can be done: The noise added by the amplifier, when referred to the input, must be at least half a quantum at the operating frequency.  This well-known quantum limit on deterministic linear amplifiers only constrains the second moments of the added noise. In~\cite{caves_quantum_2012}, we derived the quantum constraints on the entire distribution of added noise.  We showed that any phase-preserving linear amplifier is equivalent to a parametric amplifier with a physical state for the ancillary mode; the noise added to the amplified field mode is distributed according to the Wigner function of the ancilla state.

A noiseless linear amplifier takes an input coherent state to an amplified coherent state, but only works part of the time.  In~\cite{pandey_quantum_2013}, we bounded the working probabilities of probabilistic and approximate noiseless amplifiers and constructed theoretical model amplifiers that achieve some of these bounds.  Our chief conclusions were the following: (i) the working probability of any phase-insensitive noiseless amplifier is very small in the phase-plane region where the device works with high fidelity; (ii) phase-sensitive noiseless amplifiers that work only on coherent states sparsely distributed on a phase-plane circle centered at the origin can have a reasonably high working probability.

Any evolution described by a completely positive trace-preserving linear map can be imagined as arising from the interaction of the evolving system with an initially uncorrelated ancilla.  The interaction is given by a joint unitary operator, acting on the system and the ancilla.  In~\cite{jiang_ancilla_2013}, we determined the properties such a unitary operator must have in order to force the choice of a physical---that is, positive---state for the ancilla if the end result is to be a physical---that is, completely positive---evolution of the system.  Thus Ref.~\cite{jiang_ancilla_2013} finds the general solution to this problem, which reveals a surprising and previously unsuspected structure in the joint unitary operators.  The problem, in a more restricted setting, arose in our work on the general limits to the noise added by deterministic linear amplifiers~\cite{caves_quantum_2012}.

In quantum optics a pure state is considered classical, relative to the statistics of photodetection, if and only if it is a coherent state.  A different and newer notion of nonclassicality is based on modal entanglement.  One example that relates these two notions is the Hong-Ou-Mandel effect, where modal entanglement is generated by a beamsplitter from the nonclassical photon-number state $\ket{1}\otimes\ket{1}$.  This suggests that beamsplitters or, more generally, linear-optical networks are mediators of the two notions of nonclassicality. In~\cite{jiang_mixing_2013}, we showed the following: Given a nonclassical pure-product-state input to an $N$-port linear-optical network, the output is almost always mode entangled; the only exception is a product of squeezed states, all with the same squeezing strength, input to a network that does not mix the squeezed and antisqueezed quadratures. Our work thus gives a necessary and sufficient condition for a linear network to generate modal entanglement from pure-product inputs, a result that is of immediate relevance to the boson-sampling problem.

In~\cite{jiang_quantum_2014}, I derived explicit expressions for the quantum Fisher information and the Symmetric Logarithmic Derivative (SLD) of a quantum state that is expressed in exponential form $\rho=\exp(G)$; the SLD is expressed in terms of the generator $G$.  Applications include quantum-metrology problems with Gaussian states, including the effects of photon losses or other decoherence, and general thermal states.  Specifically, I gave the SLD for a Gaussian state in two forms, first, in terms of its generator and, second, in terms of its moments; the Fisher information was calculated for both forms.

Probabilistic metrology attempts to improve parameter estimation by doing a measurement and post-selecting states that are especially sensitive to the parameter in question.  Such probabilistic protocols thus occasionally report an excellent estimate and the rest of the time either guess or do nothing at all.  In~\cite{combes_quantum_2014}, we showed that such post-selected probabilistic protocols can never improve quantum limits on estimation of a single parameter, both on average and asymptotically in number of trials, if performance is judged relative to the mean-square estimation error.  We extended the result by showing that for a finite number of trials, the probability of obtaining better estimates using probabilistic metrology, as measured by mean-square error, decreases exponentially with the number of trials.

\renewcommand{\lb}[1]{\label{basics_bec:#1}}
\renewcommand{\rf}[1]{\ref{basics_bec:#1}}

\chapter{Basics of Bose-Einstein Condensates}
\label{ch:basics_bec}
\chaptermark{Basics about BECs}

\begin{quote}
From a certain temperature on, the molecules ``condense'' without attractive forces; that is, they accumulate at zero velocity. The theory is pretty, but is there some truth in it. \\[4pt]
-- Albert Einstein\ai{Einstein, Albert}
\end{quote}

\noindent Bose-Einstein condensation has attracted many theoretical efforts since its first realizations~\cite{anderson_observation_1995, davis_bose-einstein_1995}.  Perhaps one reason is that BECs are much ``cleaner'' than other many-body systems and thus can be modeled by simple Hamiltonians.  The ``simpleness'' of BECs makes possible theoretical investigations of important concepts such as nonlinear effects, elementary excitations, and macroscopic quantum coherence~\cite{penrose_bose-einstein_1956, yang_concept_1962}.  In this chapter, I review the theoretical frameworks for describing BECs, which include the definition of a condensate, the Gross-Pitaevskii Equation (GPE)~\cite{gross_structure_1961, pitaevsk_vortex_1961}, and relative phases of BECs~\cite{andrews_observation_1997}.  In particular, the GPE is derived by projecting the evolved many-body state back to the manifold of product states; a similar method is also used in Chap.~\chref{ch:pcs_2} to derive the time-dependent equations for the so called Pair-Correlated State (PCS). \footnote{For situations where depletion (noncondensate fraction) is large, neither the GPE nor the Bogoliubov approximation works, and one usually needs to rely on methods such as quantum Monte Carlo.  In Chap.~\chref{ch:pcs_2}, we discuss how to apply the PCS ansatz introduced in Chap.~\chref{ch:pcs_1} to fragmented BECs.  Compared to other approaches, PCS is much less numerically demanding while capturing interparticle correlations.}

In 1924 Einstein received a letter from Bose\ai{Bose, Satyendra Nath} on the derivation of Planck's law using a new statistics for photons, later known as Bose-Einstein statistics.  He soon realized the importance of that work, applied Bose's idea to massive particles, and predicted a new phase of matter---Bose-Einstein condensation---where almost all the particles condensate to a single quantum state when the temperature drops beneath a critical temperature $T_\mathrm{\ssC}$.  Interestingly, the condensate is a consequence of Bose-Einstein statistics, which dramatically suppress the number of low-energy excited states, rather than due to attractive particle-particle interactions.  Although the idea of BEC was initially proposed for noninteracting particles, it also applies when there exist interparticle interactions, but great care needs to be taken, because the interparticle interactions deplete part of the condensate even at zero temperature.  To strictly define BECs for interacting bosons at zero or finite temperature, one introduces the single-particle Reduced Density Matrix (1RDM),\si{1RDM}
\begin{align}
\rho^{(1)}\big(\xbf\, \vert\, \mathbf{x'}\big)
&=\brab{\varPsi}\,\uppsi^\dagger(\mathbf{x'})\, \uppsi(\xbf)\,\ketb{\varPsi}\;,
\end{align}
where we use upright Greek letters to denote field operators and slanted capital Greek letters to denote many-body states.  The Penrose-Onsager criterion~\cite{penrose_bose-einstein_1956}\si{Penrose-Onsager criterion}\ai{Penrose, Oliver}\ai{Onsager, Lars} states that a BEC occurs when the largest eigenvalue of $\rho^{(1)}$ is of order $N$, where $N$ is the total number of particles.  This criterion remains meaningful for interacting systems and corresponds to the existence of Off-Diagonal Long Range Order (ODLRO)~\cite{yang_concept_1962}\si{Off-diagonal long range order}\ai{Yang, Chen-Ning}.  In this chapter, we only consider the case of weakly interacting particles with small depletion, in which case the largest eigenvalue of $\rho^{(1)}$ does approach $N$.  When there are interparticle interactions, however, the bosons no longer condense to the ground state of the single-particle Hamiltonian, but rather condense to the ground state of the time-independent Gross-Pitaevskii equation\si{Gross-Pitaevskii equation}
\begin{align}\lb{eq:GP_equation_time_independent}
 \mu\, \phi(\xbf)= \Big(\mathord{-}\frac{{\hbar}^2}{2m}\boldsymbol\nabla^2 +V(\xbf)+g(N-1)\norm{\phi(\xbf)}^2\,\Big) \phi(\xbf)\;,
\end{align}
where $\phi(\xbf)$ is the condensate wavefunction, $V(\xbf)$ is the trapping potential, $N$ is the number of particles, and $g$ represents the strength of the inter-particle interactions.  Since the atoms are very cold, their interaction can be described by the single parameter $g=4\pi\hbar^2 a_\mathrm{s}/m$, where $a_\mathrm{s}$ is the $s$-wave scattering length\si{S-wave scattering length} and $m$ is the mass per particle.  This situation, in stark contrast to the case of liquid helium, is a simplification that collisions can be described in terms of the lowest-energy scattering length~\cite{fermi_motion_1936, huang_statistical_1987}.\ai{Fermi, Enrico}  Notice also that in the Gross-Pitaevskii equation, the interaction term in Eq.~(\rf{eq:GP_equation_time_independent}) is ``amplified'' $N-1$ times by the Bose statistics.  Thus, even for weakly interacting dilute Bose gases, the interaction term in the GPE is considerable; the effects of interactions are less crucial when temperature is increased towards the critical temperature.

\section{Derivation of the Time-Dependent GPE}

Generally we need to rely on approximations to deal with interacting many-body quantum systems.  The inner product between the approximated and the actual state, whose absolute value is called the fidelity, is usually considered a measure of merit of the approximation.  As the dimension of the Hilbert space becomes large, however, this inner product goes to zero.  One example of this is that the inner product between the GP ground state, which is a  product state, and the more accurate Bogoliubov ground state is quite small.  Such a situation urges one to find a more useful measure of the merit of approximations that does not go to zero as the number of particles becomes large.  Often it is the case that the important physical properties of many-body systems can be determined by the correlation functions (also called Green's functions) instead of the whole many-body state vector, and an approximation may be considered good if it gives the correct low-order correlation functions.  We emphasis that this criterion is weaker than the inner product criterion; the low-order correlation functions of two states can be almost identical, while the whole many-body wavefunctions are vastly different.

Of particular interest are the equal-time $2q$-point correlation matrices,\si{Correlation matrix} i.e., the $q$-particle Reduced Density Matrices ($q$RDM),\si{Reduced density matrix}
\begin{align}\lb{eq:2q_correlation_function}\hspace{-2em}
 \rho^{(q)}\big(\xbf_1,\ldots,\xbf_q\, ,\, \xbf'_1,\ldots,\xbf'_q\,;\, t\big)&\equiv
 \brab{\varPsi(t)}\,\uppsi^\dagger(\xbf'_1)\cdots \uppsi^\dagger(\xbf'_q)\; \uppsi(\xbf_q)\cdots \uppsi(\xbf_1)\,\ketb{\varPsi(t)}\;,
\end{align}
where the field operators are time independent (Schr\"odinger picture). The time derivative of the $q$RDM is
\begin{align}\lb{eq:derivative_2q_rdm}
 i\hbar\, \dt{\rho}^{(q)}(t) &=\brab{\varPsi(t)}\,\commutb{\uppsi^\dagger(\xbf_1)\cdots  \uppsi^\dagger(\xbf_q)
 \,\uppsi(\xbf_q')\cdots \uppsi(\xbf_1')}{\sH(t)}\,\ketb{\varPsi(t)} \;.
\end{align}
If the Hamiltonian only contains terms up to two-body interactions (quartic in the field operators), the commutators $\commut{\sH}{\uppsi^\dagger(\xbf_j)}$ and $\commut{\sH}{\uppsi(\xbf_j)}$ are at most cubic functions of the field operators, which suggests that $\dt{\rho}^{(q)}$ is only a function of $\rho^{(q+1)}$.  Generally, the $k$th time derivative of the $q$RDM, $\dif^k\!\rho^{(q)}\!/\dif\ssp t^k$, is a function of $\rho^{(q+k)}$~\cite{BBGKY}\si{BBGKY hierarchy}, and the short time evolutions of the low-order RDMs are immune from errors of higher-order RDMs.

Before going further, let us review a general procedure for approximating state evolution in a Hilbert space $\dsH$ by a set of ansatz states $\dsA\subset \dsH$, where the subset $\dsA$ is not necessarily a subspace of $\dsH$ (it can be a submanifold of $\dsH$).  Initially, the state of the system is assumed to be in the ansatz subset, $\ket{\varPsi(0)}\in \dsA$.  After evolving it for a short period of time $\dif t$, we ``project'' the evolved state $\ket{\varPsi(\dif t)}=\exp(-i\sH \dif t/\hbar)\ssp\ket{\varPsi(0)}$ back to $\dsA$.  The ``projection'' step simply means to find the normalized state $\ket{\varPsi_\dsA(\dif t)}\in \dsA$ such that the inner product $\braket{\varPsi_\dsA(\dif t)}{\varPsi(\dif t)}$ is real and its norm is maximized.   Repeating the same procedure for $n=t/\dif t$ steps, we find a state $\ket{\varPsi_\dsA(t)}\in \dsA$ within $\dsA$ that approximates the actual state $\ket{\varPsi(t)}=\exp(-i \sH t/\hbar)\ket{\varPsi(0)}$.  Note that this procedure does not in any way guarantee a large overlap, $\braket{\varPsi_\dsA(t)}{\varPsi(t)}$, between the actual state and the approximated state.  Nevertheless, at each step, it is the best approximation to the evolution within $\dsA$, and we are satisfied if it gives good approximations for the low-order RDMs.  The adequacy of the approximation is determined by the system Hamiltonian and a judicious choice of the subset $\dsA$.

I now discuss how to derive the Gross-Pitaevskii equation using the procedure described in the preceding paragraph; moreover, the question why the GP ansatz works so well is investigated.  The GP ansatz consists of choosing $\dsA$ to be all of the product states; in the second-quantized picture, these product states have the form\si{Product ansatz}
\begin{equation}\lb{eq:product_ansatz}
 \ket{\varPsi_\mathrm{gp}(t)}=\frac{1}{\sqrt{N !}}\, \big[\a_{\smash{\psinot(t)}}^\dagger\big]^N\, \ket{\vac}\;,
\end{equation}
where $\a_{\smash{\psinot(t)}}^\dagger$ is the creation operator for the condensate mode,
\begin{equation}
\a_{\smash{\psinot(t)}}^\dagger=\int\uppsi^\dagger(\mathbf{x})\,\psinot(\mathbf{x},t)\,\dif \mathbf{x}
=\braket{\uppsi}{\psinot(t)}=\braket{\psinot^*(t)}{\uppsi^\dagger}\;.
\end{equation}
Here we introduce a shorthand notation for the integral as a bra-ket inner product between a single-particle state and the field operator.  The bra-ket notation introduced here, though \emph{ad hoc}, is useful for manipulating the complicated expressions that arise as we proceed.   We define the error vector of the product ansatz as
\begin{subequations}
\begin{align}\lb{eq:errors_product_ansatz}
 \ket{\dt\varPsi_\mathrm{err}(t)}&=\frac{\sH(t)}{i\hbar}\, \ket{\varPsi_\mathrm{gp}(t)}-\, \ket{\dt\varPsi_\mathrm{gp}(t)}\\[3pt]
 &= \frac{1}{\sqrt{N !}}\,  \Big(\, \frac{\sH(t)}{i\hbar}\, \big[\a_{\smash{\psinot(t)}}^{\dagger}\big]^N- N\,   \dt{\a}_{\smash{\psinot(t)}}^\dagger \big[\a_{\smash{\psinot(t)}}^{\dagger}\big]^{N-1}\,\Big)\, \ket{\vac}\;,
\end{align}
\end{subequations}
where $\dt{\a}_{\smash{\psinot(t)}}^{\dagger}=\int\uppsi^\dagger(\xbf)\,\dt{\psinot}(\xbf,t)\,\dif \xbf=\braket{\uppsi}{\dt\psinot(t)}$ and the many-body Hamiltonian $\sH(t)$ takes the form
\begin{equation}
\sH(t)=\int\uppsi^\dagger (\xbf)  \Big(\mathord{-}\frac{{\hbar}^2}{2m}\boldsymbol\nabla^2 +V(\xbf,t)\Big) \uppsi(\xbf)+\frac{g}{2}\, [\uppsi^\dagger (\xbf)]^2\,\uppsi^2(\xbf)\,\dif \xbf\;.
\end{equation}
We notice that
\begin{subequations}
\begin{align}\hspace{-2em}
\sH\,  \big(\a_{\smash \psinot}^{\dagger}\big)^N\,\ket{\vac}&=N\, \braB{\uppsi}\,\mathord{-}\frac{{\hbar}^2}{2m}\boldsymbol\nabla^2 +V+g(N-1)\norm{\psinot}^2\, \ketB{\psinot}\,   \big(\a_{\smash \psinot}^{\dagger}\big)^{N-1}\,\ket{\vac}\lb{eq:evolv_actual_state_expansion_a}\\[4pt]
&\quad+N\, (N-1)\bigg(\frac{g}{2}\int \big(\uppsi^\dagger_\perp\big)^2\,\psinot^2\, \dif \xbf\bigg)\; \big(\a_{\smash{\psinot}}^{\dagger}\big)^{N-2}\,\ket{\vac} \lb{eq:evolv_actual_state_expansion_b}\\[4pt]
&\quad-\frac{\eta}{2}\,N(N-1)\,\big(\a_{\smash{\psinot}}^{\dagger}\big)^N\,\ket{\vac}\;,\lb{eq:evolv_actual_state_expansion_c}
\end{align}
\end{subequations}
where $\uppsi^\dagger_\perp(\xbf,t)=\uppsi^\dagger(\xbf)-\a_{\smash{\psinot(t)}}^\dagger\psinot^*(\xbf,t)$ is the field operator for the noncondensate modes, and
\begin{align}
\eta(t)=g\int \norm{\psinot(\xbf,t)}^4\,\dif \xbf\;.
\end{align}
The term Eq.~(\rf{eq:evolv_actual_state_expansion_c}) contributes only an overall phase and will be neglected hereafter.  By letting $\psinot(t)$ satisfy the Gross-Pitaevskii equation,\si{Gross-Pitaevskii equation}
\begin{align}\lb{eq:GP_equation_time_dependent}
 i \hbar\,\dt{\psinot}(\xbf,t)= \Big(\mathord{-}\frac{{\hbar}^2}{2m}\boldsymbol\nabla^2 +V(\xbf,t)+g(N-1)\norm{\psinot(\xbf,t)}^2\,\Big) \psinot(\xbf,t)\;,
\end{align}
we have
\begin{align}\hspace{-2em}\lb{eq:evolution_product_ansatz}
 \dt{\a}_{\smash{\psinot(t)}}^\dagger &\equiv \braket{\uppsi}{\dt{\psinot}(t)}
 = \frac{1}{i\hbar}\,\braB{\uppsi} \Big(-\frac{{\hbar}^2}{2m}\boldsymbol\nabla^2 +V(t)+g(N-1)\norm{\psinot(t)}^2\,\Big) \ketB{\psinot(t)}\;,
\end{align}
which cancels the term~(\rf{eq:evolv_actual_state_expansion_a}).  Thus, the error vector takes the form,
\begin{align}
 \ket{\dt\varPsi_\mathrm{err}(t)}= \frac{N\, (N-1)}{i\hbar\,\sqrt{N !}}\, \bigg(\frac{g}{2}\int \big[\uppsi^\dagger_\perp(\xbf,t)\big]^2\, \psinot^2(\xbf,t)\,\dif \xbf\bigg) \, \big[\a_{\smash{\psinot(t)}}^{\dagger}\big]^{N-2}\,\ket{\vac}\;,
\end{align}
which comes solely from the term~(\rf{eq:evolv_actual_state_expansion_b}).  This kind of error, which involves two particles excited out of the condensate mode, takes the state outside the product ansatz~(\rf{eq:product_ansatz}), where only one particle is allowed to be excited.  Thus the GP solution~(\rf{eq:GP_equation_time_dependent}), by taking into account the term~(\rf{eq:evolv_actual_state_expansion_a}), is optimal in minimizing the length of the error vector $\ket{\dt\varPsi_\mathrm{err}(t)}$.  

For $gN\sim 1$, we have
\begin{align}\lb{eq:error_norm}                                                                                                                                                                                                                                        \norm{\dt\varPsi_\mathrm{err}(t)}^2\sim \frac{1}{i\hbar}\sim \norm{\dt\varPsi_\mathrm{\!gp}(t)}^2\;,
\end{align}
which says that the error is not small at all, and the GP ansatz $\ket{\varPsi_\mathrm{\!gp}(t)}$ will deviate substantially from the actual state.  The error of the 1RDM, however, vanishes,
\begin{align}
\dt\rho_\mathrm{err}^{(1)}(\xbf, \xbf'\ssp)
=\brab{\varPsi_\mathrm{gp}}\,\uppsi^\dagger(\xbf')\,\uppsi(\xbf)\,\ketb{\dt\varPsi_\mathrm{err}}+\mathrm{H.c.}=0\;,
\end{align}
where $\mathrm{H.c.}$ means exchanging $\xbf$ and $\xbf'$ and taking the complex conjugate. 

The error in the two-particle Reduced Density Matrix (2RDM) is\si{2RDM}
\begin{subequations}
\begin{align}
&\hspace{-2em}\dt\rho_\mathrm{err}^{(2)}\big(\xbf_1,\xbf_2\, ,\, \xbf'_1,\xbf'_2\big)\nonumber\\[2pt]
&=\brab{\varPsi_\mathrm{gp}}\,\uppsi^\dagger(\xbf'_1)\uppsi^\dagger(\xbf'_2)\,\uppsi(\xbf_2)\uppsi(\xbf_1)\,\ketb{\dt\varPsi_\mathrm{err}}+\mathrm{H.c.}\\[4pt]
&=\frac{N(N-1)}{\sqrt{N !}}\:\Big(\psinot^*(\xbf'_1)\psinot^*(\xbf'_2)\,\brab{\vac}\, \a_{\smash{\psinot}}^{N-2}\, \uppsi(\xbf_2)\uppsi(\xbf_1)\,\ketb{\dt\varPsi_\mathrm{err}}+\mathrm{H.c.}\Big)\;.\lb{eq:2_matrix_a}
\end{align}
\end{subequations}
To evaluate Eq.~(\rf{eq:2_matrix_a}), we notice
\begin{subequations}\hspace{-1em}
\begin{align}
&\hspace{-2em}\frac{i\hbar}{\sqrt{N !}}\, \brab{\vac}\, \a_{\smash \psinot}^{N-2}\: \uppsi(\xbf_2)\uppsi(\xbf_1)\,\ketb{\dt\varPsi_\mathrm{err}}\\
&= \frac{g}{2}\int \frac{\psinot^2(\xbf)}{(N-2)!}\, \brab{\vac}\, \a_{\smash \psinot}^{N-2}\: \uppsi(\xbf_2)\uppsi(\xbf_1)\,[\uppsi^\dagger_\perp(\xbf)]^2\, \big(\a_{\smash \psinot}^{\dagger}\big)^{N-2}\,\ketb{\vac}\: \dif \xbf\\[3pt]
&= \frac{g}{2}\int \psinot^2(\xbf)\, \brab{\vac}\,  \uppsi(\xbf_2)\uppsi(\xbf_1)\,[\uppsi^\dagger_\perp(\xbf)]^2\, \ketb{\vac}\: \dif \xbf\\[3pt]
&= g\int \psinot^2(\xbf)\, \brab{\vac}\,  \uppsi(\xbf_1)\,\uppsi^\dagger_\perp (\xbf)\, \ketb{\vac}\brab{\vac}\,  \uppsi(\xbf_2)\,\uppsi^\dagger_\perp (\xbf)\, \ketb{\vac}\: \dif \xbf\\
&= g\int \psinot^2(\xbf)\,\Big(\delta(\xbf_1, \xbf) -\psinot(\xbf_1) \psinot^*(\xbf)\Big) \Big(\delta(\xbf_2, \xbf)-\psinot(\xbf_2)\psinot^*(\xbf)\Big) \: \dif \xbf\\[3pt]
&= g\,\psinot(\xbf_1)\psinot(\xbf_2)\Big(\delta(\xbf_1, \xbf_2)-\norm{\psinot(\xbf_1)}^2- \norm{\psinot(\xbf_2)}^2+\eta/g \,\Big)\;.\lb{eq:2_matrix_midstep}
\end{align}
\end{subequations}
Putting Eq.~(\rf{eq:2_matrix_midstep}) into Eq.~(\rf{eq:2_matrix_a}), we have
\begin{subequations}
\begin{align}
&\hspace{-2em}\dt\rho_\mathrm{err}^{(2)}\big(\xbf_1,\xbf_2\, ,\, \xbf'_1,\xbf'_2\big)\nonumber\\
&\hspace{-0.8em}=\frac{N(N-1)}{\sqrt{N !}}\,\Big(\psinot^*(\xbf'_1)\psinot^*(\xbf'_2)\,\brab{\vac}\, \a_{\smash \psinot}^{N-2}\, \uppsi(\xbf_2)\uppsi(\xbf_1)\,\ketb{\dt\varPsi_\mathrm{err}}+\mathrm{H.c.}\Big)\\[2pt]
&\hspace{-0.8em}=\frac{gN(N-1)}{i\hbar}\,\psinot^*(\xbf'_1)\psinot^*(\xbf'_2)\psinot(\xbf_1)\psinot(\xbf_2)\nonumber\\[2pt]
&\quad \times\Big(\delta(\xbf_1, \xbf_2)-\norm{\psinot(\xbf_1)}^2-\norm{\psinot(\xbf_2)}^2+\eta/g \,\Big)+\mathrm{H.c.}\;.\lb{eq:2_matrix_b}
\end{align}
\end{subequations}
The expression~(\rf{eq:2_matrix_b}) says that $\dt\rho_\mathrm{err}^{(2)}$ is of order $gN^2$ (or $N$) and is negligible compared to $\rho^{(2)}$ which is of order $N^2$.  

Similarly, for the error in the $q$RDM, we have
\begin{subequations}
\begin{align}
&\hspace{-2em}\dt\rho_\mathrm{err}^{(q)}\big(\xbf_1,\ldots,\xbf_q\, ,\, \xbf'_1,\ldots,\xbf'_q\big)\big/N(N-1)\cdots(N-q+1)\nonumber\\[2pt]
&\hspace{-0.7em}=\frac{1}{\sqrt{N!}}\,\Big(\psinot^*(\xbf'_1)\cdots\psinot^*(\xbf'_q)\,\brab{\vac}\, \a_{\smash \psinot}^{N-q}\, \uppsi(\xbf_q)\cdots\uppsi(\xbf_1)\,\ketb{\dt\varPsi_\mathrm{err}}+\mathrm{H.c.}\Big)\\[2pt]
&\hspace{-0.7em}= \frac{g}{i\hbar}\,\prod_{l=1}^q \psinot^*(\xbf'_{l})\ssp \psinot(\xbf_l)\sum_{\substack{j,k=1\\ k>j}}^{q}\! \Big(\delta(\xbf_j, \xbf_k)- \norm{\psinot(\xbf_j)}^2-\norm{\psinot(\xbf_k)}^2+\eta/g \Big)+\mathrm{H.c.} \lb{eq:q_matrix_b}
\end{align}
\end{subequations}
From Eq.~(\rf{eq:q_matrix_b}), we observe that $\dt\rho_\mathrm{err}^{(q)}\sim N^{-1}\rho^{(q)}$ for $q\ll \sqrt N$.  This justifies that the product ansatz is a good approximation provided that the evolution time is not too long.\footnote{In practice, the product ansatz probably also works for very long time $t$, but to prove it rigorously is hard.}

\section{The Relative Phase of Two BECs}\si{Relative phase}

Although it is not possible to attribute an overall phase to a BEC with a fixed number of particles, the relative phase\si{Relative phase} of two BECs can be well defined when there is an uncertainty in the relative number of bosons.  This is linked closely with the well-known uncertainty relationship between phase and number~\cite{caves_quantum-mechanical_1981}.  The many-body wavefunction is therefore a superposition of states with different numbers of particles distributed in the two BECs; such states evolve at different rates due to their different chemical potentials caused by the particle-particle interactions.  This effect, called phase diffusion,\si{Phase diffusion} degrades the relative phase of the two BECs.  In  App.~\chref{ch:mgpes_vs_bogoliubov}, I discuss the equivalence of two seemingly different approaches to phase diffusion.  

Many authors have discussed the relative phase of two BECs from a variety of perspectives.  For example, Imamo\=glu~\cite{imamoglu_inhibition_1997} and Villain \emph{et al.}~\cite{villain_quantum_1997} showed that the phase memories of BECs are lost on a relatively short time scale, in some cases vanishing in the large $N$ limit.\footnote{The coherence time is less than \unit{50}\milli\second\ for a typical million-atom BEC (with diluteness parameter $na^3 \simeq 10^{-4}$).}  Preparation and detection of the relative phase of two BECs using optical means were discussed in~\cite{javanainen_optical_1996, ruostekoski_nondestructive_1997, hall_measurements_1998}. Horak and Barnett~\cite{horak_creation_1999} showed that coherence between two BECs can be generated by just measuring several atoms, but collapsing to the phase state requires order $N$ measurements.   Zapata \emph{et al.}~\cite{zapata_phase_2003} studied the dynamics of the relative phase following the connection of two independently formed BECs.   Saba \emph{et al.}~\cite{saba_light_2005} performed the first nondestructive measurement of the relative phase of two spatially separated BECs by stimulated light scattering.  Chwede\'nczuk \emph{et al.}~\cite{chwedenczuk_phase_2011} showed that the sensitivity of phase estimation by measuring the position of atoms saturates the bound set by the quantum Fisher information.

Condensed-matter physicist Philip W. Anderson\ai{Anderson, Philip W.} once asked the question: ``Do two superfluids which have never seen one another possess a relative phase?''  This intricate question was partially answered by the MIT experiment~\cite{andrews_observation_1997}, where two BECs of sodium atoms are formed separately in a double-well trapping potential.  The double-well potential is a result of a combination of a magnetic trap and sheet of blue-detuned far-off-resonant laser light in the middle.  The magnetic trap and the laser-light sheet are suddenly switched off, and after about \unit{40}\milli\second\ time-of-flight, the two expanding condensates overlapped and were observed by absorption imagining.  Very clear fringe patters were observed, and the fringe period was the de Broglie wavelength associated with the relative motion of atoms.  The centers of the fringe patterns, however, were different from shot to shot.  This experiment implies that two separate BECs possess a relative phase, although this phase might be completely random.  Many authors discussed this phenomenon using different theoretical approaches; see~\cite{javanainen_quantum_1996, naraschewski_interference_1996, cirac_continuous_1996, castin_relative_1997, pethick_bose-einstein_2008}.  Here, I briefly go over the argument in~\cite{naraschewski_interference_1996}, which makes use of the correlation functions.

The quantum state of two separated BECs, each with $n$ atoms, can be approximated by the double-Fock state
\begin{align}\lb{eq:double_Fock_state}
\ket{\varPsi_\mathrm{dfs}}=\frac{1}{n!}\,\big(\a_\mathrm{\ssR}^\dagger\big)^{n}\big(\a_\mathrm{\ssL}^\dagger\big)^{n}\, \ket{\vac}\;,
\end{align}
where $\a_\mathrm{\ssL}^\dagger$ and $\a_\mathrm{\ssR}^\dagger$ are creation operators of the left and the right single-particle states $\psi_\mathrm{\ssL}(x)$ and $\psi_\mathrm{\ssR}(x)$.\footnote{In this section, we assume the BECs are quasi one-dimensional.}  One simple way to explain the experimental results is by calculating the correlation of particle\si{Correlation function} densities~\cite{naraschewski_interference_1996},
\begin{subequations}
\begin{align}\hspace{-2em}
\varrho\ssp (x_1, x_2)&=\frac{1}{(n!)^2}\,\brab{\vac}\, \a_\mathrm{\ssL}^{n}\a_\mathrm{\ssR}^{n}\,\uppsi^\dagger(x_1) \uppsi^\dagger(x_2)\, \uppsi(x_2) \uppsi(x_1)\,\big(\a_\mathrm{\ssR}^\dagger\big)^{n}\big(\a_\mathrm{\ssL}^\dagger\big)^{n}\, \ketb{\vac}\\[3pt]
\begin{split}
&=n(n-1)\,\norm{\psi_\mathrm{\ssL}(x_1)}^2\norm{\psi_\mathrm{\ssL}(x_2)}^2+n(n-1)\,\norm{\psi_\mathrm{\ssR}(x_1)}^2\norm{\psi_\mathrm{\ssR}(x_2)}^2\\[4pt]
&\quad +n^2\, \normb{\psi_\mathrm{\ssL}(x_2)\psi_\mathrm{\ssR}(x_1)+\psi_\mathrm{\ssL}(x_1)\psi_\mathrm{\ssR}(x_2)}^2
\end{split}\\[3pt]
\begin{split}
&\simeq n^2\big(\norm{\psi_\mathrm{\ssL}(x_1)}^2+\norm{\psi_\mathrm{\ssR}(x_1)}^2\big)\big(\norm{\psi_\mathrm{\ssL}(x_2)}^2+\norm{\psi_\mathrm{\ssR}(x_2)}^2\big)\\[3pt]
&\quad +n^2\, \big(\psi_\mathrm{\ssL}^*(x_2)\psi_\mathrm{\ssR}^*(x_1)\psi_\mathrm{\ssL}(x_1)\psi_\mathrm{\ssR}(x_2)+\mathrm{c.c.}\big)\;, \lb{eq:interference_term}
\end{split}
\end{align}
\end{subequations}
where the Fock terms in Eq.~(\rf{eq:interference_term}) describe the interference of the two BECs. With the correlation function, we have the expectation of the squared norm of the Fourier component of the particle density distribution,
\begin{subequations}
\begin{align}
 \hspace{-2.3em}\avB{\normB{\int e^{ikx}\, \uppsi^\dagger(x)\uppsi(x) \,\dif x}^2} &=
 \int e^{ik(x_1-x_2)}\, \avb{\uppsi^\dagger(x_1)\uppsi(x_1) \uppsi^\dagger(x_2)\uppsi(x_2)} \,\dif x_1\dif x_2\\
 &= 2n+\int e^{ik(x_1-x_2)}\, \varrho\ssp (x_1,x_2) \,\dif x_1\dif x_2\;,
\end{align}
\end{subequations}
where the constant $2n$ comes from different ordering of the creation and annihilation operators.  Neglecting this constant, we define
\begin{align}
 \varrho_k = \int e^{ik(x_1-x_2)}\, \varrho\ssp (x_1,x_2) \,\dif x_1\dif x_2\;.
\end{align}
Consider the case where the left and right states are plane waves in an interval of length $L$, i.e., $\psi_\mathrm{\ssL} = e^{ik_0\ssp x}/\sqrt L$ and $\psi_\mathrm{\ssR} =  e^{-ik_0\ssp x}/\sqrt L$; we have the following for $k\neq 0$,
\begin{align}
 \hspace{-1.5em}\varrho_k &= \frac{n^2}{L^2}\int e^{ik(x_1-x_2)}\, \big(e^{2ik_0(x_1-x_2)}+\mathrm{c.c.}\big) \,\dif x_1\dif x_2= n^2\big(\delta_{k,\,2k_0}+\delta_{k,\,-2k_0}\big)\;.
\end{align}
The quantity $\varrho_k$ is nonzero for $k=\pm 2k_0$, and thus one only finds a fringe pattern with period $\pi/k_0$; the fluctuation of the contrast of the fringe patterns is zero from shot to shot, a result that can be proved by calculating higher-order correlation matrices.

\renewcommand{\lb}[1]{\label{n_conserving:#1}}
\renewcommand{\rf}[1]{\ref{n_conserving:#1}}

\chapter[Number-Conserving Bogoliubov Approximation Using ECS]{Number-Conserving Bogoliubov Approximation Using Extended Catalytic States}
\label{ch:n_conserving}
\chaptermark{N-Conserving Bogoliubov Approximation Using ECS}

\begin{quote}
With every passing year, BEC proves that it still has surprises left for us.\\[4pt]
-- Eric A. Cornell\ai{Cornell, Eric A.}
\end{quote}

\noindent
In this chapter, we consider the ground state and dynamics of a dilute-gas BEC of $N$ bosonic atoms trapped in an arbitrary external potential.  In order to describe how interparticle correlations modify the Gross-Pitaevskii equation, we go to the next level of approximation, the Bogoliubov approximation.  The Bogoliubov approximation~\cite{bogoliubov_theory_1947, fetter_nonuniform_1972, huang_statistical_1987} is important for several reasons: (i)~it tells when the Gross-Pitaevskii (mean-field) approach begins to break down; (ii)~it describes small deviations from the Gross-Pitaevskii equation and can be used to study the stability of a BEC; (iii)~it enables the calculation of how impurities change the behavior of a BEC; and (iv)~it is useful for studying phase coherence between BECs.

\begin{figure}[ht] 
   \centering
   \includegraphics[width=0.9\textwidth,natwidth=610,natheight=642]{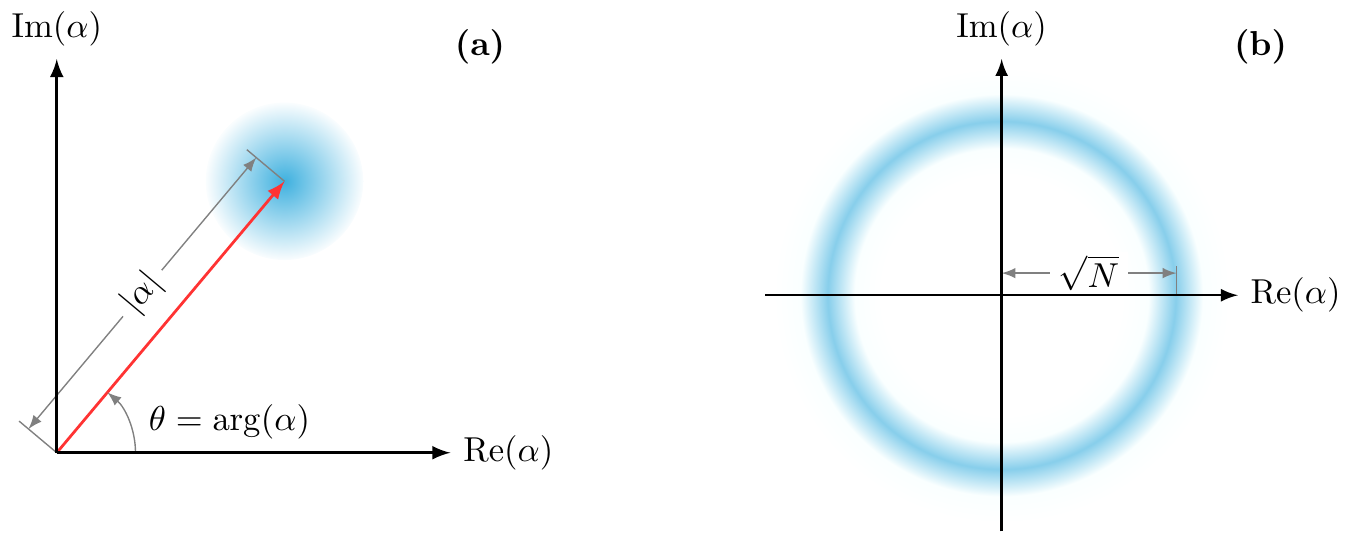}
   \caption[Coherent vs Fock States: Phase-Space Representation]{Phase-space representations for (a)~a coherent state with complex amplitude $\alpha$ and (b)~a Fock (number) state with particle number $N$.  A Fock state is distributed phase-symmetrically on the phase space, and no definite phase can be attributed to it; in contrast, a coherent state has a well defined phase.}
   \label{fig:coherent_vs_number_state}
\end{figure}

\begin{figure}[ht] 
   \centering
   \includegraphics[width=0.6\textwidth,natwidth=610,natheight=642]{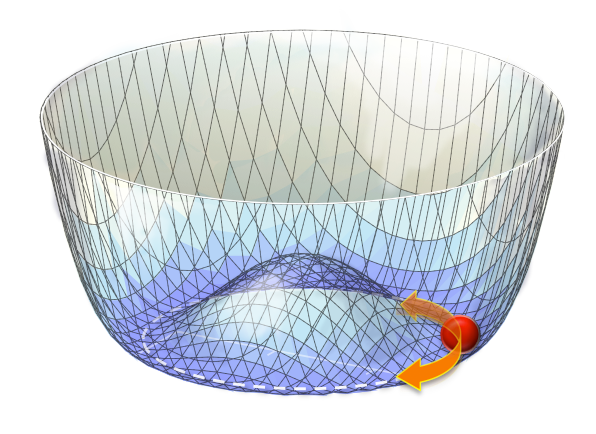}
   \caption[Goldstone Bosons]{The breaking of $U(1)$ symmetry of the condensate wavefunction causes a Goldstone boson in the Bogoliubov Hamiltonian. Because there is no restoring force on the Goldstone mode, the Bogoliubov ground state is not well defined.}
   \label{fig:goldstone_boson}
\end{figure}

Conventionally, in the Bogoliubov approximation, the condensate is treated as a small perturbation of the state where all the bosons occupy a coherent state of a particular condensate mode (i.e., a single-particle state).  When particle loss is negligible, however, the real condensate is much closer to a number state than to a coherent state (see Fig.~\hyperref[fig:coherent_vs_number_state]{\ref*{fig:coherent_vs_number_state}}).\si{Coherent state vs number state}
Since a coherent states has a well-defined phase, the conventional Bogoliubov approximation breaks the $U(1)$ symmetry possessed by the condensate; consequently, a fictitious Goldstone mode~\cite{nambu_quasi-particles_1960, goldstone_field_1961}\si{Goldstone boson}\ai{Nambu, Yoichiro}\ai{Goldstone, Jeffrey} is present in the Bogoliubov Hamiltonian (see Fig.~\hyperref[fig:goldstone_boson]{\ref*{fig:goldstone_boson}}).  Because there is no restoring force on the Goldstone mode, the Bogoliubov ground state is not well defined; worse, the Goldstone mode causes the condensate state to deviate linearly in time from a single condensate in a coherent state (i.e., this is a secular deviation, not an oscillation).
This problem is particularly pesky when the condensate is in a trapping potential, where the Goldstone mode is a mixture of the condensate mode and modes orthogonal to it and thus cannot be removed easily.  The way to get rid of the unphysical Goldstone mode is to adhere to the fact that the condensate has a fixed number of particles, i.e., by using a Bogoliubov approximation where particle number is conserved.

Many authors have already considered the number-conserving Bogoliubov approximation.  Girardeau and Arnowitt~\cite{girardeau_theory_1959}\ai{Girardeau, Marvin D.} were the first to propose a theory for the ground state and excited states of many bosons based on a particle-number-conserving ($N$-conserving) formulation of the Bogoliubov quasiparticles and quasiparticle Hamiltonian.  Gardiner~\cite{gardiner_particle-number-conserving_1997}\ai{Gardiner, Crispin W.} introduced a somewhat similar approach to Girardeau and Arnowitt's, but with an emphasis on the time-dependent case; Gardiner \emph{et al.}~\cite{gardiner_kinetics_1997} then applied this approach to the kinetics of a BEC in a trap.  Castin and Dum~\cite{castin_instability_1997, castin_low-temperature_1998, castin_bose-einstein_2001}\ai{Castin, Yvan} gave a modified form of the Bogoliubov Hamiltonian where the terms that break the $U(1)$ symmetry are removed by a projection operator.  S{\o}rensen~\cite{sorensen_bogoliubov_2002}\ai{S{\o}rensen, Anders S.} generalized the Castin-Dum result to the two-component case.  Gardiner and Morgan~\cite{gardiner_number-conserving_2007} considered an expansion in powers of the ratio of noncondensate to condensate particle numbers, rather than in the inverse powers of the total number of particles.  Leggett~\cite{leggett_bose-einstein_2001, leggett_relation_2003}\ai{Leggett, Anthony J.} used a number-conserving BCS-like ansatz to discuss the properties of the ground state of a homogeneous BEC.  The so-called truncated Wigner approximation~\cite{steel_dynamical_1998, sinatra_truncated_2002}\ai{Sinatra, Alice} is an equivalent way to implement the number-conserving Bogoliubov approximation in a phase space.

A number-conserving Bogoliubov approximation yields qualitatively different results from one that fails to conserve particle number; among these are the following. Villain \emph{et al.}~\cite{villain_quantum_1997} showed that the collapse time of the phase of a BEC is relatively short and, in some cases, vanishes in the limit of a large number of atoms.  Danshita \emph{et al.}~\cite{danshita_collective_2005} investigated collective excitations of BECs in a box-shaped double-well trap.  Trimborn \emph{et al.}~\cite{trimborn_exact_2008, trimborn_beyond_2009} demonstrated that artificial number fluctuations lead to ambiguities and large deviations within the Bose-Hubbard model.  Ole\'s \emph{et al.}~\cite{oles_n-conserving_2008} predicted considerable density fluctuations in finite systems close to the phase-separation regime.  Schachenmayer \emph{et al.}~\cite{schachenmayer_atomic_2011} studied the collapse and revival of interference patterns in the momentum distribution of atoms in optical lattices.   Gaul and M\"uller~\cite{gaul_bogoliubov_2011} studied BECs in spatially correlated disorder potentials.   Billam \emph{et al.}~\cite{billam_second-order_2013} derived equations of motion describing the coupled dynamics of the condensate and noncondensate fractions.

We return to the number-conserving Bogoliubov approximation here to provide a particularly transparent method of deriving the relevant equations.  Our approach to a number-conserving Bogoliubov approximation is to ``encode'' the many-body wavefunction of the BEC in the $N$-particle sector of a state we call an Extended Catalytic State (ECS),\si{Extended catalytic state} by which we mean a coherent state for the condensate mode and a state to be determined by the dynamics for the orthogonal modes of the atoms.  Using a time-dependent interaction picture, we move the coherent state to the vacuum, thus making all the field operators formally small compared to ${N}^{1/2}$.  The resulting Hamiltonian can then be organized by powers of ${N}^{-1/2}$.  Requiring the terms of order ${N}^{1/2}$ to vanish in the interaction-picture evolution equation gives the GP equation for the condensate wavefunction.  Going to the next order in the evolution equation, $N^0$, we derive equations equivalent to those found by Castin and Dum~\cite{castin_low-temperature_1998} for a number-conserving Bogoliubov approximation.  In contrast to other approaches, ours allows one to calculate the state evolution in the Schr\"{o}dinger picture, and it also has advantages in considering higher-order corrections and extensions to multi-component cases.

\section{Encoding the State of a BEC}

What is true in quantum optics, that coherent states are easier to deal with than number states, is often true elsewhere.  Indeed, the usual mean-field approximation to BEC evolution is based on the assumption that the BEC is in a coherent state of a condensate mode~\cite{pitaevskii_bose-einstein_2003}.  A problem with this approach is that the number of particles in a BEC is usually fixed, whereas coherent states are superpositions of states with different numbers of particles.  A related problem is that assigning a coherent state to a BEC breaks its phase symmetry, thus causing problems in developing the Bogoliubov approximation.

Our philosophy for dealing with a BEC that has $N$ particles is to extend the BEC state $\ket{\varPsi_{\! N}}$ to a state $\ket{\varPsi_\mathrm{ecs}}$, for which the condensate mode is in a coherent state, but the $N$-particle sector is the same as $\ket{\varPsi_{\!N}}$ within a constant.  Consider an arbitrary state $\ket{\varPsi_{\!N}}$ with $N$ particles, for which we have the number-state decomposition,
\begin{gather}\lb{eq:PsiN}
\ket{\varPsi_N}=
\sum_{M=0}^N\, \ket{N-M}_0\otimes \ket{\varOmega_M}_\perp\;,\\
\sN_\perp\, \ket{\varOmega_M}_\perp=M\, \ket{\varOmega_M}_\perp\;,
\end{gather}
where the kets labeled by 0 and $\perp$ apply to the condensate mode and to all the modes orthogonal to the condensate mode, respectively.  The operator $\sN_\perp$ is the particle-number operator for the orthogonal modes.  The state $\ket{\varOmega_M}$ for the orthogonal modes, which has $M$ particles in the orthogonal modes, is not necessarily normalized.  The key to our approach is that the state~(\rf{eq:PsiN}) can be written as
\begin{subequations}
\begin{align}
\ket{\varPsi_N}
&=e^{\norm{\alpha}^2/2}\,\sum_{M=0}^N\, \frac{\sqrt{(N-M) !}}{\alpha^{N-M}}\; \sP_N\,\Big(\ket{\alpha}_0\otimes
\ket{\varOmega_M}_\perp\Big) \\
&=e^{\norm{\alpha}^2/2}\,\frac{\sqrt{N!}}{\alpha^N}\;
\sP_N \Bigg(\ket{\alpha}_0\otimes
\bigg(\sum_{M=0}^N\;\alpha^M\sqrt{\frac{(N-M)!}{N!}}\;\ket{\varOmega_M}_\perp\bigg)\,\Bigg)\\
&=e^{\norm{\alpha}^2/2}\,\frac{\sqrt{N!}}{\alpha^N}
\;\sP_N\Big(\ket{\alpha}_0\otimes\ket{\varOmega}_\perp\Bigr)\;,
\end{align}
\end{subequations}
where $\sP_{N}$ is the projection operator onto the $N$-particle sector and
\begin{equation}\lb{eq:Omegaperp}
\ket{\varOmega}_\perp=
\sum_{M=0}^N\;\alpha^M\sqrt{\frac{(N-M)!}{N!}}\;\ket{\varOmega_M}_\perp
\end{equation}
is an (unnormalized) state of the modes orthogonal to the condensate mode.

We now introduce the \emph{extended catalytic state},\si{Extended catalytic state}
\begin{equation}
\ket{\varPsi_\mathrm{ecs}}
=\ket{\alpha}_0\otimes \ket{\varOmega}_\perp \;,\lb{eq:catalytic_state}
\end{equation}
which is related to the physical state by
\begin{equation}
\ket{\varPsi_N}=e^{\norm{\alpha}^2/2}\,\frac{\sqrt{N!}}{\alpha^N}\;
\sP_N \ket{\varPsi_\mathrm{ecs}}\;.
\end{equation}
The extended catalytic state is a direct product of a coherent state $\ket{\alpha}_0$ in the condensate mode and an unnormalized state $\ket{\varOmega}_\perp$ of the orthogonal modes.  Notice that once $\alpha$ is specified, the extended catalytic state has a one-to-one correspondence with the physical state.  The structure of the extended catalytic state allows us to study the dynamics of a BEC in the Schr\"{o}dinger picture, and we will see that the structure is preserved throughout the evolution in the Bogliubov approximation.

For a pure condensate with no depletion of the condensate mode, the modes orthogonal to the condensate mode are in vacuum, and the overall state has the form
\begin{equation}
\ket{\varPsi_{N}}=\ket{N}_0\otimes \ket{\vac}_\perp\;.
\end{equation}
In this case we have
\begin{equation}
 \ket{\varPsi_\mathrm{ecs}}=\ket{\alpha}_0\otimes \ket{\vac}_\perp\;.
\end{equation}
Generally one expects that a dilute-gas BEC has a state close to that of a pure condensate, in which case the noncondensate state $\ket{\varOmega}_\perp$ is close to the vacuum; we want to develop an approximate description based on this expectation.  To do so, notice that the encoding into an extended catalytic state works for any value of $\alpha$.  In other words, one has the freedom to choose $\alpha$ at will; after the projection, all values of $\alpha$ yield the same physical state.   Nonetheless, we stick to the choice $|\alpha|=N^{1/2}$, for the reason that we make approximations in deriving the dynamics of $\ket{\varPsi_\mathrm{ecs}}$ and the projection into the $N$-particle sector can amplify the errors due to these approximations.  To keep these errors under control, we need the number distribution of the coherent state to be centered at the actual atomic number $N$.  The phase of $\alpha$ is yet another matter, which we discuss further below.

The BEC Hamiltonian conserves particle number and thus commutes with the particle-number operator.  As a consequence, the evolution operator $\sU(t)$ commutes with $\sP_N$, allowing us to move the evolution operator through the projection into the $N$-particle sector so that it acts directly on the extended catalytic state:
\begin{subequations}
\begin{align}\lb{eq:projection}
\ket{\varPsi_{\! N}(t)}
&=\sU(t)\,\ket{\varPsi_{\! N}(0)}\\
&=e^{\norm{\alpha}^2/2}\,\frac{\sqrt{N!}}{\alpha^N}\;
\sU(t)\,\sP_{\! N}\ket{\varPsi_\mathrm{ecs}(0)}\\
&=e^{\norm{\alpha}^2/2}\,\frac{\sqrt{N!}}{\alpha^N}\;
\sP_{\! N}\,\sU(t)\, \ket{\varPsi_\mathrm{ecs}(0)}\\
&=e^{\norm{\alpha}^2/2}\,\frac{\sqrt{N!}}{\alpha^N}\;
\sP_{N} \ket{\varPsi_\mathrm{ecs}(t)}\;.
\end{align}
\end{subequations}
To find $\ket{\varPsi_{\! N}(t)}$, one solves for $\ket{\varPsi_\mathrm{ecs}(t)}$ and then projects into the $N$-particle sector.

The first step in developing our approximation is to go to an interaction picture in which the condensate mode is displaced from a coherent state to vacuum.  To do this, we start with a time-dependent condensate mode defined by the single-particle state $\ket{\psinot(t)}$, which has wave function
\begin{equation}
\psinot(\mathbf{x},t)=\braket{\mathbf{x}}{\psinot(t)}\;.
\end{equation}
The annihilation operator for the condensate mode is related to the Schr\"odinger-picture field operator $\uppsi(\mathbf{x})$ by
\begin{equation}\lb{eq:aphi}
\a_{\psinot(t)}=\int \psinot^*(\mathbf{x},t)\,\uppsi(\mathbf{x})\,\dif \mathbf{x}
=\braket{\psinot(t)}{\uppsi}=\braket{\uppsi^\dagger}{\psinot^*(t)}\;.
\end{equation}
Here, in the final two equalities, we introduce a shorthand notation for the integral as a bra-ket inner product between a single-particle state and the field operator.  The creation operator for the condensate mode is
\begin{equation}\lb{eq:aphidagger}
\a_{\smash{\psinot(t)}}^\dagger=\int\uppsi^\dagger(\mathbf{x})\,\psinot(\mathbf{x},t)\,\dif \mathbf{x}
=\braket{\uppsi}{\psinot(t)}=\braket{\psinot^*(t)}{\uppsi^\dagger}\;.
\end{equation}
The bra-ket notation introduced here, though \emph{ad hoc}, is useful for manipulating the complicated expressions that arise as we proceed, more so once we get to the two-component case in the next section.  The annihilation and creation operators have two different bra-ket forms, both of which we use in our treatment.  The field operator can be written as
\begin{equation}\lb{eq:fieldperp}
\uppsi(\mathbf{x})=\a_{\psinot(t)}\ssp \psinot(\mathbf{x},t)+\uppsi_{\perp}(\mathbf{x},t)\;,
\end{equation}
where $\uppsi_{\perp}(\mathbf{x},t)$, the field operator with the condensate mode excluded, can be expanded in terms of single-particle states orthogonal to $\ket{\psinot(t)}$.  In the Schr\"odinger picture, $\uppsi_{\perp}(\mathbf{x},t)$ has a time dependence because the condensate mode is changing in time.  Notice that in terms of our shorthand notation, we can write
\begin{subequations}
\begin{align}
\ket{\uppsi_\perp}&=\ket\uppsi-\ket\psinot\a_\psinot=\ket\uppsi-\ket\psinot\braket{\psinot}{\uppsi}\;,\\
\bra{\uppsi_\perp}&=\bra\uppsi-\a_\psinot^\dagger\bra\psinot=\bra\uppsi-\braket{\uppsi}{\psinot}\bra\psinot\;.
\end{align}
\end{subequations}

The extended catalytic state for a pure condensate in the time-dependent condensate mode $\ket{\psinot(t)}$ is
\begin{equation}\lb{eq:ecs_td}
\sD\big(\alpha,\psinot(t)\big)\ket{\vac}=
\ket{\alpha,\psinot(t)}_0\otimes\ket{\vac}_\perp\;,
\end{equation}
where the displacement operator $\sD\big( \alpha, \psinot(t) \big)$ for the condensate mode, which we usually abbreviate as $\sD(t)$, is defined as\si{Displace operator}
\begin{equation}
\sD\big( \alpha, \psinot(t) \big)= \sD(t)
=\exp \big(\alpha\ssp \a^\dagger_{\smash{\psinot(t)}} - \alpha^* \a_{\psinot(t)} \big)\;,
\end{equation}
The state~(\rf{eq:ecs_td}), which describes a pure condenstate with no depletion, is the one we perturb about in developing our approximate description.

We can now introduce the desired interaction picture as the one where the condensate mode is displaced to vacuum; i.e., states transform to\si{Interaction picture}
\begin{equation}\lb{eq:intpic}
\ket{\varPsi_\mathrm{int}(t)}=\sD^\dagger
\big( \alpha, \psinot(t) \big) \ket{\varPsi_{\rm ecs}(t)}
=\overbrace{\mathcal{D}^\dagger\big( \alpha, \psinot(t) \big)\,
\mathcal{U}(t)\,\mathcal{D}\big( \alpha, \psinot(0) \big)}^{\displaystyle{\mathcal{U}_\mathrm{int}(t)}}\, \ket{\varPsi_\mathrm{int}(0)}\;,
\end{equation}
where $\mathcal{U}_\mathrm{int}(t)$ is the evolution operator in the interaction picture.  The Schr\"odinger-picture evolution operator $\mathcal{U}(t)$ obeys the Schr\"odinger equation
\begin{equation}
i\hbar\:\dt{\mathcal{U}}(t)=\sH(t)\,\mathcal{U}(t)\;,
\end{equation}
where $\sH(t)$ is the (possibly time-dependent) BEC Hamiltonian.  The time dependence of the condensate wave function, which enters into the displacement operator $\mathcal{D}\big(\alpha, \psinot(t) \big)$ through the annihilation and creation operators, $\a_{\psinot(t)}$ and $\a_{\smash{\psinot(t)}}^\dagger$, is to be determined. The evolution operator $\mathcal{U}_\mathrm{int}(t)$ obeys the equation
\begin{subequations}
\begin{align}
i \hbar\: \dt{\mathcal U}_\mathrm{int}(t)&=i \hbar\,\dt{\mathcal D}^\dagger(t)\, \mathcal{U}(t) \,\mathcal{D}(0)+i \hbar\,\mathcal{D}^\dagger(t)\, \dt{\mathcal{U}}(t)\,\mathcal{D}(0)\\[3pt]
&=\Big(\,i \hbar\,\dt{\mathcal D}^\dagger(t)\,
\mathcal{D}(t)+\mathcal{D}^\dagger(t)\,\sH(t)\,\mathcal{D}(t)\Big)\, \mathcal{U}_\mathrm{int}(t)\;.\lb{eq:evolution_Uint}
\end{align}
\end{subequations}
The time derivative of the displacement operator is
\begin{subequations}
\begin{align}
\dt{\mathcal D}^\dagger(t)&=\di{}{t}\,\Big(e^{\alpha^* \a_{\psinot(t)}-\alpha\, \a^\dagger_{\smash{\psinot(t)}}}\Big)\\
&=e^{\half\,\norm{\alpha}^2}\,\di{}{t}\,\Big(\,e^{\alpha^* \a_{\psinot}}\,
e^{\mathord{-}\alpha\, \a_{\smash \psinot}^\dagger}\Big)\\
&=\Big(\alpha^*\,\dt{\a}_\psinot-\alpha\,e^{\alpha^*\a_{\psinot}}\,\dt{\a}^\dagger_{\smash \psinot}\,
e^{\mathord{-}\alpha^*\a_{\psinot}}\Big)\:\mathcal{D}^\dagger
(t)\\
&=\Big(\alpha^*\, \dt{\a}_\psinot-\alpha\,\dt{\a}^\dagger_{\smash \psinot}-\,
\norm{\alpha}^2\, \commutb{\a_\psinot}{
\dt{\a}^\dagger_{\smash \psinot}}\Big)\:\mathcal{D}^\dagger(t)\\
&=\Big(\alpha^*\,\dt{\a}_\psinot-\alpha\,
\dt{\a}^\dagger_{\smash \psinot}-\norm{\alpha}^2\,
\braket{\psinot}{\dt{\psinot}}\Big)\:
\mathcal{D}^\dagger(t)\;.
\end{align}
\end{subequations}
Putting the above expression into Eq.~(\rf{eq:evolution_Uint}), we have
\begin{equation}
i \hbar\; \dt{\mathcal{U}}_\mathrm{int}(t)=\sH_\mathrm{int}(t)\: \mathcal{U}_\mathrm{int}(t)\;,
\end{equation}
where the interaction picture Hamiltonian reads
\begin{equation}\lb{eq:intpicH}
\sH_\mathrm{int}(t)=\mathord{-}i \hbar\,\Big(\,\norm{\alpha}^2\,
\braket{\psinot(t)}{\dt{\psinot}(t)}+\alpha\,
\dt{\a}^\dagger_{\smash{\psinot(t)}}-\alpha^*\, \dt{\a}_{\psinot(t)}\Big)+
\mathcal{D}^\dagger(t)\,\sH(t)\,\mathcal{D}(t)\;.
\end{equation}
Equivalently, we have
\begin{equation}
i \hbar\,\ket{\dt{\varPsi}_\mathrm{int}(t)}=\sH_\mathrm{int}(t)\,
\ket{\varPsi_\mathrm{int}(t)}\;.
\end{equation}

In the interaction picture the field operator takes the form
\begin{equation}
\mathcal{D}^\dagger(t)\,\uppsi(\mathbf{x})\,\mathcal{D}(t)=\uppsi(\mathbf{x})+\alpha\, \psinot(\mathbf{x},t)\;.
\end{equation}
An expansion of $\sH_\mathrm{int}(t)$ in powers of $1/|\alpha|=1/N^{1/2}$ is a good approximation as long as the field operator is small relative to the interaction picture, i.e., more formally, as long as the one-particle density matrix is small in the sense that
\begin{equation}
\rho_\mathrm{int}(\mathbf{x},\mathbf{x}')=\bra{\varPsi_\mathrm{int}}\, \uppsi^\dagger(\mathbf{x}')\,\uppsi(\mathbf{x}) \ket{\varPsi_\mathrm{int}}\sim N^0\;.
\end{equation}
This requirement is satisfied as long as the system is a condensate.  In second-quantized form, the model Hamiltonian for the BEC is
\begin{equation}\lb{BECHt}
\sH(t)
=\int\left(\uppsi^\dagger(\mathbf{x}) \Big(\mathord{-}\frac{{\hbar}^2}{2m}\boldsymbol{\nabla}^2 +V(\mathbf{x},t)\Big) \uppsi(\mathbf{x})
+ \frac{g}{2}\,\uppsi^\dagger(\mathbf{x})\uppsi^\dagger(\mathbf{x})\uppsi(\mathbf{x})\uppsi(\mathbf{x})\right)\dif \mathbf{x} \;,
\end{equation}
where the first term is the second-quantized Hamiltonian for particles trapped in a potential $V(\mathbf{x},t)$ and the second term represents the two-body scattering energy.  The only explicit time dependence in the Hamiltonian~(\rf{BECHt}) comes from a possible time dependence in the trapping potential $V(\mathbf x,t)$.  For our expansion in powers of $1/\norm\alpha$ to work, we must have that $g\norm\alpha^2$ is of order $N^0$; i.e., $g$ is of order $N^{-1}$.

Going to the interaction picture, we have
\begin{subequations}
\begin{align}
&\hspace{-2.5em}\sH_\mathrm{int}(t)\nonumber\\
&\hspace{-2.2em}= - i \hbar\,\Big(\,\norm{\alpha}^2
\braket{\psinot(t)}{\dt{\psinot}(t)}
+\alpha\,\dt{\a}^\dagger_{\smash{\psinot(t)}}-\alpha^*\,
\dt{\a}_{\psinot(t)}\,\Big)+ \mathcal{D}^\dagger(t)\,\sH(t)\,\mathcal{D}(t)\lb{eq:Hint}\\
&\hspace{-2.2em}\simeq\norm{\alpha}^2 \int\psinot^*
\Big(\mathord{-}i \hbar\,\pa{}{t}-\frac{{\hbar}^2}{2m}\boldsymbol{\nabla}^2
+V+\frac{g}{2}\,\norm{\alpha}^2 \norm{\psinot}^2 \Big)
\psinot\: \dif \mathbf{x} \lb{eq:Hint_meanfield_energy} \\
&\hspace{-2.2em}\quad +\bigg(\alpha\int
\uppsi^\dagger\Big(-i \hbar\,\pa{}{t}-\frac{{\hbar}^2}{2m}\boldsymbol{\nabla}^2 +V+g\norm{\alpha}^2\norm{\psinot}^2\Big)\psinot\,\dif \mathbf{x}+\mathrm{H.c.}\bigg) \lb{eq:Hint_linear}\\
&\hspace{-2.2em}\quad
+\int\hspace{-0.25em}\bigg[\uppsi^\dagger\Big(-\frac{{\hbar}^2}{2m}\boldsymbol{\nabla}^2
+V+2 g \norm{\alpha}^2\norm{\psinot}^2\Big) \uppsi+\frac{g}{2}\Big(\alpha^2\uppsi^\dagger\uppsi^\dagger\,\psinot^2
+(\alpha^*)^2\uppsi\uppsi\,(\psinot^*)^2\Big)\bigg]\dif \mathbf{x}\;, \lb{eq:Hint_quadratic}
\end{align}
\end{subequations}
where we neglect terms of order $N^{-1/2}$ or less. The $c$-number term~(\rf{eq:Hint_meanfield_energy}), of order $N$, is, in the time-independent case, the mean-field energy of the BEC.  By requiring the linear term~(\rf{eq:Hint_linear}), of order ${N}^{1/2}$, to vanish, we get\si{Gross-Pitaevskii equation}
\begin{equation}\lb{eq:GPE}
i \hbar\,\dt{\psinot}(\mathbf{x},t)=\Big(\,\mathord{-}\frac{{\hbar}^2}{2m}\boldsymbol{\nabla}^2 +V(\mathbf{x},t)+g\norm{\alpha}^2
\norm{\psinot(\mathbf{x},t)}^2\,\Big)\,\psinot(\mathbf{x},t)\;,
\end{equation}
which is the celebrated Gross-Pitaevskii equation. Thus the structure of our approach is clear: By going to the interaction picture, the mean-field evolution is removed, and then by neglecting the terms of higher order than $N^0$, we are left with the quadratic Bogoliubov Hamiltonian\si{Bogoliubov Hamiltonian}
\begin{align}\lb{eq:Hbog}
\begin{split}
\sH_\mathrm{bog}=\int\bigg[&\uppsi^\dagger\Big(\mathord{-}\frac{{\hbar}^2}{2m}\boldsymbol{\nabla}^2
+V+2 g \norm{\alpha}^2\norm{\psinot}^2\Big)\uppsi\\
&+\frac{g}{2}\Big(\alpha^2\,\uppsi^\dagger \uppsi^\dagger\psinot^2
+(\alpha^*)^2\,\uppsi\uppsi (\psinot^*)^2\Big)\bigg]\dif \mathbf{x}\;,
\end{split}
\end{align}

To reveal the symplectic structure of the Bogoliubov Hamiltonian, it is useful to write it in the matrix form
\begin{equation}\lb{Hbog_matrixform}
\sH_\mathrm{bog}=\half\,
\COLON
\begin{pmatrix}
\bra{\uppsi} & \bra{\uppsi^\dagger}
\end{pmatrix}
\begin{pmatrix}
H_\mathrm{gp} + g \norm{\alpha}^2  \norm{\psinot}^2  & g \alpha^2\,  \psinot^2 \\[6pt]
g (\alpha^*)^2  (\psinot^*)^2   & H_\mathrm{gp} + g \norm{\alpha}^2  \norm{\psinot}^2
\end{pmatrix}
\begin{pmatrix}
\ket{\uppsi}\\[5pt] \ket{\uppsi^\dagger}
\end{pmatrix}
\COLON\;,
\end{equation}
where the colons denote normal ordering of annihilation and creation operators and the GP single-particle Hamiltonian takes the form
\begin{equation}
H_\mathrm{gp}(t)=\mathord{-}\frac{{\hbar}^2}{2m}\boldsymbol{\nabla}^2
+V(t)+g \norm{\alpha}^2\, \norm{\psinot(t)}^2\;.
\end{equation}
Notice that the normal ordering has an effect only on the lower-right corner of the matrix.

Like all the matrices of symplectic structures in this dissertation, we denote the $2\times 2$ matrix in Eq.~(\rf{Hbog_matrixform}) by a special sans serif character:
\begin{equation}
\mathsp{H}_\mathrm{bog}
=\begin{pmatrix}
H_\mathrm{gp} + g \norm{\alpha}^2  \norm{\psinot}^2  & g \alpha^2\,  \psinot^2 \\[6pt]
g (\alpha^*)^2  (\psinot^*)^2   & H_\mathrm{gp} + g \norm{\alpha}^2  \norm{\psinot}^2
\end{pmatrix}\;.
\end{equation}
As shown by Lewenstein and You~\cite{lewenstein_quantum_1996},\ai{Lewenstein, Maciej} $\mathsp{H}_\mathrm{bog}$ has a nilpotent subspace, where phase diffusion takes place.  Such phase diffusion is not physical, but rather is a consequence of the arbitrary phase assigned to the condensate wavefunction, i.e., to $\alpha$.  This problem was addressed by introducing the so-called number-conserving approaches~\cite{girardeau_theory_1959, gardiner_particle-number-conserving_1997, castin_low-temperature_1998}.\ai{Girardeau, Marvin D.}\ai{Gardiner, Crispin W.}\ai{Castin, Yvan}  Particularly in the work of Castin and Dum, a systematic expansion of the field operators was used in deriving the equations for the number-conserving approach.  The aim is to eliminate the artificial nilpotent subspace that gives rise to the phase diffusion.  Here we solve the same problem by introducing an additional contribution to the Hamiltonian, an auxiliary Hamiltonian $\sF(t)$,\si{Auxiliary Hamiltonian} which does not affect the $N$-particle sector of $\ket{\varPsi_\mathrm{ecs}(t)}$ and thus keeps the physical state $\ket{\varPsi_{\! N}(t)}$ unchanged, i.e.,
\begin{equation}
\mathcal{P}_{\! N}\, \sF(t)\,\ket{\varPsi_{\rm ecs}(t)}=0\;.
\lb{eq:F_condition}
\end{equation}
With this term $\sF(t)$, we can solve the phase diffusion problem by eliminating the nilpotent subspace of $\mathsp{H}_\mathrm{bog}$.

To determine the form of $\sF(t)$, we must go to the Bogoliubov level of approximation, but for now let us suppose $\sF(t)$ takes the form
\begin{equation}\lb{eq:F}
\sF(t)=-\frac{\eta(t)}{2}\, (\mathcal{N}-N)^2
+  \big(\alpha\ssp \a^\dagger_{\smash{\psinot(t)}}+\mathcal{N}_{\perp}(t)-N\big)\, \mathcal{F}_\perp(t)+   \big(\alpha^* \a_{\psinot(t)}-N\big)\, \mathcal{F}_\perp^\dagger(t)\;.
\end{equation}
Here
\begin{equation}
\mathcal{N}=
\int \uppsi^\dagger(\mathbf{x})\uppsi(\mathbf{x})\,\dif \mathbf{x}
\end{equation}
is the total particle-number operator, and $\mathcal{N}_{\perp}=\mathcal{N}-\a_{\smash \psinot}^\dagger\a_{\psinot}$ is the particle-number operator for all the modes orthogonal to the condensate mode, i.e., the depletion number operator.  The time-dependent parameter $\eta(t)$, which is to be determined, is of order $N^{-1}$.  The operator $\mathcal{F}_\perp$, also to be determined, is of the order $N^{-1/2}$, is a linear function of the annihilation and creation operators of the modes orthogonal to the condensate mode, and thus commutes with $\a_\psinot$ and $\a_{\smash \psinot}^\dagger$.  The first term in Eq.~(\rf{eq:F}) clearly satisfies Eq.~(\rf{eq:F_condition}). For the other two terms, we have
\begin{gather}\hspace{-2em}
\sP_N \big(\alpha \a_{\smash \psinot}^\dagger+\mathcal{N}_{\perp}-N\big)\mathcal{F}_\perp\,\ket{\alpha}_0\otimes \ket{\varOmega}_\perp
=\sP_N \big(\mathcal{N}-N\big)\,\ket{\alpha}_0\otimes \mathcal{F}_\perp\ket{\varOmega}_\perp=0\;,\\[2pt]
\big(\alpha^* \a_{\psinot}-N\big)\mathcal{F}_\perp^\dagger\,\ket{\alpha}_0\otimes \ket{\varOmega}_\perp=\big(\,\norm{\alpha}^2-N\big)\,\ket{\alpha}_0\otimes \mathcal{F}_\perp^\dagger\ket{\varOmega}_\perp=0\;,
\end{gather}
where in the first equation we use $\alpha \a_{\smash \psinot}^\dagger+\mathcal{N}_{\perp}-N=\a_{\smash \psinot}^\dagger(\alpha-a_\psinot)+\mathcal{N}-N$.
As long as the condensate mode stays in the coherent state with amplitude $\alpha$, these two terms do not affect the physical state $\ket{\varPsi_{\! N}(t)}$.  We show that the condensate mode does remain in a coherent state at Bogoliubov order at the end of this section.

An astute reader might object at this point that the auxiliary Hamiltonian~(\rf{eq:F}) is not Hermitian.  This is not a problem at Bogoliubov order, however, because the only nonHermitian term in $\mathcal{F}(t)$ is $\sN_\perp\sF_\perp$, which, being of order $N^{-1/2}$, can be neglected in the Bogoliubov approximation (order $N^0$).  Going now to the interaction picture, we have
\begin{subequations}
\begin{align}\lb{eq:Fint}
\hspace{-1em}
\sF_\mathrm{int}(t)
&=\mathcal{D}^\dagger\big(\alpha, \psinot(t) \big)\,\sF(t)\,\mathcal{D}\big( \alpha, \psinot(t) \big)\\
\begin{split}
&=-\frac{\eta}{2}\,
\bigg(
\int\big(\alpha^*\psinot^*+\uppsi^\dagger\big)\big(\alpha \psinot+\uppsi\big)\,\dif\mathbf{x}-N
\bigg)^2  \\
&\qquad
+\Big( \alpha\, ( \alpha^*+\a_{\smash \psinot}^\dagger )+\mathcal{N}_{\perp}-N \Big)\,\mathcal{F}_\perp
+\,\Big(\alpha^* ( \alpha+\a_{\psinot} )-N \Big)\,\mathcal{F}_\perp^\dagger
\end{split}\\[3pt]
&=-\frac{\eta}{2}\Big(\alpha\a_{\smash \psinot}^\dagger+\alpha^*\a_\psinot+\mathcal{N}\Big)^2
+\big(\alpha\a_{\smash \psinot}^\dagger+\mathcal{N}_\perp\big)\sF_\perp
+\alpha^*\a_{\psinot}\,\mathcal{F}_\perp^\dagger\\[5pt]
\begin{split}
&\simeq \mathord{-}\frac{\eta}{2}\,\Big(\,\norm{\alpha}^2\,(\a_{\smash \psinot}^\dagger \a_{\psinot}+\a_\psinot\a_{\smash \psinot}^\dagger)+\alpha^2  \a_{\smash \psinot}^\dagger \a_{\smash \psinot}^\dagger
+(\alpha^*)^2\a_\psinot\a_\psinot\Big)
+\alpha \a_{\smash \psinot}^\dagger\,\mathcal{F}_\perp+\alpha^*\a_{\psinot}\,\mathcal{F}_\perp^\dagger\;,
\end{split}
\end{align}
\end{subequations}
where the identity $\norm{\alpha}^2=N$ is used to cancel several terms and where terms of order $N^{-1/2}$ or smaller are neglected in the final form (recall that interaction-picture field operators are order $N^0$).  Adopting this final, approximate form, we have
\begin{equation}\lb{eq:FintBog}
\sF_\mathrm{int}(t)
=\mathord{-}\frac{\eta}{2}\,\Big(\,2\norm{\alpha}^2\,\a_{\smash \psinot}^\dagger \a_{\psinot}+\alpha^2  \a_{\smash \psinot}^\dagger \a_{\smash \psinot}^\dagger
+(\alpha^*)^2\a_\psinot\a_\psinot\Big)
+\alpha\a_{\smash \psinot}^\dagger\,\mathcal{F}_\perp+\alpha^*\a_{\psinot}\,\mathcal{F}_\perp^\dagger
-\frac{\eta}{2}|\alpha|^2\;,
\end{equation}
where we normally order the creation and annihilation operators of the condensate mode in preparation for incorporating $\sF_{\mathrm{int}}$ into the main Bogoliubov Hamiltonian.  This normal ordering introduces a $c$-number term, $-\eta|\alpha|^2/2$, which could be important as a second-order correction to the condensate energy, but which only adds an overall phase to the evolving quantum state.  Thus we neglect this term henceforth.  The modified (number-conserving) Bogoliubov Hamiltonian then takes the form
\begin{subequations}
\begin{align}\lb{modHbog}
\sH_\mathrm{ncb}
&=\sH_\mathrm{bog}+\sF_{\mathrm{int}}\\
&=\sH_\mathrm{bog}-\frac{\eta}{2}\,\Big(\,2\norm{\alpha}^2 \a_{\smash \psinot}^\dagger \a_{\psinot}+\alpha^2 \a_{\smash \psinot}^\dagger \a_{\smash \psinot}^\dagger
+(\alpha^*)^2\a_\psinot\a_\psinot\Big)
+\alpha\a_{\smash \psinot}^\dagger\,\mathcal{F}_\perp+\alpha^*\a_{\psinot}\,\mathcal{F}_\perp^\dagger\;.
\end{align}
\end{subequations}

To eliminate the phase diffusion, we choose
\begin{subequations}
\begin{align}
\eta(t)&= g \int \norm{\psinot(\mathbf{x},t)}^4\,\dif\mathbf{x}\\
&=g\bra\psinot\norm{\psinot}^2\ket\psinot
=g\bra{\psinot^*}\norm{\psinot}^2\ket{\psinot^*}
=g\bra{\psinot^*}(\psinot^*)^2\ket{\psinot}
=g\bra{\psinot}\psinot^2\ket{\psinot^*}
\lb{eq:eta}
\end{align}
\end{subequations}
and
\begin{subequations}
\begin{align}
\mathcal{F}_\perp(t)
=\mathcal{F}_\perp^\dagger(t)
&=-g\alpha^*\!\int \psinot^*(\mathbf{x}, t)\norm{\psinot(\mathbf{x},t)}^2
\uppsi_{\perp}(\mathbf{x},t)\,\dif \mathbf{x}+\mathrm{H.c.}\\
&=\eta(t)\alpha^*a_\psinot
-g\alpha^*\!\int \psinot^*(\mathbf{x}, t)\norm{\psinot(\mathbf{x},t)}^2\uppsi(\mathbf{x})\,\dif \mathbf{x}+\mathrm{H.c.}\\
&=g\alpha^*\Big(\bra{\psinot}\norm\psinot^2\ket{\psinot}\braket{\psinot}{\uppsi}
-\bra{\psinot}\norm{\psinot}^2\ket{\uppsi}\Big)+\mathrm{H.c.}\;,
\lb{eq:Fperp}
\end{align}
\end{subequations}
where $\uppsi_{\perp}(\mathbf{x},t)$ is the field operator of Eq.~(\rf{eq:fieldperp}), which has the condensate mode excluded.  It is now a tedious calculation to show that
\begin{subequations}
\begin{align}
\begin{split}
\sF_\mathrm{int}
&=\frac{\eta}{2}\,\Big(\,2\norm{\alpha}^2\,\a_{\smash \psinot}^\dagger \a_{\psinot}+\alpha^2  \a_{\smash \psinot}^\dagger \a_{\smash \psinot}^\dagger
+(\alpha^*)^2\a_\psinot\a_\psinot\Big)\\
&\qquad-g\bigg(\Big(|\alpha|^2\a_{\smash \psinot}^\dagger+(\alpha^*)^2\a_\psinot\Big)
\int \psinot^*(\mathbf{x},t)\norm{\psinot(\mathbf{x},t)}^2\uppsi(\mathbf{x})\,\dif \mathbf{x}+\mathrm{H.c.}\bigg)
\end{split}\\[3pt]
\begin{split}
&=\frac{g}{2}\,\Big(\,
\norm{\alpha}^2\,\braket{\uppsi}{\psinot}\bra{\psinot}\norm\psinot^2\ket\psinot\braket{\psinot}{\uppsi}\\
&\qquad\quad
+(\alpha^*)^2\,\braket{\uppsi^\dagger}{\psinot^*}\bra{\psinot^*}(\psinot^*)^2\ket\psinot\braket{\psinot}{\uppsi}
+\mathrm{H.c.}\Big)\\
&\qquad-g\Big(
\norm{\alpha}^2\,\braket{\uppsi}{\psinot}\bra{\psinot}\norm\psinot^2\ket\uppsi
+(\alpha^*)^2\,\braket{\uppsi^\dagger}{\psinot^*}\bra{\psinot^*}(\psinot^*)^2\ket\uppsi
+\mathrm{H.c.}\Big)\;.
\end{split}
\end{align}\lb{eq:Fint2}
\end{subequations}

Translating this into matrix notation, we get
\begin{equation}
\sF_\mathrm{int}
=\frac{1}{2}
\,\COLON
\begin{pmatrix}
\bra{\uppsi} & \bra{\uppsi^\dagger}
\end{pmatrix}
\mathsp{F}_{\mathrm{int}}
\begin{pmatrix}
\ket{\uppsi}\\[5pt] \ket{\uppsi^\dagger}
\end{pmatrix}\COLON\;,
\end{equation}
where the symplectic matrix is
\begin{align}
\mathsp{F}_{\mathrm{int}}=g
\begin{pmatrix}
|\alpha|^2\big(Q|\psinot|^2Q-|\psinot|^2\big)
&\alpha^2\big(Q\psinot^2 Q^*-\psinot^2\big)\\[6pt]
(\alpha^*)^2\big(Q^*(\psinot^*)^2 Q-(\psinot^*)^2\big)
&|\alpha|^2\big(Q^*|\psinot|^2 Q^*-|\psinot|^2\big)
\end{pmatrix}\;.
\end{align}
Here
\begin{equation}\lb{eq:Q}
Q(t)=\identity-P(t)=\identity-\proj{\psinot(t)}
\end{equation}
is the projector onto the single-particle space orthogonal to the condensate mode, with $Q^*(t)=\identity-\proj{\psinot^*(t)}$.  In Eqs.~(\rf{eq:eta}), (\rf{eq:Fperp}), and (\rf{eq:Fint2}), we use the bra-ket notation, which is the easiest way to carry out the algebraic manipulations and which also generalizes straightforwardly to the two-component case considered in the next section.  The modified (number-conserving) Bogoliubov Hamiltonian~(\rf{modHbog}) now reads\si{Number-conserving Bogoliubov}
\begin{equation}\lb{Hncb_matrixform}
\sH_\mathrm{ncb}=\half\,\COLON
\begin{pmatrix}
\bra{\uppsi} & \bra{\uppsi^\dagger}
\end{pmatrix}\!
\mathsp{H}_\mathrm{ncb}
\begin{pmatrix}
\ket{\uppsi}\\[5pt] \ket{\uppsi^\dagger}
\end{pmatrix}\COLON\;,
\end{equation}
with
\begin{equation}
\mathsp{H}_\mathrm{ncb}
=\mathsp{H}_{\mathrm{bog}}+\mathsp{F}_{\mathrm{int}}
=\begin{pmatrix}
H_\mathrm{gp} + g \norm{\alpha}^2 Q \norm{\psinot}^2 Q & g \alpha^2\, Q \psinot^2 Q^*\\[6pt]
g (\alpha^*)^2  Q^* (\psinot^*)^2  Q & H_\mathrm{gp} + g \norm{\alpha}^2 Q^* \norm{\psinot}^2 Q^*
\end{pmatrix}\;.
\end{equation}

We return now to showing that the condensate mode remains in a coherent state under the evolution at Bogoliubov order.  This is a bit different from saying that the condensate mode is decoupled from the orthogonal modes, as would be the case if we were considering the time-independent ground state of the condensate.  Rather, because $\ket{\phi(t)}$ is changing in time, what we need is that the coupling to the orthogonal modes is of just right sort to maintain the condensate mode in a coherent state.

To show this formally is quite easy.  We begin by noting that the interaction-picture evolution equation at Bogoliubov order,
\begin{equation}
i \hbar  \di{}{t}\,\ket{\varPsi_\mathrm{int}(t)}= \sH_\mathrm{ncb}(t)\,\ket{\varPsi_\mathrm{int}(t)}\;,
\end{equation}
allows us to write
\begin{subequations}
\begin{align}
i\hbar\,\di{}{t}\,\big(\a_{\psinot}\, \ket{\varPsi_\mathrm{int}} \big)
&=i\hbar\,{\dot\a_{\psinot}}\,\ket{\varPsi_\mathrm{int}}
+\a_{\psinot}\ssp\sH_\mathrm{ncb}\,\ket{\varPsi_\mathrm{int}}\\
&=\Big(i\hbar\,{\dot\a_{\psinot}}
-\commutb{\sH_\mathrm{ncb}}{\a_{\psinot}}\Big)\ket{\varPsi_\mathrm{int}}
+\sH_\mathrm{ncb} \big( \a_{\psinot} \ket{\varPsi_\mathrm{int}}\big)\;.
\lb{eq:evolution_condensate_mode}
\end{align}
\end{subequations}
If we could show that the first term in Eq.~(\rf{eq:evolution_condensate_mode}) vanished, then we could conclude that if $\a_{\psinot(0)}\,\ket{\varPsi_\mathrm{int}(0)}=0$, i.e., the condensate mode starts in vacuum in the interaction picture, then $\a_{\psinot(t)}\, \ket{\varPsi_\mathrm{int}(t)}=0$, i.e., the condensate mode remains in vacuum at all times in the interaction picture.  That the first term in Eq.~(\rf{eq:evolution_condensate_mode}) vanishes is thus what we mean by saying that the time-dependent coupling to the orthogonal modes is of the right sort.  Showing this involves using the GP equation~(\rf{eq:GPE}), the commutator $\commutb{\uppsi^\dagger(\mathbf x)}{\a_\psinot}=-\psinot^*(\mathbf x)$, and the fact that the only term in Eq.~(\rf{Hncb_matrixform}) that contains the annihilation and creation operators of the condensate mode is $\bra{\uppsi} H_\mathrm{gp}(t) \ket{\uppsi}=\Colon\,\bra{\uppsi^\dagger} H_\mathrm{gp}(t)\ket{\uppsi^\dagger}\ssp\ssp\Colon\,$.  Pulling all this together, we get
\begin{align}\lb{eq:comm}
i \hbar\, \dt{\a}_{\psinot(t)}
=i \hbar\,\braket{{\dt\psinot(t)}}{\uppsi}
&=-\bra{\psinot(t)} H_\mathrm{gp}(t)\ket{\uppsi}\nonumber\\
&=\commutb{\bra{\uppsi} H_\mathrm{gp}(t) \ket{\uppsi}}{\a_{\psinot(t)}}
=\commutb{\sH_\mathrm{ncb}(t)}{\a_{\psinot(t)}}\;,
\end{align}
which shows that the first term in Eq.~(\rf{eq:evolution_condensate_mode}) vanishes.  Equivalent to the statement that in the interaction picture the condensate mode stays in vacuum if it begins in vacuum is the statement that  in the Schr\"{o}dinger picture the condensate mode is always in the coherent state $\mathcal{D}\big(\alpha,\psinot(t)\big)\ket{\vac}=
\ket{\alpha,\psinot(t)}$.  As a result, Eq.~(\rf{eq:F_condition}) is always satisfied, and the auxiliary Hamiltonian $\sF$ does not affect the physical state.

Notice that this means that in the extended catalytic state, the condensate mode is never entangled with the other modes.  When we project the extended catalytic state to the $N$-particle sector to obtain the physical state of the BEC, however, entanglement makes its appearance.

We can see the effect of the Bogoliubov Hamiltonian~(\rf{Hncb_matrixform}) on the condensate mode more directly by dividing the field operators in the Bogoliubov Hamiltonian into the contribution from the condensate mode and the contribution from the orthogonal modes, as in Eq.~(\rf{eq:fieldperp}).  The condensate mode does not notice the terms with projection operators $Q$ and $Q^*$ in them, so we get
\begin{subequations}\lb{eq:Hbogcond}
\begin{align}\lb{eq:Hbogcond1}
\sH_\mathrm{ncb}
&=\a_{\smash \psinot}^\dagger\a_\psinot\bra{\psinot}H_{\mathrm{gp}}\ket\psinot
+\a_{\smash \psinot}^\dagger\bra\psinot H_{\mathrm{gp}}\ket{\uppsi_\perp}
+\bra{\uppsi_\perp}H_{\mathrm{gp}}\ket\psinot\a_\psinot
+\sH_{\mathrm{ncb}\ssp\perp}\\
&=\a_{\smash \psinot}^\dagger\bra\psinot H_{\mathrm{gp}}\ket{\uppsi}
+\bra{\uppsi_\perp}H_{\mathrm{gp}}\ket\psinot\a_\psinot
+\sH_{\mathrm{ncb}\ssp\perp}\;,
\lb{eq:Hbogcond2}
\end{align}
\end{subequations}
where
\begin{equation}
\sH_{\mathrm{ncb}\ssp\perp}
=\half\,\COLON
\begin{pmatrix}
\bra{\uppsi_\perp} & \bra{\uppsi_\perp^\dagger}
\end{pmatrix}\!
\mathsp{H}_{\mathrm{bog}}
\begin{pmatrix}
\ket{\uppsi_\perp}\\[5pt] \ket{\uppsi_\perp^\dagger}
\end{pmatrix}\COLON
\end{equation}
is the Hamiltonian for the orthogonal modes, in which we can use $\mathsp{H}_{\mathrm{bog}}$ instead of $\mathsp{H}_\mathrm{ncb}$ because the projectors $Q$ and $Q^*$ have no effect.  Using the GP equation~(\rf{eq:GPE}), we can rewrite Eq.~(\rf{eq:Hbogcond2}) as
\begin{align}
\sH_\mathrm{ncb}
=-i\hbar\ssp \a_{\smash \psinot}^\dagger\dt\a_\psinot
+i\hbar\braket{\uppsi_\perp}{\dt\psinot}\a_\psinot
+\sH_{\mathrm{ncb}\ssp\perp}\;,
\end{align}
from which we can immediately verify the commutator identity~(\rf{eq:comm}).

\section{The Two-Component Case}\si{Two-component BEC}
\lb{sec:two-comp}

In the preceding section we discussed how to derive the number-conserving Bogoliubov approximation for a single-component BEC by going to an interaction picture where the condensate mode is displaced to vacuum.  In this section we show that it is a simple task to generalize our method to multi-component BECs.  We do the two-component case as an example, but the generalization to many components is straightforward.  In the two-component case the condensate wavefunction, which is generally a single-particle state that is entangled between the translational and internal degrees of freedom, takes the form
\begin{subequations}\lb{eq:condensate_state_two_component}
\begin{align}
\ket{\psinot(t)}
&=\frac{1}{\alpha}\sum_\sigma\alpha_\sigma(t)\ket{\psinot_\sigma(t)}\otimes\ket\sigma\\
&=\frac{1}{\alpha}\,\Big(\,\alpha_1(t)\,
\ket{\psinot_1(t)}\otimes \ket{1}+\alpha_2(t)\, \ket{\psinot_2(t)}\otimes \ket{2}\,\Big)\;,
\end{align}
\end{subequations}
where $\sigma$ labels the hyperfine levels, taking on values 1 and 2 in the two-component case, and
where  $\norm{\alpha_1}^2+\norm{\alpha_2}^2=\norm{\alpha}^2=N$, with $N$ being the total number of particles. The states $\ket1$ and $\ket2$ are internal states of the bosonic atoms, which we refer to as hyperfine levels because that would be a typical situation.  In the subspace spanned by $\ket{\psinot_1}\otimes \ket{1}=\ket{1,\psinot_1}$ and $\ket{\psinot_2}\otimes \ket{2}=\ket{2,\psinot_2}$, the single-particle state that is orthogonal to the condensate mode is
\begin{equation}\lb{eq:barpsinot}
\ket{\bar\psinot(t)}
=\frac{1}{\alpha^*}\,\Big(\,\alpha_2^*(t)\, \ket{\psinot_1(t)}\otimes \ket{1}-\alpha_1^*(t)\, \ket{\psinot_2(t)}\otimes \ket{2}\,\Big)\;.
\end{equation}
Notice that
\begin{subequations}
\begin{align}
\ket{1,\psinot_1(t)}
=\ket{\psinot_1(t)}\otimes\ket1
&=
\bigg(\frac{\alpha_1^*(t)}{\alpha^*}\ket{\psinot(t)}
+\frac{\alpha_2(t)}{\alpha}\ket{\bar\psinot(t)}\bigg)\;,\\[3pt]
\ket{2,\psinot_2(t)}=\ket{\psinot_2(t)}\otimes\ket2
&=\bigg(\frac{\alpha_2^*(t)}{\alpha^*}\ket{\psinot(t)}
-\frac{\alpha_1(t)}{\alpha}\ket{\bar\psinot(t)}\bigg)\;.
\end{align}
\end{subequations}

The field operator that destroys a particle in internal level~$\sigma$ at position $\mathbf x$ is $\uppsi_\sigma(\mathbf x)$.  In our shorthand bra-ket notation for field operators, we have
\begin{equation}
\uppsi_\sigma(\mathbf x)=\braket{\mathbf x}{\uppsi_\sigma}=\braket{\sigma,\mathbf x}{\uppsi}\;.
\end{equation}
In the final form we extend our notation by introducing a total field operator
\begin{equation}
\ket{\uppsi}=\sum_\sigma\ket{\uppsi_\sigma}\ket\sigma\;,
\end{equation}
which is a spinor field operator, including both spatial and internal degrees of freedom.  It gives the hyperfine-level field operators according to $\braket{\sigma}{\uppsi}=\ket{\uppsi_\sigma}$; notice that since $\uppsi_\sigma^\dagger(\mathbf x)=\braket{\uppsi_\sigma}{\mathbf x}=\braket{\uppsi}{\sigma,\mathbf x}$, we also have $\braket{\uppsi}{\sigma}=\bra{\uppsi_\sigma}$.  The spinor representation is
\begin{equation}
\uppsi(\mathbf x)=\braket{\mathbf x}{\uppsi}=\sum_\sigma\uppsi_\sigma(\mathbf x)\ket\sigma\;.
\end{equation}

The annihilation and creations operators that destroy or create a particle in internal level $\sigma$ and spatial wave function $\psi(\mathbf x)$ are
\begin{subequations}
\begin{align}\lb{eq:achi}
\a_{\sigma,\psi}&=\int \psi^*(\mathbf{x})\,\uppsi_\sigma(\mathbf{x})\,\dif\mathbf{x}
=\braket{\psi}{\uppsi_\sigma}=\braket{\sigma,\psi}{\uppsi}\;,\\
\a_{\sigma,\psi}^\dagger&=\int \psi(\mathbf{x})\,\uppsi_\sigma^\dagger(\mathbf{x})\,\dif\mathbf{x}
=\braket{\uppsi_\sigma}{\psi}=\braket{\uppsi}{\sigma,\psi}\;.
\end{align}
\end{subequations}
The annihilation operators for the states $\ket\psinot$ and $\ket{\bar\psinot}$ are thus
\begin{gather}\hspace{-2em}
\a_{\psinot}=\braket{\psinot}{\uppsi}
=\frac{1}{\alpha^*}\,\Big(\alpha_1^* \braket{\psinot_1}{\uppsi_1}+\alpha_2^*\braket{\psinot_2}{\uppsi_2}\Big)
=\frac{1}{\alpha^*}\,\big(\alpha_1^*\ssp\a_{1,\psinot_1}+\alpha_2^*\ssp \a_{2,\psinot_2}\big)\;,\\[4pt] \hspace{-2em}
\bar{\a}_{\psinot}=\braket{\bar\psinot}{\uppsi}
=\frac{1}{\alpha}\,\Big(\alpha_2\braket{\psinot_1}{\uppsi_1}-\alpha_1\braket{\psinot_2}{\uppsi_2}\Big)
=\frac{1}{\alpha}\, \big(\alpha_2\ssp  \a_{1,\psinot_1}-\alpha_1\ssp \a_{2,\psinot_2}\big)\;,
\end{gather}
where $\a_{\sigma,\psinot_\sigma(t)}$ is the annihilation operator that destroys a particle in hyperfine state~$\sigma$ with spatial wavefunction $\psinot_\sigma(\mathbf x,t)$.  The field operator for the atoms in hyperfine level~$\sigma$ can be written as
\begin{align}
\uppsi_\sigma(\mathbf x)&=a_{\sigma,\psinot_\sigma(t)}\psinot_\sigma(\mathbf x,t)+\uppsi_{\sigma\perp}(\mathbf x,t)\;,
\end{align}
where $\uppsi_{\sigma\perp}(\mathbf x,t)$ excludes the mode with wave function $\psinot_\sigma(\mathbf x,t)$.  The total field operator can be written in a variety of forms,
\begin{subequations}
\begin{align}
\ket{\uppsi}
&=\sum_\sigma\a_{\sigma,\psinot_\sigma(t)}\ket{\psinot_\sigma(t)}\otimes\ket\sigma+\ket{\uppsi_\doubleperp(t)}\\
&=\a_{\psinot(t)}\ket{\psinot(t)}+\bar\a_{\psinot(t)}\ket{\bar\psinot(t)}+\ket{\uppsi_\doubleperp(t)}\lb{eq:uppsidoubleperp}\\[7pt]
&=\a_{\psinot(t)}\ket{\psinot(t)}+\ket{\uppsi_\perp(t)}\;,\lb{eq:uppsiperp2}
\end{align}
\end{subequations}
where
\begin{subequations}
\begin{align}
\ket{\uppsi_\doubleperp(t)}&=\sum_\sigma\ket{\uppsi_{\sigma\perp}(t)}\ket\sigma\\
&=\ket{\uppsi}-\a_{\psinot(t)}\ket{\psinot(t)}-\bar\a_{\psinot(t)}\ket{\bar\psinot(t)}\vphantom{\sum_\sigma}\\
&=\ket{\uppsi}-\big(\,\ket{\psinot(t)}\bra{\psinot(t)}+\ket{\bar\psinot(t)}\bra{\bar\psinot(t)}\,\big)\ket\uppsi
\end{align}
\end{subequations}
is the total field operator with modes $\ket\psinot$ and $\ket{\bar\psinot}$ removed and
\begin{equation}\lb{eq:psiperptwo}
\ket{\uppsi_\perp(t)}
=\bar\a_{\psinot(t)}\ket{\bar\psinot(t)}+\ket{\uppsi_\doubleperp(t)}
=\ket\uppsi-\a_{\psinot(t)}\ket{\psinot(t)}
=\ket\uppsi-\ket{\psinot(t)}\braket{\psinot(t)}{\uppsi}
\end{equation}
is the total field operator with only the condensate mode removed.  By using our bra-ket shorthand, all the manipulations for two components can be made identical to that for a single component.

Just as in the single-component case, we perturb about the extended catalytic state for a pure condensate
that is in a coherent state for the condensate mode:
\begin{equation}
\mathcal{D}\big(\alpha,\psinot(t)\big)\ket{\vac}=
\ket{\alpha,\psinot(t)}_0\otimes\ket{\vac}_\perp\;,
\end{equation}
The physical state is obtained by projecting into the $N$-particle sector.

In the two-component case the model Hamiltonian for the $N$ atoms is
\begin{equation}\lb{eq:H_two_component}
\begin{split}
\sH(t)
&=\sum_\sigma\int \uppsi^\dagger_\sigma\Big(\mathord{-}
\frac{{\hbar}^2}{2m_\sigma}\boldsymbol{\nabla}^2 +V_\sigma(t)\Big)
\uppsi_\sigma\,\dif \mathbf{x}
+\sum_{\sigma,\tau}\hbar\omega_{\sigma\tau}\!
\int\uppsi^\dagger_\sigma\uppsi_\tau\,\dif \mathbf{x}\\
&\qquad\quad+\frac{1}{2}\,\sum_{\sigma,\tau}g_{\sigma\tau}
\int\uppsi^\dagger_\sigma\uppsi^\dagger_\tau\uppsi_\tau\uppsi_\sigma\,\dif\mathbf{x} \;.
\end{split}
\end{equation}
The diagonal terms of the Hermitian matrix $\hbar\omega_{\sigma\tau}$ give the energies of the internal levels, and the off-diagonal terms give the single-particle coupling between the two levels.  The real, symmetric matrix $g_{\sigma\tau}$ describes the scattering of the atoms in each component off one another and the cross-scattering between components.  The single-particle terms are trivial to treat, so the only new effect here comes from the cross scattering described by $g_{12}$.

The next step is to go to the interaction picture where the condensate mode is displaced to vacuum, just as in Eqs.~(\rf{eq:intpic}).  In this interaction picture, the field operators transform according to
\begin{equation}
\mathcal{D}^\dagger\big( \alpha, \psinot(t) \big)\,\uppsi_\sigma(\mathbf{x})\,
\mathcal{D}\big( \alpha, \psinot(t) \big)
=\alpha_\sigma(t)\, \psinot_\sigma(\mathbf{x},t)+\uppsi_\sigma(\mathbf{x})\;,
\end{equation}
thus allowing an expansion in powers of $1/{N}^{1/2}=1/|\alpha|$.  We can write this transformation more abstractly as
\begin{equation}
\Big|\mathcal{D}^\dagger\big(\alpha,\psinot(t)\big)\,\uppsi\,
\mathcal{D}\big(\alpha,\psinot(t)\big)\Big\rangle
=\alpha\ket{\psinot(t)}+\ket\uppsi\;,
\end{equation}
The interaction-picture Hamiltonian, as in Eq.~(\rf{eq:intpicH}), is given by
\begin{align}
\hspace{-1.5em}\sH_\mathrm{int}(t)=- i \hbar\, \Big(\, \norm{\alpha}^2
\braket{\psinot}{\dt{\psinot}(t)}+\alpha\,
\braket{\uppsi}{\dt{\psinot}(t)}-\,\alpha^*\, \braket{\dt{\psinot}(t)}{\uppsi}\,\Big)+ \mathcal{D}^\dagger(t)\,\sH(t)\,\mathcal{D}(t)\;.
\end{align}
The time derivative of the condensate state is
\begin{equation}
\ket{\dt{\psinot}(t)}
=\frac{1}{\alpha}\left(
\frac{d}{dt}\big(\alpha_1\,\ket{\psinot_1}\big)\otimes\ket1+
\frac{d}{dt}\big(\alpha_2\,\ket{\psinot_2}\big)\otimes\ket2
\right)\;.
\end{equation}
Putting all this together, we get the interaction-picture Hamiltonian to Bogoliubov order, i.e., order $N^0$,
\begin{subequations}\lb{eq:Hbog_two_component}
\begin{align}\hspace{-2em}
\begin{split}
\sH_\mathrm{int}(t)
&=\int\sum_\sigma\alpha_\sigma^*\psinot_\sigma^*
\bigg[\bigg(
\mathord{-}i\hbar \pa{}{t}+H_\sigma+\half\,\sum_\tau g_{\sigma\tau}\norm{\alpha_\tau}^2\,\norm{\psinot_\tau}^2
\bigg)\alpha_\sigma\psinot_\sigma\\
&\hspace{8em}+\sum_\tau\hbar\omega_{\sigma\tau}\alpha_\tau\psinot_\tau
\bigg]\,\dif\mathbf{x}
\end{split}\\
\begin{split}
&\qquad+\int\Bigg(\sum_\sigma\uppsi_\sigma^\dagger
\bigg[\bigg(
\mathord{-}i\hbar \pa{}{t}+H_\sigma+\sum_\tau g_{\sigma\tau}\norm{\alpha_\tau}^2\,\norm{\psinot_\tau}^2
\bigg)\alpha_\sigma\psinot_\sigma\\
&\hspace{8em}+\sum_\tau\hbar\omega_{\sigma\tau}\alpha_\tau\psinot_\tau
\bigg]
+\mathrm{H.c.}\Bigg)\,\dif\mathbf{x}
\end{split}\\
\begin{split}
&\qquad+\int\Bigg(\sum_\sigma\uppsi_\sigma^\dagger
\bigg(H_\sigma+\sum_\tau g_{\sigma\tau}\norm{\alpha_\tau}^2\norm{\psinot_\tau}^2\bigg)\uppsi_\sigma\\ &\hspace{3em}+\sum_{\sigma,\tau}\uppsi_\sigma^\dagger
\Big(\hbar\omega_{\sigma\tau}+g_{\sigma\tau}\alpha_\sigma\psinot_\sigma\alpha_\tau^*\psinot_\tau^*\Big)
\uppsi_\tau\\
&\hspace{3em}
+\half\sum_{\sigma,\tau}\Bigl(\uppsi_\sigma^\dagger\uppsi_\tau^\dagger
g_{\sigma\tau}\alpha_\sigma\psinot_\sigma\alpha_\tau\psinot_\tau
+\mathrm{H.c.}\Big)\Bigg)
\dif \mathbf{x}\;,
\end{split}
\end{align}
\end{subequations}
where the single-body translational Hamiltonians are
\begin{equation}
H_\sigma=\mathord{-}\frac{{\hbar}^2}{2m_\sigma}\boldsymbol{\nabla}^2 +V_\sigma\;.
\end{equation}

Just as for the case of a single component, we can neglect the $c$-number, mean-field-energy term.  By requiring the term of order $N^{1/2}=\norm{\alpha}$ to vanish, we get a pair of coupled GP equations,\si{Gross-Pitaevskii equation}
\begin{equation}
\bigg(
\mathord{-}i\hbar\pa{}{t}+H_\sigma+\sum_\tau g_{\sigma\tau}\norm{\alpha_\tau}^2\,\norm{\psinot_\tau}^2
\bigg)
\alpha_\sigma\psinot_\sigma
+\sum_\tau\hbar\omega_{\sigma\tau}\alpha_\tau\psinot_\tau
=0\;.
\end{equation}
Notice that these are best thought of as coupled equations for the unnormalized wave functions, $\alpha_1\psinot_1$ and $\alpha_2\psinot_2$.  It is often convenient to have the two GP equations written out separately as
\begin{gather}\lb{eq:GP_two_component}
\Big(\mathord{-}i \hbar \pa{}{t}+H_{\mathrm{gp}}^{(1)}\,\Big)\,\alpha_1 \psinot_1+\hbar\omega_{12}\,\alpha_2\psinot_2=0\;,\\[6pt]
\Big(\mathord{-}i \hbar \pa{}{t}+H_{\mathrm{gp}}^{(2)}\,\Big)\,\alpha_2 \psinot_2+\hbar\omega_{21}\,\alpha_1\psinot_1=0\;,
\end{gather}
where the GP Hamiltonians are
\begin{gather}
H_\mathrm{gp}^{(1)}
=H_1+\hbar\omega_{11}+g_{1\nsp 1}\norm{\alpha_1}^2 \norm{\psinot_1}^2
+g_{1\nsp 2}\norm{\alpha_2}^2\norm{\psinot_2}^2\;,\\[6pt]
H_\mathrm{gp}^{(2)}
=H_2+\hbar\omega_{22}+g_{22}\norm{\alpha_2}^2\norm{\psinot_2}^2
+g_{2\nsp 1}\norm{\alpha_1}^2\norm{\psinot_1}^2
\end{gather}
(remember that $\omega_{21}=\omega_{12}^*$ and $g_{21}=g_{12}$).  It is also convenient to compactify the equations in terms of spinors relative to the two hyperfine levels so that we can take advantage of our bra-ket notation,
\begin{equation}\lb{eq:spinorGP}
\Big(\mathord{-}i\hbar\pa{}{t}+H_{\mathrm{gp}}\,\Big)
\begin{pmatrix}
\alpha_1\psinot_1\\\alpha_2\psinot_2
\end{pmatrix}
=0\;,
\end{equation}
where
\begin{equation}\lb{eq:Hgpmatrix}
H_\mathrm{gp}=\begin{pmatrix}
H_\mathrm{gp}^{(1)} &
\hbar\omega_{12}\\[3pt]
\hbar\omega_{21} &
H_\mathrm{gp}^{(2)}
\end{pmatrix}
=H_\mathrm{gp}^{(1)}\proj{1}+H_\mathrm{gp}^{(2)}\proj{2}+
\hbar\omega_{12}\ket1\bra2+\hbar\omega_{21}\ket2\bra1\;.
\end{equation}
Recognizing that the spinor in Eq.~(\rf{eq:spinorGP}) is the spinor representation of the state $\alpha\ket\psinot$,
we can write the coupled GP equations in the very compact form
\begin{equation}\lb{eq:GP2}
\Big(\mathord{-}i\hbar\pa{}{t}+H_{\mathrm{gp}}\,\Big)\ket\psinot=0\;,
\end{equation}
where it is assumed, as our formalism requires, that $\alpha$ does not change in time.

The Bogoliubov Hamiltonian governing the dynamics in the interaction picture is given by Eq.~(\rf{eq:Hbog_two_component}).  In $4\times4$ matrix form, we have
\begin{equation}
\sH_\mathrm{bog}= \half\,
\COLON\,
\begin{pmatrix}
\,\bra{\uppsi_1} & \bra{\uppsi_2} & \bra{\uppsi_1^\dagger} & \bra{\uppsi_2^\dagger}\,
\end{pmatrix}
\mathsp{H}_\mathrm{bog}
\begin{pmatrix}
\ket{\uppsi_1} \\[4pt] \ket{\uppsi_2} \\[4pt]\, \ket{\uppsi_1^\dagger}\\[4pt] \ket{\uppsi_2^\dagger}
\end{pmatrix}
\,\COLON\;,
\end{equation}
where the symplectic-style matrix $\mathsp{H}_\mathrm{bog}$ takes the form\\[-12pt]
\begin{equation}
\lb{bog_matrixform_two_component}
\hspace{-0.7em}
\scalebox{0.85}{\mbox{$\left(\begin{array}{cc|cc}
H_\mathrm{gp}^{(1)}+g_{1\nsp 1} \norm{\alpha_1}^2\norm{\psinot_1}^2
&\hbar\omega_{12}+g_{1\nsp2}\alpha_1 \alpha_2^*\, \psinot_1 \psinot_2^*
&g_{1\nsp 1}\alpha_1^2\, \psinot_1^2
&g_{1\nsp 2}\alpha_1 \alpha_2\,\psinot_1 \psinot_2\\[3pt]
\hbar\omega_{21}+g_{2\nsp 1}\alpha_1^*\alpha_2\,\psinot_1^*\psinot_2
&H_\mathrm{gp}^{(2)}+g_{22}\norm{\alpha_2}^2 \norm{\psinot_2}^2
&g_{2\nsp1}\alpha_1\alpha_2\,\psinot_1\psinot_2
&g_{22}\alpha_2^2\, \psinot_2^2\\[3pt]
\hline&&&\\[-18pt]
g_{1\nsp1} \big(\alpha_1^*)^2 \big(\psinot_1^*\big)^2
&g_{2\nsp1}\alpha_1^*\alpha_2^*\,\psinot_1^*\psinot_2^*
&H_\mathrm{gp}^{(1)}+g_{1\nsp 1} \norm{\alpha_1}^2 \norm{\psinot_1}^2
&\hbar\omega_{21}+g_{2\nsp1}\alpha_1^*\alpha_2\,\psinot_1^*\psinot_2\\[3pt]
g_{1\nsp 2} \alpha_1^*\alpha_2^*\,\psinot_1^*\psinot_2^*
&g_{22}\big(\alpha_2^*)^2 \big(\psinot_2^*\big)^2
&\hbar\omega_{12}+g_{1\nsp 2}\alpha_1 \alpha_2^*\,\psinot_1\psinot_2^* &H_\mathrm{gp}^{(2)}+g_{22}\norm{\alpha_2}^2 \norm{\psinot_2}^2
\end{array}\right)$}}
\;.
\end{equation}
\\[-18pt]
To get back to the compact spinor notation, we introduce, along with the matrix~(\rf{eq:Hgpmatrix}), two other matrices that operate in the spinor space defined by the hyperfine levels $\ket1$ and $\ket2$:
\begin{equation}\lb{eq:definition_matrices_two_component}
\Phi=\frac{1}{\alpha}
\begin{pmatrix}
\alpha_1 \psinot_1&
0\\[3pt]
0 &
\alpha_2 \psinot_2
\end{pmatrix}\;,\qquad
G=\begin{pmatrix}
g_{1\nsp 1} &
g_{1\nsp 2}\\[3pt]
g_{2\nsp 1} &
g_{22}
\end{pmatrix}\;.
\end{equation}
With these matrices, we have
\begin{equation}
\mathsp{H}_\mathrm{bog}=
\begin{pmatrix}
H_\mathrm{gp}+\norm\alpha^2\Phi G \Phi^{\nsp *} &
\alpha^2\Phi G \Phi\\[3pt]
(\alpha^*)^2\Phi^{\nsp *} G \Phi^{\nsp *} &
H_\mathrm{gp}^*+\norm\alpha^2\Phi^{\nsp *} G \Phi
\end{pmatrix}\;.
\end{equation}
Notice that since $\Phi$ is diagonal and $G$ is real and symmetric, $\Phi G\Phi^*$ and $\Phi^*G\Phi$ are both Hermitian, and they are transposes and complex conjugates of one another; $\Phi G\Phi$ and $\Phi^* G\Phi^*$ are both symmetric, and they are complex conjugates and Hermitian conjugates of one another.  Using our total field operator and interpreting the $2\times2$ submatrices as operators in the space of the internal levels, we can write the Bogoliubov Hamiltonian in the suggestive form, identical to that for a single component,
\begin{equation}
\sH_\mathrm{bog}=\half\,
\COLON\,
\begin{pmatrix}
\bra{\uppsi} & \bra{\uppsi^\dagger}
\end{pmatrix}
\mathsp{H}_\mathrm{bog}
\begin{pmatrix}
\ket{\uppsi}\\[5pt] \ket{\uppsi^\dagger}
\end{pmatrix}
\,\COLON
\;.
\end{equation}

To eliminate phase diffusion in the condensate mode, we now introduce the auxiliary (nonHermitian) Hamiltonian $\sF$ in exactly the same form it has in the single-component case [see Eq.~(\rf{eq:F})],
\begin{equation}\lb{eq:F_two_component}
\sF(t)=-\frac{\eta(t)}{2}\, (\mathcal{N}-N)^2
+ \big(\alpha\ssp \a^\dagger_{\smash{\psinot(t)}}+\mathcal{N}_{\perp}(t)-N\big)\mathcal{F}_\perp
+ \big(\alpha^* \a_{\psinot(t)}-N\,\big) \mathcal{F}_\perp^\dagger\;,
\end{equation}
where $\mathcal{N}_{\perp}=\mathcal{N}-\a_{\smash \psinot}^\dagger \a_{\psinot}$ and where the coefficient $\eta$ and the operator $\sF_\perp$ are defined in analogy to the single-component case,
\begin{subequations}
\begin{align}
\hspace{-1.5em}
\eta(t)
&=\bra{\psinot}\Phi G \Phi^{\nsp*}\ket{\psinot}
=\bra{\psinot^*}\Phi^{\nsp*}G \Phi\ket{\psinot^*}
=\bra{\psinot^*}\Phi^{\nsp*}G \Phi^{\nsp*}\ket{\psinot}
=\bra{\psinot}\Phi G \Phi\ket{\psinot^*}\\[3pt]
&=\frac{1}{\norm{\alpha}^4}
\int
\sum_{\sigma,\tau} g_{\sigma\tau}\norm{\alpha_\sigma}^2\norm{\alpha_\tau}^2 \norm{\psinot_\sigma}^2\norm{\psinot_\tau}^2
\,\dif\mathbf{x}\;,
\end{align}
\end{subequations}
and
\begin{subequations}
\begin{align}
\sF_\perp=\sF_\perp^\dagger
&=-\alpha^*\bra{\psinot}\Phi G\Phi^{\nsp*}\ket{\uppsi_{\perp}}
-\alpha\bra{\uppsi_\perp}\Phi G \Phi^{\nsp*}\ket{\psinot}\\
&=\eta\alpha^*\a_\psinot+\eta\alpha\a^\dagger_\psinot
-\alpha^*\bra{\psinot}\Phi G\Phi^{\nsp*}\ket{\uppsi}
-\alpha\bra{\uppsi}\Phi G \Phi^{\nsp*}\ket{\psinot}
\\
&=-\frac{1}{\norm\alpha^2}
\int\bigg(
\sum_{\sigma,\tau}g_{\sigma\tau}\norm{\alpha_\sigma}^2\norm{\psinot_\sigma}^2
\alpha_\tau^*\psinot_\tau^*\uppsi_{\tau\perp}
+\mathrm{H.c.}\bigg)\,\dif\mathbf x
\;,
\end{align}
\end{subequations}
where $\ket{\uppsi_\perp}$ is the total field operator with the condensate mode excluded [see Eq.~(\rf{eq:psiperptwo})].  Just as in the single-component case, $\eta$ is of order $1/N$ and $\sF_\perp$ is of order $1/N^{1/2}$.

The transition to the interaction picture goes exactly as in the single-component case, yielding Eq.~(\rf{eq:FintBog}) at Bogoliubov order $N^0$.  Dropping the $c$-number term from the result, we have
\begin{subequations}
\begin{align}
\sF_\mathrm{int}(t)
&=\mathord{-}\frac{\eta}{2}\,\Big(\,2\norm{\alpha}^2\,\a_{\smash \psinot}^\dagger \a_{\psinot}+\alpha^2  \a_{\smash \psinot}^\dagger \a_{\smash \psinot}^\dagger
+(\alpha^*)^2\a_\psinot\a_\psinot\Big)
+\alpha\a_{\smash \psinot}^\dagger\,\mathcal{F}_\perp+\alpha^*\a_{\psinot}\,\mathcal{F}_\perp^\dagger\\[3pt]
\begin{split}
&=\frac{\eta}{2}\,\Big(\,
2\norm{\alpha}^2\,\a_{\smash \psinot}^\dagger \a_{\psinot}+\alpha^2 \a_{\smash \psinot}^\dagger \a_{\smash \psinot}^\dagger
+(\alpha^*)^2\a_\psinot\a_\psinot\Big)\\
&\qquad+\Big(
\big(|\alpha|^2\a_{\smash \psinot}^\dagger+(\alpha^*)^2\a_\psinot\big)
\bra{\phi}\Phi G\Phi^*\ket{\uppsi}+\mathrm{H.c.}
\Big)
\end{split}\\[3pt]
\begin{split}
&=\frac{1}{2}\,\Big(\,
\norm{\alpha}^2\,\braket{\uppsi}{\psinot}
\bra{\psinot}\Phi G\Phi^{\nsp*}\ket\psinot\braket{\psinot}{\uppsi}\\
&\qquad\quad+(\alpha^*)^2\,\braket{\uppsi^\dagger}{\psinot^*}
\bra{\psinot^*}\Phi^{\nsp*}G\Phi^{\nsp*}\ket\psinot\braket{\psinot}{\uppsi}
+\mathrm{H.c.}\Big)\\
&\quad-\Big(
\norm{\alpha}^2\,\braket{\uppsi}{\psinot}\bra{\psinot}\Phi G\Phi^{\nsp*}\ket\uppsi\\
&\qquad\quad+(\alpha^*)^2\,\braket{\uppsi^\dagger}{\psinot^*}\bra{\psinot^*}\Phi^{\nsp*}G\Phi^{\nsp*}\ket\uppsi
+\mathrm{H.c.}\Big)\;.
\end{split}
\end{align}
\end{subequations}

In symplectic form, we have
\begin{equation}
\sF_\mathrm{int}(t)
=\frac{1}{2}
\,\COLON
\begin{pmatrix}
\bra{\uppsi} & \bra{\uppsi^\dagger}
\end{pmatrix}
\mathsp{F}_{\mathrm{int}}(t)
\begin{pmatrix}
\ket{\uppsi}\\[5pt] \ket{\uppsi^\dagger}
\end{pmatrix}\COLON\;.
\end{equation}
Here the matrix of symplectic structure is given by
\begin{align}
\mathsp{F}_{\mathrm{int}}(t)=
\begin{pmatrix}
|\alpha|^2\big(Q\Phi G\Phi^{\nsp*}Q-\Phi G\Phi^{\nsp*}\big)
&\alpha^2\big(Q\Phi G\Phi Q^*-\Phi G\Phi\big)\\[6pt]
(\alpha^*)^2\big(Q^*\Phi^{\nsp*}G\Phi^{\nsp*} Q-\Phi^{\nsp*}G\Phi^{\nsp*}\big)
&|\alpha|^2\big(Q^*\Phi^{\nsp*}G\Phi Q^*-\Phi^{\nsp*}G\Phi\big)
\end{pmatrix}\;,
\end{align}
where $Q(t)$ is the projector onto the single-particle subspace orthogonal to the condensate mode, as in Eq.~(\rf{eq:Q}).  The modified (number-conserving) Bogoliubov Hamiltonian matrix assumes the form\si{Number-conserving Bogoliubov}
\begin{equation}\lb{eq:Hbogtwofinal}
\sH_\mathrm{ncb}=\half\,
\COLON\,
\begin{pmatrix}
\bra{\uppsi} & \bra{\uppsi^\dagger}
\end{pmatrix}
\mathsp{H}_\mathrm{ncb}
\begin{pmatrix}
\ket{\uppsi}\\[5pt] \ket{\uppsi^\dagger}
\end{pmatrix}
\,\COLON
\;,
\end{equation}
with
\begin{equation}\lb{eq:number_conserving_bogliubov_hamiltonian_matrix}
\mathsp{H}_\mathrm{ncb}
=\mathsp{H}_{\mathrm{bog}}+\mathsp{F}_{\mathrm{int}}
=\begin{pmatrix}
H_\mathrm{gp}+\norm\alpha^2Q\ssp \Phi G \Phi^{\nsp *} Q &
\alpha^2Q\ssp \Phi G \Phi\ssp Q^*\\[3pt]
(\alpha^*)^2Q^* \Phi^{\nsp *} G \Phi^{\nsp *} Q &
H_\mathrm{gp}^*+\norm\alpha^2 Q^*\Phi^{\nsp *} G \Phi\ssp Q^*
\end{pmatrix}\;.
\end{equation}

The related demonstrations that at Bogoliubov order, if the condensate mode begins in a coherent state, it remains in a coherent state and that the auxiliary Hamiltonian $\sF(t)$ does not change the evolution in the $N$-particle sector can be repeated word for word from the single-component case.  There is, however, an important difference from the single-component case, which involves the orthogonal mode $\ket{\bar\psinot}$ of Eq.~(\rf{eq:barpsinot}).  We get at this difference by first plugging into the number-conserving Bogoliubov Hamiltonian~(\rf{eq:Hbogtwofinal}) the field-operator decomposition~(\rf{eq:uppsiperp2}); this separates out the condensate mode $\ket\psinot$ and brings $\sH_\mathrm{ncb}$ into the form~(\rf{eq:Hbogcond1}).  By further separating out the mode $\ket{\bar\psinot}$ using the field-operator decomposition~(\rf{eq:uppsidoubleperp}), we arrive at
\begin{align}\hspace{-2.5em}
\begin{split}
\sH_\mathrm{ncb}
&=\a_{\smash\psinot}^\dagger\a_\psinot\bra{\psinot}H_{\mathrm{gp}}\ket\psinot
+\Big(\a_{\smash\psinot}^\dagger \bar\a_\psinot\bra{\psinot}H_{\mathrm{gp}} \ket{\bar\psinot}
+\a_{\smash \psinot}^\dagger\bra\psinot H_{\mathrm{gp}}\ket{\uppsi_\doubleperp}+\mathrm{H.c.}\Big)\\[3pt]
&\hspace{1.3em}
+\bar\a_{\smash\psinot}^\dagger \bar\a_\psinot\ssp
\bra{\bar\psinot}(H_{\mathrm{gp}}+\norm{\alpha}^2\ssp \Phi G \Phi^{\nsp *})\ket{\bar\psinot}
+\Big(\bar\a_{\smash \psinot}^\dagger\bra{\bar\psinot}
(H_{\mathrm{gp}}+\norm{\alpha}^2\ssp \Phi G \Phi^{\nsp*})\ket{\uppsi_\doubleperp}+\mathrm{H.c.}\Big)\\[3pt]
&\hspace{1.3em}
+\half\Big(\alpha^2\ssp \bar\a_{\smash\psinot}^\dagger \bar\a_{\smash\psinot}^\dagger\ssp\bra{\bar\psinot}\Phi G \Phi\ket{\bar\psinot^*}
+\alpha^2\ssp\bar\a_{\smash\psinot}^\dagger \bra{\bar\psinot\nsp} \Phi G \Phi \ket{\uppsi_\doubleperp^\dagger}+\mathrm{H.c.}\Big)\\
&\hspace{1.3em}+\sH_{\mathrm{ncb}\ssp\doubleperp}\;,
\end{split}
\lb{eq:Hbogcond_two_component}
\end{align}
where
\begin{equation}
\sH_{\mathrm{ncb}\ssp\doubleperp}
=\half\,\COLON
\begin{pmatrix}
\bra{\uppsi_\doubleperp} & \bra{\uppsi_\doubleperp^\dagger}
\end{pmatrix}\!
\mathsp{H}_{\mathrm{bog}}
\begin{pmatrix}
\ket{\uppsi_\doubleperp}\\[5pt] \ket{\uppsi_\doubleperp^\dagger}
\end{pmatrix}\COLON\;.
\end{equation}
By the same calculation as for a single component, this Hamiltonian conserves $\a_\psinot$.

We want to study the evolution of the mode $\ket{\bar\psinot}$; for that purpose, we need only keep the parts of the Hamiltonian~(\rf{eq:Hbogcond_two_component}) that depend on $\bar\a_{\psinot}$ and $\bar\a_{\smash\psinot}^\dagger$.  Re-organizing the terms a bit, we have
\begin{subequations}
\begin{align}\hspace{-2.5em}
\overbar\sH_\mathrm{ncb}
&=\norm{\alpha}^2\bar\a_{\smash\psinot}^\dagger \bar\a_\psinot\ssp
\bra{\bar\psinot}\ssp \Phi G \Phi^{\nsp *}\ket{\bar\psinot}
+\half\Big(\alpha^2\ssp \bar\a_{\smash\psinot}^\dagger \bar\a_{\smash\psinot}^\dagger\ssp\bra{\bar\psinot}\Phi G \Phi\ket{\bar\psinot^*}
+\mathrm{H.c.}\Big)\lb{eq:spin_squeeze_term}\\[3pt]
&\hspace{1.3em}
+\Big(\bar\a_{\smash\psinot}^\dagger \a_\psinot\bra{\bar\psinot}H_{\mathrm{gp}} \ket{\uppsi}+\mathrm{H.c.}\Big)\lb{eq:second_mode_bogoliubov_gp}\\[3pt]
&\hspace{1.3em}
+\Big(\norm{\alpha}^2\bar\a_{\smash \psinot}^\dagger\bra{\bar\psinot}
\ssp \Phi G \Phi^{\nsp*}\ket{\uppsi_\doubleperp}+\half\,\alpha^2\ssp\bar\a_{\smash\psinot}^\dagger \bra{\bar\psinot\nsp} \Phi G \Phi \ket{\uppsi_\doubleperp^\dagger}+\mathrm{H.c.}\Big)\;,\lb{eq:second_component_other_modes}
\end{align}
\end{subequations}
where the term~(\rf{eq:second_mode_bogoliubov_gp}) evolves the state $\ket{\bar\psinot}$ according to the GP equation, while the term~(\rf{eq:second_component_other_modes}) expresses the coupling to modes orthogonal to both $\ket\psinot$ and $\ket{\bar\psinot}$.  The remaining  term~(\rf{eq:spin_squeeze_term}) takes the form
\begin{subequations}\lb{eq:spin_squeeze}
\begin{align}\lb{eq:spin_squeezea}
\sH_\mathrm{ss}
&= \frac{\bar\eta(t)}{2}\ssp
\Big(\ssp 2\ssp\norm{\alpha}^2\ssp \bar\a_{\smash\psinot}^\dagger\ssp\bar\a_{\psinot}
+e^{2\ssp i\ssp\theta}\ssp \alpha^2\ssp \bar\a_{\smash\psinot}^\dagger\ssp \bar\a_{\smash\psinot}^\dagger
+e^{-2\ssp i\ssp\theta}\ssp (\alpha^*)^2\ssp \bar\a_{\psinot}\ssp  \bar\a_{\psinot}\ssp\Big)\\[3pt]
&= \frac{\bar\eta}{2}\, \big(\ssp e^{ i\ssp\theta}\ssp\alpha\ssp  \bar\a_{\smash\psinot}^\dagger + e^{- i\ssp\theta}\ssp\alpha^*\ssp  \bar\a_{\psinot}\ssp\big)^2- \frac{\bar\eta\ssp \norm{\alpha}^2}{2}\;,\lb{eq:spin_squeezeb}
\end{align}
\end{subequations}
where
\begin{subequations}
\begin{align}
\bar\eta(t)
&=\bra{\bar\psinot}\Phi G \Phi^{\nsp *}\ket{\bar\psinot}
=e^{-2i\theta}\,\bra{\bar\psinot}\Phi G \Phi\ket{\bar\psinot^*}
=e^{2i\theta}\,\bra{\bar\psinot^*}\Phi^* G \Phi^*\ket{\bar\psinot}\\
&=\frac{\norm{\alpha_1}^2\ssp \norm{\alpha_2}^2}{\norm{\alpha}^4}
\int \Big( g_{1\nsp 1}\ssp  \norm{\psinot_1}^4+g_{22}\ssp  \norm{\psinot_2}^4-2\ssp g_{1 2}\ssp  \norm{\psinot_1}^2\ssp \norm{\psinot_2}^2\Big)\,\dif \mathbf x\;,
\end{align}
\end{subequations}
with $\theta = \arg(\alpha_1\alpha_2/\alpha^2)$.

Presuming that $\alpha_1$ and $\alpha_2$ do not change in time---and, hence, that $\theta$ is also constant in time---we can choose $\alpha\,e^{i\theta}=\beta$ to be real and positive, thus putting the Hamiltonian in the form (after discarding the irrelevant $c$-number that comes from operator ordering)
\begin{equation}\lb{eq:Hss}
\sH_\mathrm{ss}
=\frac{\bar\eta\ssp\beta^2}{2}\,
\big(\ssp\bar\a_{\smash\psinot}^\dagger + \bar\a_{\psinot}\ssp\big)^2
=\bar\eta\ssp\beta^2\bar x_{\smash\psinot}^2\;,
\end{equation}
where $\bar x_\psinot=(\bar\a_{\smash\psinot}^\dagger + \bar\a_{\psinot})/\sqrt2$ is the position quadrature corresponding to $\bar\a_\psinot$.  The Hamiltonian~(\rf{eq:Hss}) produces shearing and squeezing in the direction of the momentum quadrature at a variable rate given by $2\bar\eta(t)\beta^2$.

Another way to think about the Hamiltonian~(\rf{eq:spin_squeeze}) is to recall that it arises, in the Bogoliubov approximation, from replacing $\a_\psinot$ and $\a_{\smash\psinot}^\dagger$ by $\alpha$ and $\alpha^{\nsp*}$ in the original Schr\"odinger-picture Hamiltonian.  Restoring, in normal order, the creation and annihilation operators for the condensate mode to the Hamiltonian~(\rf{eq:spin_squeezea}) gives a Kerr-like interaction
\begin{subequations}\lb{eq:Kerr}
\begin{align}\lb{eq:Kerra}
\sH_\mathrm{ss}
&=\frac{\bar\eta}{2}\,
\Big(\ssp 2\ssp\a_{\smash\psinot}^\dagger\bar\a_{\smash\psinot}^\dagger\ssp\a_\psinot\ssp\bar\a_{\psinot}
+e^{2\ssp i\ssp\theta}\ssp\bar\a_{\smash\psinot}^\dagger\ssp\bar\a_{\smash\psinot}^\dagger\a_\psinot\ssp\a_\psinot
+e^{-2\ssp i\ssp\theta}\ssp\a_{\smash\psinot}^\dagger\ssp\a_{\smash\psinot}^\dagger\bar\a_{\psinot}\ssp  \bar\a_{\psinot}\ssp\Big)\\
&=\frac{\bar\eta}{2}\,
\big(\ssp e^{ i\ssp\theta}\ssp\bar\a_{\smash\psinot}^\dagger\ssp \a_\psinot
+e^{-i\ssp\theta}\ssp\a_{\smash\psinot}^\dagger \ssp \bar\a_\psinot\ssp\big)^2
-\frac{\bar\eta}{2}\big(\a_{\smash\psinot}^\dagger\a_\psinot+\bar\a_{\smash\psinot}^\dagger\bar\a_\psinot\big)\;.
\lb{eq:Kerrb}
\end{align}
\end{subequations}
The first term in Eq.~(\rf{eq:Kerra}) comes from scattering of $\psinot$- and $\bar\psinot$-particles off one another, the second term from scattering of two $\psinot$-particles into the $\bar\psinot$-mode, and the last term from scattering of two $\bar\psinot$-particles into the $\psinot$-mode.

As an example of how the Hamiltonian~(\rf{eq:Kerr}) works, consider the situation where the condensate wavefunction $\psinot$ is an equal superposition of the two hyperfine levels, i.e., $\alpha_1=\alpha_2=1/\sqrt2$, $\theta=0$, and
\begin{equation}
\a_\psinot=\frac{1}{\sqrt{2}}\, \big(\ssp \a_1+\a_2\ssp\big)\;, \qquad
\bar\a_\psinot=\frac{1}{\sqrt{2}}\, \big(\ssp \a_1-\a_2\ssp\big)\;,
\end{equation}
where $\a_1$($\a_2$) is a shorthand for $\a_{1,\psinot_1}$($\a_{2,\psinot_2}$).  By introducing the Schwinger operator\si{Schwinger operators}
\begin{align}
\sJ_z\equiv \frac{1}{2}\, \big(\ssp \a_1^\dagger \a_1-\a_2^\dagger \a_2\ssp \big) =\frac{1}{2}\, \big(\ssp \bar\a_{\smash\psinot}^\dagger\ssp \a_\psinot+ \a_{\smash\psinot}^\dagger\ssp \bar\a_\psinot\ssp\big)\;,
\end{align}
we bring the Hamiltonian~(\rf{eq:Kerrb}) into the form
\begin{align}\lb{eq:one_axis_twist}
\sH_\mathrm{ss}= 2\ssp \bar\eta(t)\sJ_z^2
-\frac{\bar\eta}{2}\big(\a_{\smash\psinot}^\dagger\a_\psinot+\bar\a_{\smash\psinot}^\dagger\bar\a_\psinot\big) \;.
\end{align}
The $\sJ_z^2$ term is the so-called one-axis-twisting~\cite{kitagawa_squeezed_1993}\si{One-axis-twisting} Hamiltonian and is widely used to generate spin squeezing\si{Spin squeezing} in BECs~\cite{esteve_squeezing_2008, grond_optimizing_2009, riedel_atom-chip-based_2010, gross_nonlinear_2010}.  The interplay of spatial and spin dynamics is considered in~\cite{sinatra_binary_2000, sorensen_many-particle_2001, li_spin_2009}; Sinatra \emph{et al.}~\cite{sinatra_limit_2011} showed that the amount of squeezing is bounded from above by the initial noncondensed fraction at finite temperature.

\renewcommand{\lb}[1]{\label{pcs_1:#1}}
\renewcommand{\rf}[1]{\ref{pcs_1:#1}}

\chapter{Bosonic \PCS{}s}\label{ch:pcs_1}

\begin{quote}
It would indeed be remarkable if nature fortified herself against
further advances in knowledge behind the analytical difficulties
of the many-body problem.\\[4pt]
-- Max Born\ai{Born, Max}
\end{quote}

\section{Introduction}
\lb{sec:intro}

Quantum many-body problems are notoriously hard.  This is partly because the Hilbert space becomes exponentially large with the number of particles $N$.  As a consequence, one needs an exponentially large number of parameters merely to record an arbitrary state, not to say computing its time evolution.  While exact solutions are often considered intractable, numerous approximate approaches have been proposed.  A common trait of these approaches is to use an ansatz such that the number of parameters either does not depend on $N$ or is proportional to $N$, e.g., the matrix-product state for spin chains~\cite{white_density_1992, white_density-matrix_1993, verstraete_density_2004, daley_time-dependent_2004}, the BCS wave function for superconductivity~\cite{cooper_bound_1956, bardeen_microscopic_1957, bardeen_theory_1957}, the Laughlin wavefunction for the fractional quantum Hall effects~\cite{laughlin_anomalous_1983}, and the Gross-Pitaevskii theory for BECs~\cite{gross_structure_1961, pitaevsk_vortex_1961}.

The Gross-Pitaevskii theory, which uses the product ansatz, has precisely predicted many useful properties of Bose gases at ultra-low temperature. As particle-particle correlations become important, however, it begins to fail. To capture the quantum correlations, we propose a new set of states, which constitute a natural generalization of the product-state ansatz,\si{Bosonic particle-correlated state}
\begin{align}
 &\rho\ssp\big(\xbf^{(l)}_1,\xbf^{(l)}_2,\ldots,\xbf^{(l)}_n\, ;\, \ybf^{(l)}_1,\ybf^{(l)}_2,\ldots,\ybf^{(l)}_n \big)\nonumber\\[4pt]
 &\quad= \frac{ \sP_S\,  \sigma (\xbf^{(l)}_1, \,\ybf^{(l)}_1)\otimes \sigma(\xbf^{(l)}_2, \,\ybf^{(l)}_2)\otimes \cdots \otimes \sigma(\xbf^{(l)}_n, \,\ybf^{(l)}_n )\, \sP_S}{\tr\Big(\sP_S\,  \sigma(\xbf^{(l)}_1,  \,\ybf^{(l)}_1)\otimes \sigma(\xbf^{(l)}_2, \,\ybf^{(l)}_2)\otimes \cdots \otimes \sigma(\xbf^{(l)}_n, \,\ybf^{(l)}_n)\, \sP_S\Big)}\;,\lb{eq:pcs}
\end{align}
where $\xbf^{(l)}_{j} = \big(\xbf_{j,1},\,\xbf_{j,2},\ldots, \xbf_{j,l}\big)$ and $\ybf^{(l)}_{j} = \big(\ybf_{j,1},\,\ybf_{j,2},\ldots, \ybf_{j,l}\big)$ denotes the coordinates of $l$ particles, $\sigma$ is an arbitrary state (density matrix) of the $l$ particles,\footnote{The $l$-particle states $\sigma$ can be restricted to symmetrized states, or they can be left arbitrary, with $\sP_S$ taking care of the symmetrization when the BPCS state is constructed.} and $\sP_S$ is the projection operator onto the symmetric subspace of all the $N=l\times n$ particles,\si{Symmetric subspace}
\begin{align}
 \sP_S\, \ketb{\psi_1,\psi_2,\ldots,\psi_N} = \frac{1}{N!}\sum_{\pi\in S_N}\, \ketb{\psi_{\pi(1)},\psi_{\pi(2)},\ldots,\psi_{\pi(N)}}\;,
\end{align}
where the sum is over the permutations $\pi$ in the symmetric group $S_N$.  The state~(\rf{eq:pcs}) is derived by symmetrizing the $n$-fold tensor product of the $l$-particle state $\sigma$; we call the resulting state a Bosonic \PCS{}~(BPCS).\footnote{For the fermion case, we can simply substitute the anti-symmetrizing operator $\sP_A$ for the symmetrizing operator $\sP_S$.  The resulting state can be called a Fermionic~\PCS{} (FPCS).}  As a consequence of symmetrization, the quantum correlations existing in the $l$-boson state $\sigma$ ``spread out'' to any subset of the $l\times n$ bosons. The parameter space of the BPCS does not grow with $n$; it equals to that of the bosonic states for $l$ particles.

In this dissertation, I pay most attention to the special case that $\sigma$ is pure and $l=2$.  When $l=2$, \mbox{PCS} can also be read as Pair-Correlated State\si{Pair-correlated state}.

In this chapter, I show that BPCS is a many-body ansatz which is both quantumly correlated and computationally efficient. One advantage of BPCS is that it can represent quantum states with or without Off-Diagonal Long Range Order (ODLRO)~\cite{yang_concept_1962}\si{Off-diagonal long range order}. For example, both the superconducting and the Mott insulating phases in the Bose-Hubbard model can be described by the BPCS ansatz. An interesting question is whether the BPCS ansatz can faithfully interpolate the two phases. In chapter~\chref{ch:pcs_1}, we will show that the answer is yes even if we restrict ourselves to the case $l=2$; there we compare both the ground state and the dynamics predicted by BPCS to fully numerical results.

\section{Historical Remarks}

A central topic in condensed matter physics is to study the particle-particle correlations existing in many-body wavefunctions.  Here I briefly review some existing approaches that are used to capture the particle correlations in ultracold bosonic gases, with an emphasis on the relations to the BPCS ansatz.

One of the first and most influential approaches to BECs that take interparticle correlations into consideration is the Bogoliubov approximation~\cite{bogoliubov_theory_1947, fetter_nonuniform_1972, huang_statistical_1987}\si{Bogoliubov approximation}. Although the Bogoliubov approach has made many precise predictions, it is a perturbative method; i.e., the depletion of the condensate must be small for it to work. I will argue that the nonperturbative BPCS recovers the number-conserving Bogoliubov approximation in the limit of small depletion.

Inspired by the BCS wavefunction proposed by Bardeen, Cooper, and Schrieffer for superconductivity~\cite{bardeen_microscopic_1957, bardeen_theory_1957}, Valatin and Butler~\cite{valatin_collective_1958}\ai{Valatin, J. G.}\ai{Butler, D.}\si{Valatin-Butler wavefunction} introduced a similar pairing wavefunction for bosons,
\begin{align}\lb{eq:valatin_butler_wavefunction}
\ket{\varPsi_\mathrm{vb}} = \frac{1}{\sqrt \NFB}\, \exp\Big(\, \lambda_0\ssp \a_0^\dagger\ssp \a_0^\dagger + 2\sum_k \lambda_k\ssp \a^\dagger_k \a^\dagger_{-k}\Big)\,\ket{\vac}\;,
\end{align}
where $\NFB$ is a normalization factor. This state, with a quadratic form of the creation operators in the exponential, is very different from a coherent state.  The coherent state has a Poissonian number distribution which is peaked around $N = \norm{\alpha}^2$, whereas the Valatin-Butler state~(\rf{eq:valatin_butler_wavefunction}) satisfies an exponential number distribution. Consequently, it might not be suitable to describe situations where the total number of particles is conserved. One easy way to see the exponential number distribution is by setting $\lambda_j = 0$ for all $j\neq 0$, which gives
\begin{subequations}
\begin{align}
 \ket{\varPsi_\mathrm{vb}} &= \frac{1}{\sqrt \NFB}\, \exp\big(\lambda_0\ssp \a_0^\dagger\ssp \a_0^\dagger \big)\,\ket{\vac}\\
 &= \frac{1}{\sqrt \NFB}\, \sum_{n=0}^\infty \frac{\lambda_0^n}{n!}\, \big(\a_0^\dagger\big)^{2n}\,\ket{\vac}\\
 &= \frac{1}{\sqrt \NFB}\, \sum_{n=0}^\infty \frac{\lambda_0^n}{n!}\,\sqrt{(2n)!}\;\ket{2n}_0\otimes \ket{\vac}_\perp\lb{eq:number_state_representation_c}\\
 &\sim \frac{1}{\sqrt \NFB}\,\sum_{n=0}^\infty \frac{\;(2 \lambda_0)^n}{\sqrt[4]{\pi n}}\:\ket{2n}_0\otimes \ket{\vac}_\perp\;,\lb{eq:number_state_representation_d}
\end{align}
\end{subequations}
where we use Stirling's formula in the last step. It turns out this exponential distribution is more general; it is valid even when more than one $\lambda_j$ is nonzero.\footnote{In the large $n$ limit, the normalization factor of the $2n$-particle sector scales as $\big(2^n n!\, \lambda^n\big)^2$ [see Eq.~(\rf{eq:nfb_upsilon})]. By replacing $\sqrt{(2n)!}$ in Eq.~(\rf{eq:number_state_representation_c}) with $2^n n!$, we have the same exponential distribution as in Eq.~(\rf{eq:number_state_representation_d}).} Unlike the pairing wavefunction for fermions, which is always normalizable, the Valatin-Butler wavefunction~(\rf{eq:valatin_butler_wavefunction}) can only be normalized when $\norm{\lambda_j}< 1/2$ for all $j\in \{0,1,\ldots\}$. The Valatin-Butler wavefunction has been used to investigate the transition from a single condensate to a multicondensate~\cite{coniglio_condensation_1967, evans_pairing_1969, nozieres_particle_1982}.

A number-conserving version of the Valatin-Butler wavefunction was introduced by Leggett~\cite{leggett_bose-einstein_2001, leggett_relation_2003}\ai{Leggett, Anthony J.},
\begin{align}\lb{eq:leggett_wavefunction}
\ket{\varPsi_\mathrm{legg}} = \frac{1}{\sqrt{N!}}\, \Big(\,\a_0^\dagger\ssp \a_0^\dagger + 2\sum_k \lambda_k \a^\dagger_k \a^\dagger_{-k}\Big)^{N/2}\,\ket{\vac}\;,
\end{align}
where the normalization factor is not exact. This state is a special case of the BPCS with $\sigma$ pure and $l=2$, but a systematic treatment seems to be lacking in the literature. I will discuss the single-particle and two-particle reduced density matrices of this $N$-particle state in the following sections; a generalized Gross-Pitaevskii equation is derived using this ansatz in the next chapter. It is worth mentioning that the ground state of a spin-1 Bose gas with an antiferromagnetic interaction takes a form similar to Eq.~(\rf{eq:leggett_wavefunction})~\cite{ho_fragmented_2000}\si{Spin-1 Bose gas}. The fermion version of this state was also used by Leggett \emph{et al.} to treat the BEC-BCS crossover\si{BEC-BCS crossover} problem~\cite{leggett_diatomic_1980,leggett_becbcs_2012} and by others~\cite{combescot_n-exciton_2003, law_quantum_2005, combescot_many-body_2008, tichy_how_2013} to discuss the composite boson problem\si{Composite boson problem}.

\section{The Second-Quantized Picture}\si{Second-quantized picture}

It is difficult to do any calculation with the form~(\rf{eq:pcs}), because of the need to do an explicit symmetrization. By going to the second-quantized picture, symmetrization is done automatically.  For a pure PCS with $l=2$, the PCS is specified by a two-boson wavefunction $\varPsi^{(2)}(\xbf_1,\, \xbf_2)$, and the PCS is given by
\begin{equation}\lb{eq:pcs_rank2_pure_first_quantization}
\varPsi_\mathrm{pcs}(\xbf_1,\xbf_2,\ldots,\xbf_{2n})\,\propto\,
\sP_S\Bigl(\varPsi^{(2)}(\xbf_1,\, \xbf_2)\,\varPsi^{(2)}(\xbf_3,\, \xbf_4)\cdots \varPsi^{(2)}(\xbf_{2n-1},\,\xbf_{2n})\Bigr)\;.
\end{equation}
Such a PCS can be regarded as constructed by a mapping of the two-boson Hilbert space into a submanifold of the Hilbert space of $2n$ bosons.  Note that the two-boson wavefunction always has a Schmidt decomposition of the form~\cite{paskauskas_quantum_2001}\si{Two-boson state}\si{Schmidt decomposition},
\begin{equation}\lb{eq:schmidt_first_quantization}
\varPsi^{(2)}(\xbf_1,\, \xbf_2)=\sum_{j=1}^\rank \normalized\lambda_j\, \psi_j(\xbf_1)\psi_j(\xbf_2) \;,
\end{equation}
where $\rank$ is the Schmidt rank and the $\normalized\lambda_j$s are the square roots of the Schmidt coefficients ($\ssp\sum_{j=1}^\rank \normalized\lambda_j^2=1$). The coincidence of the Schmidt bases of the two bosons is a consequence of the symmetry of the wavefunction.

It is instructive to perform the Schmidt decomposition in the second-quantized picture, where an arbitrary two-boson state takes the form
\begin{equation}
 \ket{\varPsi^{(2)}}=\frac{1}{\sqrt 2}\,\sum_{j,k=1}^{\rank} \normalized\Lambda_{jk}\, \b^\dagger_k \b^\dagger_j\,\ket{\vac}\,.
\end{equation}
Here $\b^\dagger_j$ is the creation operator of the $j$th single particle state, and $\normalized\Lambda_{jk}=\normalized\Lambda_{kj}$ is a symmetric\footnote{We can always make $\Lambda$ symmetric by redefining $\Lambda\rightarrow (\Lambda+\Lambda^T)/2$, without changing $\ket{\varPsi^{(2)}}$.} matrix, normalized according to $\mbox{tr}(\Lambda\Lambda^\dagger)=1$ to make $\ketb{\varPsi^{(2)}}$ normalized.  The Autonne-Takagi factorization theorem\si{Autonne-Takagi factorization} [see Corollary 4.4.4 (c) of~\cite{horn_matrix_2013}, or App.~\chref{ch:at_factorization} of this dissertation] says that any complex symmetric matrix $\normalized\Lambda$ can be diagonalized by a unitary matrix $U$ in the following way,
\begin{equation}\lb{eq:Lambda_diagonalized}
 U \normalized\Lambda\, U^T = \diag\big(\normalized\lambda_1, \normalized\lambda_2,\ldots, \normalized\lambda_\rank\big) \;,
\end{equation}
where the $\lambda_j$s are real and positive and normalized as $\ssp\sum_{j=1}^\rank \normalized\lambda_j^2=1$.  The $\lambda_j$s are the singular values of $\Lambda$, and the diagonalization~(\rf{eq:Lambda_diagonalized}) is a special case of the singular-value decomposition, specialized to symmetric matrices.  We adopt the following convention throughout this dissertation,
\begin{equation}\lb{eq:lambda_convention}
\lambda_1\geq \lambda_2\geq \cdots \geq \lambda_\rank\;.
\end{equation}
By introducing a new set of creation operators $\a_j^\dagger=\sum_{k=1}^\rank U_{jk}^*\, \b_k^\dagger$, we have\si{Schmidt decomposition}
\begin{equation}\lb{eq:schmidt_second_quantization}
 \ket{\varPsi^{(2)}}=\frac{1}{\sqrt 2}\,\sum_{j=1}^{\rank} \normalized\lambda_j \ssp \big(\a^{\dagger}_j\big)^2 \,\ket{\vac}\;;
\end{equation}
the corresponding wave function $\braket{\xbf_1,\xbf_2}{\varPsi^{(2)}}=\varPsi^{(2)}(\xbf_1,\, \xbf_2)$ has the Schmidt form~(\rf{eq:schmidt_first_quantization}), with $\psi_j(\xbf)$ being the single-particle wavefunction that goes with the annihilation operator $\a_j$, i.e.,
\begin{equation}
\a_j=\int\psi_j^*(\xbf)\,\uppsi(\xbf)\,d\xbf\;.
\end{equation}

We now define the pair creation operator\si{Pair creation operator}
\begin{equation}
\sA^\dagger\equiv \sum_{j=1}^\rank \normalized\lambda_j \ssp \big(\a^{\dagger}_j\big)^2
=\int\varPsi^{(2)}(\xbf_1,\,\xbf_2)\uppsi^\dagger(\xbf_1)\uppsi^\dagger(\xbf_2)\,d\xbf_1\,d\xbf_2\;.
\end{equation}
Using
\begin{subequations}
\begin{align}
\Psi^\dagger(\xbf_1)\cdots\Psi^\dagger(\xbf_N)\,\ketb{\vac}
&=\sqrt{N!}\;\sP_S\ketb{\xbf_1,\ldots,\xbf_N}\\
&=\frac{1}{\sqrt{N!}}\sum_{\pi\in S_N}\ketb{\xbf_{\pi(1)},\ldots,\xbf_{\pi(N)}}\;,
\end{align}
\end{subequations}
where the sum is over the permutations $\pi$ in the symmetric group $S_N$, we have
\begin{align}\lb{eq:Anvac}
\hspace{-2em}
\big(\sA^\dagger\big)^n\,&\ketb{\vac}\nonumber\\
&=\sqrt{N!}\int\ketb{\xbf_1,\ldots,\xbf_{2n}}\,
\sP_S\Bigl(\varPsi^{(2)}(\xbf_1,\, \xbf_2) \cdots \varPsi^{(2)}(\xbf_{2n-1},\, \xbf_{2n})\Bigr) \,d\xbf_1\cdots d\xbf_{2n}\;,
\end{align}
where
\begin{align}
\sP_S\Bigl(\varPsi^{(2)}&(\xbf_1,\, \xbf_2)\cdots \varPsi^{(2)}(\xbf_{2n-1},\, \xbf_{2n})\Bigr)\nonumber\\
&=\frac{1}{N!}\sum_{\pi\in S_N}
\varPsi^{(2)}\big(\xbf_{\pi(1)},\,\xbf_{\pi(2)}\big) \cdots \varPsi^{(2)}\big(\xbf_{\pi(2n-1)},\, \xbf_{\pi(2n)}\big)\;.
\end{align}
Equation~(\rf{eq:Anvac}) is the relation between the first- and second-quantized pictures; it can be written in the equivalent form
\begin{equation}\lb{eq:12rel}
 \hspace{-2em}\frac{1}{\sqrt{N!}}\,\brab{\xbf_1,\ldots,\xbf_{2n}}\big(\sA^\dagger\big)^n \ketb{\vac}=\sP_S\Bigl(\varPsi^{(2)}(\xbf_1,\, \xbf_2) \cdots \varPsi^{(2)}(\xbf_{2n-1},\, \xbf_{2n})\Bigr)\;,
\end{equation}
which states that one gets the PCS state~(\rf{eq:pcs_rank2_pure_first_quantization}) by applying the pair creation operator $n$ times to the vacuum.

The state in Eq.~(\rf{eq:12rel}) is not normalized.  The properly normalized PCS state is given by
\begin{equation}\lb{eq:pcs_rank2_pure_second_quantization}
\ket{\varPsi_\mathrm{pcs}}\equiv \frac{1}{\sqrt{\NFB}}\, \big(\sA^\dagger\big)^n\,  \ket{\vac}\;,
\end{equation}
where $\NFB$ is a normalization factor\si{Normalization factor} that plays an important role in our consideration of PCS states and $n=N/2$ is the number of pairs.  We abandon the normalization restriction on the Schmidt coefficients from now on, requiring only that the $\lambda_j$s be real and positive, since an overall scaling of the $\lambda_j$s is automatically absorbed into $\NFB$.

The form~(\rf{eq:pcs_rank2_pure_second_quantization}) is  convenient for calculations, but let us first build some intuition by considering the state decompositions in the first-quantized picture.  The relative-state decomposition\si{Relative-state decomposition} of the particle $\xbf_1$ relative to all the other particles is
\begin{subequations}
\begin{align}
\varPsi_\mathrm{pcs} &\propto\, \sP_S\Bigl(\varPsi^{(2)}(\xbf_1,\, \xbf_2)\,\varPsi^{(2)}(\xbf_3,\, \xbf_4)\cdots \varPsi^{(2)}(\xbf_{2n-1},\, \xbf_{2n})\Bigr)\\
 &\propto\, \sum_j\lambda_j\; \psi_j(\xbf_1)\,\sP_S\Bigl(\psi_j(\xbf_2)\,\varPsi^{(2)}(\xbf_3,\, \xbf_4)\cdots \varPsi^{(2)}(\xbf_{2n-1},\, \xbf_{2n})\Bigr)\;.
\end{align}
\end{subequations}
This is a Schmidt decomposition\si{Schmidt decomposition}, and the Schmidt basis of the particle $\xbf_1$ consists of all the single-particle wavefunctions $\psi_j(\xbf_1)$.  The Schmidt coefficients, however, are not given by the $\lambda_j$s, because the norms of the relative states of the other particles are different for different values of $j$.

More interestingly, we have the following relative-state decomposition for particles $\xbf_1$ and $\xbf_2$,
\begin{subequations}
\begin{align}
\varPsi_\mathrm{pcs}
&\,\propto\, \varPsi^{(2)}(\xbf_1,\, \xbf_2)\,\sP_S\Bigl(\varPsi^{(2)}(\xbf_3,\, \xbf_4)\cdots \varPsi^{(2)}(\xbf_{2n-1},\, \xbf_{2n})\Bigr)\\
\begin{split}
&\qquad+(N-2)\sum_{j,k} \lambda_{\smash j}\lambda_{\smash k}\,\psi_j(\xbf_1)\psi_k(\xbf_2)\\
&\qquad\qquad\times\,\sP_S\Bigl(\psi_j(\xbf_3)\psi_k(\xbf_4)\,\varPsi^{(2)}(\xbf_5,\, \xbf_6)\cdots \varPsi^{(2)}(\xbf_{2n-1},\, \xbf_{2n})\Bigr) \;.
\end{split}
\end{align}
\end{subequations}
What this shows it that the reduced (marginal) two-particle state consists of two terms: In the first term, the two particles $\xbf_1$ and $\xbf_2$ can be perfectly correlated, whereas in the second term, they are only partially correlated.  To determine the pairwise quantum correlations in the PCS, we need to find the two-particle reduced density matrices (2RDMs), and to do that, we will find it useful to investigate the normalization factor $\NFB$.  Before getting to that, however, we detour into showing how the Bogoliubov approximation arises from the \hbox{PCS}.

\section{Relation to Bogoliubov's Wavefunction}
\lb{sec:pcs_bog_relation}

When the Schmidt rank of the two-particle state~(\rf{eq:schmidt_second_quantization}) is one, the PCS~(\rf{eq:pcs_rank2_pure_second_quantization}) is a product state (this is the Gross-Pitaevskii ansatz).  Here, I show that the PCS also reproduces the particle-number-conserving ($N$-conserving) Bogoliubov approximation~\cite{girardeau_theory_1959, gardiner_particle-number-conserving_1997, castin_low-temperature_1998}\si{Number-conserving Bogoliubov} to the BEC ground state when there is one dominant and other minor Schmidt coefficients.  This result validates using PCS for BECs for small depletion, and it also provides a different way to write the state of an $N$-conserving Bogoliubov approximation.

Consider a Bose-Einstein condensate of $N$ particles in a trapping potential.  The Gross-Pitaevskii (GP) ground state takes the form
\begin{align}\lb{eq:gp_ground}
 \ket{\varPsi_\mathrm{gpgs}} &=  \ket{N}_0\otimes\ket{\vac}_\perp\;,
\end{align}
where the subscript $\perp$ denotes all the modes that are orthogonal to the condensate mode.  The essence of the $N$-conserving Bogoliubov approximation is to perturb about the state~(\rf{eq:gp_ground}) by introducing the operators \begin{equation}
\tilde\a_j^\dagger = \a_j^\dagger\a_0/\sqrt{N}
\quad\mbox{for $j=1,2,\ldots, \rank-1$,}
\end{equation}
where $\a_0$ is the annihilation operator of the condensate mode and $\a_j^\dagger$ is the creation operator of the $j$th orthogonal mode.  Note that even though the operators $\tilde\a_j^\dagger$ and $\tilde\a_j$ are not exactly creation and annihilation operators, they satisfy the canonical commutation relations approximately in the limit of large $N$ and small depletion (small number of particles not in the condensate mode).  For a condensate with small depletion, $\a_0$ is of order $\sqrt N$ and $\tilde\a_j$s are of order $1$; the modification to the total energy in the Bogoliubov approximation is of order $1$.

The $N$-conserving Bogoliubov Hamiltonian, with the mean field removed, is a quadratic function of $\tilde\a_j^\dagger$ and $\tilde\a_j$,
\begin{align}\lb{eq:ncb_hamiltonian}
 \sH_\mathrm{ncb}= \sum_{j,k=1}^{\rank-1} M_{jk}\tilde\a_j^\dagger \tilde\a_k+\half\Big(M_{jk}' \tilde\a_j^\dagger \tilde\a_k^\dagger+\mathrm{H.c.}\Big)\;,
\end{align}
where $M^\dagger = M$ and $(M')^T=M'$.  This Hamiltonian can always be diagonalized by a Bogoliubov transformation
\begin{align}
 \sB^\dagger\, \sH_\mathrm{ncb}\, \sB = \sum_{k=1}^{\rank-1} \epsilon_k\, \tilde\a_k^\dagger \tilde\a_k\;,
\end{align}
where $\sB$ is a Gaussian unitary inducing a symplectic transformation on the operators $\tilde\a_j^\dagger$ and $\tilde\a_j$ for $j=1,2,\ldots, \rank-1$.  The Bogoliubov ground state\si{Bogoliubov ground state} of the Hamiltonian~(\rf{eq:ncb_hamiltonian}) thus takes the form
\begin{subequations}\lb{eq:bgs_a}
\begin{align}
 \ket{\varPsi_\mathrm{bgs}} &= \sB\, \ket{N}_0\otimes\ket{\vac}_\perp\\
 &= \sU\,\bigg(\prod_{k=1}^{\rank-1}\, \sS_k(\gamma_k,\tilde\a_k)\bigg)\, \sV^\dagger\, \ket{N}_0\otimes\ket{\vac}_\perp\\
 &= \sU\,\bigg(\prod_{k=1}^{\rank-1}\, \sS_k(\gamma_k,\tilde\a_k)\bigg)\, \sU^\dagger\, \ket{N}_0\otimes\ket{\vac}_\perp\;,
\end{align}
\end{subequations}
where we use the Bloch-Messiah reduction theorem~\cite{bloch_canonical_1962}\si{Bloch-Messiah reduction} to decompose the Gaussian unitary $\sB$ into a multiport beam splitter $\sV^\dagger$, followed by a set of single-mode squeezers $\sS_k(\gamma_k,\tilde\a_k)= e^{\half(\gamma_k \tilde\a_k^2-\gamma_k \tilde\a_k^{\dagger\,2})}$ with $\gamma_k$ real, followed by yet another multiport beam splitter $\sU$.  Also note that $\sU^\dagger$ and $\sV^\dagger$ have no effect on the GP ground state $\ket{N}_0\otimes\ket{\vac}_\perp$; thus we can use either of them (or neither).

The action of the multiport beam splitter $\sU$ is given by
\begin{align}
 \sU\, \tilde\a_j\, \sU^\dagger = \sum_{k=1}^{\rank-1} \tilde\a_k U_{kj} = \tilde\b_j = \a_0^\dagger\ssp \b_j/\sqrt N\;,
\end{align}
where $U$ is some unitary matrix that specifies the multiport splitter.  Note that $\b_j$ is the annihilation operator for a new set of modes orthogonal to the condensate mode.  The state~(\rf{eq:bgs_a}) thus takes the form
\begin{subequations}
\begin{align}
 \ketb{\varPsi_\mathrm{bgs}} &= \sU\bigg(\prod_{k=1}^{\rank-1} \, \sS_k(\gamma_k,\tilde\a_k)\bigg)\, \sU^\dagger\, \ketb{N}_0\otimes\ketb{\vac}_\perp\\
 &= \bigg(\prod_{k=1}^{\rank-1} \, \sS_k(\gamma_k,\tilde\b_k)\bigg)\, \ketb{N}_0\otimes\ketb{\vac}_\perp\\
&= \bigg(\prod_{k=1}^{\rank-1} \frac{1}{\sqrt{\cosh \gamma_k}}\;\exp\Big(\mathord{-}\frac{\tanh
\gamma_k}{2}\,\tilde\b_k^{\dagger\, 2}\,\Big) \bigg)\, \ketb{N}_0\otimes\ketb{\vac}_\perp\\
&= \frac{1}{\sqrt{\prod_{k=1}^{\rank-1} \cosh \gamma_k }}\, \exp\bigg(-\sum_{k=1}^{\rank-1}\frac{\tanh
\gamma_k}{2}\,\tilde\b_k^{\dagger\, 2}\,\bigg)\, \ketb{N}_0\otimes\ketb{\vac}_\perp\;.\lb{eq:nc_bog_a}
\end{align}
\end{subequations}
Here we use the ``quasi-normal-ordered'' factored form of the squeeze operator~\cite{perelomov_generalized_1977, hollenhorst_quantum_1979}\si{Squeeze operator}:
\begin{equation}\lb{eq:quasi_normal_ordered}
\sS(\gamma, \a)= \frac{1}{\sqrt{\cosh \gamma}}\, \exp\Bigl(\mathord{-}\frac{\tanh \gamma}{2}\, a^{\dagger\, 2} \Bigr)\,\bigl(\cosh \gamma\bigr)^{- \a^{\dagger} \a}\,\exp\Bigl(\frac{\tanh \gamma}{2}\, \a^2 \Bigr)\;.
\end{equation}
Note that we can always make $\gamma_k$ negative, by redefining the phase of $\tilde\b_k$, so that the coefficient $-\half\tanh \gamma_k$ in Eq.~(\rf{eq:nc_bog_a}) is positive.

On the other hand, when there is one dominant Schmidt coefficient $\lambda_0$, the PCS~(\rf{eq:pcs_rank2_pure_second_quantization}) of $N=2n$ particles takes the form
\begin{subequations}\lb{eq:pcs_exponential}
\begin{align}
 \ketb{\varPsi_\mathrm{pcs}} &= \frac{1}{\sqrt \NFB}\,\bigg(\lambda_0\big(\a_0^\dagger\big)^2+\sum_{k=1}^{\rank-1} \lambda_k\big(\b_k^\dagger\big)^2\bigg)^n\,\ketb{\vac}\\
 &\simeq \sqrt{\frac{(2n)!}{\NFB}}\,\bigg(\lambda_0+\frac{1}{2n}\sum_{k=1}^{\rank-1} \lambda_k\big(\tilde\b_k^\dagger \big)^2\bigg)^n \,\ketb{2n}_0\otimes\ketb{\vac}_\perp\\
 &\simeq \sqrt{\frac{N!\,\lambda_0^N}{ \NFB}}\,\exp\bigg(\frac{1}{2\lambda_0}\sum_{k=1}^{\rank-1} \lambda_k\big(\tilde\b_k^\dagger \big)^2\bigg)\, \ketb{N}_0\otimes\ketb{\vac}_\perp\;,\lb{eq:nc_bog_b}
\end{align}
\end{subequations}
where the approximations are good when $N$ is large and the depletion from the condensate mode is small.  Comparing Eqs.~(\rf{eq:nc_bog_a}) and~(\rf{eq:nc_bog_b}), we find that they can be made the same by choosing
\begin{gather}
  \NFB = N!\,\lambda_0^N \prod_{k=1}^{\rank-1} \cosh \gamma_k\;,\\
  \lambda_j/\lambda_0 = -\tanh \gamma_j\,,\;\;\mbox{for $j=1,2,\ldots,\rank-1$.}
\end{gather}
Hence, as promised, the PCS~(\rf{eq:pcs_rank2_pure_second_quantization}) encompasses the Bogoliubov approximation to the BEC ground state.

Another way to prove the same result is by noticing that
\begin{align}
 \ketb{\varPsi_\mathrm{pcs}} &\sim \sP_N\,\exp\bigg(\frac{1}{2\lambda_0}\sum_{k=1}^{\rank-1} \lambda_k\big(\b_k^\dagger \big)^2\bigg)\, \ketb{\alpha}_0\otimes\ketb{\vac}_\perp\;,
\end{align}
where $\sP_N$ is the projection operator onto the $N$-particle sector and $\ket{\alpha}_0$ is a coherent state for the condensate mode with $\alpha=\sqrt N$.  This is nothing but the extended catalytic state\si{Extended catalytic state}~(\chref{n_conserving:eq:catalytic_state}) that we have introduced for the number-conserving Bogoliubov approximation.

\section{The Normalization Factor}\si{Normalization factor}

The importance of the normalization factor $\NFB$ to PCS states is analogous to the utility of the partition function in statistical physics.  By taking derivatives of the normalization factor with respect to the Schmidt coefficients $\lambda_j$, one can calculate the reduced density matrices (RDMs) in the Schmidt basis, and these, in turn, give all the physical observables.

In the second-quantized picture, the normalization factor $\NFB$ introduced in Eq.~(\rf{eq:pcs_rank2_pure_second_quantization}) takes the form
\begin{subequations}
\begin{align}
 \NFB &=\brab{\vac} \sA^{n} \big(\sA^\dagger\big)^n  \ketb{\vac}\\
 &=\frac{1}{\pi^\rank}\int  \brab{\vac} \sA^{n} \projb{\vec{\alpha}} \big(\sA^\dagger\big)^n  \ketb{\vac}\; \dif^2 \alpha_1\cdots\dif^2 \alpha_\rank\\
 &=\frac{1}{\pi^\rank}\int  e^{-\norm{\vec{\alpha}}^2}\: \biggl| \sum_{j=1}^\rank \lambda_j\,\alpha_j^2 \biggr|^{2n}\,\dif^2 \alpha_1\cdots\dif^2 \alpha_\rank\;,\lb{eq:nfb_integral}
\end{align}
\end{subequations}
where we insert a complete basis of coherent states. Expanding the monomial in Eq.~(\rf{eq:nfb_integral}), we have
\begin{subequations}
\begin{align}\hspace{-2em}
 \NFB  &=\sum_{\{\vec{\idu},\, \vec{\idv}\}}\frac{n!}{\idu_1!\, \idu_2!\cdots\idu_\rank!}\: \frac{n!}{\idv_1!\, \idv_2!\cdots\idv_\rank!}\,\biggl(\prod_{j=1}^\rank \lambda_j^{\idu_j+\idv_j}\biggr) \nonumber\\[3pt]
 &\qquad \times \frac{1}{\pi^\rank} \int e^{-\norm{\vec{\alpha}}^2} \bigg(\prod_{j=1}^\rank \alpha_j^{2\idu_j} \big(\alpha_j^*\big)^{2\idv_{j}}\bigg)\,\dif^2 \alpha_1\cdots\dif^2 \alpha_\rank\\[2pt]
 &=\sum_{\{\vec{\idu}\}}\frac{(n!\ssp)^2}{\idu_1!\, \idu_2!\cdots\idu_\rank!}\:
 \bigg(\prod_{j=1}^\rank\frac{(2\idu_j)!\,\lambda_j^{2\idu_j}}{\idu_j!}\bigg)\;,\lb{eq:nfb_a}
\end{align}
\end{subequations}
where $\idu_j$ ($\idv_j$) are non-negative integers satisfying $\sum_{j=1}^\rank \idu_j=\sum_{j=1}^\rank \idv_j=n$.  Although the sum in Eq.~(\rf{eq:nfb_a}) appears to requires exponential time to evaluate, it can be evaluated in polynomial time by using an iterative algorithm.  It is still, however, computationally demanding for large $N$, in addition to not being intuitive.

To make progress in interpreting and evaluating $\NFB$, we take what might be construed as a backward step by writing $\NFB$ in a different integral form.  To do so, we use $(2\idu_j)!/\idu_j!=2^{\idu_j}(2\idu_j-1)!!$ to write
\begin{equation}\hspace{-0.8em}
\prod_{j=1}^\rank\frac{(2\idu_j)!\,\lambda_j^{2\idu_j}}{\idu_j!}=
\frac{2^{n}}{(\sqrt{2\pi}\,)^{\rank}  } \int_{-\infty}^\infty
e^{-\norm{\vec{y}\ssp}^2/2}\bigg(
\prod_{j=1}^\rank \big(\lambda_j^2\,y_j^2\big)^{\idu_j}\bigg)
\,\dif y_1\cdots \dif y_\rank
\end{equation}
and then use this to put Eq.~(\rf{eq:nfb_a}) in the form
\begin{equation}
 \NFB
 =\frac{2^{n}n!}{(\sqrt{2\pi}\,)^{\rank}}
 \int_{-\infty}^\infty  e^{-\norm{\vec{y}\ssp}^2/2}\biggl( \sum_{j=1}^\rank \lambda_j^2\, y_j^2\biggr)^{\! n}\,\dif y_1\cdots \dif y_\rank\;. \lb{eq:nfb_b}
\end{equation}
We use the above expression to evaluate the normalization factor in the large-$N$ limit in the next section.  For now, however, we employ it to derive several exact expressions.

The first use of Eq.~(\rf{eq:nfb_b}) is to derive a generating function\si{Generating function},
\begin{subequations}
\begin{align}
 \NFB
 &=\frac{2^{{2n}}n!}{(\sqrt{2\pi}\,)^{\rank}  }\:
 \frac{\partial^{n}}{\partial \tau^{n}}
 \bigg(\int_{-\infty}^\infty  e^{-\norm{\vec{y}\ssp}^2/2}
 \exp\bigg(\half\tau\sum_{j=1}^\rank\!\lambda_j^2 y_j^2\bigg)
 \,\dif y_1\cdots \dif y_\rank\bigg)\bigg|_{\tau=0}\\[3pt]
 &=2^{{2n}}n!\, \frac{\partial^{n}}{\partial \tau^{n}} \bigg(\prod_{j=1}^\rank\frac{1}{\sqrt{1-\tau\lambda_{\smash j}^2}}\bigg)\bigg|_{\tau=0}
 \lb{eq:nfb_c}\;.
 \end{align}
\end{subequations}
An equivalent way to obtain the generating function~(\rf{eq:nfb_c}) is to evaluate the quantity $\bra{\vac}e^{\sqrt{\tau}\sA}e^{\sqrt{\tau}\sA^\dagger}\ket{\vac}$, which can be calculated using the ``quasi-normal-ordered'' factored form of the squeeze operator found in Eq.~(\rf{eq:quasi_normal_ordered}).

The second use is to derive expressions for the diagonal elements of RDMs.\footnote{The $q$-particle RDM is normalized to $N!/(N-q)!=N(N-1)\cdots (N-q+1)$ in this dissertation unless stated otherwise; normalized in this way, the RDMs are equal to correlation functions.}  As a first example, we are able to represent, using Wick's theorem\si{Wick's theorem}, the $j$th diagonal element of the 1RDM $\rho^{(1)}$ with the normalization factor,
\begin{subequations}\lb{eq:diagonal_elements_rdm_single_particle}
\begin{align}\hspace{-2em}
 \rho_{j j}^{(1)} &=
 \frac{1}{\NFB}\: \brab{\vac} \sA^{n}\, \a_j^\dagger \a_j\, \big(\sA^\dagger\big)^n  \ketb{\vac}\\[3pt]
 &= \frac{n\,\lambda_j}{\NFB}\,\Big(\brab{\vac} \sA^{n-1}\, \a_j^2\, \big(\sA^\dagger\big)^n \ketb{\vac}+\brab{\vac} \sA^n \, \big(\a_j^\dagger\ssp\big)^2\, \big(\sA^\dagger\big)^{n-1}\ketb{\vac}\Big)\lb{eq:1rdm_derivative_b}\\[3pt]
 &=\frac{\lambda_j}{\NFB}\, \bigg( \brab{\vac} \,\pa{\sA^{n}}{ \lambda_j}\,\big(\sA^\dagger\big)^n  \ketb{\vac}+\brab{\vac} \,\sA^{n}\, \pa{\big(\sA^\dagger\big)^n }{ \lambda_j}\ketb{\vac}\bigg)\\[3pt]
 &= \frac{\lambda_j}{\NFB}\, \pa{\, \NFB}{\lambda_j}\;.
\end{align}
\end{subequations}
The two terms in Eq.~(\rf{eq:1rdm_derivative_b}), which correspond to contracting $\a^\dagger_j$ and $\a_j$ with the pair annihilation and creation operators, are equal.  In addition, by using Wick's theorem, it is not hard to prove that all the off-diagonal elements of $\rho^{(1)}$ in the Schmidt basis are zero and, therefore, the normalization factor and its first derivative determine the 1RDM.

More generally, we have the following result for the diagonal elements of the $q$-particle RDM $\rho^{(q)}$:
\begin{subequations}
\begin{align}
\rho_{j_1\cdots j_q,\, j_1 \cdots j_q}^{(q)}
&=\frac{1}{\NFB}\:
\brab{\vac}\sA^{n}\,
\a_{j_1}^\dagger\cdots\a_{j_q}^\dagger\ssp\a_{j_q}\cdots\a_{j_1}
\,\big(\sA^\dagger\big)^n  \ketb{\vac}\\[3pt]
&=\frac{\lambda_{j_1}\cdots\lambda_{j_q}}{\NFB}\:
\frac{\partial^q \NFB}{\partial\lambda_{j_1} \cdots \partial\lambda_{j_q}}\;,\lb{eq:diagonal_elements_rdm_a}
\end{align}
\end{subequations}
This result can be proved by mathematical induction.  We already have that Eq.~(\rf{eq:diagonal_elements_rdm_a}) holds for $q=1$, so to show that it holds for all positive integers $q$ is that if it holds for $q$, it is satisfied for $q+1$.  The inductive hypothesis is thus that
\begin{align}\lb{eq:diagonal_elements_rdm_b}
\frac{\partial^q \NFB}{\partial\lambda_{j_1} \cdots\partial\lambda_{j_q}}
=\frac{1}{\lambda_{j_1}\cdots\lambda_{j_q}}\,\brab{\vac} \sA^{n}\,
\a_{j_1}^\dagger\cdots\a_{j_q}^\dagger\ssp\a_{j_q}\cdots\a_{j_1}
\,\big(\sA^\dagger\big)^n  \ketb{\vac}\;.
\end{align}
By taking derivatives with respect to $\lambda_{j_{q+1}}$ of both sides of Eq.~(\rf{eq:diagonal_elements_rdm_b}), we have
\small
\begin{subequations}\lb{eq:diagonal_elements_rdm_c}
\begin{align}\hspace{-2.3em}
\frac{\partial^{q+1} \NFB}{\partial\lambda_{j_1} \cdots \partial\lambda_{j_{q+1}}}
&=\frac{1}{\lambda_{j_1}\cdots\lambda_{j_q}}\,\bigg(n\,
\brab{\vac}\sA^{n-1}\,
\a_{j_{q+1}}^2\,\a_{j_1}^\dagger\cdots\a_{j_q}^\dagger\ssp\a_{j_q}\cdots\a_{j_1}
\,\big(\sA^\dagger\big)^n \ketb{\vac}+\mathrm{c.c.}\nonumber\\[3pt]
&\qquad-\frac{1}{\lambda_{j_{q+1}}}\,
\brab{\vac} \sA^{n}\,
\a_{j_1}^\dagger\cdots\a_{j_q}^\dagger\ssp\a_{j_q}\cdots\a_{j_1}
\,\big(\sA^\dagger\big)^n\ketb{\vac}\sum_{k=1}^q \delta(j_k,\, j_{q+1})\bigg)\\[3pt]
&=\frac{n}{\lambda_{j_1}\cdots\lambda_{j_q}}\nonumber\\
&\qquad\times\bigg(
\brab{\vac} \sA^{n-1}\,
\a_{j_{q+1}}\,\a_{j_1}^\dagger\cdots\a_{j_q}^\dagger\ssp\a_{j_q}\cdots\a_{j_1}\,\a_{j_{q+1}}
\,\big(\sA^\dagger\big)^n\ketb{\vac}+\mbox{c.c.}\bigg)\\[3pt]
&=\frac{1}{\lambda_{j_1}\cdots\lambda_{j_{q+1}}}\,
\brab{\vac} \sA^{n}\,
\a_{j_1}^\dagger\cdots\a_{j_{q+1}}^\dagger\ssp\a_{j_{q+1}}\cdots\a_{j_1}
\,\big(\sA^\dagger\big)^n \ketb{\vac} \;,
\end{align}
\end{subequations}
\normalsize
which is the required result.  We note that for $q>1$, the $q$-particle RDM is generally not diagonalized in the Schmidt basis.  In Sec.~\rf{sec:2rdms}, we show how to construct the entire $q$-particle RDM using only the diagonal elements calculated by the above method.

Although we can avail ourselves of the power of Wick's theorem to derive the results of this section, we can demonstrate the same results using only the commutators $[a_j,(\sA^\dagger)^n]=2n\lambda_j\a_j^\dagger(\sA^\dagger)^{n-1}$ and $[\sA^n,\a_j^\dagger]=2n\lambda_j\sA^{n-1}\a_j$, which imply that
\begin{subequations}
\begin{align}
\a_j(\sA^\dagger)^n\ket\vac&=2n\lambda_j\a_j^\dagger(\sA^\dagger)^{n-1}\ket\vac\;,\\
\bra\vac\sA^n\a_j^\dagger&=2n\lambda_j\bra\vac\sA^{n-1}\a_j\;.
\end{align}
\end{subequations}

\section{The Large-\texorpdfstring{$N$}{\textit{N}} Limit}\si{Large-$N$ limit}
\lb{sec:large_N}

Often there are thousands to millions of atoms in a BEC, and it is sufficient to work with results that are valid in the large-$N$ limit.  In this section, we discuss how to derive an asymptotic form of the normalization factor for large $N$.  To get the desired analytical results, terms that are of order $1/N$ smaller than the leading terms are neglected.  This means that the following results do not include the Bogoliubov approximation;\footnote{The depletion predicted by the Bogoliubov approximation is of order $1$, i.e., it modifies the 1RDM by order $1$, which is $1/N$ times smaller than the 1RDM itself.} instead we should think of them as a generalization of the Gross-Pitaevskii approximation.

\begin{figure}[ht] 
   \centering
   \includegraphics[width=0.76\textwidth,natwidth=610,natheight=642]{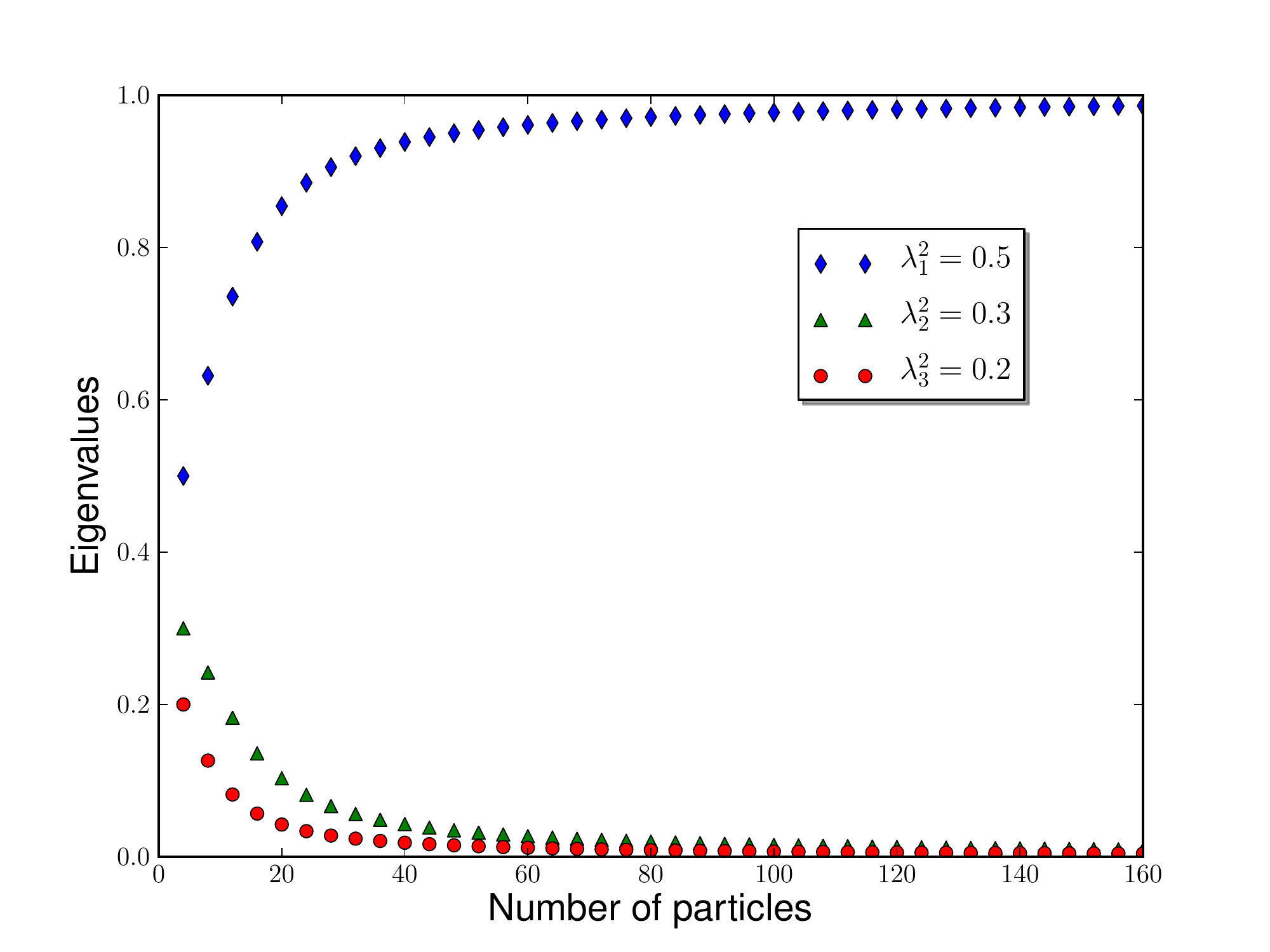}
   \caption[Eigenvalues of 1RDMs as Functions of $N$ I]{The three eigenvalues of the single-particle RDM (normalized to unity) plotted as a function of the number of particles~$N$ for the case $\lambda_1^2=0.5$, $\lambda_2^2=0.3$, $\lambda_3^2=0.2$.   As $N$ gets large, the eigenvalue corresponding to the biggest $\lambda$ approaches one while the other eigenvalues become negligible.}
   \label{fig:eigenv_a}
\end{figure}

Before the analytical calculation, let us look first at some numerical examples to get some intuition. In Fig.~\ref{fig:eigenv_a}, the eigenvalues of the 1RDM (normalized to unity instead of to $N$) are plotted as a function of the number of particles $N$ for $\lambda_1^2=0.5$, $\lambda_2^2=0.3$, and $\lambda_3^2=0.2$.  For $N=2$, the eigenvalues of the 1RDM are, of course, equal to the $\lambda_j$s, but as $N$ gets larger, the eigenvalues become further apart.  Eventually, the biggest eigenvalue approaches one, leaving the other two eigenvalues negligible; thus, as far as the 1RDM can tell, the PCS becomes an uninteresting product state for large $N$.

\begin{figure}[ht]
   \centering
   \includegraphics[width=0.76\textwidth,natwidth=610,natheight=642]{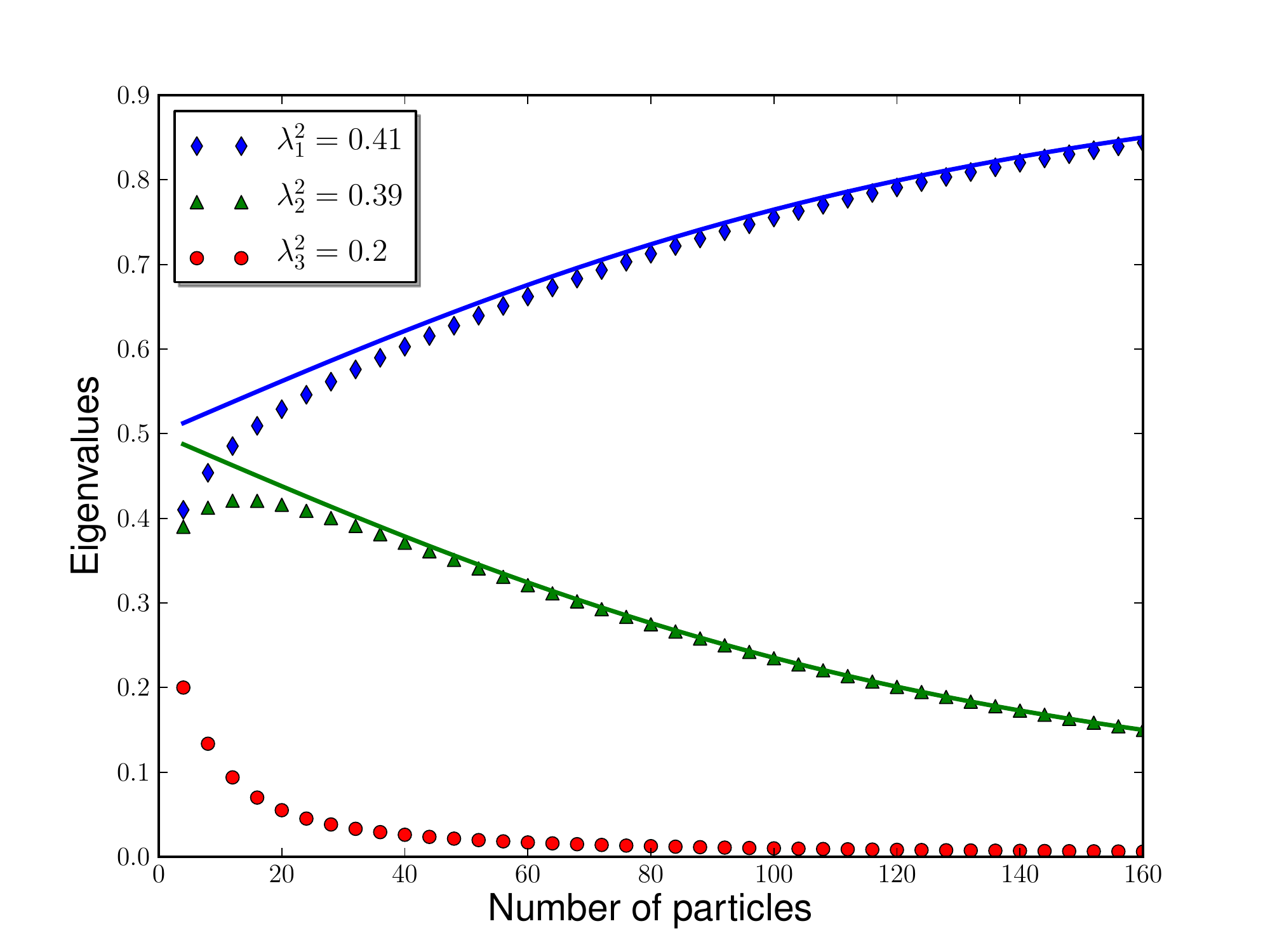}
   \caption[Eigenvalues of 1RDMs as Functions of $N$ II]{The three eigenvalues of the single-particle RDM (normalized to unity) plotted as a function of the number of particles~$N$ for the case $\lambda_1^2=0.41$, $\lambda_2^2=0.39$, $\lambda_3^2=0.2$.  The solid lines are the eigenvalues calculated by setting $\lambda_3=0$ and keeping $\lambda_{1}$ and $\lambda_2$ unchanged except for renormalizing; they conform pretty well with the other results for large $N$.}
   \label{fig:eigenv_c}
\end{figure}

In the second numerical example, plotted in Fig.~\ref{fig:eigenv_c}, we consider the situation where there are two $\lambda_j$s that are nearly degenerate, $\lambda_1^2=0.41$ and $\lambda_2^2=0.39$, and a third smaller value, $\lambda_3^2=0.2$.  When $N$ is of the order $1/(\lambda_1-\lambda_2)=40$ or larger, the third eigenvalue has died out, and only the two biggest eigenvalues play much of a role in determining the 1RDM.  Our numerics suggest that in the large-$N$ limit, only those $\lambda_j$s that are within order $1/N$ of $\lambda_1$ survive.

Generally, we speculate that one only need to keep those $\lambda_j$s that are within $1/N$ of $\lambda_1$, the largest eigenvalue given our ordering convention~(\rf{eq:lambda_convention}); the other $\lambda_j$s can be omitted without affecting the PCS---i.e., the relevant low-order RDMs are not affected.  This speculation is already supported by the above numerical results, and I will argue further for it based on the analytical results for the large~$N$ limit.  The important $\lambda_j$s, being very close to each other, can be rescaled (i.e., they are no longer normalized to one) and parameterized as
\begin{equation}\lb{eq:lambda_zeta}
\lambda_j^2\equiv 1+\frac{\pcss_j}{n}\quad \mbox{or}\quad\lambda_j\simeq 1+\frac{\pcss_j}{2n}\;,
\end{equation}
where the $\pcss_j$s are real parameters of order unity.  Because of the rescaling, all the $\lambda_j$s are very close to $1$, and their differences are of order $1/N$.  Putting Eq.~(\rf{eq:lambda_zeta}) into Eq.~(\rf{eq:nfb_b}), we manipulate the normalization factor $\NFB$ through the following sequence of steps:
\begin{subequations}
\begin{align}
\NFB_{\smash{\vec\pcss},\ssp n}
&=\frac{2^{n}n!}{(\sqrt{2\pi}\,)^{\rank}  } \int_{-\infty}^\infty  e^{-\norm{\vec{y}\ssp}^2/2}\biggl( \sum_{j=1}^\rank \lambda_j^2\ssp y_j^2\biggr)^{\! n}\,\dif y_1\cdots \dif y_\rank\\
&=\frac{2^{n}n!}{(\sqrt{2\pi}\,)^{\rank}  } \int_{-\infty}^\infty  e^{-\norm{\vec{y}\ssp}^2/2}\biggl( \sum_{j=1}^\rank \Big(1+\frac{\pcss_j}{n}\Big) y_j^2\biggr)^{\! n}\,\dif y_1\cdots \dif y_\rank\\
&=\frac{2^{n}n!}{(\sqrt{2\pi}\,)^{\rank}  } \int_{-\infty}^\infty  e^{-\norm{\vec{y}\ssp}^2/2} \norm{\vec{y}\ssp}^{2n} \biggl(1+\frac{1}{n\norm{\vec{y}\ssp}^{2}} \sum_{j=1}^\rank \pcss_j\ssp y_j^2\biggr)^{\! n}\,\dif y_1\cdots \dif y_\rank\\[3pt]
&\simeq \frac{2^{n}n!}{(\sqrt{2\pi}\,)^{\rank}}
\int_{-\infty}^\infty e^{-\norm{\vec{y}\ssp}^2/2} \norm{\vec{y}\ssp}^{2n} \exp\bigg(\frac{1}{\norm{\vec{y}\ssp}^2}\sum_{j=1}^\rank \pcss_j\ssp y_j^2\bigg)\,\dif y_1\cdots \dif y_\rank\lb{eq:large_n_approximation}\\
&= \frac{2^{n}n!}{(\sqrt{2\pi}\,)^{\rank}}
\int_{0}^\infty e^{\mathord{-}r^2/2} r^{2n+\rank-1}\,\dif r\int_{\norm{\vec y}=1}  \exp\Big(\sum_{j=1}^\rank \pcss_j y_j^2\Big)\:\dif \Omega\\
&=\frac{4^{n}n!}{2 \pi^{\rank/2}} \; \Gamma\Big(n+\frac{\rank}{2}\Big)\int_{\norm{\vec y}=1}  \exp\Big(\sum_{j=1}^\rank \pcss_j y_j^2\Big)\:\dif \Omega\;,\lb{eq:nfb_d}
\end{align}
\end{subequations}
where $\dif \Omega$ denotes the area element on the unit $(\rank-1)$-dimensional sphere $\norm{\vec{y}\ssp}=1$.  The only approximation here is to replace, in Eq.~(\rf{eq:large_n_approximation}), the power function by the exponential function.  For each low-degree monomial of $\pcss_j$s the error in its expansion coefficient as a result of this replacement is of order $1/n$; such error only becomes substantial when the degree of the monomial approaches $n$.  This is an excellent approximation for our purpose of calculating low-order RDMs, because the high-degree monomials only affect high-order RDMs.

Denoting the Gaussian integral in Eq.~(\rf{eq:nfb_d}) by
\begin{subequations}
\begin{align}\lb{eq:upsilon_definition}
\Upsilon\big(\vec{\pcss}\, \big)
&\equiv\frac{1}{2 \pi^{\rank/2}}
\int_{\norm{\vec y}=1}\exp\bigg(\sum_{j=1}^\rank \pcss_j\ssp y_j^2\bigg)\:\dif\Omega\\
&=\frac{1}{2 \pi^{\rank/2}}
\int_{-\infty}^\infty \delta\big(\norm{\vec y}-1\big)
\exp\bigg(\sum_{j=1}^\rank \pcss_j\ssp y_j^2\bigg)\,\dif y_1\cdots \dif y_\rank\;,
\end{align}
\end{subequations}
the normalization factor takes the form
\begin{align}\lb{eq:nfb_upsilon}
\NFB_{\smash{\vec\pcss},\ssp n}&\simeq 4^{n}n!\: \Gamma\Big(n+\frac{\rank}{2}\Big)\,\Upsilon\big(\vec{\pcss}\,\big)\;.
\end{align}
The greatest significance of this expression is that the dependences on $n$ and on $\vec\pcss$ (or~$\vec\lambda$) factorize.  According to Eq.~(\rf{eq:diagonal_elements_rdm_a}), the diagonal elements of the RDMs can now be expressed approximately, with errors of order $1/N$, as
\begin{subequations}
\begin{align}
\rho_{j_1\cdots j_q,\, j_1 \cdots j_q}^{(q)}
&\simeq\frac{\lambda_{j_1}\cdots\lambda_{j_q}}{\Upsilon\big(\vec{\lambda}\,\big)}
\:\frac{\partial^q \Upsilon\big(\vec{\lambda}\,\big)}{\partial\lambda_{j_1} \cdots \partial\lambda_{j_q}}\\[3pt]
&=\frac{(2n)^q\lambda^2_{j_1}\cdots\lambda^2_{j_q}}{\Upsilon\big(\vec{\lambda}\,\big)}
\:\frac{\partial^q \Upsilon\big(\vec{\lambda}\,\big)}{\partial\pcss_{j_1} \cdots \partial\pcss_{j_q}}\\[3pt]
&\simeq \frac{(2n)^q}{\Upsilon\big(\vec{\pcss}\,\big)}\: \frac{\partial^q \Upsilon\big(\vec{\pcss}\,\big)}{\partial\pcss_{j_1} \cdots \partial\pcss_{j_q}}\;. \lb{eq:upsilon_diagonal_elements_rdm}
\end{align}
\end{subequations}
The function $\Upsilon\big(\vec{\pcss}\, \big)$ determines the PCS in the large-$N$ limit, with all the $n$-dependence removed.  Relative to exact expressions like Eq.~(\rf{eq:nfb_a}), the complexity of evaluating $\Upsilon\big(\vec{\pcss}\, \big)$ is dramatically reduced because of the removal of the $n$-dependence.

As an example, consider the $k$th eigenvalue of the 1RDM, which in the large-$N$ limit takes the form
\begin{align}\lb{eq:single_particle_upsilon}
 \rho_{kk}^{(1)}= \frac{2n}{2 \pi^{\rank/2} \Upsilon(\vec\pcss)}
\int_{\norm{\vec y}=1}y_k^2 \exp\bigg(\sum_{j=1}^\rank \pcss_j\ssp y_j^2\bigg)\:\dif\Omega\;.
\end{align}
Consider the case where $\lambda_\rank$ is substantially less than $1$ and all the other $\lambda_j$s are close to $1$.  In this situation $\pcss_\rank$ is a negative number with large magnitude.  Because of the suppression of the exponential function in Eq.~(\rf{eq:single_particle_upsilon}), the magnitude of $y_\rank$ must be very small to contribute to the integral, which tells us that $\rho_{\rank\rank}^{(1)} \ll 2n$.  On the other hand, we have $\sum_{j=1}^{\rank-1} y_j^2 \simeq 1$ for the other dimensions, so $\rho_{jj}^{(1)}$ for $j=1,2,\ldots,\rank-1$ can be calculated by neglecting the last dimension $y_\rank$.  In effect, the integral in Eq.~(\rf{eq:single_particle_upsilon}) is reduced to an integral over a hypersphere of one less dimension.  This argument can be easily generalized to higher-order RDMs, and it confirms our speculation that we need only keep those $\lambda_j$s that are within $1/N$ of $\lambda_1$.

Although we have already made life easier by introducing $\Upsilon\big(\vec{\pcss}\, \big)$, it is still a difficult task to evaluate the Gaussian integral~(\rf{eq:upsilon_definition}) over the hypersphere.  Fortunately, we can reduce the expression for $\Upsilon\big(\vec\pcss\,)$ to a single-variable integral.  To do so, notice that
\begin{subequations}
\begin{align}\hspace{-2.5em}
 \int_0^{\infty}\nsp \chi^{\rank/2-1}\, e^{-\tau \chi}\, \Upsilon\big(\chi\ssp \vec{\pcss}\, \big)\,\dif \chi  \lb{eq:upsilon_a}
 &=\frac{2}{2^{\rank/2}}\int_0^{\infty} r^{\rank-1}\, e^{-\tau r^2/2}\, \Upsilon\Big(\,\frac{r^2\ssp \vec{\pcss}}{2}\,\Big)\, \dif r\\
 &= \frac{1}{(\sqrt{2\pi}\,)^\rank}\int_{-\infty}^\infty  \exp\nsp \bigg(\mathord{-}\half \sum_{j=1}^\rank \big(\tau-\pcss_j\big) y_j^2\bigg)\,\dif y_1\cdots \dif y_\rank\\
 &= \prod_{j=1}^\rank \frac{1}{\sqrt{\tau-\pcss_j}}\;,\lb{eq:upsilon_b}
\end{align}
\end{subequations}
where we do the substitution $\chi=r^2/2$ in Eq.~(\rf{eq:upsilon_a}) and  where $\tau > \pcss_1$ for convergence ($\pcss_1$ is the largest of the $\pcss_j$s). Because Eq.~(\rf{eq:upsilon_b}) is the Laplace transformation of the function  $\chi^{\rank/2-1}\, \Upsilon\big(\chi\ssp \vec{\pcss}\, \big)$, we have
\begin{subequations}
\begin{align}
\Upsilon\big(\chi\ssp \vec{\pcss}\, \big)
&= \chi^{1-\rank/2} \sL^{-1}\bigg(\prod_{j=1}^\rank \frac{1}{\sqrt{\tau-\pcss_j}}\bigg)\\
&= \frac{\chi^{1-\rank/2}}{2\pi i}\, \int_{\Delta-i\infty}^{\Delta+i\infty} e^{\tau \chi}\bigg(\prod_{j=1}^\rank \frac{1}{\sqrt{\tau-\pcss_j}}\bigg)\,\dif \tau\;,\lb{eq:upsilon_laplace}
\end{align}
\end{subequations}
where $\sL^{-1}$ stands for the inverse Laplace transformation and the real parameter $\Delta>\pcss_1$ for convergence.  We have thus succeeded in reducing the high-dimensional integral~(\rf{eq:upsilon_definition}) to the one-dimensional integral~(\rf{eq:upsilon_laplace}).

For numerical calculations, one might find a straightforward series expansion of the function $\Upsilon(\vec\pcss\,)$ to be useful:
\begin{subequations}
\begin{align}
 \Upsilon\big(\vec{\pcss}\, \big)
 &= \frac{1}{2\pi^{\rank/2}}\int_{\norm{\vec y}=1} \prod_{j=1}^\rank e^{\pcss_j\ssp y_j^2}\:\dif \Omega\\
 &= \frac{1}{2\pi^{\rank/2}} \sum_{m_1,\ldots, m_\rank=0}^\infty\, \frac{\pcss_1^{m_1}\pcss_2^{m_2}\cdots\pcss_\rank^{m_\rank}}{m_1!\, m_2!\cdots m_\rank!}\,
 \int_{\norm{\vec y}=1} y_1^{2 m_1} y_2^{2 m_2}\cdots y_\rank^{2 m_\rank}\:\dif \Omega \lb{eq:series_expression_b}\;.
\end{align}
\end{subequations}
The integral in Eq.~(\rf{eq:series_expression_b}) can be manipulated by a change of variables into the form
\begin{subequations}
\begin{align}\hspace{-2.3em}
\int_{\norm{\vec y}=1} y_1^{2 m_1}\cdots y_\rank^{2 m_\rank}\:\dif \Omega
&=\int_{-\infty}^\infty\,\delta\big(\norm{\vec y}-1\big)
y_1^{2 m_1}\cdots y_\rank^{2 m_\rank}\,\dif y_1\cdots \dif y_\rank\\
&=2\int_0^\infty\,\delta\bigg(\sum_{j=1}^\rank z_j-1\bigg)
z_1^{m_1-1/2}\cdots z_\rank^{m_\rank-1/2}\,\dif z_1\cdots \dif z_\rank\\
&=2B\big(m_1+1/2,\ldots,m_\rank+1/2\big)\;,
\end{align}
\end{subequations}
where
\begin{subequations}
\begin{align}
B\big(m_1+1/2,\ldots,m_\rank+1/2\big)
&=\frac{\prod_{j=1}^\rank\Gamma(m_j+1/2)}{\Gamma(m+\rank/2)}\\
&=\frac{\pi^{\rank/2}}{2^m\Gamma(m+\rank/2)}\prod_{j=1}^\rank(2m_j-1)!!
\end{align}
\end{subequations}
is the multivariable Beta function, with $m=\sum_{j=1}^\rank m_j$ [notice that $(-1)!!=1$].  Putting this back into Eq.~(\rf{eq:series_expression_b}) gives
\begin{align}\lb{eq:UpsilonBetaFunctionExpansion}
\Upsilon\big(\vec{\pcss}\, \big)
=\frac{1}{\pi^{\rank/2}} \sum_{m_1,\ldots, m_\rank=0}^\infty\, \frac{\pcss_1^{m_1}\pcss_2^{m_2}\cdots\pcss_\rank^{m_\rank}}{m_1!\, m_2!\cdots m_\rank!}\,
B\big(m_1+1/2,\ldots,m_\rank+1/2\big)\;.
\end{align}

For interested readers we also show here how to represent the function $\Upsilon\big(\chi\ssp \vec{\pcss}\, \big)$ as a convolution.  Note that
\begin{subequations}
\begin{align}
 \frac{1}{\sqrt{\tau-\pcss_j}}
 &= \frac{1}{\sqrt{\pi}}\int_0^{\infty} \chi^{-1/2} \, e^{-(\tau-\pcss_j) \chi}  \,\dif \chi\\
 &=\frac{1}{\sqrt{\pi}}\int_0^{\infty} e^{-\tau \chi}\, \idG_{\pcss_j}(\chi) \,\dif \chi\;, \lb{eq:upsilon_c}
\end{align}
\end{subequations}
where
\begin{align}
 \idG_{\pcss_j}(\chi)=\left\{
 \begin{aligned}
  &\chi^{-1/2}\, e^{\pcss_j \chi},\; &\mbox{for $\chi>0$}\;,\\
  &\quad 0, \; &\mbox{for $\chi\leq 0$}\;.
 \end{aligned}
 \right.
\end{align}
Putting Eq.~(\rf{eq:upsilon_c}) into Eq.~(\rf{eq:upsilon_b}), we have
\begin{subequations}\lb{eq:upsilon_d}
\begin{align}\hspace{-2em}
\int_0^{\infty} \chi^{\rank/2-1}\, e^{-\tau \chi}\, \Upsilon\big(\chi\ssp \vec{\pcss}\, \big)\:\dif \chi
& = \prod_{j=1}^{\rank} \bigg(\frac{1}{\sqrt{\pi}}\int_0^{\infty}\,e^{-\tau \chi} G_{s_j}(\chi) \:\dif \chi\bigg)\\
&=\frac{1}{\pi^{\rank/2}} \int_0^{\infty} e^{-\tau \chi} \big(\idG_{\pcss_1}*\idG_{\pcss_2}*\cdots * \idG_{\pcss_{\rank}}\big)(\chi)\:\dif \chi\;,\lb{eq:upsilon_e}
\end{align}
\end{subequations}
where $*$ stands for the convolution,
\begin{align}
\big(\idG_{\pcss_j}*\idG_{\pcss_k}\big)(\chi)&=\int_{0}^{\chi} \idG_{\pcss_j}(\chi-\chi')\, \idG_{\pcss_k}(\chi') \,\dif \chi'\;.
\end{align}
Doing the inverse Laplace transformation, we have
\begin{align}\lb{eq:upsilon_f}
\Upsilon\big(\chi\ssp \vec{\pcss}\, \big)
=\frac{\chi^{1-\rank/2}}{\pi^{\rank/2}}\, \big(\idG_{\pcss_1}*\idG_{\pcss_2}*\cdots *\idG_{\pcss_{\rank}}\big)(\chi)\;.
\end{align}

We now have four representations of $\Upsilon\big(\ssp \vec{\pcss}\, \big)$, expressed in Eqs.~(\rf{eq:upsilon_definition}), (\rf{eq:upsilon_laplace}), (\rf{eq:UpsilonBetaFunctionExpansion}), and~(\rf{eq:upsilon_f}).  All of these turn out to be useful, and we use whichever is most convenient.

We turn now to an exploration of relations among the RDMs that can be derived and expressed through the function $\Upsilon\big( \vec{\pcss}\,\big)$.  First, from the definition~(\rf{eq:upsilon_definition}) or from the Laplace transform~(\rf{eq:upsilon_laplace}), we have
\begin{align}
\Upsilon\big(\vec{\pcss}+\delta\ssp\vec{1}\big)
&=e^{\delta}\,\Upsilon\big(\vec{\pcss}\, \big)\;,\lb{eq:upsilon_scale}
\end{align}
where $\vec{1}=(1,1,\ldots,1)^T$ and $\delta$ is a $c$-number.  The only effect of adding a constant $\delta$ to all the $\pcss_j$s, which changes the normalization of the $\lambda_j^2$s by $\rank\delta/n$, is to change $\Upsilon\big(\vec{\pcss}\big)$ and, hence, $\NFB_{\vec s,n}$ by multiplying by a factor $e^\delta$.  This trivial fact implies that
\begin{align}\lb{eq:upsilon_normalization_single_rdm}
 \sum_{k=1}^{\rank} \frac{\partial \Upsilon\big(\vec{\pcss}\,\big)}{\partial \pcss_k}=\frac{\partial \Upsilon\big(\vec{\pcss}+\delta\ssp\vec{1}\,\big)}{\partial \delta}\bigg|_{\delta=0}=\Upsilon\big(\vec{\pcss}\,\big)\;,
\end{align}
which applied to Eq.~(\rf{eq:upsilon_diagonal_elements_rdm}), confirms the normalization condition for the 1RDM:
\begin{align}
\sum_{k=1}^\rank \rho_{k\ssp k}^{(1)}
&\simeq \frac{2n}{\Upsilon\big(\vec{\pcss}\,\big)}\: \sum_{k=1}^\rank \frac{\partial \Upsilon\big(\vec{\pcss}\,\big)}{\partial\pcss_k}=2n\;.
\end{align}
Equation~(\rf{eq:upsilon_normalization_single_rdm}) can be generalized to
\begin{align}
 \sum_{k=1}^{\rank} \pa{}{\pcss_k}\,\frac{\partial^{q} \Upsilon\big(\vec{\pcss}\,\big)}{\partial \pcss_{j_1}\cdots \partial \pcss_{j_q}}=\frac{\partial^{q} \Upsilon\big(\vec{\pcss}\,\big)}{\partial \pcss_{j_1}\cdots \partial \pcss_{j_q}}\;,
\end{align}
which corresponds to the following condition for the higher RDMs\si{Marginalization condition}:
\begin{subequations}
\begin{align}
\sum_{k=1}^\rank \rho_{j_1\cdots j_q\ssp k\ssp ,\, j_1 \cdots j_q\ssp k}^{(q+1)}
&\simeq \frac{(2n)^{q+1}}{\Upsilon\big(\vec{\pcss}\,\big)}\, \sum_{k=1}^\rank\, \pa{}{\pcss_k}\, \frac{\partial^q \Upsilon\big(\vec{\pcss}\,\big)}{\partial\pcss_{j_1} \cdots \partial\pcss_{j_q}}\\
&= \frac{(2n)^{q+1}}{\Upsilon\big(\vec{\pcss}\,\big)}\,  \frac{\partial^q \Upsilon\big(\vec{\pcss}\,\big)}{\partial\pcss_{j_1} \cdots \partial\pcss_{j_q}}\\[2pt]
&= 2n\,  \rho_{j_1\cdots j_q,\, j_1 \cdots j_q}^{(q)}\;. \lb{eq:upsilon_normalization_q_rdm}
\end{align}
\end{subequations}
Notice that we have the factor $2n$, instead of $2n-q$, in Eq~(\rf{eq:upsilon_normalization_q_rdm}); this is because of the approximations we have used, which should be fine for large $n$.

More powerful relations of the RDMs can be derived using the Laplace form~(\rf{eq:upsilon_laplace}),
\begin{align}\lb{eq:upsilon_laplace_s=1}
 \Upsilon\big(\vec{\pcss}\,\big)
   &= \frac{1}{2\pi i}\,  \int_{\Delta-i\infty}^{\Delta+i\infty} e^{\tau }\bigg(\prod_{m=1}^\rank \frac{1}{\sqrt{\tau-\pcss_m}}\bigg)\:\dif \tau\;.
\end{align}
For $\pcss_j\neq \pcss_k$, we find
\begin{subequations}
\begin{align}\hspace{-2em}
 \frac{\partial^2 \Upsilon\big(\vec{\pcss}\,\big)}{\partial \pcss_j\, \partial \pcss_k}
   &= \frac{1}{2\pi i}  \int_{\Delta-i\infty}^{\Delta+i\infty}\frac{e^{\tau }}{4\,(\tau-\pcss_j)(\tau-\pcss_k)}\bigg(\prod_{m=1}^\rank \frac{1}{\sqrt{\tau-\pcss_m}}\bigg)\:\dif \tau\\[3pt]
   &= \frac{1}{2\pi i}  \int_{\Delta-i\infty}^{\Delta+i\infty}\frac{e^\tau }{4\,(\pcss_j-\pcss_k)}\,\bigg(\frac{1}{\tau-\pcss_j}-\frac{1}{\tau-\pcss_k}\bigg)\bigg(\prod_{m=1}^\rank \frac{1}{\sqrt{\tau-\pcss_m}}\bigg)\:\dif \tau\\[3pt]
   &=\frac{1}{2\,(\pcss_j-\pcss_k)}\, \bigg(\frac{\partial \Upsilon\big(\vec{\pcss}\,\big)}{\partial \pcss_j}-\frac{\partial\Upsilon\big(\vec{\pcss}\,\big)}{\partial \pcss_k}\bigg)\;.\lb{eq:upsilon_derivative_a}
\end{align}
\end{subequations}
Equations~(\rf{eq:upsilon_diagonal_elements_rdm}) and (\rf{eq:upsilon_derivative_a}) together give the following relation between the single and two-particle RDMs,
\begin{align}\lb{upsilon_relations_rdms}
 \rho^{(2)}_{j\ssp k,\,j\ssp k}\simeq n\,\frac{\rho^{(1)}_{jj}-\rho^{(1)}_{kk}}{\pcss_j-\pcss_k}\;,
\end{align}
which we rederive below using Wick's theorem [see Eq.~(\rf{wick_relations_rdms})].   We can also write $\rho^{(2)}_{j\ssp k,\,j\ssp k}$ in terms of derivatives of the 1RDMs,
\begin{equation}
\rho^{(2)}_{j\ssp k,\,j\ssp k}
=\frac{(2n)^2}{\Upsilon}
\frac{\partial^2 \Upsilon}{\partial \pcss_j\, \partial \pcss_k}
=\frac{2n}{\Upsilon}\frac{\partial}{\partial\pcss_j}\big(\Upsilon\rho^{(1)}_{kk}\big)
=\rho^{(1)}_{jj}\rho^{(1)}_{kk}+2n\frac{\partial\rho^{(1)}_{kk}}{\partial\pcss_j}\;,
\end{equation}
which can be used to evaluate $\rho^{(2)}_{j\ssp k,\,j\ssp k}$ when $s_j=s_k$ or even when $j=k$.  Once we know $\rho^{(2)}_{j\ssp k,\,j\ssp k}$ for all $k\neq j$, an alternative way to find $\rho^{(2)}_{j\ssp j,\,j\ssp j}$ is by using the marginalization condition
\begin{equation}
\sum_k\rho^{(2)}_{j\ssp k,\,j\ssp k}=(2n-1)\rho_{jj}^{(1)}\;.
\end{equation}
Our conclusion is that the diagonal elements of the 2RDM can be calculated from the diagonal elements of the 1RDM.

The relation~(\rf{upsilon_relations_rdms}) can be generalized to higher orders; for example, we have
\begin{align}
 \prod_{d=1}^q \frac{1}{\tau-\pcss_{j_d}}=\sum_{d=1}^q\,  \Bigg(\,\frac{1}{\tau-\pcss_{j_d}}\, \prod_{\substack{d\ssp'\!=1\\d\ssp'\!\neq d}}^q \frac{1}{\pcss_{j_d}-\pcss_{j_{\smash{d\ssp'}}}}\,\Bigg)\;,
\end{align}
where $j_d\in \{1,2,\ldots,\rank\}$, and $j_d\neq j_{\smash{d\ssp'}}$ for $d\neq d\ssp'$. For the general case where the same $j$ appears multiple times, a similar procedure can be carried out if required, but it becomes increasingly complicated as $q$ increases.

In view of the results in this section, we should address the question of whether the 1RDM\si{1RDM} encodes all the information about a PCS, including the coefficients $\lambda_j$ and the Schmidt orbitals $\psi_j$.  The answer to this question is a decisive no.  One way to see this is to recall that the 1RDM is diagonal in the Schmidt basis and thus is insensitive to the phase of the orbitals.  In other words, the 1RDM remains the same under the transformation $\psi_j\rightarrow e^{i\theta_j}\psi_j$; in contrast, the 2RDM and higher-order RDMs are sensitive to this phase change.

\section{Examples}

In the following, I give some exactly solvable examples which include the case $\rank=2$, the totally degenerate case, and the case where the $\pcss_j$s come in pairs.

For the $\rank=2$ case, we notice that
\begin{subequations}
\begin{align}
 \big(\idG_{\pcss_1}*\idG_{\pcss_2}\big)(\chi)&=\int_{0}^{\chi} \frac{e^{\pcss_1 (\chi-\chi')}\, e^{\pcss_2 \chi'}}{\sqrt{\chi'(\chi-\chi')}}  \; \dif \chi'\\[3pt]
 &=\pi \exp\Big(\frac{\pcss_1 + \pcss_2}{2}\,\chi\Big)\, \BesselI_0\Big(\frac{\pcss_1 - \pcss_2}{2}\,\chi\Big)\;,\lb{eq:convolution_rank_2}
\end{align}
\end{subequations}
where
\begin{equation}
\BesselI_0(x)=\frac{1}{\pi}\int_0^\pi e^{-x\cos\theta}\,d\theta
=\frac{e^x}{\pi}\int_0^1\frac{e^{-2xu}}{\sqrt{u(1-u)}}\,du
\end{equation}
is the zeroth-order modified Bessel function\si{Modified Bessel function}.\footnote{In this dissertation, $\BesselI_j$ stands for the $j$th order modified Bessel function.} Putting Eq.~(\rf{eq:convolution_rank_2}) into Eq.~(\rf{eq:upsilon_f}), we have
\begin{align}\lb{eq:upsilon_s=2}
\Upsilon\big(\pcss_1,\pcss_2\big)
&=e^{\pcss_+}\BesselI_0(\pcss_-)\;,
\end{align}
where
\begin{equation}\lb{eq:spm}
\pcss_{\pm}=\frac{\pcss_1 \pm \pcss_2}{2}\;.
\end{equation}
Notice that if we added $\delta=-(\pcss_1+\pcss_2)/2=-\pcss_+$ to both $s_1$ and $s_2$, as in Eq.~(\rf{eq:upsilon_scale}), we would remove $\pcss_+$ from $\Upsilon\big(\pcss_1,\pcss_2\big)$.

\begin{figure}[ht]
   \centering
   \includegraphics[width=0.76\textwidth,natwidth=610,natheight=642]{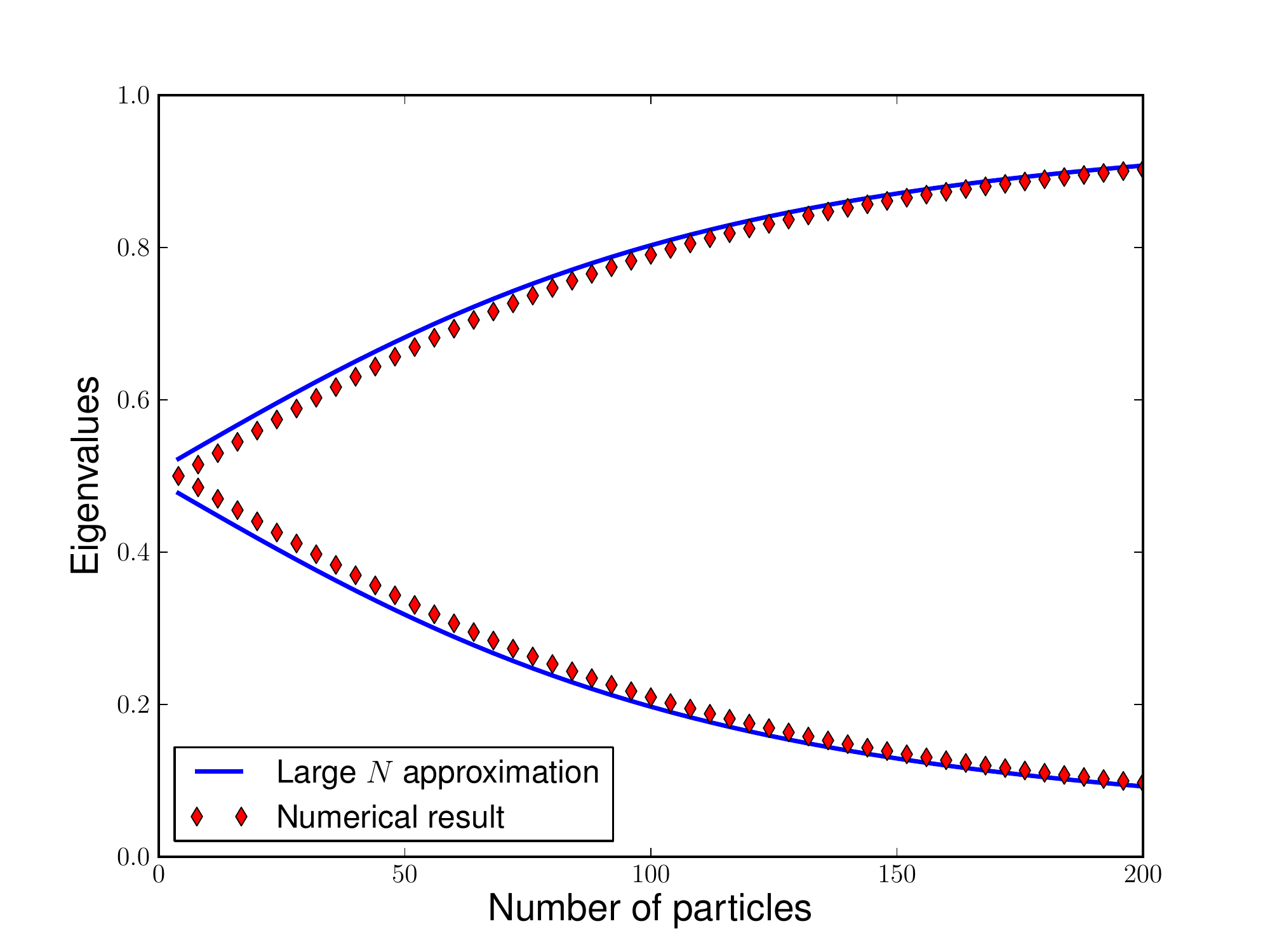}
   \caption[Large-$N$ Approximations vs Fully Numerical Results]{Eigenvalues of the 1RDM (normalized to one) as a function of the number of particles.  The coefficients $\lambda_1$ and $\lambda_2$ are fixed here; i.e., the parameter $\pcss_-$ grows linearly in $N$.  The validity of our approximation in the large-$N$ limit is confirmed by the numerical results.}
   \label{fig:eigenv_comparison}
\end{figure}

It is now straightforward to calculate the 1RDM using\si{1RDM} Eq.~(\rf{eq:upsilon_diagonal_elements_rdm}),
\begin{subequations}\lb{eq:upsilon_s=2_1rdm_a}
\begin{align}
 \rho_{11}^{(1)}
 &\simeq \frac{2n}{\Upsilon\big(\vec{\pcss}\,\big)}\:
 \frac{\partial}{\partial\pcss_1}\,\big(e^{\pcss_+}I_0(\pcss_-)\big)\\[3pt]
 &=\frac{n}{\Upsilon\big(\vec{\pcss}\,\big)}\:e^{\pcss_+}\,\Big( \BesselI_0( \pcss_-)+ \BesselI_1(\pcss_-)\Big)\\[3pt]
 &=n\,\bigg( 1+ \frac{\BesselI_1( \pcss_-)}{\BesselI_0( \pcss_-)}\,\bigg)\;.
\end{align}
\end{subequations}
Similarly, we have
\begin{align}\lb{eq:upsilon_s=2_1rdm_b}
\rho_{22}^{(1)}
&\simeq n\,\bigg(1 - \frac{\BesselI_1( \pcss_-)}{\BesselI_0( \pcss_-)}\,\bigg)\;.
\end{align}
These equations can be compared with fully numerical results, and as shown in Fig.~\ref{fig:eigenv_comparison}, the two conform quite well in the large-$N$ limit.

To see the particle-particle correlations in the $\rank=2$ PCS, we calculate the 2RDM\si{2RDM}; by putting Eq.~(\rf{eq:upsilon_s=2}) into Eq.~(\rf{eq:upsilon_diagonal_elements_rdm}), we have
\begin{subequations}\lb{eq:two_particle_rank2_a}
\begin{align}\hspace{-2em}
 \rho_{11,11}^{(2)}
 &\simeq \frac{4 n^2}{\Upsilon\big(\vec{\pcss}\,\big)}\:
 \frac{\partial^2}{\partial \pcss_1^2}\big(e^{\pcss_+}\,I_0(\pcss_-)\big)\\[3pt]
 &=\frac{2 n^2}{\Upsilon\big(\vec{\pcss}\,\big)}\:
 \frac{\partial}{\partial \pcss_1}\bigg(e^{\pcss_+}\,
 \Big( \BesselI_0( \pcss_-)+ \BesselI_1( \pcss_- )\Big)\bigg)\\[3pt]
 &=n^2\,\bigg( \frac{3}{2}+ 2\,\frac{\BesselI_1(\pcss_-)}{\BesselI_0(\pcss_-)}+\half\, \frac{\BesselI_2(\pcss_-)}{\BesselI_0(\pcss_-)}\bigg)\;.
\end{align}
\end{subequations}
Similarly, we have
\begin{gather}\lb{eq:two_particle_rank2_b}
 \rho_{12,12}^{(2)} \simeq n^2\,\bigg( \half-\half\, \frac{\BesselI_2(\pcss_-)}{\BesselI_0(\pcss_-)}\bigg)\;,\\[6pt]
 \rho_{22,22}^{(2)} \simeq n^2\,\bigg( \frac{3}{2}- 2\,\frac{\BesselI_1(\pcss_-)}{\BesselI_0(\pcss_-)}+\half\, \frac{\BesselI_2(\pcss_-)}{\BesselI_0(\pcss_-)}\bigg)\;. \lb{eq:two_particle_rank2_c}
\end{gather}
It is straightforward to check that these 2RDM elements marginalize correctly, i.e., $\rho_{11,11}^{(2)}+\rho_{12,12}^{(2)}=2n\,\rho_{11}^{(1)}$ and $\rho_{12,12}^{(2)}+\rho_{22,22}^{(2)}=2n\, \rho_{22}^{(1)}$.\footnote{We neglect the difference of $2n$ and $2n-1$ in the large-$N$ limit.}
The relation~(\rf{upsilon_relations_rdms}) can be verified by using the recurrence relations of the Bessel functions,
\begin{align}
\rho_{12,12}^{(2)}&\simeq \frac{n^2}{2}\,\frac{\BesselI_0(\pcss_-)-\BesselI_2(\pcss_-)}{\BesselI_0(\pcss_-)} =\frac{n^2}{\pcss_-}\,\frac{\BesselI_1(\pcss_-)}{\BesselI_0(\pcss_-)}=n\,\frac{\rho_{11}^{(1)}-\rho_{22}^{(1)}}{\pcss_1 - \pcss_2}\;.
\end{align}

In the totally degenerate case\si{Totally degenerate case} ($\pcss_1=\pcss_2=\cdots =\pcss_{\rank}$), all of the eigenvalues of the 1RDM are the same and thus are determined by the normalization condition,
\begin{align}
  \rho_{jj}^{(1)}=\frac{2n}{\rank}\,,\;\; \mbox{for $j=1,2,\ldots,\rank$}\;.
\end{align}
Similarly, we have that the 2RDM matrix elements $\rho_{jj,\, jj}^{(2)}$ are the same for all $j$ and the matrix elements $\rho_{j\ssp k,\, j\ssp k}^{(2)}$ are the same for all $j\ne k$; moreover, putting
the Laplace form~(\rf{eq:upsilon_laplace_s=1}) into Eq.~(\rf{eq:upsilon_diagonal_elements_rdm}) implies that
\begin{align}\lb{eq:diagonal_elements_relation_degenerated}
 \rho_{jj,\, jj}^{(2)}\simeq 3\rho_{j\ssp k,\, j\ssp k}^{(2)}\;,\quad \mbox{for $j\neq k$}\;.
\end{align}
We can now use the marginalization condition~(\rf{eq:upsilon_normalization_q_rdm}) to determine that
\begin{align}\lb{eq:diagonal_elements_degenerated}
  \rho_{j\ssp k,\, j\ssp k}^{(2)}\simeq \frac{4n^2}{\rank(\rank+2)}\:(2\delta_{jk}+1)\;.
\end{align}
This last result is a special case of the general result for diagonal matrix elements in the totally degenerate case.  In this situation, plugging Eq.~(\rf{eq:UpsilonBetaFunctionExpansion}) into Eq.~(\rf{eq:upsilon_diagonal_elements_rdm}) gives us the diagonal matrix elements of the RDMs of all orders,
\begin{subequations}
\begin{align}
\rho_{j_1\cdots j_q,\, j_1 \cdots j_q}^{(q)}
&\simeq\frac{(2n)^q}{\Upsilon\big(0\big)}
\:\bigg.\frac{\partial^q \Upsilon\big(\vec{\pcss}\,\big)}{\partial\pcss_{j_1} \cdots \partial\pcss_{j_q}}\bigg|_{\vec s=0}\\[4pt]
&=\frac{(2n)^q\Gamma(\rank/2)}{\pi^{\rank/2}}B(q_1+1/2,\ldots,q_\rank+1/2)\\[3pt]
&=\frac{n^q\,\Gamma(\rank/2)}{\Gamma(q+\rank/2)}\,\prod_{j=1}^\rank(2q_j-1)!!\\[1pt]
&=\frac{(2n)^q(\rank-2)!!}{(\rank+2q-2)!!}\,\prod_{j=1}^\rank(2q_j-1)!!\;,
\end{align}
\end{subequations}
where $q_j=\sum_{k=1}^q\delta(j,j_k)$ is the number of times $j$ appears in the list of single-particle states, $j_1,\ldots,j_q$.

The degenerate case illustrates that when the system possesses special symmetries, the expressions for matrix elements can sometimes be solved exactly.  Another example occurs when the $\pcss_j$s come in degenerate pairs, i.e., $\pcss_j=\pcss_{j+1}$ for odd $j$.  Note that Eq.~(\rf{eq:convolution_rank_2}) gives the following for $\pcss_1=\pcss_2$:
\begin{align}\lb{eq:degenerate_rank_2}
\big(\idG_{\pcss}*\idG_{\pcss}\big)(\chi)
 &=\pi e^{\chi\ssp\pcss}\;.
\end{align}
Using Eq.~(\rf{eq:upsilon_f}) and (\rf{eq:degenerate_rank_2}), we have\footnote{Another way of deriving Eq.~(\rf{eq:pair_degenerate}) is by using the residue theorem.}
\begin{subequations}\lb{eq:pair_degenerate}
\begin{align}
\Upsilon\big(\chi\ssp \vec{\pcss}\, \big)
&=\chi^{1-\rank/2} \Big(\exp\big(\chi\ssp \pcss_1 \big)*\exp\big(\chi\ssp \pcss_3 \big)*\cdots *\exp\big(\chi\ssp \pcss_{\rank-1} \big)\Big)\\[2pt]
&=\chi^{1-\rank/2}\! \sum_{j\,\in\, \mathrm{odd}}\! \bigg(e^{\chi \ssp \pcss_j} \prod_{\substack{k\,\in\, \mathrm{odd}\\[2pt] k\neq j}}\frac{1}{\pcss_j-\pcss_k}\bigg)\;,\end{align}
\end{subequations}
where $\mathrm{odd}=\{1,3,\ldots, \rank-1\}$. Note that Eq.~(\rf{eq:pair_degenerate}) corresponds to the convolution of many exponential distributions, and results from probability theory can be used.

\section{Iterative Relations}\si{Iterative relations}

The purposes of this section are the following: (i)~to find an effective way of approximating the 1RDMs; (ii)~to gain physical intuition about the 1RDMs; and (iii)~to verify and extend results already obtained. Using Wick's theorem, I will derive a relation between the 1RDMs of PCSs of $2n$ and $2n-2$ particles. In the large-$N$ limit, this iterative relation can then be turned into a differential equation, and a series solution is given for that equation.

In the Schmidt basis, we have\si{1RDM}
\begin{subequations}\lb{eq:iteration_1rdm}
\begin{align}\lb{eq:iteration_1rdm_a}
 \rho_{kj}^{(1)}(n)
 &=\frac{1}{\NFB_n}\, \brab{\vac}\sA^n \, a_j^\dagger a_k\,\big(\sA^\dagger\big)^n  \ketb{\vac}\\[3pt]
 &=\frac{4 n^2}{\NFB_n}\, \lambda_j\lambda_k\, \brab{\vac}\sA^{n-1}\, a_j a_k^\dagger\, \big(\sA^\dagger\big)^{n-1}\ketb{\vac}\\[3pt]
 &=\frac{4 n^2}{\NFB_n}\, \lambda_j\lambda_k \, \brab{\vac}\sA^{n-1}\, \big(a_k^\dagger a_j+\delta_{jk}\big) \,\big(\sA^\dagger\big)^{n-1}\ketb{\vac}\\[3pt]
 &=\frac{4 n^2\ssp \NFB_{n-1}}{\NFB_{n}}\,\Bigl( \lambda_j\lambda_k\: \rho_{jk}^{(1)}(n-1)+  \lambda_j^2\, \delta_{jk} \Bigr)\;.
\end{align}
\end{subequations}
This relation implies that $\rho^{(1)}(n)$ is diagonalized in the Schmidt basis provided that $\rho^{(1)}(n-1)$ is diagonalized.  Since $\rho^{(1)}(1)$ is diagonalized, mathematical induction allows us to conclude that $\rho^{(1)}(n)$ is diagonalized.

That $\rho^{(1)}(n)$ is diagonalized in the Schmidt basis can also be seen directly from Eq.~(\rf{eq:iteration_1rdm_a}): $\big(\sA^\dagger\big)^n\ketb{\vac}$ is a superposition of Fock states that have an even number of particles in each of the Schmidt single-particle states, so in the Fock-state superposition for $a_k\,\big(\sA^\dagger\big)^n\ketb{\vac}$, the single-particle state~$k$ always has an odd number of particles; thus $a_k\,\big(\sA^\dagger\big)^n\ketb{\vac}$ is orthogonal to $a_j\,\big(\sA^\dagger\big)^n\ketb{\vac}$ unless $j=k$.

For the diagonal elements $\rho^{(1)}_{jj}(n)\equiv\varrho_j(n)$, we have
\begin{align}\lb{eq:iteration_diagonal_1rdm}
 \varrho_j(n)=\frac{4 n^2\ssp \NFB_{n-1}}{\NFB_{n}}\, \lambda_j^2\, \big(\varrho_j(n-1)+  1\big)\;.
\end{align}
Using the condition $\sum_j \varrho_j(n) = 2n$, we can do the iteration without knowing the ratio of the normalization factor $\NFB_{n-1}/\NFB_{n}$,
\begin{align}
 \varrho_j(n)=
 2n\,\frac{\lambda_j^2\,[\varrho_j(n-1)+ 1]}{\sum_{k=1}^\rank \lambda_k^2\,[\varrho_k(n-1)+1]}\;.
\end{align}
For sufficiently large $n$, all the probability concentrates on the dominant eigenvalue $\lambda_1$.  To get useful results, we again assume the differences of the $\lambda_j$s are small and use the parametrization~(\rf{eq:lambda_zeta}), $\lambda_j^2\equiv 1+\pcss_j/n$.  To turn the iterative equation into a continuous differential equation, we introduce the parameter $\tau\in\{1/n,\, 2/n\,, \ldots\,, 1\}$ to represent the iteration steps.  In addition, we defer normalization to the end of the process, instead of imposing it at each iteration.  The new iterative equation reads
\begin{subequations}\lb{eq:iteration_1rdm_large_n}
\begin{align}
 \varrho_j(n\tau+1) & = \Big( 1+\frac{\pcss_j}{n}\,\Big)\,\Big(\varrho_j(n\tau)+\frac{1}{2 n\tau}\sum_{k=1}^\rank \varrho_k(n\tau)\Big)\\
 &\simeq \varrho_j(n\tau) +\frac{1}{n}\,\Bigl(\, \pcss_j\ssp  \varrho_j(n\tau)+\frac{1}{2 \tau }\,  \sum_{k=1}^\rank \varrho_k(n\tau)\, \Bigr) \;,
\end{align}
\end{subequations}
where the term of order $1/n^2$ is neglected, and we use $(1/2n\tau)\,\sum_k\varrho_k$ instead of $1$ to deal with the fact that $\varrho_j$ is unnormalized.  Taking the limit $n\rightarrow \infty$, we have the differential equation
\begin{align}
 \frac{\dif \varrho_j}{\dif \tau}=\pcss_j\ssp \varrho_j+\frac{1}{2 \tau}\,  \sum_{k=1}^\rank \varrho_k\;.\lb{eq:iteration_differential_a}
\end{align}
The problem with Eq.~(\rf{eq:iteration_differential_a}) is that it diverges at small $\tau$ due to the factor $1/2\tau$.  This divergence is an artificial consequence of our decision to defer the normalization to the end; one remedy is to modify the differential equation to
\begin{align}\lb{eq:iteration_differential_b}
 \frac{\dif \varrho_j}{\dif \tau}=\Bigl(\pcss_j-\frac{\rank}{2 \tau }\Bigr)\,  \varrho_j+\frac{1}{2 \tau}\,  \sum_{k=1}^\rank \varrho_k\;,
\end{align}
where the extra term, which only introduces an overall factor, keeps $\varrho_j$ from diverging from our initial condition $\varrho_j(0)=1$, for $j = 1,2,\ldots,\rank$.

For the case $\rank=2$, we can decouple the two equations in Eq.~(\rf{eq:iteration_differential_b}) by introducing $\varrho_\pm= \varrho_1 \pm \varrho_2$; the modified Bessel functions $\BesselI_0(\tau)$ and $\BesselI_1(\tau)$ solve these equations.  Thus we have recovered our former results~(\rf{eq:upsilon_s=2_1rdm_a}) and~(\rf{eq:upsilon_s=2_1rdm_b}) using an entirely different approach.

For the general case $\rank>2$, the solution of Eqs.~(\rf{eq:iteration_differential_b}), which are linear equations, can be expanded into a series.  This can be done most conveniently by introducing the matrix equation,
\begin{align}\lb{eq:matrix_iteration_differential}
 \frac{\dif\ssp T(\tau)}{\dif \tau}=\Bigl(Z-\frac{\rank}{2\tau}\, \bigl(\identity-S\bigr) \Bigr)\, T(\tau)\;,
\end{align}
where $T(\tau)$ is the transition matrix, $S_{jk}=1/\rank$ (for $j,k = 1,2,\ldots,\rank$) is the projector onto the symmetric vector $(1,1,\ldots,1)^T$; $Z=\diag(\pcss_1,\pcss_2,\ldots, \pcss_\rank)$, and $\identity$ is the identity matrix. Knowing the transition matrix allows us to solve for
\begin{align}
 \varrho_j(\tau)&=\sum_{k=1}^\rank T_{jk}(\tau)\, \varrho_k(0)=\sum_{k=1}^\rank T_{jk}(\tau)\;,
\end{align}
where the initial condition is $\varrho_k(0)=1$ (for $k = 1,2,\ldots,\rank$). By choosing the initial condition\footnote{We can choose any matrix that stabilizes the vector $(1,\,1,\ldots,1)^T$ to be $T(0)$. The choice $T(0)=S$ is most convenient, because the term of order $\tau^{-1}$ in Eq.~(\rf{eq:taylor_comparison}) disappears automatically.} of the transition matrix to be $T(0)=S$, we have
\begin{align}\lb{eq:taylor_transition_matrix}
T(\tau)=S+\sum_{m=1}^\infty \tau^m\, T_m\;,
\end{align}
where $T_m$, $m=1,2,\ldots$, are matrices to be determined. Putting Eq.~(\rf{eq:taylor_transition_matrix}) into Eq.~(\rf{eq:matrix_iteration_differential}), we get
\begin{align}
\sum_{m=1}^\infty m\, \tau^{m-1}\, T_m=Z\, S+\sum_{m=1}^\infty \tau^m\, Z\, T_m-\frac{\rank}{2}\, \sum_{m=1}^\infty \tau^{m-1} \big(\identity-S\big)\, T_m\;.\lb{eq:taylor_comparison}
\end{align}
Comparing the coefficients of different orders of $\tau$ in Eq.~(\rf{eq:taylor_comparison}), we have
\begin{align}\label{eq:taylor_solving_a}
 \Big( m\, \identity+\frac{\rank}{2}\, \big(\identity-S\big)\Big)\, T_m=Z\, T_{m-1}\,,\;\;\mbox{ for $m\geq 1$}\;.
\end{align}
Because the matrix $m\, \identity+\rank\big(\identity-S\big)/2$ is invertible, we have
\begin{align}\lb{eq:taylor_solving_b}
 T_m=\frac{\identity+\rank S/2m}{m+\rank/2}\: Z\, T_{m-1}\;.
\end{align}
Note that the solution to Eq.~(\rf{eq:taylor_solving_b}) always take the form $T_m= \poly(Z)\ssp S$, where $\poly(Z)$ is some polynomial of the $Z$ matrix.\footnote{For example, we have
$T_0=S$, $T_1=\frac{Z}{1+\rank /2}\,S$, and $T_2=\frac{Z^2+\tr(Z^2)/4}{(1+\rank/2)(2+\rank/2)}\, S$,
where we assume $\tr(Z)=0$ without loss of generality.}  This is because the identity
\begin{align}
\rank S D S=\tr(D)\,S
\end{align}
holds for any diagonal matrix $D$. Thus the matrix equation~(\rf{eq:taylor_solving_b}) can be turned into a $c$-number equation for polynomials.  In addition, we see that the matrix norm of $T_m$ begin to fall quickly after $m > \pcss_1$, which gives us an estimate of how many terms we need to include in the expansion to get a desired precision.

In this section, we have discussed an alternative way of deriving the 1RDMs with a differential equation approach. The results we have here conform with those derived previously. The solution to the differential equations can be expressed as an infinite sum.  Compared to our other approaches, this one allows one us to solve the 1RDMs efficiently, but approximately, and is suitable for numerical evaluation.

\section{Determining the 2RDMs}\si{2RDM}
\lb{sec:2rdms}

In the former sections, we only discussed the diagonal elements of all RDMs. Here, we show
how to represent the off-diagonal elements of the 2RDMs with their diagonal elements. In the Schmidt basis, the 2RDM reads
\begin{align}\lb{eq:two_particle_rdm_a}
 \rho_{k_1 k_2,\,j_1 j_2}^{(2)}= \frac{1}{\NFB}\,\brab{\vac}\sA^n \, \a^\dagger_{j_1} \a^\dagger_{j_2} \a_{k_2} \a_{k_1}\ssp \big(\sA^\dagger\big)^n \ketb{\vac}\;.
\end{align}
All the matrix elements of $\rho^{(2)}$ are real, and it satisfies the normalization condition
\begin{equation}
\tr\rho^{(2)}=\sum_{j,k}\rho_{j\ssp k,\,j\ssp k}^{(2)}=2n(2n-1)\;.
\end{equation}
Using Wick's theorem\si{Wick's theorem}, we find the 2RDM must have the form
\begin{align}\lb{eq:two_particle_rdm_b}
 \rho_{k_1 k_2,\,j_1 j_2}^{(2)}= \xi_{j_1 k_1}\delta_{j_1 j_2}\delta_{k_1 k_2}+\xi'_{j_1\ssp j_2} \delta_{j_1 k_1}\delta_{j_2 k_2}+\xi'_{j_2\ssp
 j_1}\delta_{j_1 k_2}\delta_{j_2\ssp k_1}\;,
\end{align}
where the real matrices $\xi$ and $\xi'$ are symmetric to ensure the Hermiticity of the 2RDM.

In the large-$N$ limit, this form can be further simplified by the condition $\lambda_j/\lambda_k =1+\sO(1/n)$ [see Eq.~(\rf{eq:lambda_zeta})],
\begin{subequations}
\begin{align}
 \rho_{k_1 k_2,\,j_1 j_2}^{(2)}&=\frac{2 n \lambda_k}{\NFB}\, \brab{\vac}\sA^{n}\, \a^\dagger_{j_1} \a^\dagger_{j_2} \a_{k_2} \a_{k_1}^\dagger \ssp\big(\sA^\dagger\big)^{n-1}\ketb{\vac}\\
 &=\frac{2 n \lambda_k}{\NFB}\,\brab{\vac}\sA^{n}\, \a_{k_1}^\dagger  \a^\dagger_{j_2} \a_{k_2}  \a_{j_1}^\dagger\ssp \big(\sA^\dagger\big)^{n-1}\ketb{\vac}+ \sO(n)\\
 &=\frac{1}{\NFB}\, \frac{\lambda_k}{\lambda_j}\: \brab{\vac}\sA^{n}\, \a_{k_1}^\dagger  \a^\dagger_{j_2} \a_{k_2}  \a_{j_1}\ssp \big(\sA^\dagger\big)^{n}\ketb{\vac}+ \sO(n)\\
 &=\frac{1}{\NFB}\: \brab{\vac}\sA^{n}\, \a_{k_1}^\dagger  \a^\dagger_{j_2} \a_{k_2}  \a_{j_1}\ssp \big(\sA^\dagger\big)^{n}\ketb{\vac}+ \sO(n)\\[2pt]
 &=\rho_{j_1 k_2,\,k_1 j_2}^{(2)}+ \sO(n)\;,
\end{align}
\end{subequations}
which means $\rho_{k_1 k_2,\,j_1 j_2}^{(2)}$ is symmetric at the leader order, $n^2$, under the exchange of $j_1$ and $k_1$.  Thus we can further simplify Eq.~(\rf{eq:two_particle_rdm_b}) into
\begin{align}\lb{eq:two_particle_rdm_c}
 \rho_{k_1 k_2,\,j_1 j_2}^{(2)} &\simeq \xi_{j_1 k_1} \delta_{j_1 j_2}\delta_{k_1 k_2}
+\xi_{j_1 j_2}\bigl( \delta_{j_1 k_1}\delta_{j_2\ssp k_2}+\delta_{j_1 k_2}\delta_{j_2\ssp k_1}\bigr)\;,
\end{align}
where the matrix $\xi$ determines $\rho^{(2)}$. Note that all the elements of $\xi$ are nonnegative due to the positiveness of the Schmidt coefficients.  Moreover, the positiveness of $\rho^{(2)}$ is equivalent to the positiveness of the matrix $\xi_{jk} + 2 \delta_{jk} \xi_{jj}$.  Conversely, the matrix $\xi$ can be determined by the diagonal elements of $\rho^{(2)}$,
\begin{align}\lb{eq:two_particle_rdm_diagonal}
 \rho_{j\ssp k,\,j\ssp k}^{(2)} \simeq (2\delta_{jk}+1)\,  \xi_{j k}\;,
\end{align}
and these diagonal elements can be determined from Eq.~(\rf{eq:upsilon_diagonal_elements_rdm}).

In Sec.~\rf{sec:large_N}, we derived a relation~(\rf{upsilon_relations_rdms}) between the diagonal elements of $\rho^{(2)}$ and those of $\rho^{(1)}$.   Here we rederive this relation using Wick's theorem\si{Wick's theorem}, without making the large-$N$ approximation.  By doing contractions of the single annihilation and creation operators with the pair creation and annihilation operators, we have for $j\neq k$,
\begin{subequations}
\begin{align}
 \NFB\, \rho_{j\ssp k,\,j\ssp k}^{(2)}&= \brab{\vac}\sA^n \, \a_j^\dagger \a_k^\dagger \a_k \a_j\, \big(\sA^\dagger\big)^n \ketb{\vac}\\[3pt]
 &= 4n^2 \lambda_j^2\, \brab{\vac}\sA^{n-1}\, \a_j \a_k^\dagger \a_k \a_j^\dagger\, \big(\sA^\dagger\big)^{n-1}\ketb{\vac}\\[3pt]
 &= 4n^2 \lambda_j^2\, \brab{\vac}\sA^{n-1}\, \big(\a_j^\dagger \a_k^\dagger \a_k \a_j+\a_k^\dagger \a_k\big)\, \big(\sA^\dagger\big)^{n-1}\ketb{\vac}\;.\lb{eq:diagonal_element_contraction_a}
\end{align}
\end{subequations}
Another way of contracting gives us
\begin{subequations}
\begin{align}
 \NFB\, \rho_{j\ssp k,\,j\ssp k}^{(2)}
 &= 4n^2 \lambda_k^2\, \brab{\vac}\sA^{n-1}\, \a_k \a_j^\dagger \a_j \a_k^\dagger\, \big(\sA^\dagger\big)^{n-1}\ketb{\vac}\\[3pt]
 &= 4n^2 \lambda_k^2\, \brab{\vac}\sA^{n-1}\, \big(\a_j^\dagger \a_k^\dagger \a_k \a_j+\a_j^\dagger \a_j\big)\, \big(\sA^\dagger\big)^{n-1}\ketb{\vac}\;.\lb{eq:diagonal_element_contraction_b}
\end{align}
\end{subequations}
Multiplying Eqs.~(\rf{eq:diagonal_element_contraction_b}) and (\rf{eq:diagonal_element_contraction_a}) by $\lambda_j^2$ and $\lambda_k^2$, respectively, and subtracting the results gives
\begin{subequations}
\begin{align}\hspace{-1.5em}\lb{eq:diagonal_element_contraction_c}
 \big(\lambda_j^2-\lambda_k^2\big)\, \NFB\, \rho_{j\ssp k,\,j\ssp k}^{(2)}
 &= 4n^2 \lambda_j^2 \lambda_k^2\, \brab{\vac}\sA^{n-1}\, \big(\a_j^\dagger \a_j-\a_k^\dagger \a_k \big)\, \big(\sA^\dagger\big)^{n-1}\ketb{\vac}\\[2pt]
 &= \brab{\vac}\sA^n \, \big(\lambda_k^2\, \a_j^\dagger \a_j-\lambda_j^2\, \a_k^\dagger \a_k \big)\, \big(\sA^\dagger\big)^n \ketb{\vac}\;,
\end{align}
\end{subequations}
which leads to
\begin{align}\lb{wick_relations_rdms}
 \rho_{j\ssp k,\,j\ssp k}^{(2)}
 = \frac{\lambda_k^2\,\rho_{jj}^{(1)}-\lambda_j^2\,\rho_{kk}^{(1)}}{\lambda_j^2-\lambda_k^2}\;.
\end{align}
With the large-$N$ parametrization~(\rf{eq:lambda_zeta}), Eq.~(\rf{wick_relations_rdms}) reproduces Eq.~(\rf{upsilon_relations_rdms}) in the large-$N$ limit.  This relation allows us to determine $\rho_{j\ssp k,\,j\ssp k}^{(2)}$, for $j\neq k$, from the 1RDM, and then $\rho_{j\ssp j,\,j\ssp j}^{(2)}$ is determined by the marginal condition,
\begin{align}
 \rho^{(2)}_{j j,\,j j}=(2n-1)\rho_{jj}^{(1)}-\sum_{k\neq j} \rho^{(2)}_{j\ssp k,\,j\ssp k}\;.
\end{align}

In the large-$N$ limit, we have determined the 2RDM by using the 1RDM in the Schmidt basis, which suggests that the 2RDM is a function of the 1RDM.  This conclusion assumes, however, that the Schmidt basis can be calculated from the 1RDM.  To be explicit, given the 1RDM, one can diagonalize it and thus find the Schmidt coefficients and Schmidt orbitals \emph{up to an arbitrary phase for each orbital}.  The 2RDM~(\rf{eq:two_particle_rdm_c}), written out explicitly in the Schmidt basis, takes the form
\begin{subequations}
\begin{align}\lb{eq:rho2explicit}
\rho^{(2)}&=
\sum_{j,k}\xi_{jk}\Big(
\ket{\psi_k}\bra{\psi_j}\otimes\ket{\psi_k}\bra{\psi_j}\\
&\qquad+\ket{\psi_{j}}\bra{\psi_{j}}\otimes\ket{\psi_{k}}\bra{\psi_{k}}
+\ket{\psi_{k}}\bra{\psi_{j}}\otimes\ket{\psi_{j}}\bra{\psi_{k}}
\Big)\;.
\end{align}
\end{subequations}
The terms in Eq.~(\rf{eq:rho2explicit}) are sensitive to the phases of the Schmidt orbitals when $j\ne k$ and thus cannot be determined from the 1RDM.

In the case $\rank=2$, we can solve the 2RDM\si{2RDM} exactly in the large-$N$ limit. Using the results~(\rf{eq:two_particle_rank2_a}), (\rf{eq:two_particle_rank2_b}), and~(\rf{eq:two_particle_rank2_c}) and also Eq.~(\rf{eq:two_particle_rdm_c}), we have the following expression for the 2RDM in the Schmidt basis, $\{\,\ket{11},\, \ket{12},\, \ket{21},\, \ket{22}$\,\},
\begin{align}\lb{eq:two_particle_rdm_matrix_form}
 \rho^{(2)}\simeq \frac{n^2}{2\, I_0}
 \begin{pmatrix}
  3I_0+4I_1+I_2  &0        &0         & I_0-I_2\\[3pt]
  0              &I_0-I_2  & I_0-I_2  & 0 \\
  0              &I_0-I_2  & I_0-I_2  & 0 \\
  I_0-I_2        & 0       &0         & 3 I_0-4 I_1+ I_2
 \end{pmatrix}\;.
\end{align}
where $I_0$, $I_1$, and $I_2$ are the zeroth, first, and second order modified Bessel functions with argument $\pcss_-=(\pcss_1-\pcss_2)/2$. Equivalently, we can write $\rho^{(2)}$ in the Pauli basis
\begin{align}\hspace{-2em}
 \rho^{(2)}
 &\simeq n^2 \bigg(\identity\otimes \identity +\frac{I_1}{I_0}\: \big(\identity\otimes Z+Z\otimes \identity\big) +\frac{I_0+I_2}{2\, I_0}\: Z\otimes Z+\frac{I_0-I_2}{2\, I_0}\:  X\otimes X\bigg)\;.
\end{align}
The two-particle state~(\rf{eq:two_particle_rdm_matrix_form}) is not entangled; i.e., it has zero concurrence. It is known that all pair-wise entanglement vanishes in large bosonic systems due to the monogamy of entanglement~\cite{koashi_entangled_2000}\si{Monogamy of entanglement}.

Another case that can be solved analytically is the totally degenerate case where $\pcss_1=\pcss_2=\cdots=\pcss_\rank$. Using Eqs.~(\rf{eq:diagonal_elements_degenerated}) and (\rf{eq:two_particle_rdm_c}), we have
\begin{align}
 \rho_{k_1 k_2,\,j_1 j_2}^{(2)} &\simeq \frac{(2n)^2}{\rank(\rank+2)} \Bigl(\delta_{j_1 j_2}\delta_{k_1 k_2}+ \delta_{j_1 k_1}\delta_{j_2\ssp k_2}+\delta_{j_1 k_2}\delta_{j_2 k_1}\Bigr)\;.
\end{align}

\renewcommand{\lb}[1]{\label{pcs_2:#1}}
\renewcommand{\rf}[1]{\ref{pcs_2:#1}}

\chapter{Applying Pair-Correlated States to Fragmented Condensates}
\label{ch:pcs_2}
\chaptermark{Applying PCS to Fragmented Condensates}

\begin{quote}
The fundamental laws necessary for the mathematical treatment of a large part of physics and the whole of chemistry are thus completely known, and the difficulty lies only in the fact that application of these laws leads to equations that are too complex to be solved.\\[4pt]
--  Paul Dirac\ai{Dirac, Paul}
\end{quote}

\noindent In the last chapter, we introduced the Bosonic Particle-Correlated State (BPCS), a state of $N=l\times n$ bosons that is derived by symmetrizing the $n$-fold tensor product of an arbitrary $l$-boson (pure or mixed) state $\sigma^{(l)}$.  In particular, we are interested in the pure-state case for $l=2$, i.e., $\sigma^{(2)}=\ket{\varPsi^{(2)}}\bra{\varPsi^{(2)}}$, which we call a Pair-Correlated State (PCS).  Note that PCS can also be generated by projecting a multimode squeezed vacuum to the $N$-particle sector.\footnote{With the ``quasi-normal-ordered'' factored form of the squeeze operator~\cite{perelomov_generalized_1977, hollenhorst_quantum_1979}, see Eq.~(\chref{pcs_1:eq:quasi_normal_ordered}), the squeezed vacuum takes the form $\ket{\varPsi_\mathrm{sv}}\sim \exp\big(\mathord{-}\half\sum_k\tanh
\gamma_k\,\a_k^{\dagger\, 2}\,\big)\ket{\vac}$.  Projecting this state to the $N$ particle sector gives the \hbox{PCS}.}  When there is one dominant squeezing parameter---or equivalently, a dominant Schmidt coefficient in the two-particle wavefunction---PCS reproduces the number-conserving Bogoliubov approximation.   For the case where many squeezing parameters are of the same size, PCS describes a fragmented state with nearly maximized two-particle quantum correlations.

In this chapter, I will discuss how to use the results from the last chapter to derive equations that determine the PCS ground state.  I also formulate time-dependent equations for the evolution of PCS using the technique described in Chap.~\chref{ch:basics_bec}.  For the two-site Bose-Hubbard model, we calculate the trace distance between the 2RDMs given by the full numerical solution and the closest \hbox{PCS}.  The results confirm that PCS provides a good representation of the ground state over the entire parameter space; more interestingly, for the time-dependent case our numerical simulations suggest that the error in the 2RDM given by PCS does not get larger with increasing evolution time.

\section{PCS Ground State}\si{PCS ground state}

In this section, we discuss how to solve for the PCS ground state using a variational method.  By fixing the PCS parameters $\vec\pcss$ (the eigenvalues of the 1RDM are thus fixed), we find equations that determine the orbitals.  By fixing the orbitals, we write the energy as a function of $\vec\pcss$, and we find the $\vec\pcss$ that minimizes the PCS energy.  Repeating this whole procedure several times allows one to find the PCS ground state approximately.  We also derive a condition for fragmentation using the PCS ansatz.

Consider the state of $N$ bosons modeled by the following Hamiltonian,
\begin{align}
\begin{split}\lb{eq:hamiltonian}
\sH&=\sH_1+\sH_2\\
&=\int\uppsi^\dagger (\xbf)  \Big(\mathord{-}\frac{{\hbar}^2}{2m}\boldsymbol\nabla^2 +V(\xbf)\,\Big) \uppsi(\xbf)\, \dif \xbf+\frac{g}{2}\int [\uppsi^\dagger(\xbf)]^2\,\uppsi^2(\xbf)\, \dif \xbf\;,
\end{split}
\end{align}
where $\sH_1$ and $\sH_2$ are the single- and two-particle Hamiltonians respectively.  The energy expectation value of the Hamiltonian~(\rf{eq:hamiltonian}) for a PCS reads
\begin{align}
 E_\mathrm{pcs}&=\bra{\varPsi_\mathrm{pcs}}\ssp \sH\, \ket{\varPsi_\mathrm{pcs}}=\brab{\varPsi_\mathrm{pcs}} \big(\sH_1 + \sH_2\big)\ketb{\varPsi_\mathrm{pcs}}\;.
\end{align}
The contribution from the single-particle Hamiltonian is
\begin{align}\hspace{-2em}\lb{eq:energy_H1}
\bra{\varPsi_\mathrm{pcs}}\sH_1\ket{\varPsi_\mathrm{pcs}}&=\tr \big(\rho^{(1)} H_1\big) =\sum_{j=1}^\rank \varrho_j\int \psi_j^*(\xbf)\Big(\mathord{-}\frac{i\hbar^2}{2m}\boldsymbol\nabla^2+V(\xbf)\ssp\Big)\psi_j(\xbf)\,\dif \xbf\;,
\end{align}
where $\varrho_j\equiv \rho^{(1)}_{jj}$ is the occupation number of the $j$th orbital $\psi_j(\xbf)$; the contribution from the two-particle Hamiltonian is
\begin{subequations}\lb{eq:energy_H2}
\begin{align}\hspace{-1em}
\bra{\varPsi_\mathrm{pcs}}\sH_2\ket{\varPsi_\mathrm{pcs}}
&=\frac{g}{2}\,\int \brab{\varPsi_\mathrm{pcs}}\, [\uppsi^\dagger(\xbf)]^2\,\uppsi^2(\xbf)\,\ketb{\varPsi_\mathrm{pcs}}\,\dif \xbf\\
&=\frac{g}{2}\sum_{j,k,l,m=1}^\rank  \rho^{(2)}_{l m,\,jk} \int \psi^*_j(\xbf)\psi^*_k (\xbf)\, \psi_m(\xbf) \psi_l(\xbf)\,\dif \xbf \\
&=\frac{g}{2}\sum_{j,k=1}^\rank \xi_{jk}\nsp \int
\Big([\psi^*_j (\xbf)]^2\, \psi_k^2(\xbf)+2\,\norm{\psi_j (\xbf)}^2\, \norm{\psi_k(\xbf)}^2\Big)
\:\dif \xbf\;,\lb{eq:energy_H2_final}
\end{align}
\end{subequations}
where we used the relation~(\chref{pcs_1:eq:two_particle_rdm_c}).  Note that the overall phases of the Schmidt orbitals are important in the first term of Eq.~(\rf{eq:energy_H2_final}).  Combining Eqs.~(\rf{eq:energy_H1}) and (\rf{eq:energy_H2}), we have
\begin{align} \lb{eq:energy}\hspace{-2em}
 E_\mathrm{pcs} &=\sum_{j=1}^\rank \varrho_j\int\Big[ \psi_j^*\Big(\mathord{-}\frac{i\hbar^2}{2m}\boldsymbol\nabla^2+V+\frac{R_j}{2}\ssp\Big)\psi_j+\frac{1}{4}\,\Big( Q_j\ssp \psi_j^*\psi_j^* +Q_j^*\ssp \psi_j\psi_j\ssp\Big)\Big]\,\dif \xbf\;,
\end{align}
where the effective potentials are given by
\begin{align}\lb{eq:effective_potentials}
 R_j(\xbf)&= \frac{g}{\varrho_j}\,\sum_{k=1}^\rank \big(2+\delta_{jk}\big)\ssp \xi_{jk}\,\norm{\psi_k(\xbf)}^2\,,\quad
 Q_j(\xbf)= \frac{g}{\varrho_j}\,\sum_{\substack{k=1,\\k\neq j}}^\rank \xi_{jk} \,\psi_k^2(\xbf)\;.
\end{align}
Note that $R$ and $Q$ are of order $N^0$, because $g$ is of order $N^{-1}$, $\xi_{jk}$ is of order $N^2$, and $\varrho_j$ is of order $N$.

The energy expectation value $E_\mathrm{pcs}$ depends both on the Schmidt orbitals and the PCS parameters $\vec\pcss$.  As a first step, we only allow the Schmidt orbitals to vary by fixing $\vec\pcss$.  To keep the orbitals orthonormal, we introduce the Lagrange multiplier $L= \sum_{j,k}\vartheta_{jk}\braket{\psi_j}{\psi_k}$, where $\vartheta_{jk} = \vartheta_{kj}^*$. Varying the $j$th orbital, we have the following equation for the PCS ground state,
\begin{align}
\hspace{-1.2em}
 0 &= \frac{1}{\varrho_j}\frac{\delta (E_\mathrm{pcs}-L)}{\delta \psi_j^*(\xbf)}\nonumber\\[2pt]
 &= \Big(\mathord{-}\frac{i\hbar^2}{2m}\boldsymbol\nabla^2+ V(\xbf) +R_j(\xbf) \Big)\psi_j(\xbf)+Q_j(\xbf)\ssp\psi_j^*(\xbf)-\sum_{k=1}^\rank \vartheta_{jk} \psi_k(\xbf)\;.
\end{align}
Combining the above equations for $j=1,2,\ldots,\rank$, we can determine the optimal orbitals; a promising candidate to solve these differential equations numerically is the imaginary-time evolution method, where at each time step $\vartheta$ is chosen so that the evolved orbitals remain orthonormal.

After the orbitals are fixed, the energy of PCS takes the simple form
\begin{align}\lb{eq:energy_simple}
 E_\mathrm{pcs} = \sum_{j=1}^\rank \epsilon_j \varrho_j+\frac{1}{2}\sum_{j,k=1}^\rank (1+2\delta_{jk}) \xi_{jk} \eta_{jk}\;,
\end{align}
where
\begin{gather}
\epsilon_j = \int \psi_j^*(\xbf)\Big(\mathord{-}\frac{i\hbar^2}{2m}\boldsymbol\nabla^2+V(\xbf)\ssp\Big)\psi_j(\xbf)\,\dif \xbf\;,\\[3pt]
\eta_{jk}= \frac{g}{1+2\delta_{jk}}\int\Big([\psi^*_j (\xbf)]^2\, \psi_k^2(\xbf)+2\,\norm{\psi_j (\xbf)}^2\, \norm{\psi_k(\xbf)}^2\Big)\,\dif \xbf \;.
\end{gather}
Note that the elements of $\eta_{jk}$ are determined by the Schmidt orbitals and generally are sensitive,
through the first term in the integral, to the overall phases of the Schmidt orbitals.  For the PCS ground state, the chemical potentials of different orbitals must be the same
\begin{align}\lb{eq:equal_chemical_potential}
 \frac{\partial E_\mathrm{pcs}}{\partial \varrho_j} &= \epsilon_j + \frac{1}{2}\sum_{j,k=1}^\rank (1+2\delta_{jk}) \bigg( \frac{\partial \xi_{jk}}{\partial \varrho_j}\, \eta_{jk}+\xi_{jk}\frac{\partial \eta_{jk}}{\partial \varrho_j} \bigg)=\mu\;,
\end{align}
where the partial derivative $\partial /\partial \varrho_j$ is taken by fixing $\varrho_k$ for all $k\neq j$.

Here we discuss the case $\rank=2$ in detail.  In this case we can assume that $\eta_{12}$ is real for the following reason: with everything else held fixed, we can lower the energy of PCS by varying the overall phases of the two orbitals so that the integral $\int  [\psi^*_1 (\xbf)]^2\, \psi_2^2(\xbf)\,\dif \xbf$ is negative, thus making $\eta_{12}$ always real for the PCS ground state and equal to
\begin{align}\lb{eq:eta12}
 \eta_{12}= g\int  \,\norm{\psi_1 (\xbf)}^2\, \norm{\psi_2(\xbf)}^2 \,\dif \xbf\;.
\end{align}
Note that this argument does not work for higher $\rank$ because there are several competing $\eta_{jk}$s to consider.   Moreover, for $\rank=2$, one can argue that the phases of the two orbitals do not depend on $\xbf$, because the kinetic energy term and the integral $\int  [\psi^*_1 (\xbf)]^2\, \psi_2^2(\xbf)\,\dif \xbf$ can be minimized simultaneously by making the phases homogeneous.\footnote{This also might not be true for $\rank>2$, since the minimization of the integral $\int  [\psi^*_1 (\xbf)]^2\, \psi_2^2(\xbf)\,\dif \xbf$ can become $\xbf$-dependent and thus compete with the kinetic term.}  Using the Cauchy-Schwarz inequality, we have
\begin{align}\lb{eq:eta_condition}
 \eta_{12}= g\int  \,\norm{\psi_1 (\xbf)}^2\, \norm{\psi_2(\xbf)}^2 \,\dif \xbf \;\leq\: \sqrt{\eta_{11}\eta_{22}}\;.
\end{align}

With the results~(\chref{pcs_1:eq:upsilon_s=2_1rdm_a}) and~(\chref{pcs_1:eq:upsilon_s=2_1rdm_b}) from Chap.~\chref{ch:pcs_1}, we have
\begin{align}
 \varrho_1=n\,(1 + \chi_1)\;,\qquad
 \varrho_2=n\,(1 - \chi_1)\;,
\end{align}
where $\chi_1(\pcss_-) =\BesselI_1(\pcss_-)/\BesselI_0(\pcss_-)\in [0,\;1)$ is the ratio of the first and zeroth order modified Bessel functions, with $\pcss_-= (\pcss_1-\pcss_2)/2$ being the single relevant PCS parameter.  Without loss of generality, we assume that the occupation number of the first mode is greater or equal to that of the second, i.e., $\pcss_-\ge0$.  Using Eqs.~(\chref{pcs_1:eq:upsilon_s=2_1rdm_a})--(\chref{pcs_1:eq:two_particle_rank2_c}) and the relation~(\chref{pcs_1:eq:two_particle_rdm_diagonal}), we have
\begin{gather}\hspace{-2em}
 \xi_{11} = \frac{n^2}{6}\,\big( 3+ 4\,\chi_1+ \chi_2\big)\;,\quad
 \xi_{12} = \frac{n^2}{2}\,\big( 1- \chi_2\big)\;,\quad
 \xi_{22} = \frac{n^2}{6}\,\big( 3- 4\,\chi_1+ \chi_2\big)\;,
\end{gather}
where $\chi_2(\pcss_-) =\BesselI_2(\pcss_-)/\BesselI_0(\pcss_-)\in [0,\;1)$.  Note that the elements of $\xi_{jk}$, as functions of $\pcss_-$, are properties of how the Schmidt orbitals are populated within the \hbox{PCS}.

The energy expectation value thus reads
\begin{subequations}
\begin{align}\hspace{-2em}
 E_\mathrm{pcs} &= \epsilon_1 \varrho_1 + \epsilon_2 \varrho_2 +\frac{1}{2}\Big( 3\ssp \xi_{11} \eta_{11}+3\ssp \xi_{22} \eta_{22}+\xi_{12} \big(\eta_{12}+\eta_{12}^*\big)\Big)\\
 &= E_0 + \Big(n(\epsilon_1-\epsilon_2)+ n^2\big(\eta_{11}-\eta_{22}\big)\Big)\chi_1+\frac{n^2}{4}\big(\eta_{11}+\eta_{22}-2\eta_{12}\big)\chi_2\\[3pt]
 &= E_0 + n\ssp \Big(\mathord{-} c_1\chi_1(\pcss_-) +\frac{c_2}{2}\,\chi_2(\pcss_-)\Big)\;.\lb{eq:pcs_energy_two_mode}
\end{align}
\end{subequations}
Here
\begin{equation}
E_0=n\,(\epsilon_1+\epsilon_2)+\frac{n^2}{4}(3\eta_{11}+3\eta_{22}+2\eta_{12})
\end{equation}
is the energy expectation value for equal occupation numbers, $\varrho_1=\varrho_2$ (i.e., $\pcss_- = 0$); \begin{equation}
c_1=(\epsilon_2+n\ssp \eta_{22})-(\epsilon_1+n\ssp \eta_{11})
\end{equation}
is the difference in chemical potentials at equal occupation numbers (with the orbital wavefunction fixed);
and
\begin{equation}
c_2= \frac{n}{2}(\eta_{11}+\eta_{22}-2\eta_{12})
\end{equation}
measures the difference between the strengths of intra- and inter-orbital interactions.  From the condition~(\rf{eq:eta_condition}), we have that $c_2\geq 0$.  Note that in the second term in Eq.~(\rf{eq:pcs_energy_two_mode}), $c_1$ and $c_2$ contain the dependence on the Schmidt orbitals, and $\chi_1$ and $\chi_2$ contain the dependence on how the orbitals are populated within the \hbox{PCS}.

Since $\partial E_\mathrm{pcs}/\partial\pcss_-=n\,\big(\partial E_\mathrm{pcs}/\partial\varrho_1-\partial E_\mathrm{pcs}/\partial\varrho_2\big)\,\partial\chi_1/\partial\pcss_-$, the condition~(\rf{eq:equal_chemical_potential}) that the two orbitals have the same chemical potential is simply the extremal condition, $\partial E_\mathrm{pcs}/\partial\pcss_-=0$, satisfied by the minimal energy.  Since $\chi_1(\pcss_-)$ is an odd function and $\chi_2(\pcss_-)$ is an even function, the minimum of $E_\mathrm{pcs}(\pcss_-)$ occurs when $\pcss_-$ has the same sign as $c_1$.  If $c_1$ is negative, the minimum has $\pcss_-<0$, violating our assumption that $\varrho_1\ge\varrho_2$; this is telling us that we should switch the roles of the two Schmidt orbitals to make $c_1$ nonnegative.  In what follows, we assume that $c_1\ge0$ consistent with our assumption that $\varrho_1\ge\varrho_2$.

\begin{figure}[ht]
   \centering
   \includegraphics[width=0.55\textwidth]{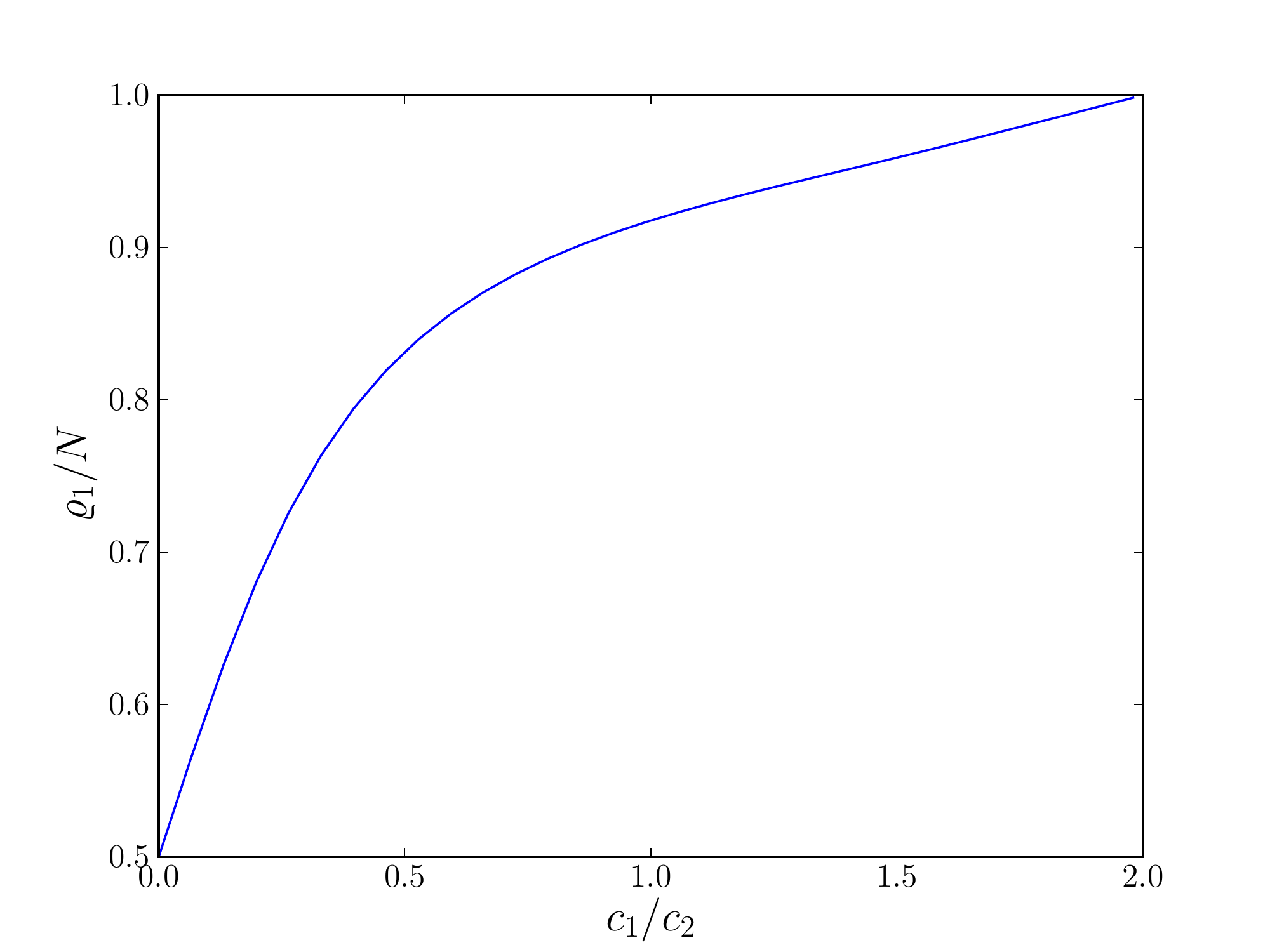}
   \caption[Properties of PCS Ground State]{The occupation ratio $\varrho_1/N$ for the PCS ground state plotted as a function of $c_1/c_2$.  We have equally populated orbitals for $c_1/c_2=0$ and a pure condensate for $c_1/c_2= 2$.}
   \label{fig:pcs_gs_two_mode}
\end{figure}

The function $E_\mathrm{pcs}(\pcss_-)$ seems to be monotonically decreasing when $c_1 \geq 2 c_2$,\footnote{For large $\pcss_-$ we can prove this rigorously by using the following asymptotic expansion of the modified Bessel functions: $I_\alpha(x) \simeq \frac{e^x}{\sqrt{2\pi x}} \left(1 - \frac{4 \alpha^{2} - 1}{8 x} + \frac{(4 \alpha^{2} - 1) (4 \alpha^{2} - 9)}{2! (8 x)^{2}}  + \cdots \right)$.} so the PCS ground state is a pure condensate, i.e., $\pcss_-=+\infty$.  When $c_1 < 2 c_2 $, the terms with coefficients $c_1$ and $c_2$ in Eq.~(\rf{eq:pcs_energy_two_mode}) compete with each other, and the PCS ground state is a fragmented condensate.  In Fig.~\ref{fig:pcs_gs_two_mode}, the occupation ratio $\varrho_1/N$ for the PCS ground state is plotted as a function of $c_1/c_2$; the data are calculated numerically.
From the figure, we see that the particles are distributed evenly between the two orbitals when $c_1/c_2=0$, and there is a single condensate when $c_1/c_2 = 2$.  Note that this single-condensate result does not conflict with the claim made in Sec.~\chref{pcs_1:sec:pcs_bog_relation} that PCS reproduces the number-conserving Bogoliubov approximation, because the depletion of a single condensate predicted by the Bogoliubov approximation is of order $1/N$ and goes away in the large-$N$ limit.  Given these considerations, we have the following condition for fragmentation,
\begin{align}\lb{eq:fragmentation_condition}
 c_1-2c_2=\epsilon_2-\epsilon_1 + N (\eta_{12}-\eta_{11})< 0\;;
\end{align}
fragmentation happens when the two orbitals are close to degenerate, and their wavefunctions have little overlap.

If we choose the dominant orbital $\psi_1$ to be the GP ground state $\psinot$, then there is no $\psi_2$ such that the condition~(\rf{eq:fragmentation_condition}) can be satisfied for a single-component BEC in the large-$N$ limit.  This can be proved by noticing that the quantity in Eq.~(\chref{gp_stable:eq:2nd_order_perturbation}) is nonnegative; by choosing $\psi_1(\xbf) = \psinot(\xbf)$ to be real and $\psi_2(\xbf) = \psinot_\perp(\xbf)$ to be pure imaginary for all $\xbf$, one has that $\epsilon_2-\epsilon_1 + N (\eta_{12}-\eta_{11})\geq 0$.  Since $\psinot_\perp$ can be any single-particle state orthogonal to the condensate wavefunction, the condition~(\rf{eq:fragmentation_condition}) cannot be satisfied.  The dominant orbital of the PCS ground state, however, does not have to be the GP ground state, and a different choice could lead to a fragmented BEC.\cmc{I am not convinced by this argument.  If one had a double-well system and used the GP ground state and the first (odd) excited state, then the condition could not be satisfied, but if one switched to right and left wavefunctions, then it could be.}

The two intertwined aspects of describing a many-body system such as a BEC are, first, which single-particle states to populate and, second, how to populate them.  Within the PCS formalism, the choice of single-particle states is taken up by the choice of Schmidt orbitals; PCS makes a particular choice for how to populate these orbitals, with the remaining freedom within the formalism being the choice of the Schmidt coefficients.  For the case $\rank=2$, the dependence on orbitals is contained in the quantities $\eta_{11}$, $\eta_{22}$, and $\eta_{12}$, and the freedom in populating the orbitals is described by the single parameter $\pcss_-$.  We can study the effectiveness of the PCS ansatz as a way of populating single-particle states by considering contexts in which all the relevant properties of the single-particle states are taken into account by just a few parameters appearing in a many-body Hamiltonian.

One such context, which we turn to now, is a two-site Bose-Hubbard model\si{Bose-Hubbard model} with Hamiltonian
\begin{align}
 \sH_\mathrm{tbh} = -\tau\ssp (\a_1^\dagger \a_2 + \a_2^\dagger \a_1) + \frac{u}{2}\ssp \Big(\sN_1(\sN_1 -1)+\sN_2(\sN_2 -1)\Big)\;,
\end{align}
where $\tau$ is the tunneling strength, $\sN_j = \a_j^\dagger \a_j$ is the particle number operator of the $j$th mode, and $u>0$ describes the on-site repulsion.  We numerically calculate the exact ground-state vector $\ket{\varPsi_\mathrm{exc}(u/\tau)}$ as a function of $u/\tau$ for $N=200$ bosons using the imaginary-time evolution method; we then calculate the exact single- and two-particle RDMs $\rho^{(1,2)}_\mathrm{exc}$ as a function of $u/\tau$ using the ground-state vector.  We construct a PCS 2RDM $\rho^{(2)}_\mathrm{pcs}$ from Eq.~(\chref{pcs_1:eq:two_particle_rdm_matrix_form}) by assuming that its marginal is the exact 1RDM $\rho^{(1)}_\mathrm{exc}$; the relative phase of the two Schmidt orbitals, which $\rho^{(1)}_\mathrm{exc}$ does not provide, are chosen to minimize the energy.

\begin{figure}[ht]
   \centering
   \includegraphics[width=0.55\textwidth]{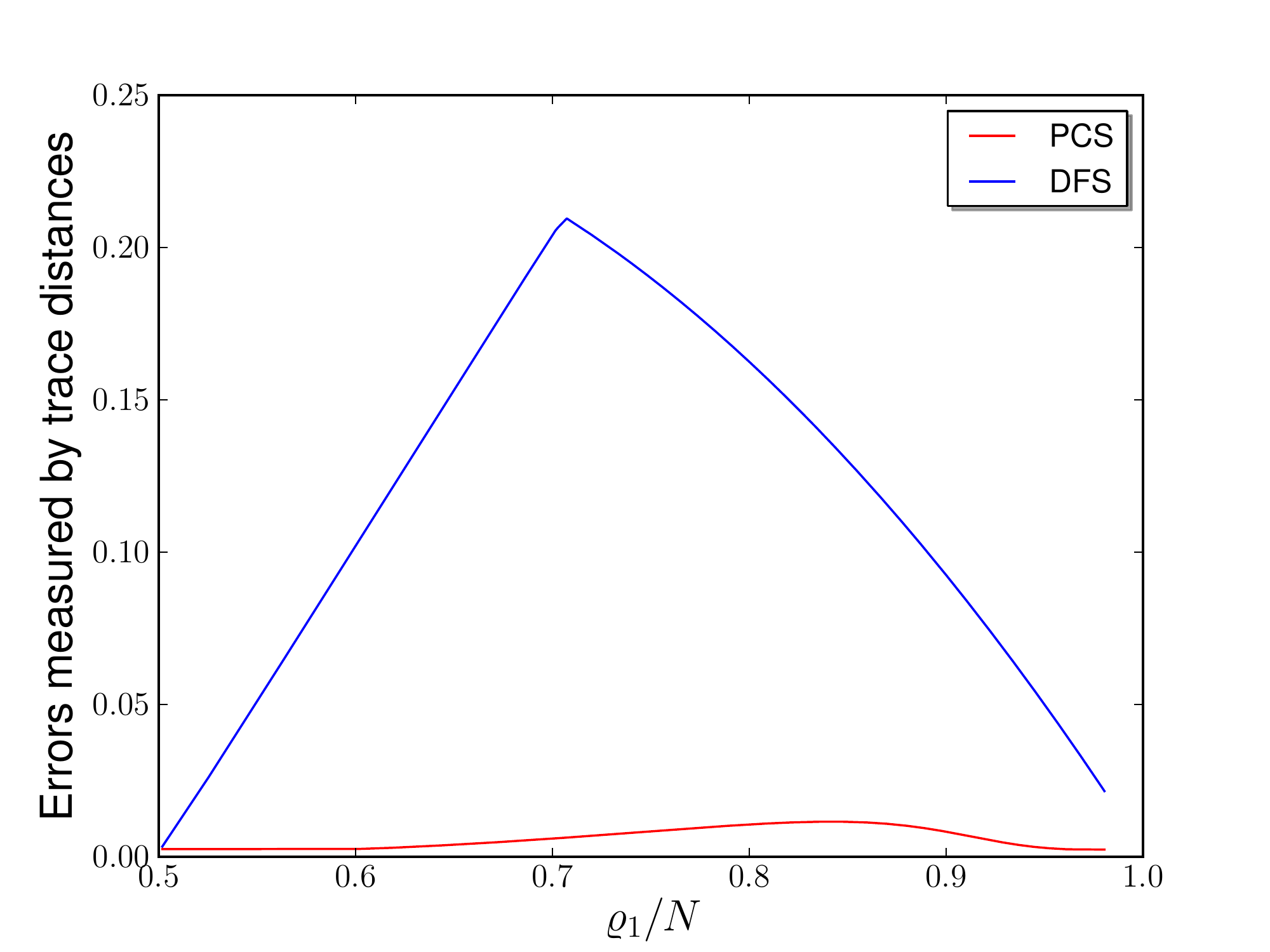}
   \caption[Errors to the 2RDMs of PCS: Ground State]{For the ground state of the 2-site Bose-Hubbard model of $N= 200$ atoms, we plot the trace distances (normalized to one) of the exact 2RDM to the closest 2RDMs given by PCS and DFS as functions of the occupation ratio $\varrho_1/N$.  The particles are split into two halves when $\varrho_1/N = 0.5$ ($u/\tau \rightarrow \infty$), whereas all the particles are condensed into a single mode when $\varrho_1/N = 1$ ($u/\tau \rightarrow 0$).  The small error at $\varrho_1/N = 0$ is due to the finite number of particles and goes away for $N\rightarrow \infty$. }
   \label{fig:pcs_dfs_gs_err}
\end{figure}

The error of the PCS 2RDM, measured by the trace distance $\half \tr\norm{\rho^{(2)}_\mathrm{pcs}-\rho^{(2)}_\mathrm{exc}}$, is plotted in Fig.~\ref{fig:pcs_dfs_gs_err} as a function of the occupation ratio $\varrho_1/N$ (which depends only on $u/\tau$); the error is smaller than two percent over the whole parameter space (hardly any error would be seen had we used fidelity instead of trace distance).  We compare the PCS result to a widely used ansatz, the  double-Fock state (DFS), where $N_1$ and $N_2$ particles occupy the two orbitals.  We also plot, in Fig.~\ref{fig:pcs_dfs_gs_err}, the trace distance $\half \tr\norm{\rho^{(2)}_\mathrm{dfs}-\rho^{(2)}_\mathrm{exc}}$ minimized over all  $\rho^{(2)}_\mathrm{dfs}$ that are consistent with the DFS ansatz, as a function of the ratio $\varrho_1/N$.
This comparison shows that PCS provides a much more faithful representation of the ground state of the 2-site Bose-Hubbard model than DFS.

\begin{figure}[ht]
   \centering
   \includegraphics[width=0.55\textwidth]{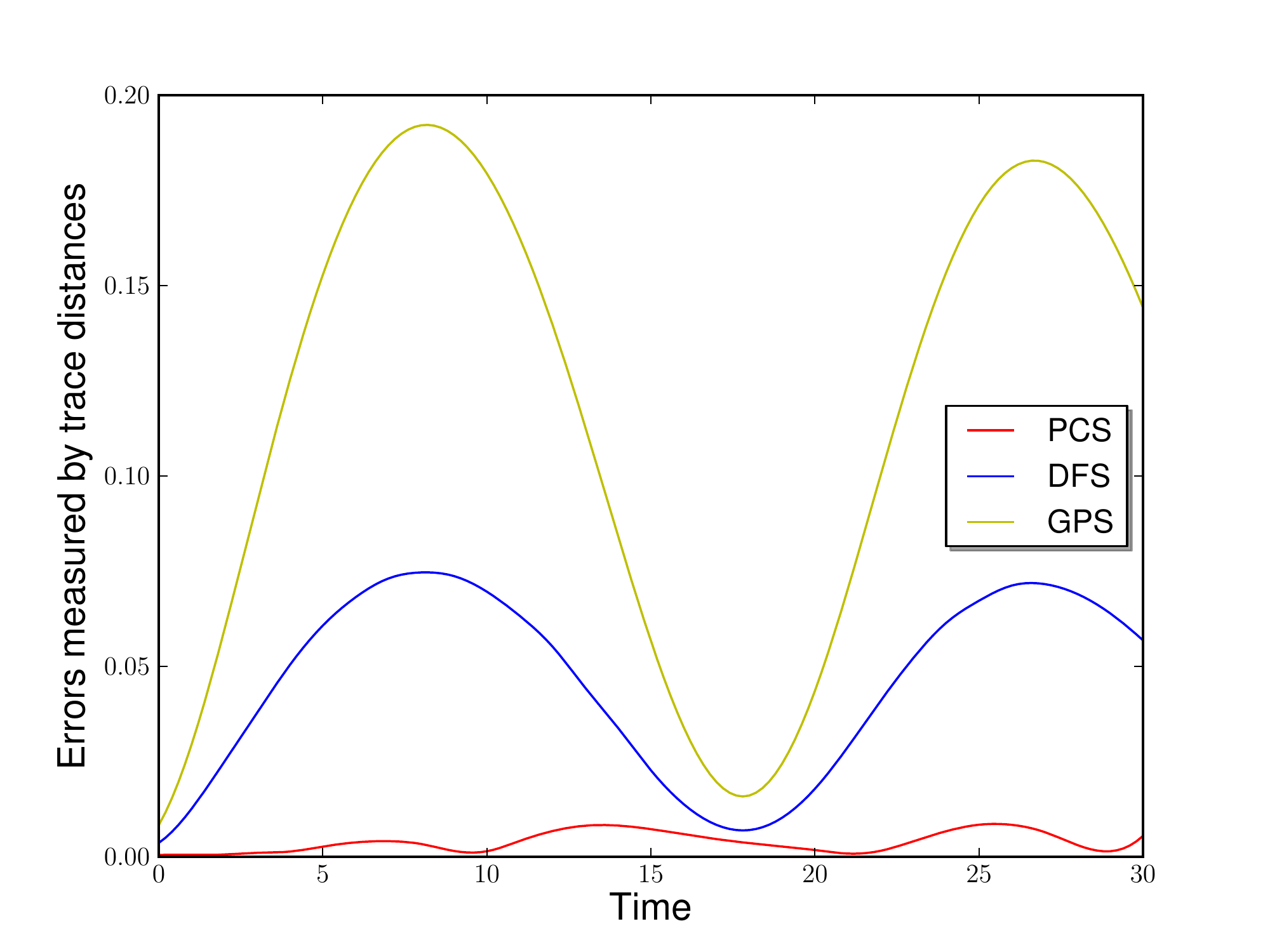}
   \caption[Errors to the 2RDMs of PCS: Time Evolution]{We consider the time evolution of the 2-site Bose-Hubbard model of $N= 1000$ atoms.  Initially, all the atoms are in the ground state of the noninteracting Hamiltonian ($u=0$, $\tau\ne0$), and then a strong on-site interaction is suddenly turned on.  We plot the trace distances (normalized to one) of the exact 2RDM to the closest 2RDMs given by PCS, DFS, and GPS as functions of evolution time.}
   \label{fig:pcs_dfs_td_err}
\end{figure}

Before diving into the time-dependent equations for the PCS ansatz, let us look at how faithfully PCS can represent the time evolution of the two-site Bose-Hubbard model.  We consider the time evolution of the 2-site Bose-Hubbard model of $N= 1000$ atoms.  Initially, all the atoms are in the ground state of the noninteracting Hamiltonian ($u=0$, $\tau\neq 0$), and then strong on-site interactions are suddenly turned on (i.e., $u$ is made very large compared to $\tau$).  We calculate the evolution of the whole state vector numerically using the Crank-Nicolson method, which has second-order precision and is always numerically stable; the exact 2RDM is derived from the state vector.  We then optimize the parameter spaces of PCS, DFS, and a Gross-Pitaevskii State (GPS) to minimize their trace distances to the exact 2RDM.  The errors of the 2RDMs, measured by the trace distances normalized to one, are plotted as functions of evolution time in Fig.~\ref{fig:pcs_dfs_td_err}.
The oscillation of the error given by GPS is a consequence of the collapse and revival of phase~\cite{greiner_collapse_2002, schachenmayer_atomic_2011}; i.e., the purity of the 1RDM oscillates.  These numerical results suggest that PCS might be useful to describing the dynamics in the strongly interacting regime.

\section{Time-Dependent Equations for PCS}
\lb{sec:time-dependent}

In this section, I discuss how to derive time-dependent equations for PCS using the ``projection'' technique developed in Chap.~\chref{ch:basics_bec}.  The key idea is to evolve the 1RDM under the full Hamiltonian and then update the PCS using the evolved 1RDM.  This method, however, is insensitive to the crucial relative phases of the orbitals, so we calculate these phases separately by directly evolving the state vector.  The result is a set of coupled equations that determines the evolution of the orbitals as well as the occupation numbers for these orbitals.  Our results, a generalization of the time-dependent GPE, enables one to solve for the dynamics of a fragmented BEC.

Suppose we have a PCS at time $t$,
\begin{align}\lb{eq:PCS_ansatz}
\ket{\varPsi_\mathrm{pcs}(t)}=\frac{1}{\sqrt{\NFB}}\,[\sA^{\dagger}(t)]^n\,  \ket{{\rm vac}}\;,\qquad \sA^\dagger(t)=\sum_{j=1}^\rank \lambda_j(t)\, [\a^{\dagger}_j(t)]^2\;,
\end{align}
where $\a_{j}^\dagger(t)=\braket{\uppsi}{\psi_j(t)}$ is the creation operator of the $j$th orbital; these orbitals $\psi_j(\xbf,t)$, for $j=1,2,\ldots,\infty$, form a complete, orthonormal basis of single-particle states (note that we are allowing $j>\rank$).  We do not put any time dependence into $\NFB$, because such time dependence can be absorbed into $\sA^\dagger(t)$ as a rescaling of the Schmidt coefficients $\lambda_j(t)$.

Consider now the state $\ket{\varPsi_\mathrm{pcs}(t)}$ evolving under the BEC Hamiltonian
\begin{subequations}\lb{eq:full_hamiltonian}
\begin{align}\hspace{-2em}
\sH(t)&=\sH_1(t)+\sH_2(t)\\
&=\int\uppsi^\dagger (\xbf)  \Big(\mathord{-}\frac{{\hbar}^2}{2m}\boldsymbol\nabla^2 +V(\xbf,t)\,\Big) \uppsi(\xbf)\, \dif \xbf+\frac{g(t)}{2}\int [\uppsi^\dagger(\xbf)]^2\,\uppsi^2(\xbf)\, \dif \xbf\;,
\end{align}
\end{subequations}
where $\sH_1$ and $\sH_2$ are the single- and two-particle Hamiltonians respectively.  After letting the state~(\rf{eq:PCS_ansatz}) evolve for a time $\dif t$ under the Hamiltonian~(\rf{eq:full_hamiltonian}), we ``project'' it back into the PCS submanifold; the rate of error in the state vector that the ``projecting'' step introduces is
\begin{align}\lb{eq:error_PCS_ansatz}
 \ket{\dt \varPsi_\mathrm{err}(t)} &=  \frac{\sH(t)}{i\hbar}\, \ket{\varPsi_\mathrm{pcs}(t)}-\ket{\dt \varPsi_\mathrm{pcs}(t)}\;,
\end{align}
where $\ket{\dt\varPsi_\mathrm{pcs}(t)}$ stands for the time evolution within the manifold of PCS states.

As in the product ansatz developed in Chap.~\chref{ch:basics_bec}, minimizing the error introduced by the projection means requiring that the error vector $\ket{\dt \varPsi_\mathrm{err}(t)}$ be perpendicular to any variation within the PCS manifold.  Thus the evolution of the PCS state is determined by the condition
\begin{equation}\lb{eq:PCS_variation_a}
 \braket{\delta{\varPsi_\mathrm{pcs}}(t)}{\dt \varPsi_\mathrm{err}(t)}=0\quad \mbox{for all\; $\delta{\varPsi_\mathrm{pcs}}(t)$}\;.
\end{equation}
For a small variation of the two-particle creation operator $\sA^\dagger(\epsilon)= \sA^\dagger+\epsilon\ssp \sV^\dagger/n$, where $\sV$ can be any quadratic function of the field operator $\uppsi(\xbf)$, we have
\begin{equation}\lb{eq:PCS_variation_b}
 \ket{\delta{\varPsi_\mathrm{pcs}}}=\frac{\partial}{\partial \epsilon}\, \ket{\varPsi_\mathrm{pcs}(\epsilon)}\Big\vert_{\epsilon=0} = \frac{1}{\sqrt \NFB}\, \sV^\dagger \,(\sA^{\dagger})^{n-1}\, \ket{\vac}\;. 
\end{equation}
Note that we treat the normalization factor $\NFB$ as constant when $\sA$ changes; i.e., we allow the perturbed state to be unnormalized.  Putting Eq.~(\rf{eq:PCS_variation_b}) into Eq.~(\rf{eq:PCS_variation_a}) gives
\begin{align}\lb{eq:orthogonal_condition_PCS_ansatz}
 \brab{\vac}\,[\sA(t)]^{n-1}\,\sV \,\ketb{\dt \varPsi_\mathrm{err}(t)}=0 \quad \mbox{for all $\sV$ quadratic in $\uppsi\,$.}
\end{align}
Putting Eq.~(\rf{eq:error_PCS_ansatz}) into Eq.~(\rf{eq:orthogonal_condition_PCS_ansatz}) determines the evolution of the PCS state $\ket{\dt \varPsi_\mathrm{pcs}}$.

In particular, for $\sV=\sA$, we have $\ket{\delta{\varPsi_\mathrm{pcs}}}= \ket{\varPsi_\mathrm{pcs}}$;
putting this into Eq.~(\rf{eq:PCS_variation_a}) and keeping only the real part gives
\begin{subequations}\lb{eq:normalization_constant}
\begin{align}
 0&=\re\Big(\frac{1}{i\hbar}\,\bra{\varPsi_\mathrm{pcs}(t)}\,\sH(t)\, \ket{\varPsi_\mathrm{pcs}(t)}-\braket{\varPsi_\mathrm{pcs}(t)}{\dt \varPsi_\mathrm{pcs}(t)}\, \Big)\\[4pt]
 &= -\re\Big(\braket{\varPsi_\mathrm{pcs}(t)}{\dt\varPsi_\mathrm{pcs}(t)}\Big)\;,
\end{align}
\end{subequations}
which says that our procedure preserves the norm of $\ket{\varPsi_\mathrm{pcs}}$,
\begin{align}
 \braket{\varPsi_\mathrm{pcs}(t)}{\varPsi_\mathrm{pcs}(t)}=\mathrm{constant}\;.
\end{align}

We can write Eq.~(\rf{eq:orthogonal_condition_PCS_ansatz}) equivalently as
\begin{align}\lb{eq:orthogonal_condition_PCS_ansatz_equivalent}
 \brab{\vac}\,[\sA(t)]^{n-1}\,\a_j(t)\ssp \a_k(t) \,  \ketb{\dt \varPsi_\mathrm{err}(t)}=0\;,\quad \mbox{for all\; $j,k\in\{1,2,\ldots,\infty\}$}\;,
\end{align}
which will be used throughout the rest of this dissertation.  Directly evaluating $\ket{\dt\varPsi_\mathrm{pcs}(t)}$ with Eq.~(\rf{eq:orthogonal_condition_PCS_ansatz_equivalent}) is hard, because it usually involves calculating the three-particle RDM arising from the product of the $\sH_2$ and the term $\a_j(t)\ssp \a_k(t)$.  To avoid this complication, we notice that for all $j,k\in\{1,2,\ldots,\infty\}$,
\begin{align}\hspace{-2em}\lb{eq:orthogonal_condition_PCS_ansatz_alternative}
 \brab{\varPsi_\mathrm{pcs}(t)}\,\a_j^\dagger(t)\, \a_k(t) \,  \ketb{\dt \varPsi_\mathrm{err}(t)}&=\frac{2n}{\sqrt \NFB}\,\lambda_j(t)\,\brab{\vac}\,[\sA(t)]^{n-1}\,\a_j(t)\, \a_k(t)\,\ketb{\dt \varPsi_\mathrm{err}(t)}=0\;,
\end{align}
where $\lambda_j=0$ for all $j>\rank$.  This equation can be recast in terms of the field operators as
\begin{subequations}
\begin{align}
0&=
\sum_{j,k=1}^\infty \psi_j^*(\mathbf{x'},t)\, \psi_k(\xbf,t)\,\brab{\varPsi_\mathrm{pcs}(t)}\,a_j^\dagger(t)\, a_k(t)\,\ketb{\dt \varPsi_\mathrm{err}(t)}\\[2pt]
&=\brab{\varPsi_\mathrm{pcs}(t)}\,
\uppsi^\dagger(\mathbf{x'})\,\uppsi(\xbf)\,
\ketb{\dt \varPsi_\mathrm{err}(t)}\;.
\lb{eq:orthogonal_condition_PCS_ansatz_alternative2}
\end{align}
\end{subequations}
The physical content of Eq.~(\rf{eq:orthogonal_condition_PCS_ansatz_alternative2}) is that the 1RDM,
\begin{subequations}\lb{eq:single_particle_correlation_function}
\begin{align}
 \rho^{(1)}\big(\xbf, \mathbf{x'}\,;\, t\big)
 &=\brab{\varPsi_\mathrm{pcs}(t)}\,\uppsi^\dagger(\mathbf{x'})\, \uppsi(\xbf)\,\ketb{\varPsi_\mathrm{pcs}(t)}\\[2pt]
 &=\sum_{j,k=1}^\infty \psi_j^*(\mathbf{x'},t)\, \psi_k(\xbf,t)\,\brab{\varPsi_\mathrm{pcs}(t)}\,a_j^\dagger(t)\, a_k(t)\,\ketb{\varPsi_\mathrm{pcs}(t)}\\
 &=\sum_{j=1}^\rank\, \varrho_j\, \psi_j(\xbf,t)\, \psi_j^*(\xbf',t)\;,
\end{align}
\end{subequations}
is left unchanged by the projection procedure; i.e., the change in the 1RDM under the evolution within the PCS manifold is the same as the change under the exact evolution.  Thus a promising procedure for determining the dynamics of PCS is to evolve $\rho^{(1)}(t)$ exactly for a short time $\dif t$ and then to update $\ket{\varPsi_\mathrm{pcs}(t+\dif t)}$ using $ \rho^{(1)}(t+\dif t)$.  The problem with this procedure is that the 1RDM does not encode the phases of the orbitals; i.e., $\rho^{(1)}$ remains the same for the transformation $\psi_j(\xbf) \rightarrow e^{i\ssp \theta_j}\psi_j(\xbf)$ with $\theta_j$ real.  These phases, however, play a crucial role in determining $\rho^{(2)}$; for example, the process of annihilating two particles in the $j$th orbital while creating two particles in the $k$th orbital feels the relative phase $e^{i(\theta_j-\theta_k)}$.

Despite this problem, we can make substantial progress by considering the evolution of the 1RDM.  According to the BBGKY hierarchy\si{BBGKY hierarchy}, $\dt\rho^{(1)}$ is a function of $\rho^{(2)}$ when there are only up to two-particle interactions; luckily, we anticipated this situation and considered the 2RDM in detail in Chap.~\chref{ch:pcs_1}.  The exact evolution of the 1RDM reads
\begin{align}\lb{eq:time_evolution_single_particle_rdm}
 i\hbar\, \dt{\rho}^{(1)}\big(\xbf, \mathbf{x'}\,;\, t\big)
  &=\brab{\varPsi_\mathrm{pcs}(t)}\,\commutb{\uppsi^\dagger(\mathbf{x'})\, \uppsi(\xbf)}{\sH(t)}\, \ketb{\varPsi_\mathrm{pcs}(t)}\;.
\end{align}
The commutator in Eq.~(\rf{eq:time_evolution_single_particle_rdm}) can be divided into two parts:
\begin{align}
 \commutb{\uppsi^\dagger(\mathbf{x'})\, \uppsi(\xbf)}{\sH(t)}=\commutb{\uppsi^\dagger(\mathbf{x'})\, \uppsi(\xbf)}{\sH_1(t)}+\commutb{\uppsi^\dagger(\mathbf{x'})\, \uppsi(\xbf)}{\sH_2(t)}\;.
\end{align}
The first part, due to the single-particle energy, corresponds to a common unitary evolution of the orbitals.
The second part, due to the interparticle interaction, assumes the form
\begin{subequations}\lb{eq:commutator_h2}
\begin{align}
\commutb{\uppsi^\dagger(\mathbf{x'})\, \uppsi(\xbf)}{\sH_2}&=\frac{g}{2}\int\commutB{\uppsi^\dagger(\mathbf{x'})\, \uppsi(\xbf)}{ \big[\uppsi^\dagger(\mathbf{x''})\big]^2\,\uppsi^2(\mathbf{x''})}\,\dif \mathbf{x''}\\[3pt]
&=g\,\Big(\uppsi^\dagger(\mathbf{x'})\,\uppsi^\dagger(\xbf)\,\uppsi^2(\xbf)-\big[\uppsi^\dagger(\mathbf{x'})\big]^2\,\uppsi(\mathbf{x'})\,\uppsi(\xbf)\Big)\;.
\end{align}
\end{subequations}
Putting Eq.~(\rf{eq:commutator_h2}) into Eq.~(\rf{eq:time_evolution_single_particle_rdm}), we have the following time derivative of the 1RDM induced by $\sH_2$,
\begin{subequations}
\begin{align}\hspace{-1.5em}
i\hbar\, \dt{\rho}_{\sH_2}^{(1)}&=\brab{\varPsi_\mathrm{pcs}}\,\commutb{\uppsi^\dagger(\mathbf{x'})\, \uppsi(\xbf)}{\sH_2}\, \ketb{\varPsi_\mathrm{pcs}}\\[5pt]
&=g\: \braB{\varPsi_\mathrm{pcs}}\,\Big(\uppsi^\dagger(\mathbf{x'})\,\uppsi^\dagger(\xbf)\,\uppsi^2(\xbf)-\big[\uppsi^\dagger(\mathbf{x'})\big]^2\,\uppsi(\mathbf{x'})\,\uppsi(\xbf)\Big)\, \ketB{\varPsi_\mathrm{pcs}}\\[2pt]
&=g\!\!\sum_{j,k,l,m=1}^\rank  \rho^{(2)}_{l m,\,jk}\: \psi^*_j(\mathbf{x'})\Big(\psi^*_k (\xbf)\, \psi_m(\xbf) -\psi^*_k (\mathbf{x'})\, \psi_m(\mathbf{x'})\Big)\psi_l(\xbf)\;.
\end{align}
\end{subequations}
Using the large-$N$ expression~(\chref{pcs_1:eq:two_particle_rdm_c}) for the 2RDM, we have
\begin{subequations}
\begin{align}\hspace{-1em}
\begin{split}
i\hbar\, \dt{\rho}_{\sH_2}^{(1)}&\simeq g\!\!\! \sum_{j,k,l,m=1}^\rank \Big(\xi_{jl}\, \delta_{jk}\delta_{lm}+ \xi_{jk}\big( \delta_{jl}\delta_{km}+\delta_{jm}\delta_{kl}\big)\Big) \\
&\hspace{5em} \times \psi^*_j(\mathbf{x'})\Big(\psi^*_k (\xbf)\, \psi_m(\xbf) -\psi^*_k (\mathbf{x'})\, \psi_m(\mathbf{x'})\Big)\psi_l(\xbf)
\end{split}\\
\begin{split}\lb{eq:evolution_single_particle_rdm_H2}
&=g \sum_{j,k=1}^\rank \xi_{jk}\,  \Big( \psi_k^2(\xbf)\,\psi^*_j (\xbf)\, \psi^*_j(\mathbf{x'}) - \psi_j(\xbf)\,\psi_j(\mathbf{x'})\,[\psi^*_k(\mathbf{x'})]^2\\
&\hspace{5.6em} + 2\,\norm{\psi_k (\xbf)}^2\, \psi_j(\xbf)\,\psi^*_j(\mathbf{x'})-2\,\psi_j(\xbf)\,\psi^*_j (\mathbf{x'}) \, \norm{\psi_k(\mathbf{x'})}^2\Big)\;.
\end{split}
\end{align}
\end{subequations}
Introducing the same effective potentials as were defined in Eq.~(\rf{eq:effective_potentials}) for the time-independent case,
\begin{align}
 R_j(\xbf)&= \frac{g}{\varrho_j}\,\sum_{k=1}^\rank \big(2+\delta_{jk}\big)\ssp \xi_{jk}\,\norm{\psi_k(\xbf)}^2\,,\qquad
 Q_j(\xbf)= \frac{g}{\varrho_j}\,\sum_{\substack{k=1,\\k\neq j}}^\rank \xi_{jk} \,\psi_k^2(\xbf)\;,
\end{align}
we have the following time-dependent equation for the evolution of the Schmidt orbitals:
\begin{align}\lb{eq:evolution_single_particle_correlation_matrix}
 i\hbar\, \dt{\psi}_j=\Big(\mathord{-}\frac{\hbar^2}{2m}\boldsymbol\nabla^2 +V+ R_j\,\Big) \ssp \psi_j+ Q_j\ssp \psi^*_j\quad \mbox{for $j\in \{1,2,\ldots,\rank\}$}\;.
\end{align}
Note that Eq.~(\rf{eq:evolution_single_particle_correlation_matrix}) reduces to the GP equation when $\rank=1$; for $\rank > 1$ we have $\rank$ coupled nonlinear equations describing the dynamics of the orbitals.

As mentioned before, our procedure of using the 1RDM to update the PCS fails to capture the dynamics of the phases, $\im\braket{\psi_j}{\dt{\psi}_j}$.  Moreover, Eq.~(\rf{eq:evolution_single_particle_correlation_matrix}) does not maintin the orthonormality of the orbitals.  I turn now to discussing how to fix these two glitches.

To evaluate the dynamics of phases of the orbitals, we use the condition~(\rf{eq:orthogonal_condition_PCS_ansatz_equivalent}),
\begin{align}\lb{eq:orthogonal_condition_PCS_ansatz_equivalent_b}
 0 = \bra{\vac}\,\sA^{n-1}\,\a_j\a_k\,  \ket{\dt \varPsi_\mathrm{err}}= \braB{\vac}\,\sA^{n-1}\,\a_j\a_k\, \Big(\frac{\sH}{i\hbar}-\frac{\dif}{\dif t}\Big)\, \ketB{\varPsi_\mathrm{pcs}}\;,
\end{align}
which is stronger than the approach of evolving the 1RDM and can be used to determine the orbital phases.  In the second-quantized picture, a change of the phase of the $j$th orbital takes the form $i\hbar\ssp\ssp\dt \a_j^\dagger = \mu_j\ssp \a_j^\dagger$, where the real number $\mu_j$ is to be determined.  The dynamics of the pair creation operator for this restricted change reads
\begin{align}
 \dt\sA = \frac{d}{\dif t}\sum_{j=1}^\rank \lambda_j \a_j^\dagger \a_j^\dagger = 2\sum_{j=1}^\rank \lambda_j \dt\a_j^\dagger \a_j^\dagger = \frac{2}{i\hbar}\sum_{j=1}^\rank \mu_j\lambda_j \a_j^\dagger \a_j^\dagger\;,
\end{align}
which is diagonal in the Schmidt basis.  Equation~(\rf{eq:orthogonal_condition_PCS_ansatz_equivalent_b}) can thus be simplified to
\begin{subequations}\hspace{-3em}
\begin{align}
&\hspace{-2em}\im\,\braB{\vac}\, \sA^{n-1}\,\a_j\a_j\, \frac{1}{i\hbar}\,\Big(\sH\, \big(\sA^{\dagger}\big)^n-2\ssp n\sum_{k=1}^\rank  \mu_k\lambda_k\a_k^\dagger\a_k^\dagger \big(\sA^{\dagger}\big)^{n-1}\Big)\,\ketB{\vac}=0\\[3pt]
&\hspace{-2em}\Longrightarrow\; \re\,\braB{\vac}\, \sA^{n-1}\,\a_j\a_j\, \Big(\sH\, -\sum_{k=1}^\rank  \mu_k\a_k^\dagger\a_k\Big)\big(\sA^{\dagger}\big)^n\,\ketB{\vac}=0\\[3pt]
&\hspace{-2em}\Longrightarrow\; \frac{\partial}{\partial \lambda_j}\,\braB{\vac}\,\sA^n\, \Big(\sH - \sum_{k=1}^\rank  \mu_k\a_k^\dagger\a_k\Big)\big(\sA^{\dagger}\big)^n\ketB{\vac}=0\;.\lb{eq:derivative_zero}
\end{align}
\end{subequations}
Thus the expectation value in Eq.~(\rf{eq:derivative_zero}) remains unchanged for any infinitesimal increment $\dif\vec\lambda$ of the coefficients $\vec\lambda$.  When $\dif\vec\lambda$ is parallel to $\vec\lambda$, this expectation value is simply rescaled, which gives the condition
\begin{align}\lb{eq:scale_invariance}
 \braB{\vac}\,\sA^n\, \Big(\sH - \sum_{k=1}^\rank  \mu_k\a_k^\dagger\a_k\Big)\big(\sA^{\dagger}\big)^n\ketB{\vac}=0\;.
\end{align}
For any $\dif\vec\lambda$ keeping the normalization factor $\NFB_{\vec\lambda}$ unchanged,\footnote{We can always make an arbitrary $\dif\vec\lambda$ conserve the normalization by adding a term that is proportional to $\vec\lambda$.} we have
\begin{align}
\hspace{-2em}\dif\, \braB{\varPsi_\mathrm{pcs}(\vec \lambda)}\, \Big(\sH - \sum_{k=1}^\rank  \mu_k\a_k^\dagger\a_k\Big)\ketB{\varPsi_\mathrm{pcs}(\vec \lambda)}= \dif\,\Big(E(\vec \lambda)-\sum_{k=1}^\rank \mu_k\varrho_k(\vec \lambda)\Big)=0\;.\lb{eq:chemical_potential}
\end{align}
Because we have the constraint $\sum_{k=1}^\rank \dif \varrho_k = 0$, this result implies that $\mu_k$ equals the chemical potential of the $k$th orbital up to a constant, which can be determined by using Eq.~(\rf{eq:scale_invariance}).  The constant, however, only adds an overall phase to the PCS state, and we can neglect it by letting $\mu_j$ be the chemical potential of the $j$th orbital,
\begin{align}
 \mu_j = \frac{\partial E(\vec\varrho)}{\partial \varrho_j}\bigg\vert_{\,\mbox{\footnotesize fixing all $\varrho_k$s for $k\neq j$}\;,}
\end{align}
where $E(\vec\varrho) = E(n,\vec\lambda)$ is the energy expectation value.  According to Eq.~(\rf{eq:evolution_single_particle_correlation_matrix}), the rate that the phase of $\psi_j(\xbf)$ changes is $\re\,\bra{\psi_j}\ssp\ssp i\hbar\,\ket{\dt{\psi}_j}$, which generally does not equal to the chemical potential $\mu_j$; the difference is
\begin{subequations}
\begin{align}
 \Delta &= \re\,\bra{\psi_j}\ssp\ssp i\hbar\,\ket{\dt{\psi}_j}-\mu_j\\
 &= \braB{\psi_j}\mathord{-}\frac{\hbar^2}{2m}\boldsymbol\nabla^2 +V+ R_j\ketB{\psi_j}+\re\,\brab{\psi_j}Q_j\ketb{\psi_j^*} -\mu_j\\[3pt]
 &= \brab{\psi_j}R_j\ketb{\psi_j}+\re\,\brab{\psi_j}Q_j\ketb{\psi_j^*} -\mu_j^{(2)}\;,
\end{align}
\end{subequations}
where
\begin{align}
 \mu_j^{(2)} = \frac{\partial}{\partial \varrho_j}\, \brab{\varPsi_\mathrm{pcs}}\sH_2 \ketb{\varPsi_\mathrm{pcs}}\;.
\end{align}
The difference $\Delta$ vanishes if we have $\xi_{jk} = \varrho_j\varrho_k$, but this is generally not true.

Another problem of Eq.~(\rf{eq:evolution_single_particle_correlation_matrix}), that it does not conserve the norm of $\psi_j(\xbf)$, can be fixed by absorbing the extra factor into the occupation number $\varrho_j$.  Assuming that $\varrho_j\neq \varrho_k$ for $j\neq k$, we have\footnote{For the degenerate case, we also need to consider the contributions to $\dt\varrho_j$ from $\braket{\psi_k}{\dt \psi_j}$ and $\braket{\dt \psi_k}{\psi_j}$ for $k\neq j$.}
\begin{align}\lb{eq:dynamics_occupation_number}\hspace{-0em}
 \frac{\dt\varrho_j}{\varrho_j} = \frac{\dif }{\dif t}\, \braket{\psi_j}{\psi_j}= 2\re\big( \braket{\psi_j}{\dt{\psi}_j}\big)=\frac{2}{\hbar}\,\im\, \bra{\psi_j}Q_j\ssp \ket{\psi^*_j}\;,
\end{align}
and thus we have
\begin{align}\lb{eq:occupation_number_dynamics}
 \dt\varrho_j = \frac{2 g}{\hbar}\,\sum_{\substack{k=1,\\k\neq j}}^\rank \xi_{jk} \int \im \Big([\psi^*_j(\xbf)]^2  \psi_k^2(\xbf)\Big)\:\dif \xbf\;.
\end{align}
The particle number-conserving condition, $\sum_{j=1}^\rank \dt\varrho_j = 0 $, follows by noticing that $\xi_{jk}=\xi_{kj}$ and $\im \big([\psi^*_j(\xbf)]^2  \psi_k^2(\xbf)\big) = -\im \big([\psi^*_k(\xbf)]^2  \psi_j^2(\xbf)\big)$.  Note that the occupation number $\varrho_j$ can be calculated using Eq.~(\chref{pcs_1:eq:upsilon_diagonal_elements_rdm}),
\begin{align}
 \varrho_j = \frac{(2n)}{\Upsilon\big(\vec{\pcss}\,\big)}\: \frac{\partial \Upsilon\big(\vec{\pcss}\,\big)}{\partial\pcss_j }\;.
\end{align}
Taking derivative with respect to the PCS parameter $\pcss_k$, we have
\begin{align}
 \frac{\partial \varrho_j}{\partial \pcss_k} = \frac{(2n)}{\Upsilon}\, \frac{\partial^2 \Upsilon}{\partial\pcss_j \partial\pcss_k} - \frac{(2n)}{\Upsilon^2}\, \frac{\partial \Upsilon}{\partial\pcss_j} \frac{\partial \Upsilon}{\partial\pcss_k} = \frac{1}{2n} \big(\rho^{(2)}_{jk,\,jk}- \varrho_j\varrho_k\big) = \frac{c_{jk}}{2n} \;,
\end{align}
where $c_{jk}$ denotes the covariance of the number operators $a_j^\dagger a_j$ and $a_k^\dagger a_k$.  Using this condition, we have
\begin{align}\lb{eq:derivative_occupation_number_pcss}
 \dt\varrho_j= \frac{1}{2n}\sum_{k=1}^\rank c_{jk}\ssp \dt\pcss_k\;.
\end{align}
The matrix $c$ is not invertible
\begin{align}
 &\sum_{k=1}^\rank c_{jk} = \sum_{k=1}^\rank \rho^{(2)}_{jk,\,jk}- \varrho_j\varrho_k = 0\;\; \Longrightarrow\;\; c\, (1,1,\ldots,1)^T = 0\;,
\end{align}
but we can invert it in the subspace orthogonal to the vector $(1,1,\ldots,1)^T$ and thus solve for $\dt\pcss_j$.

Finally, we come to the point that the orbitals do not remain orthogonal under the evolution~(\rf{eq:evolution_single_particle_correlation_matrix}).  To fix this problem, we notice that any ensemble decomposition of the form $\big\{\sum_{k=1}^\rank U_{jk}\sqrt{\varrho_k}\, \ket{\psi_k},\;j=1,2,\ldots,\rank\,\}$ gives the same 1RDM, where $U$ is a $\rank\times\rank$ unitary matrix.  For a unitary $U$ generated by $-i\vartheta\dif t /\hbar$, where $\vartheta_{jk}=\vartheta_{kj}^*$ and $\dif t$ the infinitesimal parameter, we have
\begin{align}\lb{eq:ensemble_equivalence}
 \psi_j'(\xbf)= \psi_j(\xbf)-\frac{\dif t}{i\hbar \sqrt{\varrho_j}} \sum_{\substack{k=1,\\ k\neq j}}^\rank \vartheta_{jk} \sqrt{\varrho_k}\ssp \psi_k(\xbf)\;.
\end{align}
By properly choosing the Hermitian matrix $\vartheta$, we can make the orbitals orthonormal to each other under the time evolution.  Consider the following modified equation for the orbitals
\begin{align}\hspace{-2em}
 i\hbar\ssp\ssp\dt \psi_j^{(\mathrm{mod})} =  \Big(\mathord{-}\frac{\hbar^2}{2m}\boldsymbol\nabla^2 +V+ R_j-\vartheta_j\,\Big) \ssp \psi_j+ Q_j\ssp \psi^*_j- \frac{1}{\sqrt{\varrho_j}} \sum_{\substack{k=1,\\ k\neq j}}^\rank \vartheta_{jk} \sqrt{\varrho_k}\ssp \psi_k\;,\lb{eq:modified_pcs_orbital_dynamics}
\end{align}
where
 \begin{align}
 \im \vartheta_j &= \im\, \bra{\psi_j}Q_j\ssp \ket{\psi^*_j} = \frac{g}{\varrho_j}\,\sum_{\substack{k=1,\\k\neq j}}^\rank \xi_{jk} \int \im \Big([\psi^*_j(\xbf)]^2  \psi_k^2(\xbf)\Big)\:\dif \xbf
 \end{align}
and
\begin{subequations}
\begin{align}\hspace{-2em}
 \re \vartheta_j &= \re\,\bra{\psi_j}\ssp\ssp i\hbar\,\ket{\dt{\psi}_j}-\mu_j\\
 &= \bra{\psi_j}R_j\ket{\psi_j}+\re\,\bra{\psi_j}Q_j\ket{\psi_j^*} -\mu_j^{(2)}\\
 &= \frac{g}{\varrho_j}\bigg(\sum_{k=1}^\rank \ssp \xi_{jk}\,\int 2\ssp\norm{\psi_j(\xbf)}^2\ssp \norm{\psi_k(\xbf)}^2 +\re \Big([\psi^*_j(\xbf)]^2  \psi_k^2(\xbf)\Big) \dif \xbf
 \bigg)-\mu_j^{(2)}\;.
\end{align}
\end{subequations}
Requiring that the orbitals be orthogonal gives
\begin{subequations}
\begin{align}
 0 &= \braketb{\psi_k}{\dt \psi_j^{(\mathrm{mod})}} +\braketb{\dt \psi_k^{(\mathrm{mod})}}{\psi_j}\\
 &= \braket{\psi_k}{\dt \psi_j} + \braket{\dt \psi_k}{\psi_j} - \frac{\vartheta_{jk} (\varrho_k-\varrho_j)}{i\hbar\sqrt{\varrho_j\varrho_k}}\;,
\end{align}
\end{subequations}
and this means that the Hermitian matrix $\vartheta$ can be determined
\begin{subequations}\lb{eq:orthogonal_parameter}
\begin{align}
 \vartheta_{jk} &= \frac{i\hbar\ssp \sqrt{\varrho_j\varrho_k}}{\varrho_k-\varrho_j}\,\Big(\braket{\psi_k}{\dt \psi_j}+\braket{\dt \psi_k}{\psi_j}\Big)\\[3pt]
 &= \frac{\sqrt{\varrho_j\varrho_k}}{\varrho_k-\varrho_j}\, \Big(\brab{\psi_k}R_j-R_k\ketb{\psi_j}+\bra{\psi_k}Q_j\ket{\psi_j^*}-\bra{\psi_k^*}Q_k^\dagger\ket{\psi_j}\Big)\;.
\end{align}
\end{subequations}
Note that the degenerate case needs extra care, because both Eqs.~(\rf{eq:dynamics_occupation_number}) and (\rf{eq:orthogonal_parameter}) are no longer correct.  Instead of making the orbitals orthogonal by small corrections, one can just diagonalize the evolved 1RDM in the degenerate subspace (this is similar to the nondegenerate perturbation theory).

Equations~(\rf{eq:occupation_number_dynamics}) and (\rf{eq:modified_pcs_orbital_dynamics}) together determine the dynamics of the PCS ansatz; these equations are only valid, however, for cases where $\rank$ is fixed.  This problem is caused by replacing the condition~(\rf{eq:orthogonal_condition_PCS_ansatz_equivalent}) with (\rf{eq:orthogonal_condition_PCS_ansatz_alternative}).  While the first condition can determine $\ket{\dt\varPsi_\mathrm{err}}$ in the whole space, the latter one is trivially satisfied in the nullspace of the 1RDM and gives no information about $\ket{\dt\varPsi_\mathrm{err}}$ in that subspace.  To make Eq.~(\rf{eq:orthogonal_condition_PCS_ansatz_alternative}) work in the whole space, we give an infinitesimally small occupation number to the states in the nullspace.  Thus consider an extra orbital labeled by $\rank+1$; using Eqs.~(\chref{pcs_1:eq:upsilon_definition}) and (\chref{pcs_1:eq:upsilon_diagonal_elements_rdm}), we have
\begin{align}\hspace{-1em}
  \xi_{\rank+1,\,j} \propto \frac{\partial^2 \Upsilon}{\partial \pcss_{\rank+1}\partial \pcss_j}  \propto \int_0^\pi\!\int_{\norm{\vec y}=\sin\! \theta} y_j^2 \cos^2\! \theta \exp\bigg(\pcss_{\rank+1} \cos^2\! \theta+\sum_{j=1}^\rank \pcss_j\ssp y_j^2\bigg)\: \dif \theta\,\dif \Omega\;.\lb{eq:occupation_correlation_extra}
\end{align}
Supposing that the occupation number of the extra orbital is small, $\varrho_{\rank+1}\ll N$, we have $\pcss_{\rank+1} \ll \pcss_j$ for $j=1,2,\ldots,\rank$.  The main contribution to the integral in Eq.~(\rf{eq:occupation_correlation_extra}) thus comes from $\theta \simeq \pi/2$, because of the exponential, i.e., $\norm{\vec y}\simeq 1$, and we have the following approximate relation,
\begin{align}
\xi_{\rank+1,\,j} \propto \varrho_j \int_0^\pi \cos^2\! \theta\exp\big(\pcss_{\rank+1}\ssp \cos^2\! \theta\big)\, \dif \theta\, \propto\, \varrho_j\;.
\end{align}
Using this relation and the marginal condition of the 2RDM, we have
\begin{align}\lb{eq:factorized_form}
 \xi_{\rank+1,\,j} \simeq  \varrho_{\rank+1}\ssp \varrho_j\;,
\end{align}
for $j=1,2,\ldots,\rank$.  The evolution of the occupation number $\varrho_{\rank+1}$ thus takes the form
\begin{subequations}
\begin{align}
 \dt\varrho_{\rank+1} &= \frac{2 g}{\hbar}\,\sum_{j=1}^\rank \xi_{\rank+1,\,j} \int \im \Big([\psi^*_{\rank+1}(\xbf)]^2  \psi_j^2(\xbf)\Big)\:\dif \xbf\\
 & = \frac{2 g \varrho_{\rank+1}}{\hbar}\,\sum_{j=1}^\rank \varrho_{j} \int \im \Big([\psi^*_{\rank+1}(\xbf)]^2  \psi_j^2(\xbf)\Big)\:\dif \xbf\;.\lb{eq:extra_occupation_number_dynamics}
\end{align}
\end{subequations}
From Eq.~(\rf{eq:extra_occupation_number_dynamics}) we see that the occupation number $\varrho_{\rank+1}$ grows exponentially for short time, but its initial value must be nonzero to have a nontrivial result.\footnote{The nonzero occupation number can come from two sources: (i) $\varrho_{\rank+1}$ is small but nonvanishing for the PCS ground state; (ii) by going to the next level of approximation of $\Upsilon(\vec\pcss)$, we can have a nonzero $\varrho_{\rank+1}\simeq N^0$, but $N$ must be finite for the extra level to make a difference.}  Since we have a factorized form~(\rf{eq:factorized_form}), the chemical potential of the extra orbital is
\begin{align}\hspace{-1em}
 \mu_{\rank+1}
 &= \braB{\psi_{\rank+1}}\mathord{-}\frac{\hbar^2}{2m}\boldsymbol\nabla^2 +V+ R_{\rank+1}\ketB{\psi_{\rank+1}}+\re\,\brab{\psi_{\rank+1}}Q_{\rank+1}\ketb{\psi_{\rank+1}^*}\;,
\end{align}
where
\begin{align}
 R_{\rank+1}(\xbf)&\simeq 2g\sum_{j=1}^\rank  \varrho_j\,\norm{\psi_j(\xbf)}^2\,,\quad
 Q_{\rank+1}(\xbf)\simeq g\sum_{j=1}^\rank  \varrho_j \,\psi_j^2(\xbf)\;.
\end{align}
A necessary condition for $\varrho_{\rank+1}$ to grow steadily is that $\mu_{\rank+1}\simeq \mu_j$ for at least one $j\in \{1,2,\ldots,\rank\}$; otherwise the sign of the integral in  Eq.~(\rf{eq:extra_occupation_number_dynamics}) changes rapidly.

\appendix

\renewcommand{\lb}[1]{\label{mgpes_vs_bogoliubov:#1}}
\renewcommand{\rf}[1]{\ref{mgpes_vs_bogoliubov:#1}}

\chapter{Phase Diffusion: Many GPEs vs Bogoliubov}
\label{ch:mgpes_vs_bogoliubov}
\chaptermark{Phase diffusion: Many GPEs vs Bogoliubov}

\begin{quote}
Good physicists think in many different ways so that they can pick the most efficient one.\\[4pt]
-- Carlton M. Caves\ai{Caves, Carlton M.}
\end{quote}

\noindent The relative phase of the two components of a BEC can be well defined when there is an uncertainty in the relative number of bosons.  The many-body wavefunction thus is a superposition of states with different numbers of particles distributed in the two components; such states evolve at different rates due to their different chemical potentials caused by the particle-particle interactions.  This effect, called phase diffusion\si{Phase diffusion}, degrades the relative phase of the two components.  One approach to analyzing phase diffusion is by using many Gross-Pitaevskii equations (many-GPEs)~\cite{li_spin_2009}, while another approach is to use the number-conserving Bogoliubov approximation~\cite{sorensen_bogoliubov_2002}.  Very similar results are derived using these two approaches.  Here, I discuss the equivalence of the two methods using the results in Chap.~\chref{ch:n_conserving}.

In the many-GPEs approach, we consider the state vector of a two-component condensate of $N$ bosons, with no depletion out of the two modes, expanded in the Fock basis,
\begin{align}\lb{eq:setor_dependent_gp}
 \ket{\varPsi} = \sum_{j=0}^N \frac{\chi_j}{\sqrt {j!(N-j)!}}\,\big(\a_{1,\psinot_1}^\dagger\big)^j \big(\a_{2,\psinot_2}^\dagger\big)^{N-j}\, \ket{\vac} = \sum_{j=0}^N \chi_j\, \ket{j,N-j}\;,
\end{align}
where $\a_{\sigma,\psinot_\sigma}^\dagger$ is the creation operator for the $\sigma$th hyperfine level with the spatial wavefunction $\psinot_\sigma$, and the $\chi_j$s are the complex coefficients that determine the state.  When the coupling between the two modes in the Hamiltonian~(\chref{n_conserving:eq:H_two_component}) goes to zero, i.e., $\omega_{12}=0$, each number sector $\ket{N_1,N_2}$ is decoupled from other sectors. Thus, one can solve the whole dynamics by considering the dynamics within each sector separately.  Because the interaction term is nonlinear in the number of particles, each sector evolves according to a different GP equation,
\begin{gather}
i\hbar\ssp \dt \psinot_1
=\Big(H_1+\hbar\omega_{11}+g_{1\nsp 1}N_1 \norm{\psinot_1}^2
+g_{1\nsp 2}N_2\norm{\psinot_2}^2\Big)\psinot_1\;,\\[6pt]
i\hbar\ssp \dt \psinot_2
=\Big(H_2+\hbar\omega_{22}+g_{22}N_2\norm{\psinot_2}^2
+g_{2\nsp 1}N_1\norm{\psinot_1}^2\Big)\psinot_2\;,
\end{gather}
where $H_{1,2}$ are the single-particle Hamiltonians.  This dependence of GPE on $N_1$ and $N_2$ eventually causes phase diffusion.

In the Bogoliubov approach, we consider a two-component BEC of $N$ bosons condensed in the wavefunction
\begin{subequations}\lb{eq:condensate_state_two_component}
\begin{align}
\ket{\psinot(t)}
&=\frac{1}{\alpha}\,\Big(\,\alpha_1(t)\,
\ket{\psinot_1(t)}\otimes \ket{1}+\alpha_2(t)\, \ket{\psinot_2(t)}\otimes \ket{2}\,\Big)\;,
\end{align}
\end{subequations}
which solves the Gross-Pitaevskii (GP) equations
\begin{equation}\lb{eq:spinorGP}
\Big(\mathord{-}i\hbar\pa{}{t}+H_{\mathrm{gp}}\,\Big)
\begin{pmatrix}
\alpha_1\psinot_1\\\alpha_2\psinot_2
\end{pmatrix}
=0\;.
\end{equation}
Here, the GP Hamiltonian takes the form
\begin{equation}\lb{eq:Hgpmatrix}
H_\mathrm{gp}=\begin{pmatrix}
H_\mathrm{gp}^{(1)} &
\hbar\omega_{12}\\[3pt]
\hbar\omega_{21} &
H_\mathrm{gp}^{(2)}
\end{pmatrix}
=H_\mathrm{gp}^{(1)}\proj{1}+H_\mathrm{gp}^{(2)}\proj{2}+
\hbar\omega_{12}\ket1\bra2+\hbar\omega_{21}\ket2\bra1\;,
\end{equation}
where
\begin{gather}
H_\mathrm{gp}^{(1)}
=H_1+\hbar\omega_{11}+g_{1\nsp 1}\norm{\alpha_1}^2 \norm{\psinot_1}^2
+g_{1\nsp 2}\norm{\alpha_2}^2\norm{\psinot_2}^2\;,\\[6pt]
H_\mathrm{gp}^{(2)}
=H_2+\hbar\omega_{22}+g_{22}\norm{\alpha_2}^2\norm{\psinot_2}^2
+g_{2\nsp 1}\norm{\alpha_1}^2\norm{\psinot_1}^2\;.
\end{gather}
For the case $\omega_{12}=0$, $\alpha_1$ and $\alpha_2$ do not change in time, and the following state also solves the GPE~(\rf{eq:spinorGP}),
\begin{equation}\lb{eq:barpsinot}
\ket{\bar\psinot(t)}
=\frac{1}{\alpha^*}\,\Big(\,\alpha_2^*\, \ket{\psinot_1(t)}\otimes \ket{1}-\alpha_1^*\, \ket{\psinot_2(t)}\otimes \ket{2}\,\Big)\;.
\end{equation}
Recall that the number-conserving Bogoliubov Hamiltonian matrix~(\chref{n_conserving:eq:number_conserving_bogliubov_hamiltonian_matrix}) takes the form
\begin{equation}
\mathsp{H}_\mathrm{ncb}=
\begin{pmatrix}
H_\mathrm{gp}+\norm\alpha^2Q\ssp \Phi G \Phi^{\nsp *} Q &
\alpha^2Q\ssp \Phi G \Phi\ssp Q^*\\[3pt]
(\alpha^*)^2Q^* \Phi^{\nsp *} G \Phi^{\nsp *} Q &
H_\mathrm{gp}^*+\norm\alpha^2 Q^*\Phi^{\nsp *} G \Phi\ssp Q^*
\end{pmatrix}\;,
\end{equation}
where $Q(t)=\mathbb{1}-\proj{\psinot(t)}$, and
\begin{equation}\lb{eq:definition_matrices_two_component}
\Phi=\frac{1}{\alpha}
\begin{pmatrix}
\alpha_1 \psinot_1&
0\\[3pt]
0 &
\alpha_2 \psinot_2
\end{pmatrix}\;,\qquad
G=\begin{pmatrix}
g_{1\nsp 1} &
g_{1\nsp 2}\\[3pt]
g_{2\nsp 1} &
g_{22}
\end{pmatrix}\;.
\end{equation}
To separate the part that causes phase diffusion, we divide the Bogoliubov Hamiltonian matrix into three pieces,
\begin{equation}\lb{eq:3parts}
\mathsp{H}_\mathrm{ncb}=\mathsp{H}_\mathrm{ncb\ssp \doubleperp}+\mathsp{H}_{\mathrm{ss}}+\mathsp{H}_{\mathrm{pd}}\;.
\end{equation}
The part $\mathsp{H}_\mathrm{ncb\ssp \doubleperp}$ is orthogonal to the subspace spanned by $\ket{\psinot}$ and $\ket{\bar\psinot}$,
\begin{equation}
\mathsp{H}_\mathrm{ncb\ssp\doubleperp}=
\begin{pmatrix}
H_\mathrm{gp}+\norm\alpha^2 R\ssp \Phi G \Phi^{\nsp *}\nsp R &
\alpha^2 R\ssp \Phi G \Phi R^*\\[3pt]
(\alpha^*)^2 R^* \Phi^{\nsp *} G \Phi^{\nsp *}\nsp R &
H_\mathrm{gp}^*+\norm\alpha^2 R^*\Phi^{\nsp *} G \Phi R^*
\end{pmatrix}\;,
\end{equation}
where $R(t)=\identity-\proj{\psinot(t)}-\proj{\bar\psinot(t)}$ is the projection operator on the orthogonal space.  The second term $H_{\mathrm{ss}}$ in Eq.~(\rf{eq:3parts}), the spin squeezing term, only contains contributions from the mode $\bar\psinot$
\begin{subequations}
\begin{align}\hspace{-1em}
\mathsp{H}_\mathrm{ss}&=
\begin{pmatrix}
\norm\alpha^2\ssp \proj{\bar\psinot}\ssp \Phi\ssp G\ssp \Phi^{\nsp *}\ssp \proj{\bar\psinot} &
\alpha^2\ssp \proj{\bar\psinot}\ssp \Phi\ssp G\ssp \Phi\ssp \proj{\bar\psinot^*}\\[3pt]
(\alpha^*)^2\ssp \proj{\bar\psinot^*}\ssp \Phi^{\nsp *}\ssp G\ssp \Phi^{\nsp *}\ssp \proj{\bar\psinot} &
\norm\alpha^2\ssp\proj{\bar\psinot^*}\ssp \Phi^{\nsp *}\ssp G\ssp \Phi\ssp \proj{\bar\psinot^*}
\end{pmatrix}\\[6pt]
&=\bar\eta
\begin{pmatrix}
\norm\alpha^2\ssp \ket{\bar\psinot}\bra{\bar\psinot} &
\alpha^2\ssp e^{2\ssp i\ssp\theta}\, \ket{\bar\psinot}\bra{\bar\psinot^*}\\[3pt]
(\alpha^*)^2\ssp e^{-2\ssp i\ssp\theta}\, \ket{\bar\psinot^*} \bra{\bar\psinot} &
\norm\alpha^2\ssp \ket{\bar\psinot^*} \bra{\bar\psinot^*}
\end{pmatrix}
\end{align}
\end{subequations}
where $\theta=\arg(\alpha_1\ssp \alpha_2/\alpha^2)$, and
\begin{subequations}
\begin{align}
\bar\eta&= \bra{\bar\psinot}\Phi G \Phi^{\nsp *}\ket{\bar\psinot} = e^{-2i\theta}\, \bra{\bar\psinot}\Phi G \Phi\ket{\bar\psinot^*} = e^{2i\theta}\,\bra{\bar\psinot^*}\Phi^* G \Phi^*\ket{\bar\psinot}\\
&=\frac{\norm{\alpha_1}^2\ssp \norm{\alpha_2}^2}{\norm{\alpha}^4}
\int g_{1\nsp 1}\ssp  \norm{\psinot_1}^4+g_{22}\ssp  \norm{\psinot_2}^4-2\ssp g_{1 2}\ssp  \norm{\psinot_1}^2\ssp \norm{\psinot_2}^2\; \dif \mathbf x\;.
\end{align}
\end{subequations}
The third term $H_{\mathrm{pd}}$ in Eq.~(\rf{eq:3parts}), describing the coupling of the mode $\bar\psinot$ to the orthogonal modes,  is
\begin{equation}
\mathsp{H}_{\mathrm{pd}}=
\begin{pmatrix}
\norm\alpha^2\ssp \proj{\bar\psinot}\ssp \Phi\ssp G\ssp \Phi^{\nsp *}\nsp R &
\alpha^2\proj{\bar\psinot}\ssp \Phi\ssp G\ssp \Phi\ssp R^*\\[3pt]
(\alpha^*)^2\proj{\bar\psinot^*}\ssp \Phi^{\nsp *}\ssp G\ssp \Phi^{\nsp *}\nsp R &
\norm\alpha^2\ssp \proj{\bar\psinot^*}\ssp \Phi^{\nsp *}\ssp G\ssp \Phi\ssp R^*
\end{pmatrix}+\mathrm{H.c.}\;,
\end{equation}
and the corresponding Hamiltonian reads
\begin{subequations}\lb{eq:phase_diffusion}
\begin{align}
\begin{split}\hspace{-2em}
\sH_\mathrm{pd}&=\half\, \Big(\, \norm\alpha^2\ssp \bar{a}_{\psinot}^{\dag}\ssp \bra{\bar\psinot}\ssp \Phi\ssp G\ssp \Phi^{\nsp *}\ssp \ket{\uppsi_{\doubleperp}}
+\alpha^2\ssp \bar{a}_{\psinot}^{\dag}\ssp \bra{\bar\psinot}\ssp \Phi\ssp G\ssp \Phi\ssp \ket{\uppsi_{\doubleperp}^{\dag}}\\
&\hspace{1.3em}+(\alpha^*)^2\ssp \bar{a}_{\psinot}\ssp \bra{\bar\psinot^*}\ssp \Phi^{\nsp *}\ssp G\ssp \Phi^{\nsp *}\ssp \ket{\uppsi_{\doubleperp}}+\norm\alpha^2\ssp \bar{a}_{\psinot}\ssp \bra{\bar\psinot^*}\ssp \Phi^{\nsp *}\ssp G\ssp \Phi\ssp \ket{\uppsi_{\doubleperp}^{\dag}}+\mathrm{H.c.}\,\Big)\lb{eq:phase_diffusion_a}
\end{split}\\[3pt]
&= \big(\ssp \alpha e^{i\ssp\theta}\ssp \bar{a}_{\psinot}^{\dag}+\alpha^* e^{-\ssp i\ssp\theta} \ssp \bar{a}_{\psinot}\ssp \big)\ssp \Big(\ssp \alpha^*e^{-\ssp i\ssp\theta}\bra{\bar\psinot}\ssp \Phi\ssp G\ssp \Phi^{\nsp *}\ssp \ket{\uppsi_{\doubleperp}}
+ \alpha e^{i\ssp\theta}\ssp \bra{\bar\psinot^*}\ssp \Phi^*\ssp G\ssp \Phi\ssp \ket{\uppsi_{\doubleperp}^{\dag}}\ssp\Big)
\lb{eq:phase_diffusion_b}
\end{align}
\end{subequations}
where $\uppsi_{\doubleperp}$ denotes the field operator orthogonal to the two modes $\ket{\psinot}$ and $\ket{\bar\psinot}$.  Another way to think about the Hamiltonian~(\rf{eq:phase_diffusion_b}) is to recall that it arises, in the Bogoliubov approximation, from replacing $\a_\psinot$ and $\a_{\smash\psinot}^\dagger$ by $\alpha$ and $\alpha^{\nsp*}$ in the original Schr\"odinger-picture Hamiltonian.  Restoring the creation and annihilation operators for the condensate mode to the Hamiltonian~(\rf{eq:phase_diffusion}) gives
\begin{align}\hspace{-1em}
\mathcal{H}_{\mathrm{pd}}&=  \big(\ssp e^{i\ssp\theta}\bar{a}_\psinot^{\dag}\ssp a_\psinot+e^{-i\ssp\theta}a_\psinot^{\dag}\ssp \bar{a}_\psinot\ssp\big )\ssp \Big(\ssp e^{-i\ssp\theta} a_\psinot^{\dag}\ssp \bra{\bar\psinot}\ssp \Phi\ssp G\ssp \Phi^{\nsp *}\ssp \ket{\uppsi_{\doubleperp}}
+ e^{i\ssp\theta} a_\psinot\ssp \bra{\bar\psinot}\ssp \Phi\ssp G\ssp \Phi\ssp \ket{\uppsi_{\doubleperp}^{\dag}}\ssp\Big)\;.\lb{eq:phase_diffusion_c}
\end{align}
As an example of how the Hamiltonian~(\rf{eq:phase_diffusion_c}) works, consider the situation where the condensate wavefunction $\psinot$ is an equal superposition of the two hyperfine levels, i.e., $\alpha_1=\alpha_2=1/\sqrt2$, $\theta=0$, and
\begin{equation}
\a_\psinot=\frac{1}{\sqrt{2}}\, \big(\ssp \a_1+\a_2\ssp\big)\;, \qquad
\bar\a_\psinot=\frac{1}{\sqrt{2}}\, \big(\ssp \a_1-\a_2\ssp\big)\;,
\end{equation}
where $\a_1$($\a_2$) is a shorthand for $\a_{1,\psinot_1}$($\a_{2,\psinot_2}$).  By introducing the Schwinger operator\si{Schwinger operators}
\begin{align}
\sJ_z\equiv \frac{1}{2}\, \big(\ssp \a_1^\dagger \a_1-\a_2^\dagger \a_2\ssp \big) =\frac{1}{2}\, \big(\ssp \bar\a_{\smash\psinot}^\dagger\ssp \a_\psinot+ \a_{\smash\psinot}^\dagger\ssp \bar\a_\psinot\ssp\big)\;,
\end{align}
we bring the Hamiltonian~(\rf{eq:phase_diffusion_c}) into the form
\begin{align}
\mathcal{H}_{\mathrm{pd}}&=  2\sJ_z\ssp \Big(\ssp a_\psinot^{\dag}\ssp \bra{\bar\psinot}\ssp \Phi\ssp G\ssp \Phi^{\nsp *}\ssp \ket{\uppsi_{\doubleperp}}
+ a_\psinot\ssp \bra{\bar\psinot}\ssp \Phi\ssp G\ssp \Phi\ssp \ket{\uppsi_{\doubleperp}^{\dag}}\ssp\Big)\;.\lb{eq:phase_diffusion_e}
\end{align}

Now we are in a position to see how the many-GPEs approach is related to the Bogoliubov approach.  Note that $\sJ_z$ counts the difference between the numbers of particles in the two components.  Thus, we can regard Eq.~(\rf{eq:phase_diffusion_e}) as a controlled Hamiltonian depending on the value of $N_1-N_2$, which accounts for the different GPEs~(\rf{eq:setor_dependent_gp}) for different sectors.

\renewcommand{\lb}[1]{\label{bargmann:#1}}
\renewcommand{\rf}[1]{\ref{bargmann:#1}}

\chapter{Bargmann-Fock Representation and Its Applications in BECs}\label{ch:bargmann}

\begin{quote}
God used beautiful mathematics in creating the world.\\[4pt]
-- Paul Dirac\ai{Dirac, Paul}
\end{quote}

\noindent In Chap.~\chref{ch:n_conserving}, we have introduced the extended catalytic state as a coherent state for the condensate mode and a state for the orthogonal modes.  The physical state of the BEC is encoded in the $N$-particle sector of the Extended Catalytic State (ECS) Eq.~(\chref{n_conserving:eq:catalytic_state})\si{Extended catalytic state}.  To decode, one simply projects the extended catalytic state to the $N$-particle sector.  In this appendix, I discuss how to represent the ECS and to project it into number states with the Bargmann-Fock (BF) representation~\cite{bargmann_hilbert_1961, bargmann_remarks_1962}.\ai{Bargmann, Valentine}

The BF representation, introduced to quantum optics by Glauber~\cite{glauber_coherent_1963},\footnote{The Bargmann representation also finds various applications in fields such as quantum foundations~\cite{kempf_uncertainty_1994}, quantum gravity~\cite{thiemann_gauge_2001, livine_spinor_2012}, quantum chemistry~\cite{harriman_quantum_1994, martin-fierro_derivation_2006}, nuclear physics~\cite{filippov_structure_2005}, and lattice quantum mechanics~\cite{celeghini_quantum_1995}.} is a representation where quantum states of bosonic modes are mapped onto analytic (holomorphic) functions, which are convergent with respect to the Gaussian measure\si{Gaussian measure}
\begin{equation}
\Vert f\Vert_\mathrm{gauss}\defeq \int\vert f(z)\vert^2 e^{-\vert z\vert^2}\,\dif^2 z\;.
\end{equation}
Many powerful results of analytic functions in the complex plane can thus be exploited within the quantum-mechanical context. The BF representation is so versatile because we can manipulate it by multiplying functions, taking derivatives, and doing integral transformations. Although it is used extensively by mathematical physicists, the BF representation has not attracted as much attention from the quantum optics and condensed-matter communities, with two possible reasons being: (i)~it is not very convenient for mixed states; (ii)~it is not as intuitive as the quasi-probability distributions.  We will show, however, that the Bargmann-Fock representation is ideal for certain tasks, including many of ours, for which the quasi-probability distributions are not well suited.

In the BF representation representation, the Fock states (number states) are mapped to monomials,
\begin{equation}
\ket{n_1,n_2,\ldots, n_\rank}\;\rightarrow\;\prod_{k=1}^\rank
\frac{\;z_k^{n_k}}{\sqrt{n_k !}}\;,
\end{equation}
and consequently the annihilation and creation operators take the form
\begin{equation}\lb{eq:B_annihilation_creation}
 \a_k\;\rightarrow\;  \di{}{z_k}\,,\qquad \a^\dagger_k\;\rightarrow\; z_k\;.
\end{equation}
Since any state in the Fock space can be expanded in the basis of number states,
\begin{equation}
 \ket{\varPsi}=\sum_{\{n_k\}} f_{n_1,n_2,\ldots, n_\rank}\,\ket{n_1,n_2,\ldots, n_\rank}\;,
\end{equation}
we have the BF representation for that state,\si{Bargmann-Fock representation}
\begin{equation}\lb{eq:BF_number_representation}
 \mathfrak{B}_{\ket\varPsi}(\mathbf z)=
 \mathfrak{B}_{\ket\varPsi}(z_1,z_2,\ldots, z_\rank)
 =\sum_{\{n_k\}}\bigg( f_{n_1,n_2,\ldots, n_\rank}\,\prod_{k=1}^\rank\frac{\;z_k^{n_k}}{\sqrt{n_k !}}\bigg)\;,
\end{equation}
and the state is reconstructed from the BF representation by
\begin{equation}\lb{eq:inverse_relation}
 \ket{\varPsi}=\mathfrak{B}_{\ket\varPsi}(\a_1^\dagger,\a_2^\dagger,\ldots, \a_\rank^\dagger)\,\ket{\vac}\;.
\end{equation}

Another quite useful definition of the BF representation, in terms of coherent states, is
\begin{equation}\lb{eq:B_convenient_form}
 \mathfrak{B}_{\ket\varPsi}(z_1,z_2,\ldots, z_\rank)=e^{\norm{\mathbf z}^2/2}\: \braket{z_1^*,z_2^*,\ldots, z_\rank^*}{\varPsi}\;,
\end{equation}
where $\bra{z_1^*,z_2^*,\ldots, z_\rank^*}\,$ is the bra for the coherent state that has complex amplitudes $(z_1^*,z_2^*,\ldots, z_\rank^*)$, and $\norm{\mathbf z}^2=\norm{z_1}^2+\norm{z_2}^2+ \cdots+\norm{z_\rank}^2$. Note that the BF representation should not be confused with the square root of the Husimi $Q$ function despite their similar appearance. Note also that
\begin{equation}
e^{\norm{\mathbf z}^2/2}\:\braket{\varPsi}{z_1^*,z_2^*,\ldots, z_\rank^*}
=[\mathfrak{B}_{\ket\varPsi}(z_1,z_2,\ldots, z_\rank)]^*
=\mathfrak{B}_{\ket\varPsi}^*(z_1,z_2,\ldots, z_\rank)\;,
\end{equation}
which defines what we mean by complex conjugation of the BF representation.

Using the definition~(\rf{eq:B_convenient_form}) and the properties of coherent states, it is not hard to check Eq.~(\rf{eq:B_annihilation_creation}). Also, using Eq.~(\rf{eq:B_convenient_form}) and the completeness condition of the coherence states,\footnote{\quad ${\displaystyle \frac{1}{\pi}\int \,\proj{\alpha}\, \dif^2\alpha=\frac{1}{\pi}\int \,\proj{\alpha^*}\, \dif^2\alpha=\identity}$} we have the following result for the inner product of two quantum states,
\begin{subequations}
\begin{align}
 \braket{\varPsi_2}{\varPsi_1}&=\frac{1}{\pi^\rank}\int \braket{\varPsi_2}{z_1^*,z_2^*,\ldots, z_\rank^*}\braket{z_1^*,z_2^*,\ldots, z_\rank^*}{\varPsi_1}\, \dif^2 z_1\cdots \dif^2 z_\rank\\
 &=\frac{1}{\pi^\rank}\int \mathfrak{B}^*_{\ket{\varPsi_2}}(\mathbf z)\,\mathfrak{B}_{\ket{\varPsi_1}}(\mathbf z)\, e^{-\norm{\mathbf z}^2}\, \dif^2 \mathbf z\;.\lb{eq:inner_product}
\end{align}
\end{subequations}
Thus the inner product of  $\ket{\varPsi_1}$ and $\ket{\varPsi_2}$ is given in terms of the BF representation by Eq.~(\rf{eq:inner_product}); indeed, it is easy to verify, by plugging Eq.~(\rf{eq:BF_number_representation}) into Eq.~(\rf{eq:inner_product}), that the inner product reduces to the standard Fock-space form.

Consider the operator $\sL=\exp\big(\sum_{jk} G_{jk}^{\,*}\,\a_j^\dagger \a_k)$, which is not necessarily Hermitian, i.e., $G$ is not necessarily anti-Hermitian.  This operator imposes a linear transformation on the amplitudes of a coherent state,
\begin{equation}
\sL\ssp\ssp \ket{\mathbf z^*}
= e^{(\,{\mathbf z}^T \! L^T\! L^* \mathbf z^*-\norm{\mathbf z}^2\,)/2} \ket{\!L^* \mathbf z^*}\;,
\end{equation}
where $L=e^G$.  Notice that we can obtain any invertible transformation of the coherent-state amplitudes by this method.  We have
\begin{subequations}
\begin{align}
 \bra{\varPsi_2}\sL\ssp\ssp\ket{\varPsi_1}&=\frac{1}{\pi^\rank}\int \bra{\varPsi_2}\sL\ssp\ssp\ket{\mathbf z^*}\braket{\mathbf z^*}{\varPsi_1}\, \dif^2 \mathbf z\\
 &=\frac{1}{\pi^\rank}\int \mathfrak{B}^*_{\ket{\varPsi_2}}(L\ssp \mathbf z)\,\mathfrak{B}_{\ket{\varPsi_1}}(\mathbf z)\, e^{-\norm{\mathbf z}^2}\, \dif^2 \mathbf z\;.
\end{align}
\end{subequations}
For the same operator $\sL$, we also have $\bra{\mathbf z^*}\ssp \sL = e^{(\,{\mathbf z}^T \! L^*\! L^T \mathbf z^*-\norm{\mathbf z}^2\,)/2}\,\bra{\!L^T \mathbf z^*}$, and consequently,
\begin{subequations}
\begin{align}
 \bra{\varPsi_2}\sL\ssp\ssp\ket{\varPsi_1}&=\frac{1}{\pi^\rank}\int \braket{\varPsi_2}{\mathbf z^*}\bra{\mathbf z^*}\sL\ssp\ssp\ket{\varPsi_1}\, \dif^2 \mathbf z\\
 &=\frac{1}{\pi^\rank}\int \mathfrak{B}^*_{\ket{\varPsi_2}}(\mathbf z)\,\mathfrak{B}_{\ket{\varPsi_1}}(L^\dagger \mathbf z)\, e^{-\norm{\mathbf z}^2}\, \dif^2 \mathbf z\;.
\end{align}
\end{subequations}
Thus, we have
\begin{align}
 \frac{1}{\pi^\rank}\int \mathfrak{B}^*_{\ket{\varPsi_2}}(L\ssp \mathbf z)\,\mathfrak{B}_{\ket{\varPsi_1}}(\mathbf z)\, e^{-\norm{\mathbf z}^2}\, \dif^2 \mathbf z
 &=\frac{1}{\pi^\rank}\int \mathfrak{B}^*_{\ket{\varPsi_2}}(\mathbf z)\,\mathfrak{B}_{\ket{\varPsi_1}}(L^\dagger \mathbf z)\, e^{-\norm{\mathbf z}^2}\, \dif^2 \mathbf z\;,
\end{align}
which says that doing a linear transformation on one of the BF variables in the inner product~(\rf{eq:inner_product}) is equivalent to doing the conjugate transformation on the other one.

Many useful properties of the BF representation are discussed in~\cite{vourdas_thermal_1994, vourdas_analytic_2006}; these include the relation of $\mathfrak{B}_{\ket\varPsi}$ to the wavefunction in the coordinate or momentum representation and also the operator-representation in the Bargmann-Fock space and its relation to the Glauber-Sudarshan $P$ function, the Husimi $Q$ function, and the Wigner $W$ representation. In this dissertation, particular emphasis is put on representing Gaussian states and Gaussian unitaries. We also discuss how to project an arbitrary pure state to the $N$-particle sector with the BF representation.

Pure Gaussian states have a particularly simple form in the BF representation.  For example, we have the following for the coherent state $\ket{\alpha}=\sD(\alpha)\ket{\vac}$, where \begin{equation}\lb{eq:displacement_operator}
\sD(\alpha)=e^{\alpha\a^\dagger-\alpha^*\a}=e^{-\half\norm{\alpha}^2} e^{\alpha\a^\dagger}e^{-\alpha^*a}
\end{equation}
is the displacement operator:
\begin{subequations}
\begin{align}
 \mathfrak{B}_\mathrm{coh}(\alpha,z)&=e^{\half\ssp\norm{z}^2}\,\braket{z^*}{\alpha}\\
 &=e^{\half\ssp\norm{z}^2}\,e^{-\half\ssp(\,\norm{z}^2+\norm{\alpha}^2\,)}\, e^{\alpha z}\\
 &=e^{-\half\ssp\norm{\alpha}^2}\, e^{\alpha z}\;.
\end{align}
\end{subequations}
For squeezed vacuum states, $\sS(\gamma)\ket{\vac}$, where\si{Squeezed state}
\begin{equation}\lb{eq:squeeze_operator}
\sS(\gamma)=e^{\half(\gamma \a^2-\gamma \a^{\dagger\,2})}\;,\quad\mbox{$\gamma$ real,}
\end{equation}
is the single-mode squeeze operator, we have
\begin{subequations}
\lb{eq:squeezed_vacuum}
\begin{align}
\mathfrak{B}_\mathrm{sqv}(\gamma,z)
&=e^{\half\ssp\norm{z}^2}\,\bra{z^*}\sS(\gamma)\ket{\vac}\\
&=\frac{e^{\half\ssp\norm{z}^2}}{\sqrt{\cosh \gamma}}\;\brab{z^*}
\exp\Big(\mathord{-}\frac{\tanh \gamma}{2}\,\a^{\dagger\, 2}\Big)\ketb{\vac}\\[3pt]
&=\frac{1}{\sqrt{\cosh \gamma}}\;\exp\Big(\mathord{-}\frac{\tanh
\gamma}{2}\,z^2\, \Big)\;.
\end{align}
\end{subequations}
Here we use the following ``quasi-normal-ordered'' factored form of the squeeze operator~\cite{perelomov_generalized_1977, hollenhorst_quantum_1979},
\begin{equation}\lb{eq:quasi_normal_ordered}
\sS(\gamma)= \frac{1}{\sqrt{\cosh \gamma}}\, \exp\Bigl(\mathord{-}\frac{\tanh \gamma}{2}\, a^{\dagger\, 2} \Bigr)\,\bigl(\cosh \gamma\bigr)^{- \a^{\dagger} \a}\,\exp\Bigl(\frac{\tanh \gamma}{2}\, \a^2 \Bigr)\;.
\end{equation}
Notice that the result in Eq.~(\rf{eq:squeezed_vacuum}) can also be derived by finding a differential equation for $\mathfrak{B}_\mathrm{sqv}(\gamma,z)$:
\begin{subequations}
\begin{align}
 &e^{\half \gamma \a^2-\half \gamma \a^{\dagger\,2}}\a\,\ket{\vac}=0\\
 &\Longrightarrow\;  (\a \cosh \gamma+\a^\dagger\sinh \gamma)\, e^{\half\a^2-\half \gamma \a^{\dagger\,2}}\ket{\vac}=0\\
 &\Longrightarrow\;  \Big(\di{}{z} \cosh \gamma+ z\sinh \gamma\Big)\, \mathfrak{B}_\mathrm{sqv}(\gamma,z)=0\\
 &\Longrightarrow\;  \mathfrak{B}_\mathrm{sqv}(\gamma,z)\propto \exp\Big(\mathord{-}\frac{\tanh
\gamma}{2}\,z^2\, \Big)\;.
\end{align}
\end{subequations}
The magnitude of the proportionality factor can be determined from the normalization condition
\begin{equation}
 \frac{1}{\pi}\int\, \norm{\mathfrak{B}_\mathrm{sqv}(\gamma,z)}^2\, e^{-\norm{z}^2}\,\dif^2 z=1\;,
\end{equation}
and the phase of the proportionality factor can be fixed by knowing that $\mathfrak{B}_\mathrm{sqv}(\gamma,0)=1/\sqrt{\cosh\gamma}$ is real and positive.

It turns out that the BF representation of the most general pure Gaussian states for an arbitrary number of modes (not at all restricted just to coherent and squeezed vacuum states) can be determined from the action of Gaussian unitaries on the vacuum.  Gaussian unitaries are generated by Hamiltonians containing terms linear or quadratic in the annihilation and creation operators; they take simple forms in the BF representation. For the displacement operation~(\rf{eq:displacement_operator}), we have
\begin{subequations}
\lb{eq:displacement}
\begin{align}
\mathfrak{B}_{\ket{\psi}}(z)
\;\xrightarrow{\;\mbox{\small displaced by $\alpha$}\;}\;\;
&\mathfrak{B}_{\sD(\alpha)\ket{\psi}}(z)\\
&=e^{-\half\norm{\alpha}^2} e^{\alpha
z}\,e^{-\alpha^*\nsp\,d/dz}\,\mathfrak{B}_{\ket{\psi}}(z)\\
&=\mathfrak{B}_\mathrm{coh}(\alpha,z)\,\mathfrak{B}_{\ket{\psi}}(z-\alpha^*)\;.
\end{align}
\end{subequations}
The rotation (phase-shift) operation,
\begin{equation}
\sR(\theta)=e^{i\theta \a^\dagger \a}\;,
\end{equation}
is also straightforward:
\begin{equation}
\mathfrak{B}_{\ket{\psi}}(z)\;\xrightarrow{\;\mbox{\small rotated by $\theta$}\;}\;\;
\mathfrak{B}_{\sR(\theta)\ket{\psi}}(z)
=\mathfrak{B}_{\ket{\psi}}(e^{i\theta} z)\;.\lb{eq:rotation}
\end{equation}
This result can be derived easily by using the definition~(\rf{eq:B_convenient_form}). For the squeezing operation~(\rf{eq:squeeze_operator}), we have
\begin{subequations}
\lb{eq:squeezer}
\begin{align}\hspace{-1em}
\mathfrak{B}_{\ket{\psi}}(z)
\;\xrightarrow{\;\mbox{\small squeezed by $\gamma$}\;}\;\;
&\mathfrak{B}_{\sS(\gamma)\ket{\psi}}(z)\\
&=\exp\Big(\frac{\gamma}{2}\, \di{^2}{z^2}-\frac{\gamma}{2}\, z^2\Big)\,\mathfrak{B}_{\ket{\psi}}(z)\\
&=\mathfrak{B}_\mathrm{sqv}(\gamma,z)\,\bigl(\cosh \gamma\bigr)^{- z\,\di{}{z}}\,\exp\biggl(\frac{\tanh \gamma}{2}\, \di{^2}{z^2} \biggr)\,\mathfrak{B}_{\ket{\psi}}(z)\;,
\end{align}
\end{subequations}
where the factored form~(\rf{eq:quasi_normal_ordered}) of the squeeze operator is used.
The term $(\cosh \gamma)^{- z\,d/dz}$ rescales the independent variable as $z\ssp\rightarrow\ssp z / \cosh \gamma$; the general rule here is that for $\tau$ real,  $\tau^{a^\dagger a}$ is represented by $\tau^{z\,d/dz}$ in the BF representation and that
\begin{equation}
\tau^{z\,d/dz}f(z)=f(z\tau)\;.
\end{equation}
The other term in Eq.~(\rf{eq:squeezer}), $\exp\!\Big(\frac12\tanh \gamma\,d^2/dz^2\Big)$, can be evaluated in the Fourier domain; it maps Gaussian functions to Gaussian functions without changing their centers.

Another important Gaussian unitary operation is the passive linear-optical network or multiport beamsplitter.  The beamsplitter is described by a unitary operator $\sU$, which acts on $\rank$ input modes according to\si{Multiport beamsplitter}
\begin{equation}
 \a_k\;\xrightarrow{\;\mbox{\small multiport beamsplitter}\;}\;\;
 \sU^\dagger a_k\,\sU=\sum_{j=1}^\rank\,\a_j\,U_{jk}\;.
\end{equation}
Here $U$ is a matrix that must be unitary in order to preserve the canonical commutation relations. This transformation can be written in the equivalent, inverted form
\begin{equation}
\sU\,a_k^\dagger\,\sU^\dagger=\sum_{j=1}^\rank\,U_{kj}\,\a_j^\dagger\;.
\end{equation}
Using Eq.~(\rf{eq:inverse_relation}) or Eq.~(\rf{eq:B_convenient_form}), we can find the transformation law for the BF representation,
\begin{subequations}
\lb{eq:beam_splitter}
\begin{align}\hspace{-1.5em}
\mathfrak{B}_{\ket{\varPsi}}(z_1,z_2,\ldots, z_\rank)
\;\xrightarrow{\;\mbox{\small multiport beamsplitter}\;}\;\;
&\mathfrak{B}_{\sU\ket{\varPsi}}(z_1,z_2,\ldots, z_\rank)\\
&=\mathfrak{B}_{\ket{\varPsi}}(z'_1,z'_2,\ldots, z'_\rank)\;,
\end{align}
\end{subequations}
where
\begin{equation}
z'_k=\sum_j\, U_{kj}\, z_j\;.
\end{equation}

Most generally, Gaussian unitaries are unitaries generated by Hamiltonians containing terms linear or quadratic in the annihilation and creation operators. It turns out that the simple transformation rules of displacement in Eq.~(\rf{eq:displacement}), single-mode squeezing in Eq.~(\rf{eq:squeezer}), and beamsplitters in Eq.~(\rf{eq:beam_splitter}) are enough to construct the most general Gaussian unitaries. Using displacement operations, we can remove the linear terms in the generators of a Gaussian unitary, and the resulting unitary is generated by a quadratic Hamiltonian. The Bloch-Messiah reduction theorem~\cite{bloch_canonical_1962} states that any multimode Gaussian unitary generated by a quadratic (Bogoliubov) Hamiltonian can always be decomposed into a multiport beamsplitter, followed by a set of single-mode squeezers on each mode, followed by yet another multiport beamsplitter,\si{Bloch-Messiah reduction}
\begin{align}
 \sU_\mathrm{bog}=\sU_{\ssp\mathrm{msp}}\,\bigg(\prod_{k=1}^\rank\, \sS_k(\gamma_k)\bigg)\, \sV_{\ssp\mathrm{msp}}^\dagger\;.
\end{align}
\begin{figure}
   \centering
   \hspace{-3em}
   \includegraphics[width=0.75\textwidth,natwidth=610,natheight=612]{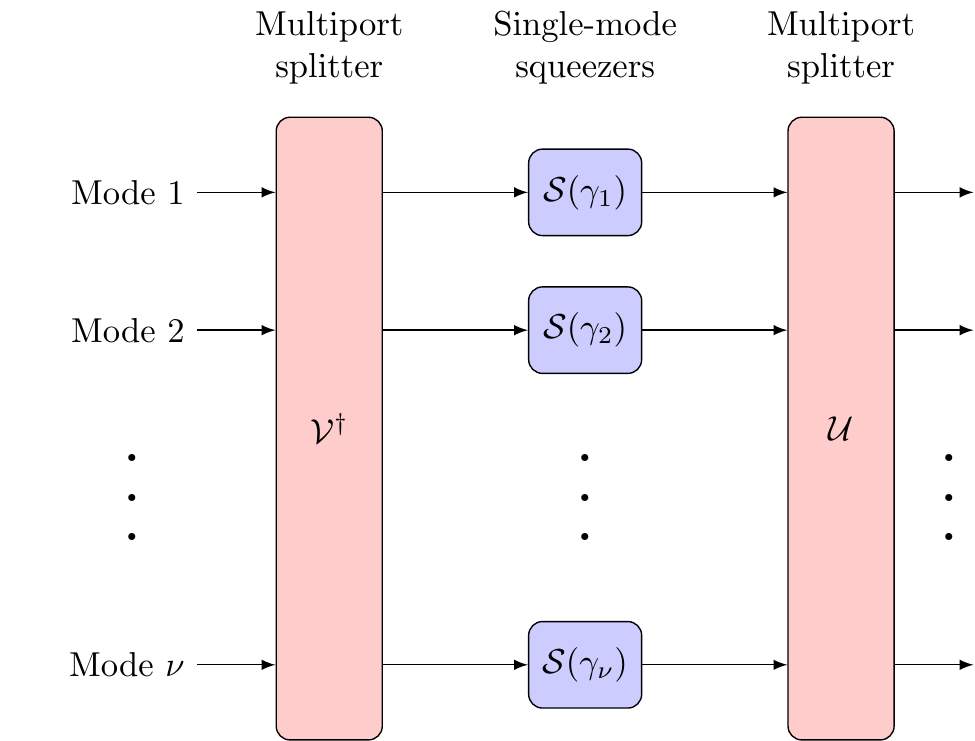}
   \caption[Bloch-Messiah Reduction Theorem]{The Bloch-Messiah reduction theorem states that any Gaussian unitary generated by a quadratic Hamiltonian can be decomposed into a multiport beamsplitter $\sV^\dagger$, followed by a set of single-mode squeezers on each mode, followed by yet another multiport beamsplitter $\sU$.}
   \label{fig:bloch_messiah_reduction}
\end{figure}
The BF representation allows a simple manipulation of all these elementary operations and thus is a suitable platform to work on Gaussian transformations.

Any pure Gaussian state can be generated by a Gaussian unitary acting on the vacuum state,\footnote{The covariance matrix of any pure Gaussian state is symplectically equivalent to that of the vacuum state, and this symplectic transformation can always be realized by a Gaussian unitary.}
\begin{subequations}
\begin{align}
 \ket{\varPsi}&=\sU\, \ket{\vac}\\
 &=\sD(\alpha_1,\alpha_2 ,\ldots, \alpha_\rank)\; \sU_{\ssp\mathrm{bog}}\, \ket{\vac}\;,
 \end{align}
 \end{subequations}
where the Gaussian unitary is broken up into a displacement operator and a unitary $\sU_{\ssp\mathrm{bog}}$ which is generated by a quadratic (Bogoliubov) Hamiltonian.
Using the Bloch-Messiah reduction theorem, we have
\begin{subequations}
\begin{align}
 \ket{\varPsi}&=\sD(\alpha_1,\alpha_2 ,\ldots, \alpha_\rank)\; \sU_{\ssp\mathrm{msp}}\,\bigg(\prod_{k=1}^\rank\, \sS_k(\gamma_k)\bigg)\, \sV_{\ssp\mathrm{msp}}^\dagger\, \ket{\vac}\\
 &=\sD(\alpha_1,\alpha_2 ,\ldots, \alpha_\rank)\; \sU_{\ssp\mathrm{msp}}\,\bigg(\prod_{k=1}^\rank\, \sS_k(\gamma_k)\bigg)\,\ketb{\vac}\;.\lb{eq:pure_gaussian_state}
\end{align}
\end{subequations}
where the initial multiport beamsplitter $\sV_{\ssp\mathrm{msp}}^\dagger$ can be discarded because it does not change the vacuum state.  All the elementary operations in Eq.~(\rf{eq:pure_gaussian_state}) have been discussed before, in Eqs.~(\rf{eq:squeezed_vacuum}), (\rf{eq:displacement}), and (\rf{eq:beam_splitter}).  Thus we have the BF representation for the most general Gaussian state,\si{Gaussian state}
\begin{equation}
 \mathfrak{B}_{\ket\varPsi}(z_1,z_2,\ldots, z_\rank)=\prod_{k=1}^\rank \frac{\mathfrak{B}_\mathrm{coh}(\alpha_k, z_k)}{\sqrt{\cosh \gamma_k}}\;\exp\bigg(\mathord{-}\frac{\tanh\gamma_k}{2}\,\Big(\sum_{j=1}^\rank U_{kj}\,\big(z_j-\alpha_j^*\big)\Big)^2\, \bigg)\;.
\end{equation}

For the Extended Catalytic State (ECS) Eq.~(\chref{n_conserving:eq:catalytic_state}),\si{Extended catalytic state} where the state of the noncondensate modes is a squeezed state centered at the origin, we have
\begin{align}\lb{eq:bargmann_ecs}
 \mathfrak{B}_\mathrm{ecs}&(z_0,z_1,\ldots, z_\rank)\nonumber\\
 &=\mathfrak{B}_\mathrm{coh}(\alpha, z_0)\prod_{k=1}^\rank \frac{1}{\sqrt{\cosh \gamma_k}}\;\exp\bigg(\mathord{-}\frac{\tanh\gamma_k}{2}\,\Big(\sum_{j=1}^\rank U_{kj}\,z_j\Big)^2\, \bigg)\;,
\end{align}
where the condensate mode is denoted by $j=0$ (and hence with the BF variable $z_0$) and the other modes, orthogonal to the condensate mode, by $j=1,2,\ldots,\rank$.  Projecting the state $\ket{\varPsi_\mathrm{ecs}}$ to the $N$-particle sector is equivalent to keeping only the terms of power $N$ in the BF variables in a power-series expansion of $\mathfrak{B}_\mathrm{ecs}(\mathbf z)$.

With the BF representation, it is also convenient to calculate the correlation matrices or, equivalently, the reduced density matrices (RDMs) of a multi-boson system. For example, the single-particle reduced matrix (1RDM) takes the form
\begin{subequations}
\begin{align}
 \rho_{kj}^{(1)}&=\bra{\varPsi}\a_j^\dagger \a_k\ket{\varPsi}\\[3pt]
 &=-\delta_{jk}+\frac{1}{\pi^\rank}\int \mathfrak{B}^*_{\ket\varPsi}(\mathbf z)\, \mathfrak{B}_{\ket\varPsi}(\mathbf z)\,z_k^* z_j \, e^{-\norm{\mathbf z}^2}\, \dif^2 \mathbf z\;.\lb{eq:reduced_density_matrix}
\end{align}
\end{subequations}
Returning to the case of the extended catalytic state, where the physical state is encoded in the $N$-particle sector of the extended catalytic state, we have
\begin{subequations}
\begin{align}
 \rho_{kj}^{(1)}&=\bra{\varPsi_\mathrm{ecs}}\sP_N\, \a_j^\dagger \a_k\, \sP_N\ket{\varPsi_\mathrm{ecs}}\\
 &=\bra{\varPsi_\mathrm{ecs}} \a_j^\dagger \a_k\, \sP_N\ket{\varPsi_\mathrm{ecs}}\;.\lb{eq:reduced_density_matrix_ecs}
\end{align}
\end{subequations}
A convenient way to evaluate Eq.~(\rf{eq:reduced_density_matrix_ecs}) is through the generating function
\begin{subequations}
\begin{align}
 G_{k,\,j}(\tau)
 &=\bra{\varPsi_\mathrm{ecs}}\a_j^\dagger \a_k\, \tau^{\,\sN}\ket{\varPsi_\mathrm{ecs}}\\[3pt]
 &=\bra{\varPsi_\mathrm{ecs}}\tau^{\sN/2}\,\a_j^\dagger \a_k\, \tau^{\,\sN/2}\ket{\varPsi_\mathrm{ecs}}\\[3pt]
 &=\frac{1}{\pi^\rank}  \int\normb{\mathfrak{B}_\mathrm{ecs}\big(\mathbf z\sqrt{\tau}\,\big)}^2\,\big(z_j z_k^*-\delta_{jk}\big)\, e^{-\norm{\mathbf z}^2}\, \dif^2 \mathbf z\;,
 \lb{eq:generating_function_ecs}
\end{align}
\end{subequations}
where $\sN=\sum_j \a_j^\dagger \a_j$ is the total particle-number operator, so that $\tau^{\,\sN}=\sum_{N=1}^\infty \tau^N\sP_N$. The value of $\rho_{kj}^{(1)}$ is encoded in the coefficient of the term $\tau^{N}$ in a power-series expansion of $G_{k,\,j}(\tau)$. For Gaussian states, the integral~(\rf{eq:generating_function_ecs}) is not hard to do, thus allowing $\rho_{kj}^{(1)}$ to be retrieved.

This procedure of deriving 1RDMs by using the BF representation can be straightforwardly generalized to the higher order RDMs.

\renewcommand{\lb}[1]{\label{symplectic:#1}}
\renewcommand{\rf}[1]{\ref{symplectic:#1}}

\chapter{Symplectic Methods for Quantum Mechanics}
\label{ch:symplectic}
\si{Symplectic method}

\begin{quote}
Mathematicians must have felt this way when they discovered that complex numbers were more than just one extra gimmick: Virtually every idea of mathematics, from the geometry of curves to the analysis of partial differential equations, was ripe for complexification. Mathematics exploded overnight.\\[4pt]
-- Ian Stewart\ai{Stewart, Ian}
\end{quote}

\section{Introduction}\lb{sec:introduction}

Quantum field theory is notoriously hard, which explains why superconductivity and superfluidity eluded people for so long.  Our knowledge of them, however, has tremendously increased with the help of the Bogoliubov transformation~\cite{bogoliubov_theory_1947, bardeen_theory_1957, kittel_quantum_1987}.\ai{Bogoliubov, Nikolay}  The Bogoliubov transformation of $\nrank$ bosonic modes is a linear transformation of the annihilation and creation operators,\si{Bogoliubov transformation}\footnote{For the purpose of this dissertation, I only discuss the bosonic case and leave out the equally important fermionic case.}
\begin{subequations}\lb{eq:bogoliubov_transformation_a}
\begin{align}
& \abf=U \bbf+V^* \bbf^\dagger\,, \lb{eq:bogoliubov_transformation_a_(a)} \\
& \abf^\dagger=U^* \bbf^\dagger+V \bbf\;, \lb{eq:bogoliubov_transformation_a_(b)}
\end{align}
\end{subequations}
where $\abf=(\a_1\;\a_2\,\cdots\, \a_\nrank)^\transp$ and $\abf^\dagger=(\a_1^\dagger\;\a_2^\dagger\,\cdots\, \a_\nrank^\dagger)^\transp$ are the column vectors of the input annihilation and creation operators, $\bbf=(\b_1\;\b_2\,\cdots\, \b_\nrank)^\transp$ and $\bbf^\dagger=(\b_1^\dagger\;\b_2^\dagger\, \cdots\, \b_\nrank^\dagger)^\transp$ are the output operators determined by implicitly Eqs.~(\rf{eq:bogoliubov_transformation_a}), and $U$ and $V$ are $\nrank\times \nrank$ matrices that specify the transformation.  The inputs, being annihilation and creation operators of different modes, satisfy the canonical commutation relations,
\begin{equation}\si{Canonical commutation relations}\lb{eq:canonical_commutators}
 \commut{\a_j}{\a_k}=\commut{\a_j^\dagger}{\a_k^\dagger}=0\,,\qquad \commut{\a_j}{\a_k^\dagger}=\delta_{jk}\,.
\end{equation}
The Bogoliubov transformation Eqs.~(\rf{eq:bogoliubov_transformation_a}) conserves these relations,
\begin{equation}\lb{eq:conserve_commutators}
 \commut{\b_j}{\b_k}=\commut{\b_j^\dagger}{\b_k^\dagger}=0\,,\qquad \commut{\b_j}{\b_k^\dagger}=\delta_{jk}\,,
\end{equation}
or equivalently, we have the following constrains on the matrices $U$ and $V$,
\begin{subequations}\lb{eq:symplectic_condition_a}
\begin{align}
& U^\dagger U-V^\dagger V = \identity\,,\lb{eq:symplectic_condition_a_(a)} \\
& U^\transp V-V^\transp U = \nullmatrix\;, \lb{eq:symplectic_condition_a_(b)}
\end{align}
\end{subequations}
where $\identity$ and $\nullmatrix$ denote the $\nrank\times \nrank$ identity matrix and null matrix, respectively.  These conditions~(\rf{eq:symplectic_condition_a_(a)}, \rf{eq:symplectic_condition_a_(b)}) are called the symplectic conditions\si{Symplectic condition}.  Therefore, the study of Bogoliubov transformations is equivalent to the study of symplectic transformations.

Symplectic structures arise naturally in classical Hamiltonian mechanics, where linear transformations of the canonical coordinates conserve the Poisson brackets.  Not restricted to Hamiltonian mechanics, it also finds applications in many other aspects of physics and engineering; the reader is referred to~\cite{berndt_introduction_2000, arnold_mathematical_1989} for pedagogical introductions to the subject, \cite{ferraro_gaussian_2005, gosson_symplectic_2006, xiao_theory_2009, adesso_continuous_2014}~for applications to quantum mechanics, and~\cite{guillemin_symplectic_1990}~for general applications to physics.

This Appendix provides a systematic introduction to the symplectic formalism to quantum mechanics in the basis of the annihilation and creation operators.\footnote{Almost all materials on symplectic structures adopt the position-momentum basis of $\hatx$ and $\hatp$.}  Doing this is especially convenient for the purpose of quantum optics and quantum many-body physics, partly because the many-body Hamiltonians are usually written in terms of the annihilation and creation operators.  Another, perhaps more important reason is that using the basis of annihilation and creation operators allows one to distinguish beamsplitters from squeezers easily; while beamsplitters are considered readily available in quantum optics, squeezers are much harder to implement and considered as a resource~\cite{braunstein_squeezing_2005}.

Following is a list of the topics covered in this Appendix: (i)~the standard symplectic structure of a quadratic Hamiltonian; (ii)~the Heisenberg time evolution under a quadratic Hamiltonian; (iii)~the relation between the ``diagonalization'' of a quadratic Hamiltonian and the symplectic eigenvalue problem; (iv)~the polar decomposition of symplectic matrices with a proof of the Bloch-Messiah theorem~\cite{bloch_canonical_1962}; and (v)~some examples and applications to quantum optics and quantum many-body physics.

\section{Quadratic Hamiltonians}
\lb{sec:the_quadratic_hamiltonian}

The class of Hamiltonians that are quadratic in annihilation and creation operators, or simply quadratic Hamiltonian, is a very special one.  Many important Hamiltonians can be reduced to this class if approximations are allowed.  Most importantly, systems evolving under quadratic Hamiltonians can be solved efficiently, and their ground states are easy to find.  Before going on, however, let us look at the notation for an $\nrank$-mode quadratic Hamiltonian,\footnote{For quadratic Hamiltonians, the difference between different orderings of the annihilation and creation operators is a $c$-number.  In this Appendix, we assume all Hamiltonians are symmetrically ordered.}\si{Quadratic Hamiltonian}
\begin{equation}\lb{eq:quadratic_Hamiltonian_a}
\sH=\half\,
\begin{pmatrix}
\abf^\dagger & \abf
\end{pmatrix}
\begin{pmatrix}
H_1 & H_3 \\ H_2 & H_4
\end{pmatrix}
\begin{pmatrix}
\abf \\  \; \abf^\dagger
\end{pmatrix}\;,
\end{equation}
where $H_1$, $H_2$, $H_3$, and $H_4$ are all $\nrank\times \nrank$ matrices, and the following conventions are used throughout this Appendix,
\begin{equation}\lb{eq:spinor_vector_convention}
\begin{pmatrix}
\abf^\dagger & \abf
\end{pmatrix}=
\begin{pmatrix}
\a_1^\dagger\:\cdots\: \a_\nrank^\dagger\:\; \a_1 \:\cdots\: \a_\nrank
\end{pmatrix}
\;, \qquad
\begin{pmatrix}
\abf \\ \;\abf^\dagger
\end{pmatrix}=
\begin{pmatrix}
\,\a_1\, \\[-4pt]\vdots \\[-4pt]\a_\nrank \\ \a_1^\dagger  \\[-4pt] \vdots \\[-4pt] \a_\nrank^\dagger
\end{pmatrix}\;.
\end{equation}
The $2\nrank\times 2\nrank$ matrix in Eq.~(\rf{eq:quadratic_Hamiltonian_a}), often called the Hamiltonian matrix\si{Hamiltonian matrix}, determines the quadratic Hamiltonian $\sH$.  To distinguish it from ordinary $\nrank\times \nrank$ matrices, we denote it by a special sans serif font,\footnote{In this dissertation, slanted sans serif fonts always denote matrices of symplectic structure.}
\begin{equation}\lb{eq:quadratic_Hamiltonian_b}
\mathsp{H}=
\begin{pmatrix}
H_1 & H_3 \\ H_2 & H_4
\end{pmatrix}\;.
\end{equation}

The canonical commutation relations~(\rf{eq:canonical_commutators}) take the form
\begin{equation}\lb{eq:canonical_commutators_matrixZ}
 \bigg[
 \begin{pmatrix}
\abf\\ \;\abf^\dagger\nsp
\end{pmatrix}
,\;
\begin{pmatrix}
\abf^\dagger\nsp & \abf
\end{pmatrix}
\bigg]
=
\begin{pmatrix}
\;\identity & \nullmatrix \\ \;\nullmatrix & -\identity\,
\end{pmatrix}
=\mathsp{Z}\;,
\end{equation}
where the commutator notation on the left means to put the commutators of the elements in the column and row vectors into the corresponding positions of the $2\nrank\times 2\nrank$ matrix.
We also have
\begin{subequations}\lb{eq:spinor_vector_conjugation} 
\begin{align}
\begin{pmatrix}
\abf & \abf^\dagger
\end{pmatrix}&=
\begin{pmatrix}
\a_1\:\cdots\: \a_\nrank\:\; \a_1^\dagger \:\cdots\: \a_\nrank^\dagger
\end{pmatrix}
=\begin{pmatrix}
\abf^\dagger & \abf
\end{pmatrix}
\mathsp{X}
\;,\\[6pt]
\begin{pmatrix}
\;\abf^\dagger \\ \abf
\end{pmatrix}&=
\begin{pmatrix}
\,\a_1^\dagger\, \\[-4pt]\vdots \\[-4pt]\a_\nrank^\dagger \\ \a_1  \\[-4pt] \vdots \\[-4pt] \a_\nrank
\end{pmatrix}
=\mathsp{X}
\begin{pmatrix}
\abf \\ \;\abf^\dagger
\end{pmatrix}\;,
\end{align}
\end{subequations}
where
\begin{equation}
\mathsp{X}=
\begin{pmatrix}
\,\nullmatrix\, &\, \identity\; \\ \,\identity\, &\, \nullmatrix\;
\end{pmatrix}\;.
\end{equation}
In words, Hermitian conjugation of the annihilation and creation operators is the same as multiplication by $\mathsp{X}$.  Equations~(\rf{eq:canonical_commutators_matrixZ}) and (\rf{eq:spinor_vector_conjugation}) introduce the matrices $\mathsp{Z}=\mathsp{Z}^\dagger=\mathsp{Z}^*=\mathsp{Z}^T$ and matrices $\mathsp{X}=\mathsp{X}^\dagger=\mathsp{X}^*=\mathsp{X}^T$ in the ways that demonstrate their importance within the symplectic formalism.

Using Eqs.~(\rf{eq:spinor_vector_conjugation}), it is easy to see that matrix $\mathsp{X}\mathsp{H}^T\mathsp{X}$ gives the same Hamiltonian $\sH$ as does $\mathsp{H}$.  Thus we can redefine the Hamiltonian matrix by
\begin{equation}\lb{eq:transpose_equivalence_b}
\mathsp{H} \,\rightarrow\, \half\,\big(\, \mathsp{H}+\mathsp{X} \mathsp{H}^\transp  \mathsp{X}\,\big)\,.
\end{equation}
The new $\mathsp{H}$, which gives the same Hamiltonian $\sH$, is now constrained by the condition
\begin{equation}\lb{eq:XHTX_condition}
\mathsp{H}^\transp=\mathsp{X} \mathsp{H} \mathsp{X}\,.
\end{equation}
With this condition, the Hamiltonian matrix $\mathsp{H}$ has a one-to-one correspondence with the quadratic Hamiltonian $\sH$.  Given the condition~(\rf{eq:XHTX_condition}), the Hermiticity of $\sH$ requires $\mathsp{H}$ to be a Hermitian matrix,
\begin{equation}\lb{eq:Hermitian_condition}
\mathsp{H}^\dagger=\mathsp{H}\,.
\end{equation}
Combining Eqs.~(\rf{eq:XHTX_condition}) and~(\rf{eq:Hermitian_condition}), we have
\begin{equation}\lb{eq:XHX_condition}
\mathsp{H^*}=\mathsp{X} \mathsp{H} \mathsp{X}\,.
\end{equation}
Only two of the three conditions, (\rf{eq:XHTX_condition}), (\rf{eq:Hermitian_condition}), and~(\rf{eq:XHX_condition}), are independent; we use the conditions~(\rf{eq:Hermitian_condition}) and~(\rf{eq:XHX_condition}) as primary (the reason for this choice should become apparent after the introduction of symplectic matrices). Imposing these two consraints, the Hamiltonian matrix $\mathsp{H}$ adopts the form
\begin{equation}\lb{eq:Hamiltonian_matrix}
\mathsp{H}=
\begin{pmatrix}
H_1 & H_2^* \\ H_2 & H_1^*
\end{pmatrix}\;,\quad
\mbox{where $H_1=H_1^\dagger$, and $H_2=H_2^\transp$.}
\end{equation}

The Hamiltonian $\sH$ therefore takes the form
\begin{equation}\lb{eq:quadratic_Hamiltonian_c}
\sH=\half\,
\begin{pmatrix}
\abf^\dagger & \abf
\end{pmatrix}
\begin{pmatrix}
H_1 & H_2^* \\ H_2 & H_1^*
\end{pmatrix}
\begin{pmatrix}
\abf \\  \; \abf^\dagger
\end{pmatrix}\;.
\end{equation}
It is clear that $H_1$ describes an $\nrank$-port beamsplitter and $H_2$ a squeezer. In contrast, had we written the Hamiltonian in the basis of $x$ and $p$, it would not be so easy to distinguish these two different sorts of transformation.  We have now prepared ourselves with the representation of the quadratic Hamiltonian $\sH$, and the next step is to solve for the time evolution.

\section{Heisenberg Time Evolution}

In the Heisenberg picture, the annihilation and creation operators undergo a linear transformation under the evolution of the quadratic Hamiltonian $\sH(t)$,
\begin{equation}\lb{eq:heisenberg_evolution_matrix}
\begin{pmatrix}
\abf(t)\\ \,\abf^\dagger\nsp (t)
\end{pmatrix}
=\mathsp{S}(t)
\begin{pmatrix}
\abf\\ \;\abf^\dagger
\end{pmatrix}\;,
\end{equation}
where $\mathsp{S}(t)$ is the evolution matrix.  Note that the initial values are used if the time arguments of the annihilation and creation operators are omitted.  Because any operator on the Fock space can be expanded as a Taylor series of the annihilation and creation operators, knowing $\mathsp{S}(t)$ allows one to solve the whole dynamical problem.\footnote{Different ordering of the annihilation and creation operators results in different expansion coefficients.}

Consider the following time-dependent quadratic Hamiltonian in the Schr\"odinger picture
\begin{equation}\lb{eq:quadratic_Hamiltonian_d}
\sH(t)=\half\,
\begin{pmatrix}
\abf^\dagger & \abf
\end{pmatrix}
\begin{pmatrix}
H_1(t) & H_2^*(t) \\ H_2(t) & H_1^*(t)
\end{pmatrix}
\begin{pmatrix}
\abf \\  \; \abf^\dagger
\end{pmatrix}\;.
\end{equation}
Going to the Heisenberg picture, we have
\begin{equation}\lb{eq:quadratic_Hamiltonian_e}
\sH_\mathrm{\ssO}(t)=\half\,
\begin{pmatrix}
\abf^\dagger\nsp(t) & \abf(t)
\end{pmatrix}
\begin{pmatrix}
H_1(t) & H_2^*(t) \\ H_2(t) & H_1^*(t)
\end{pmatrix}
\begin{pmatrix}
\abf(t) \\  \; \abf^\dagger\nsp(t)
\end{pmatrix}\;.
\end{equation}
Since the annihilation and creation operators in the Heisenberg picture also satisfy the canonical commutation relations, we have
\begin{subequations}
\begin{align}
 &\commutb{\sH_\mathrm{\ssO}(t)}{\abf(t)}=-H_1(t)\, \abf(t) - H_2^*(t)\, \abf^\dagger(t)\,, \lb{eq:commutator_H_A}\\
 &\commutb{\sH_\mathrm{\ssO}(t)}{\abf^\dagger(t)}=H_2(t)\, \abf(t) + H_1^*(t)\, \abf^\dagger(t)\;.\lb{eq:commutator_H_A*}
\end{align}
\end{subequations}
Thus, the Heisenberg evolutions of the annihilation and creation operators obey the differential equations ($\hbar=1$)
\begin{align}\lb{eq:heisenberg_evolutions_a}
\begin{pmatrix}
\dt\abf(t)\\ \,\dt\abf^\dagger\nsp (t)
\end{pmatrix}
&=i\,
\begin{pmatrix}
\commutb{\sH_\mathrm{\ssO}(t)}{\abf(t)}\\[2pt] \commut{\sH_\mathrm{\ssO}(t)}{\abf^\dagger\nsp (t)}
\end{pmatrix}
=-i\ssp\mathsp{Z}\mathsp{H}(t)
\begin{pmatrix}
\abf(t) \\  \; \abf^\dagger\nsp(t)
\end{pmatrix}\;.
\end{align}
Comparing Eqs.~(\rf{eq:heisenberg_evolution_matrix}) and (\rf{eq:heisenberg_evolutions_a}), we have
\begin{equation}\lb{eq:heisenberg_evolutions_time}
\dt{\mathsp{S}}(t)
=\mathord{-}i \mathsp{Z}\mathsp{H}(t)\mathsp{S}(t)\;,
\end{equation}
which can be solved formally if $\mathsp{H}$ is time independent,\footnote{The formal solution involves a time-ordered exponential when the Hamiltonian matrix $\mathsp{H}(t)$ is explicitly time dependent.}
\begin{equation}\lb{eq:heisenberg_evolutions_c}
\mathsp{S}(t)
=\exp\big(\mathord{-}i t\mathsp{Z}\mathsp{H}\ssp \big)\;.
\end{equation}

The evolution matrix has a symmetry that comes from the fact that $\abf(t)$ and $\abf^\dagger(t)$ are Hermitian conjugates of one another:
\begin{equation}\lb{eq:XSX_condition_a}
 \mathsp{S}^*\nsp (t)=\mathsp{X}\mathsp{S}(t)\mathsp{X}\,.
\end{equation}
More intuitively, this symmetry can be represented by the following matrix form,
\begin{equation}\lb{eq:S_matrix_b}
\mathsp{S}(t)=
\begin{pmatrix}
S_1(t) & S_2^*(t) \\ S_2(t) & S_1^*(t)
\end{pmatrix} \;,
\end{equation}
where $S_1(t)$ and $S_2(t)$ are $\nrank\times \nrank$ matrices.  The careful reader will find that this condition is similar to condition~(\rf{eq:XHX_condition}) on $\mathsp{H}$, and it can actually be derived from that condition.  In the time-independent case, the derivation is
\begin{align}\lb{eq:XSX_condition_b}\hspace{-1em}
\mathsp{S}^*\nsp (t)
&=\exp\big( it\mathsp{Z}\mathsp{H^*} \big)
=\exp\big( it\mathsp{Z}\mathsp{X}\mathsp{H} \mathsp{X}\ssp \big)
=\exp\big( \mathord{-}it\mathsp{X}\mathsp{Z}\mathsp{H} \mathsp{X}\ssp \big)
=\mathsp{X}\mathsp{S}(t)\mathsp{X}\;.
\end{align}
This explains why we choose Eq.~(\rf{eq:XHX_condition}) over Eq.~(\rf{eq:XHTX_condition}): the evolution matrix $\mathsp{S}(t)$ does not satisfy a condition similar to Eq.~(\rf{eq:XHTX_condition}).  

In the Heisenberg picture, the canonical commutation relations take the form
\begin{align}\hspace{-2em}
 \bigg[
 \begin{pmatrix}
\abf(t)\\ \,\abf^\dagger\nsp (t)
\end{pmatrix}
,\;
\begin{pmatrix}
\abf^\dagger\nsp(t) & \abf (t)
\end{pmatrix}
\bigg]
&=\mathsp{S}(t)
 \bigg[
 \begin{pmatrix}
\abf\\ \;\abf^\dagger\nsp
\end{pmatrix}
,\;
\begin{pmatrix}
\abf^\dagger\nsp & \abf
\end{pmatrix}
\bigg]\mathsp{S}^\dagger\nsp(t)=\mathsp{S}(t) \mathsp{Z} \mathsp{S}^\dagger\nsp(t)\;.
\end{align} 
Conservation of the canonical commutations means that
\begin{equation}\lb{eq:symplectic_condition_b}
\mathsp{S}(t) \mathsp{Z} \mathsp{S}^\dagger\nsp(t)=\mathsp{Z} \;.
\end{equation}
This, called the symplectic condition, is a consequence of the Hermiticity of $\mathsp{H}$,
\begin{subequations}\lb{eq:symplectic_condition_c}
\begin{align}
\mathsp{S}(t) \mathsp{Z} \mathsp{S}^\dagger\nsp(t)
&=\exp\big(\mathord{-}it \mathsp{Z} \mathsp{H}\big) \, \mathsp{Z}\, \exp\big(it \mathsp{H}^\dagger \mathsp{Z}^\dagger \big)\\
&=\exp\big(\mathord{-} i t \mathsp{Z} \mathsp{H}  \big)\, \exp\big(i t \mathsp{Z} \mathsp{H}  \big)\,\mathsp{Z}=\mathsp{Z}\;.
\end{align}
\end{subequations}
Two different, but also very useful, ways of writing the symplectic condition are
\begin{align}
&\mathsp{S}^{-1} (t) =\mathsp{Z} \mathsp{S}^\dagger\nsp(t) \mathsp{Z}\,, \lb{eq:symplectic_condition_d}\\
&\mathsp{S}^\dagger\nsp(t) \mathsp{Z} \mathsp{S}(t)= \mathsp{Z}\,.\lb{eq:symplectic_condition_e}
\end{align}
Note that the determinant of a symplectic matrix is either $1$ or $-1$ as a consequence of Eqs.~(\rf{eq:XSX_condition_a}) and (\rf{eq:symplectic_condition_b}).

We will see that the Bogoliubov transformation is a symplectic transformation of the modal operators and can be generated in the same way as $\mathsp{S}(t)$ in Eq.~(\rf{eq:heisenberg_evolutions_c}).

\section{Bogoliubov Transformations}

We have already seen how to put a quadratic Hamiltonian into a standard form and how formally to derive the Heisenberg dynamics of the annihilation and creation operators.  Questions still remain: (i)~how to solve the energy spectrum of the Hamiltonian $\sH$ and (ii)~how to derive $\mathsp{S}(t)$ beyond the formal result in Eq.~(\rf{eq:heisenberg_evolutions_c}). To answer these questions, we will need to ``diagonalize'' the Hamiltonian matrix $\mathsp{H}$ by the Bogoliubov transformation~(\rf{eq:bogoliubov_transformation_a}),\si{Bogoliubov transformation}
\begin{equation}
\begin{pmatrix}
\abf \\ \;\abf^\dagger
\end{pmatrix}=
\mathsp{B}
\begin{pmatrix}
\bbf \\  \; \bbf^\dagger
\end{pmatrix}\;, \qquad
\mbox{where}\,\: \mathsp{B}=
\begin{pmatrix}
U & V^*\\
V & U^*
\end{pmatrix}\;.
\end{equation}
The matrix $\mathsp{B}$ satisfies the condition
\begin{align}
\mathsp{B^*}=\mathsp{X}\mathsp{B}\mathsp{X}\;, \lb{eq:XBX_condition}
\end{align}
by virtue of the top and bottom halves of the vectors being Hermitian conjugates on one another, and the preservation of commutators implies that [this is equivalent the condition~(\rf{eq:symplectic_condition_a})] 
\begin{align}
\mathsp{B}^\dagger \mathsp{Z} \mathsp{B}= \mathsp{Z}\;. \lb{eq:symplectic_condition_f}
\end{align}
Thus the matrix $\mathsp{B}$ shares the same properties as $\mathsp{S}(t)$ and is also symplectic.  

In the new basis of $\bbf$ and $\bbf^\dagger$, the Hamiltonian matrix becomes 
\begin{equation}\lb{eq:symplectic_transformation}
\mathsp{H}'=\mathsp{B}^\dagger\mathsp{H}\mathsp{B}\,.
\end{equation}
The condition~(\rf{eq:XHX_condition}) is preserved by this symplectic transformation,
\begin{align}
\mathsp{X}\mathsp{H}'\mathsp{X} &=\mathsp{X}\mathsp{B}^\dagger \mathsp{H} \mathsp{B} \mathsp{X}=\mathsp{B}^\transp\mathsp{X}\mathsp{H}\mathsp{X}\mathsp{B^*}=\mathsp{H}'^{\, *}\;.
\end{align}
It is also straightforward to see that $\mathsp{H}'=\mathsp{H}'^{\,\dagger}$.   Therefore, symplectic transformations preserve the symmetries of the Hamiltonian matrix.

Often it is useful to consider conserved quantities under symplectic transformations. For example, symplectic transformations keeps the determinant of the Hamiltonian matrix unchanged,
\begin{equation}\lb{eq:symplectic_invariants_b}
\det(\mathsp{H}')=\det(\mathsp{H})\,.
\end{equation}
Also, we have
\begin{align}\lb{eq:symplectic_invariants_a}
\tr(\mathsp{Z}\mathsp{H}'\mathsp{Z} \mathsp{H}')&=\tr(\mathsp{Z}\mathsp{B}^\dagger  \mathsp{H}\mathsp{B}\mathsp{Z}\mathsp{B}^\dagger\mathsp{H}\mathsp{B})=\tr(\mathsp{Z}\mathsp{H}\mathsp{Z}\mathsp{H})\,，
\end{align}
which can be easily generalized to higher orders.  Note that the trace of any odd repetition of $\mathsp{Z}\mathsp{H}$ is zero, for example $\tr(\mathsp{Z}\mathsp{H})= \tr(\mathsp{Z}\mathsp{H}\mathsp{Z}\mathsp{H}\mathsp{Z}\mathsp{H})=0$.\footnote{One can simply use  $\mathsp{H}^*=\mathsp{X}\mathsp{H}\mathsp{X}$, $\mathsp{H}^\dagger=\mathsp{H}$, and $\mathsp{X}\mathsp{Z}\mathsp{X}=-\mathsp{Z}$ to prove this.}
We will use these invariants in Secs.~\rf{subsec:single_mode_case}, \rf{subsec:solve_two_mode}, and~\rf{subsec:fluctuation_quantum_mechanics} to solve practical problems.

The condition Eq.~(\rf{eq:symplectic_condition_f}) may not be the most convenient for the purpose of expressing the symplectic restrictions on $\mathsp{B}$; instead, one can use the exponential form
\begin{equation}
 \mathsp{B}=e^{i\mathspst{Z}\mathspst{M}}\;,
\end{equation}
where the matrix $\mathsp{M}$ has the same constraints as the Hamiltonian matrix (indeed, this the same trick as writing a unitary operator $U$ as $U=e^{-iH}$, where $H$ is Hermitian),
\begin{align}
&\mathsp{M}^*= \mathsp{X} \mathsp{M} \mathsp{X}\quad\mbox{and $\mathsp{M}^\dagger= \mathsp{M}$}\,.
\end{align}
We will use this form to parameterize all the two-dimensional symplectic matrices in Subsec.~\rf{subsec:single_mode_case}.

\section{Symplectic Form and Basis}

The symplectic form\si{Symplectic form}, which plays a role similar to the inner product, is defined as
\begin{equation}
\varpi_\mathrm{sp}(\ffq_1,\,\ffq_2)=\bra{\ffq_2}\mathsp{Z}\,\ket{\ffq_1}\,,
\end{equation}
where $\ket{\ffq_1}$ and $\ket{\ffq_2}$ are $2\nrank$-dimensional vectors in the symplectic vector space.
A symplectic transformation is a transformation that preserves the symplectic form,
\begin{align}
\bra{\ffq_2'}\mathsp{Z}\,\ket{\ffq_1'}&=\bra{\ffq_2}\mathsp{S}^\dagger \mathsp{Z}\mathsp{S}\,\ket{\ffq_1}=\bra{\ffq_2}\mathsp{Z}\,\ket{\ffq_1}\,.
\end{align}

Another useful concept is that of a symplectic basis,\si{Symplectic basis}
\begin{equation}\lb{eq:symplectic_basis_a}
\big\{\,\ket{\ffs_\sigma^{(j)}}\; \big\vert\; j=1,2,\ldots, \nrank;\, \sigma = 0,1\,\big\}\;,
\end{equation}
which satisfies
\begin{align}\lb{eq:symplectic_basis_conditions}
\bra{\ffs_{\upsilon}^{(k)}} \mathsp{Z}\,\ket{\ffs_\sigma^{(j)}}
=(-1)^\sigma \delta_{\sigma, \upsilon}\, \delta_{j,k}\,.
\end{align}
For a particular $j$, the two values of $\sigma$ label two symplectically conjugate states, which are related by
\begin{equation}
\ket{\ffs_{\sigma+1}^{(j)}}=\mathsp{X}\, \ket{\ffs_\sigma^{(j\,)}}^*
\lb{eq:symplectic_conjugate_condition}
\end{equation}
($\sigma+1$ should be interpreted as modulo $2$ addition; i.e., $\sigma$ changes to the other value).  
An example of a symplectic basis is the natural basis
\begin{equation}\lb{eq:natural_basis}
 \ket{\ffe_\sigma^{(j)}}=\big(0\; 0 \,\cdots\, 0\; 1\; 0\, \cdots \cdots\, 0\; 0 \big)^\transp\,,
\end{equation}
where there is just one nonzero element, equal to 1, located at the $(\sigma \nrank+j)$th position.  A symplectic transformation is a transformation that transforms a symplectic basis to another symplectic basis,
\begin{equation}\lb{eq:basis_transformation}
 \ket{\ffh_\sigma^{(j)}}=\mathsp{S}\,\ket{\ffs_\sigma^{(j)}}\,.
\end{equation}
It is straightforward to verify that Eqs.~(\rf{eq:symplectic_basis_conditions}) and (\rf{eq:symplectic_conjugate_condition}) hold for the new basis.  Note that conjugate states are mapped to conjugate states,
\begin{align}
\mathsp{S}\,\ket{\ffs_{\sigma+1}^{(j)}}
&=\mathsp{S}\,\mathsp{X}\,\ket{\ffs_\sigma^{(j\,)}}^* =\mathsp{X}\,\big(\mathsp{S}\,\ket{\ffs_\sigma^{(j)}}\big)^*\,.
\end{align}
Knowing the two bases in Eq.~(\rf{eq:basis_transformation}) allows one to write the symplectic transformation matrix as
\begin{equation}
 \mathsp{S}=\sum_{\sigma=0,1} (-1)^\sigma \sum_{j=1}^n\, \ket{\ffh_\sigma^{(j)}} \bra{\ffs_\sigma^{(j)}}\, \mathsp{Z}\;,
\end{equation}
which will be used to find the canonical form of the Hamiltonian matrix in Sec.~\rf{sec:canonical_form}.

\section{Symplectic Eigenvalue Problem}\si{Eigenvalue problem (symplectic)}

Knowing the eigenvalues and eigenstates of $\mathsp{Z}\mathsp{H}$ allows one to solve explicitly for the time evolution
\begin{equation}
 \mathsp{S}(t)
=\exp\big(\mathord{-}it\mathsp{Z}\mathsp{H}\ssp \big)\,.
\end{equation}
Moreover, these eigenvalues and eigenstates can be used to ``diagonalize'' the Hamiltonian $\sH$.

The symplectic eigenvalue equation reads
\begin{equation}\lb{eq:eigenvalue_a}
\mathsp{Z}\mathsp{H}\,\ket{\ffq}=\lambda_{\ffq}\,
\ket{\ffq}\;,
\end{equation}
where $\ket{\ffq}$ is an eigenvector and $\lambda_{\ffq}$ its eigenvalue. An equivalent, perhaps more useful way to write the eigenvalue equation is
\begin{equation}\lb{eq:eigenvalue_b}
\mathsp{H}\,\ket{\ffq}=\lambda_{\ffq}\,\mathsp{Z}\,
\ket{\ffq}\;.
\end{equation}
To find all the eigenvalues, one can solve the characteristic equation
\begin{equation}\lb{eq:characteristic_equation}
\det(\mathsp{H}-\lambda \mathsp{Z})=0\;.
\end{equation}
Since $\mathsp{H}$ and $\mathsp{Z}$ are both Hermitian, we have
\begin{align}\lb{eq:eigenvalue_complex_conjugate}\hspace{-2em}
\big[\det(\mathsp{H}-\lambda \mathsp{Z})\big]^*
&= \det(\mathsp{H^*}-\lambda^* \mathsp{Z^*})
=\det(\mathsp{H}^\transp-\lambda^* \mathsp{Z}^\transp)
=\det(\mathsp{H}-\lambda^* \mathsp{Z})\;,
\end{align}
which implies that if $\lambda$ is an eigenvalue, then $\lambda^*$ is also an eigenvalue. Using the condition Eq.~(\rf{eq:XHX_condition}), we have
\begin{align}\lb{eq:eigenvalue_symplectic_conjugate_a}
\mathsp{H} \mathsp{X}\,\ket{\ffq}^*&= \mathsp{X} \mathsp{H^*}\,\ket{\ffq}^*
=\lambda_{\ffq}^*\,\mathsp{X}\mathsp{Z}\, \ket{\ffq}^*
=-\lambda_{\ffq}^*\,\mathsp{Z}\mathsp{X}\, \ket{\ffq}^*\,,
\end{align}
which means that $-\lambda_{\ffq}^*$ is also an eigenvalue of $\mathsp{Z}\mathsp{H}$, corresponding to the conjugate eigenvector $\mathsp{X}\, \ket{\ffq}^*$.  Thus we have 
\begin{equation}\lb{eq:eigenvalue_symplectic_conjugate_b}
\mathsp{H}\, \ket{\thickbar{\ffq}}=-\lambda_{\ffq}^*\mathsp{Z}\,
\ket{\thickbar{\ffq}}\;,\quad\mbox{where $\ket{\thickbar{\ffq}}=\mathsp{X}\,\ket{\ffq}^*$ is the conjugate of $\ket{\ffq}$}\,.
\end{equation}
Thus, the eigenvalues appear in quadruplets
\begin{equation}\lb{eq:eigenvalues_quadruplet}
\{\lambda,\, \lambda^*,\, \mathord{-}\lambda^*,\, \mathord{-}\lambda\}\;.
\end{equation}
This is an expression of the fact that the characteristic function~(\rf{eq:characteristic_equation}) is even and real.

For two eigenvectors $\ket{\ffq_1}$ and $\ket{\ffq_2}$, we have
\begin{align}\lb{eq:symplectic_orthonormal}
\bra{\ffq_2} \mathsp{H}\,\ket{\ffq_1}
&=\lambda_{\ffq_1}\,\bra{\ffq_2}\mathsp{Z}\,\ket{\ffq_1} =\lambda_{\ffq_2}^*\ssp \bra{\ffq_2}\mathsp{Z}\,\ket{\ffq_1}\,,
\end{align}
which says that the symplectic form is zero when the eigenvalues of two eigenvectors are not a pair of complex conjugate numbers,
\begin{equation}\lb{eq:orthogonal_eigenvectors}
\bra{\ffq_2}\mathsp{Z}\,\ket{\ffq_1}=0\quad\mbox{if}\,\: \lambda^*_{\ffq_2}\neq \lambda_{\ffq_1}\,.
\end{equation}
A special case is when $\ket{\ffq_1}$ and $\ket{\ffq_2}$ are the same state, and we have
\begin{equation}\lb{eq:orthogonal_eigenvectors_self}
\bra{\ffq}\mathsp{Z}\,\ket{\ffq}=0\quad\mbox{if $\lambda_{\ffq}$ is not real.}
\end{equation}

It turns out that the eigenvalues of Eq.~(\rf{eq:eigenvalue_b}) can be real, pure imaginary, or complex; I will be treat these possibilities separately in the following subsections.  The central topic is to construct a symplectic basis from the eigenvectors.  Such symplectic bases turn out to be crucial to bringing the Hamiltonian matrix $\mathsp{H}$ to the so-called canonical form.

\subsection{Real-Eigenvalue Case}

When $\lambda$ is real, the four eigenvalues in the quadruplet Eq.~(\rf{eq:eigenvalues_quadruplet}) degenerate into $\{\lambda,\,\mathord{-}\lambda\}$, and the eigenvalue equations take the form
\begin{equation}\lb{eq:eigenvalue_eq_real}
\mathsp{H}\,\ket{\ffr_{\sigma}}
=(-1)^\sigma \lambda\, \mathsp{Z}\,
\ket{\ffr_{\sigma}}\;,
\end{equation}
where $\sigma=0, 1$ denotes the two conjugate eigenvectors, which are related by the condition~(\rf{eq:eigenvalue_symplectic_conjugate_b}),
\begin{equation}\lb{eq:eigenvalue_symplectic_conjugate_c}
\ket{\ffr_{\sigma+1}}=\mathsp{X}\,\ket{\ffr_\sigma}^*\;.
\end{equation}
As a consequence, we have
\begin{subequations}\lb{eq:real_symplectic_orthonormal_a}
\begin{align}
&\bra{\ffr_1}\mathsp{Z}\,\ket{\ffr_0}=0\,,\\
&\bra{\ffr_1}\mathsp{Z}\,\ket{\ffr_1}
=-\,\bra{\ffr_0}\mathsp{Z}\,\ket{\ffr_0}\,.
\end{align}
\end{subequations}
Note that the symplectic form of $\ket{\ffr_{\sigma}}$ with itself, $\bra{\ffr_{\sigma}}\mathsp{Z}\,\ket{\ffr_{\sigma}}$, must be nonzero because of the symplectic nondegenerate argument.\footnote{In other words, the eigenvectors defined by Eq.(\rf{eq:eigenvalue_b}) span the whole symplectic space.  Only the null vector has zero symplectic form with all vectors in a symplectic space.}  Therefore, we can ``normalize'' the two eigenvectors, so that the symplectic forms in Eqs.~(\rf{eq:real_symplectic_orthonormal_a}) become standard,
\begin{equation}\lb{eq:real_symplectic_orthonormal_b}
 \bra{\ffr_{\sigma{\smash '}}}\mathsp{Z}\, \ket{\ffr_\sigma}= (-1)^{\sigma}\delta_{\sigma,\sigma{\smash '}}\,.
\end{equation}
Note that we adopt the convention that the eigenvector that has positive symplectic form with itself is labeled as $\sigma=0$.  The two eigenvectors, $\ket{\ffr_0}$ and $\ket{\ffr_1}$, form a symplectic basis on a two-dimensional plane.

\subsection{Pure-Imaginary-Eigenvalue Case}

When $\lambda$ is pure imaginary, the four eigenvalues also degenerate into $\{\lambda,\,\mathord{-}\lambda\}$, and the eigenvalue equations are
\begin{equation}\lb{eq:imaginary_eigenvalue}
\mathsp{H}\,\ket{\ffg_{\pm}}=\pm\, \lambda\, \mathsp{Z}\,\ket{\ffg_{\pm}}\,.
\end{equation}
Since $\lambda=-\lambda^*$, the eigenvectors are now conjugate to themselves,\footnote{Strictly speaking, the two sides of Eq.~(\rf{eq:self_symplectic_conjugate}) can differ by a phase factor $e^{i\theta}$, but it always can be eliminated by redefining the eigenvector as $\ket{\ffg}\,\rightarrow\, e^{-i\theta/2}\,\ket{\ffg}$.}
\begin{equation}\lb{eq:self_symplectic_conjugate}
\ket{\ffg_{\pm}}=\mathsp{X}\,\ket{\ffg_{\pm}}^* \,,
\end{equation}
which leads to the following relation
\begin{align}
\big(\,\bra{\ffg_-}\mathsp{Z}\, \ket{\ffg_+}\,\big)^*
&=\bra{\ffg_-}\, \mathsp{X}\mathsp{Z^*} \mathsp{X}\,\ket{\ffg_+}=-\,\bra{\ffg_-}\mathsp{Z}\, \ket{\ffg_+}\,.
\end{align}
Thus $\bra{\ffg_-}\mathsp{Z}\, \ket{\ffg_+}$ is pure imaginary and can always be written in the following standard form by rescaling either $\ket{\ffg_+}$ or $\ket{\ffg_-}$,\footnote{Again, we can argue that this symplectic form must be nonzero by the symplectic nondegenerate condition.}
\begin{equation}
\bra{\ffg_-}\mathsp{Z}\,\ket{\ffg_+}=i\,.
\end{equation}
Furthermore, from the ``orthogonality'' condition Eq.~(\rf{eq:orthogonal_eigenvectors_self}), we have
\begin{equation}
 \bra{\ffg_+}\mathsp{Z}\, \ket{\ffg_+}= \bra{\ffg_-}\mathsp{Z}\, \ket{\ffg_-}=0\,.
\end{equation}
By introducing the new basis
\begin{equation}
\left\{\;
\begin{aligned}
&\ket{\ffg_0}=\frac{1}{\sqrt 2}\,\big(\,\ket{\ffg_+}
+i\,\ket{\ffg_-}\,\big) \;,\\
&\ket{\ffg_1}=\frac{1}{\sqrt 2}\,\big(\,\ket{\ffg_+} -i\,
\ket{\ffg_-}\,\big)\;,
\end{aligned}
\right.
\end{equation}
we have
\begin{subequations}\lb{eq:imaginary_symplectic_orthonormal}
\begin{align}
&\ket{\ffg_0}=\mathsp{X}\, \ket{\ffg_1}^*\;,\\
&\bra{\ffg_{\sigma{\smash'}}}\mathsp{Z}\, \ket{\ffg_\sigma}= (-1)^{\sigma}\delta_{\sigma,\sigma{\smash '}}\,,
\end{align}
\end{subequations}
where $\sigma =0,1$.  Thus $\ket{\ffg_0}$ and $\ket{\ffg_1}$ form a symplectic basis; in this basis, the eigenvalue equations~(\rf{eq:imaginary_eigenvalue}) become
\begin{equation}\lb{eq:eigenvalue_eq_imaginary}
\mathsp{H}\,\ket{\ffg_\sigma}=\lambda\, \mathsp{Z}\, \ket{\ffg_{\sigma+1}}\,.
\end{equation}

\subsection{Complex-Eigenvalue Case}

When $\lambda$ is complex, the eigenvalues appear in a group of four $\{\lambda,\, \lambda^*,\, \mathord{-}\lambda^*,\, \mathord{-}\lambda\}$, and the eigenvalue equations read
\begin{equation}\lb{eq:complex_eigenvalue_a}
\mathsp{H}\, \ket{\ffc_{\sigma, \mathord{\pm}}}
=\Big((-1)^{\sigma}\re \lambda \pm i\im \lambda\Big)\mathsp{Z}\,
\ket{\ffc_{\sigma, \mathord{\pm}}}\;,
\end{equation}
where $\pm$ denote the complex conjugate eigenvalues and $\sigma=0, 1$ labels the conjugate eigenvectors.  The only situation such that the symplectic form of two eigenvectors can be nonzero is when their eigenvalues are a pair of complex conjugate numbers,
\begin{align}
\bra{\ffc_{0,\mathord{-}}}\mathsp{Z}\, \ket{\ffc_{0,\mathord{+}}}
&=\bra{\ffc_{1,\mathord{-}}}^*\mathsp{X}\mathsp{Z}\mathsp{X}\, \ket{\ffc_{1,\mathord{+}}}^*=-\big(\,\bra{\ffc_{1,\mathord{-}}}\mathsp{Z}\, \ket{\ffc_{1,\mathord{+}}}\,\big)^*\,.
\end{align}
By multiplying the eigenvectors by appropriate complex numbers, we can always ``normalize'' these symplectic forms,
\begin{equation}
\bra{\ffc_{0,\mathord{-}}}\mathsp{Z}\, \ket{\ffc_{0,\mathord{+}}}=-\,\bra{\ffc_{1,\mathord{-}}}\mathsp{Z}\, \ket{\ffc_{1,\mathord{+}}}=1\;.
\end{equation}
We introduce the vectors $\ket{\ffc_{\sigma, \tau}}$ with $\tau=0,1$ as
\begin{equation}
\left\{\;
\begin{aligned}
&\ket{\ffc_{\sigma, 0}}=\big(\,\ket{\ffc_{\sigma,\mathord{+}}}+\ket{\ffc_{\sigma,\mathord{-}}}\,\big)/\sqrt{2}\;,\\
&\ket{\ffc_{\sigma, 1}}=\big(\,\ket{\ffc_{\sigma+1,\mathord{+}}}-\ket{\ffc_{\sigma+1,\mathord{-}}}\,\big)/\sqrt{2}\;
\end{aligned}
\right.
\end{equation}
which form a symplectic basis,
\begin{subequations}\lb{eq:complex_symplectic_orthonormal}
\begin{align}
&\ket{\ffc_{\sigma+1,\tau}}=\mathsp{X}\,\ket{\ffc_{\sigma,\tau}}^*\;,\\
&\bra{\ffc_{\sigma{\smash '},\tau{\smash '}}}\mathsp{Z}\, \ket{\ffc_{\sigma,\tau}}
=(-1)^{\sigma}\delta_{\sigma,\sigma{\smash '}}\,\delta_{\tau, \tau{\smash '}}\;.
\end{align}
\end{subequations}
In this basis, the eigenvalue equations~(\rf{eq:complex_eigenvalue_a}) become
\begin{equation}\lb{eq:eigenvalue_eq_complex}
\mathsp{H}\,\ket{\ffc_{\sigma, \tau}}=(-1)^{\sigma+\tau}\re(\lambda)\,\mathsp{Z}\,\ket{\ffc_{\sigma, \tau}}+i\im(\lambda)\,\mathsp{Z}\, \ket{\ffc_{\sigma+1, \tau+1}}\;.
\end{equation}

\section{Canonical (Normal) Forms}\si{Canonical form (symplectic)}
\lb{sec:canonical_form}

We have already seen how to construct symplectic bases from the eigenvectors of $\mathsp{Z}\mathsp{H}$ with real, pure imaginary, and complex eigenvalues.  These cases can coexist, and the whole symplectic vector space is a direct sum of the three.
From the ``orthogonality'' condition~(\rf{eq:orthogonal_eigenvectors}),
we have the following completeness condition
\begin{equation}\lb{eq:completeness_condition}\hspace{-1em}
\mathsp{I} =\sum_{\sigma=0,1}(-1)^\sigma
\bigg(
\sum_{j=1}^{\nrank_\ffr}\,\proj{\ffr_\sigma^{(j)}}+
\sum_{k=1}^{\nrank_\ffg}\proj{\ffg_\sigma^{(k)}}+
\sum_{l=1}^{\nrank_\ffc}\sum_{\tau=0,1}\proj{\ffc_{\sigma,\tau}^{(l)}}
\bigg)\,\mathsp{Z}\,,
\end{equation}
where $\mathsp{I}$ denotes the $2\nrank\times 2\nrank$ identity matrix, and $\nrank_\ffr$, $\nrank_\ffg$, and $\nrank_\ffc$ are the numbers of sets of real, pure imaginary, and complex eigenvalues, respectively. Note that these numbers must satisfy
\begin{equation}
 \nrank_\ffr+\nrank_\ffg+2 \nrank_\ffc=\nrank\,.
\end{equation}
With the completeness condition Eq.~(\rf{eq:completeness_condition}), we have the symplectic spectral decomposition for the Hamiltonian matrix
\begin{align}
\hspace{-2em}
\begin{split}
\mathsp{H} &= \mathsp{H}\sum_{\sigma=0,1}(-1)^\sigma
\bigg(
\sum_{j=1}^{\nrank_\ffr}\,\proj{\ffr_\sigma^{(j)}}+
\sum_{k=1}^{\nrank_\ffg}\proj{\ffg_\sigma^{(k)}}+
\sum_{l=1}^{\nrank_\ffc}\sum_{\tau=0,1}\proj{\ffc_{\sigma,\tau}^{(l)}}
\bigg)\,\mathsp{Z} \\
&= \mathsp{Z} \sum_{\sigma=0,1} \bigg(
\sum_{j=1}^{\nrank_\ffr}\lambda_\ffr^{(j)}\,\proj{\ffr_\sigma^{(j)}}+
\sum_{k=1}^{\nrank_\ffg}(-1)^\sigma \lambda_\ffg^{(k)} \,\ket{\ffg_{\sigma+1}^{(k)}} \bra{\ffg_\sigma^{(k)}}
+\sum_{l=1}^{\nrank_\ffc}\sum_{\tau=0,1} \\[3pt]
&\hspace{5em}(-1)^\tau \re\lambda_{\ffc}^{(l)}\, \proj{\ffc_{\sigma,\tau}^{(l)}}+ (-1)^\sigma i\im\lambda_{\ffc}^{(l)}\, \ket{\ffc_{\sigma+1,\tau+1}^{(l)}}\bra{\ffc_{\sigma,\tau}^{(l)}}
\bigg)\,\mathsp{Z}\,,
\end{split}\lb{eq:canonical_form}
\end{align}
where Eqs.~(\rf{eq:eigenvalue_eq_real}), (\rf{eq:eigenvalue_eq_imaginary}), and (\rf{eq:eigenvalue_eq_complex}) are used for actions of $\mathsp{H}$ on the symplectic basis.  If $\mathsp{Z}\mathsp{H}$ has a nilpotent subspace, its eigenvectors may not span the whole symplectic space.  In this case $\mathsp{H}$ cannot be written in the form~(\rf{eq:canonical_form}) in that subspace; we will come to this case at the end of Sec.~\rf{subsec:single_mode_case}.

To take the Hamiltonian matrix to the canonical form, we introduce the symplectic transformation
\begin{align}
\begin{split}
 \mathsp{B}&=\sum_{\sigma=0,1}\bigg(\sum_{j=1}^{\nrank_\ffr} \ket{\ffr_\sigma^{(j)}}\bra{\ffe_{\sigma}^{(j)}}+\sum_{k=1}^{\nrank_\ffg} \ket{\ffg_\sigma^{(k)}}\bra{\ffe_{\sigma}^{(\nrank_\ffr+k)}}+\sum_{l=1}^{\nrank_\ffc}\ket{\ffc_{\sigma,0}^{(l)}}\bra{\ffe_{\sigma}^{(\nrank_\ffr+\nrank_\ffg+l)}}\\[2pt]
 &\hspace{4em} +\ket{\ffc_{\sigma,1}^{(l)}}\bra{\ffe_{\sigma}^{(\nrank_\ffr+\nrank_\ffg+\nrank_\ffc+l)}}\,\bigg) \;,
\end{split}
\end{align}
where $\big\{\,\ket{\ffe_\sigma^{(j)}}\; \big\vert\; j=1,2,\ldots, \nrank;\, \sigma = 0,1\,\big\}$ is the natural basis.  The symplectic matrix $\mathsp{B}$ transforms the basis in Eq.~(\rf{eq:canonical_form}) into the natural basis, and we have
\begin{equation}\lb{eq:symplectic_transformation_b}
\mathsp{H}_\mathrm{can}=\mathsp{B}^\dagger\mathsp{H}\mathsp{B}\,.
\end{equation}
In matrix form, the symplectic matrix $\mathsp{B}$ reads
\begin{equation}
\mathsp{B}=
\begin{pmatrix}
U & V^* \\ V & U^*
\end{pmatrix}\;,
\end{equation}
where the symmetry $\mathsp{X}\mathsp{B}\mathsp{X} = \mathsp{B}^*$ is a consequence of the symmetry~(\rf{eq:symplectic_conjugate_condition}) of the symplectic basis.

It is worthwhile writing the canonical Hamiltonian of all the three cases in explicit matrix form.  For the real-eigenvalue case,
\begin{subequations}\lb{eq:canonical_form_matrix}
\begin{equation}
\mathsp{H}_{\nsp\mathrm{real}}=
\begin{pmatrix}
\lambda & 0\,  \\
0 & \lambda\\
\end{pmatrix}\;;
\end{equation}
for the pure-imaginary-eigenvalue case,
\begin{equation}\lb{eq:canonical_form_matrix_i}
\mathsp{H}_{\nsp\mathrm{imag}}=
\begin{pmatrix}
0 & \lambda\,  \\
\lambda^* & 0\\
\end{pmatrix}=
\begin{pmatrix}
0 & \lambda\,  \\
\mathord{-}\lambda & 0\\
\end{pmatrix}\;;
\end{equation}
and for the complex-eigenvalue case, we have
\begin{equation}\lb{eq:canonical_form_matrix_c}
\mathsp{H}_{\nsp\mathrm{comp}}=
\begin{pmatrix}
\re\lambda    & 0                  & 0                 & i\im\lambda \\
0                  & -\re\lambda   & i\im\lambda   & 0            \\
0                  & -i\im\lambda   & \re\lambda   & 0            \\
-i\im\lambda   & 0                  & 0                 & -\re\lambda
\end{pmatrix}\;.
\end{equation}
\end{subequations}
Note that by rephasing the basis set, one can always make the pure-imaginary form Eq.~(\rf{eq:canonical_form_matrix_i}) real,
\begin{equation}\lb{eq:canonical_form_matrix_i_b}
 \mathsp{H}_{\nsp\mathrm{imag}}=
\begin{pmatrix}
0                   & \norm{\lambda}  \\
\norm{\lambda} & 0\\
\end{pmatrix}\;.
\end{equation}

The only one of the canonical forms~(\rf{eq:canonical_form_matrix}) that can be positive definite is the real-eigenvalue case.  Because the symplectic transformation Eq.~(\rf{eq:symplectic_transformation_b}) keeps the positiveness of $\mathsp{H}$, we have
\begin{equation}\lb{eq:positive_stable}
\mathsp{H} > 0 \;\Longrightarrow\; \mbox{all the eigenvalues of $\mathsp{Z}\mathsp{H}$ are real.}
\end{equation}
Thus, any $\mathsp{H}>0$ can be turned into a collection of independent harmonic oscillators with positive frequencies by a Bogoliubov transformation.  

While the real-eigenvalue case corresponds to harmonic oscillators, the pure-imaginary-eigenvalue case corresponds to single-mode squeezers.  Perhaps the most interesting case is when the eigenvalues are complex Eq.~(\rf{eq:canonical_form_matrix_c}),
\begin{equation}\lb{eq:complex_symplectic_Hamiltonian}
\sH_\mathrm{comp} =\re(\lambda)\,\big(\a^\dagger \a-\b^\dagger \b\big)+i\im(\lambda)\,\big(\a^\dagger \b^\dagger-\a\ssp \b\big)\;.
\end{equation}
In this situation, the Hamiltonian matrix cannot be decoupled into a sum of single-mode Hamiltonians.  Physically, the Hamiltonian~(\rf{eq:complex_symplectic_Hamiltonian}) arises when two sidebands symmetrically placed about a carrier frequency, with the carrier energy removed, interact via a two-mode squeezing interaction.

\section{Polar Decompositions}\si{Polar decomposition (symplectic)}
\lb{sec:the_polar_decomposition}

The polar decomposition allows one to break a complicated symplectic transformation into several simple ones.  It can also be used to determine the squeezing ability of a unitary Gaussian operator.  Both of these have important applications in quantum information processing for continuous variables.

The polar decomposition of a symplectic matrix $\mathsp{S}$ is as follows,
\begin{equation}\lb{eq:polar_decomposition}
\mathsp{S}=\mathsp{K}\mathsp{U}=\overbar{\mathsp{U}}\ssp \overbar{\mathsp{K}}
\end{equation}
where $\mathsp{U}$ and $\overbar{\mathsp{U}}$ are unitary, and $\mathsp{K}$ and $\overbar{\mathsp{K}}$ are Hermitian.  Interestingly, the matrices $\mathsp{U}$, $\overbar{\mathsp{U}}$, $\mathsp{K}$, and $\overbar{\mathsp{K}}$ are also symplectic.  The proofs of this for the two ways of polar decomposition are similar, so I just sketch the first one here.  Since  $\mathsp{S}^{-1}=\mathsp{Z} \mathsp{S}^\dagger \mathsp{Z}$, we have $\mathsp{U}^{-1}\mathsp{K}^{-1}=\mathsp{Z}\mathsp{U}^\dagger\mathsp{Z}\mathsp{Z}\mathsp{K}^\dagger \mathsp{Z}$.  Because of the uniqueness of the polar decomposition, we conclude that $\mathsp{U}^{-1}=\mathsp{Z} \mathsp{U}^\dagger \mathsp{Z}$ and $\mathsp{K}^{-1}=\mathsp{Z} \mathsp{K}^\dagger \mathsp{Z}$, and these are nothing but the symplectic conditions for $\mathsp{U}$ and $\mathsp{K}$, respectively.

The polar decomposition can be used to break down any Gaussian unitary into a sequence of multiport beamsplitters and single-mode squeezers.  To see this, we first consider the unitary part,
\begin{equation}\lb{eq:exponential_V}
\mathsp{U}=\exp\big(\mathord{-}i \mathsp{Z} \mathsp{H}_{\nsp\ssU}\big)\;.
\end{equation}
Because $\mathsp{U}$ is both unitary and symplectic, its generator satisfies
\begin{equation}\lb{eq:generator_V}
\big(\mathsp{Z} \mathsp{H}_{\nsp\ssU}\big)^\dagger = \mathsp{Z} \mathsp{H}_{\nsp\ssU}\,,\qquad \mathsp{H}_{\nsp\ssU}^\dagger = \mathsp{H}_{\nsp\ssU}\,,\qquad \mathsp{H}_{\nsp\ssU}^* = \mathsp{X} \mathsp{H}_{\nsp\ssU} \mathsp{X}\;.
\end{equation}
These three conditions imply that $\mathsp{H}_{\nsp\ssU}$ must have the following block-diagonalized form,
\begin{equation}
\mathsp{H}_{\nsp\ssU}=
\begin{pmatrix}
H_{\nsp\ssU} & \nullmatrix\\
\nullmatrix  & H_{\nsp\ssU}^*
\end{pmatrix}\;,
\end{equation}
where $H_{\nsp\ssU}$ is an $\nrank\times \nrank$ Hermitian matrix.  Thus $\mathsp{U}$ is also block diagonalized and describes a multiport beamsplitter,
\begin{equation}
\mathsp{U}=
\begin{pmatrix}
e^{-i H_{\nsp\ssU}}  & \nullmatrix\\
\nullmatrix          & e^{it H_{\nsp\ssU}^*}
\end{pmatrix}\;.
\end{equation}
For the Hermitian part of the polar decomposition, we have
\begin{equation}\lb{eq:exponential_K}
\mathsp{K}=\exp\big(\mathord{-}i \mathsp{Z} \mathsp{H}_{\!\ssK}\big)\;.
\end{equation}
Because $\mathsp{K}$ is both Hermitian and symplectic, its generator satisfies
\begin{equation}\lb{eq:generator_K}
\big(\mathsp{Z} \mathsp{H}_{\!\ssK}\big)^\dagger = \mathord{-}\mathsp{Z} \mathsp{H}_{\!\ssK}\,,\qquad \mathsp{H}_{\!\ssK}^\dagger = \mathsp{H}_{\!\ssK}\,,\qquad \mathsp{H}_{\!\ssK}^* = \mathsp{X} \mathsp{H}_{\!\ssK} \mathsp{X}\;.
\end{equation}
These conditions imply that $\mathsp{H}_{\!\ssK}$ must take the form
\begin{equation}
\mathsp{H}_{\!\ssK}=
\begin{pmatrix}
\nullmatrix   & H_{\!\ssK}^*\\
H_{\!\ssK}    & \nullmatrix
\end{pmatrix}\;,
\end{equation}
where $H_{\!\ssK}^\transp = H_{\!\ssK}$ is an $\nrank\times \nrank$ symmetric matrix.  Symmetric matrices like $H_{\!\ssK}$ can always be diagonalized by unitary matrices in the following way (see App.~\chref{ch:at_factorization}),
\begin{equation}
 H_{\!\ssK}=V D\,  V^\transp\;,
\end{equation}
where $V^\dagger=V$ is a unitary matrix and $D$ a diagonal matrix.  Thus we have
\begin{equation}\lb{eq:diagonalize_H_k}
\mathsp{H}_{\!\ssK} = \mathsp{V}\mathsp{D}\mathsp{V}^\dagger\,,\qquad \mbox{where}\,\:
\mathsp{V}=
\begin{pmatrix}
V^*          & \nullmatrix \\
\nullmatrix  & V
\end{pmatrix}\;,\quad
\mbox{and}\,\: \mathsp{D}=
\begin{pmatrix}
\nullmatrix & D^*\\
D           &\nullmatrix
\end{pmatrix}\;.
\end{equation}
Putting Eq.~(\rf{eq:diagonalize_H_k}) into Eqs.~(\rf{eq:exponential_K}) and~(\rf{eq:polar_decomposition}), we have
\begin{align}
 &\mathsp{K}=\mathsp{V}  \exp\big(\mathord{-}i\mathsp{Z}\mathsp{D}\big) \mathsp{V}^\dagger\,.
\end{align}
Defining $\mathsp{W}^\dagger= \mathsp{V}^\dagger \mathsp{U}$ gives us
\begin{equation}\lb{eq:bolch_messiah}
 \mathsp{S}=\mathsp{V}  \exp\big(\mathord{-}i\mathsp{Z}\mathsp{D}\big) \mathsp{V}^\dagger \mathsp{U} =\mathsp{V}  \exp\big(\mathord{-}i\mathsp{Z}\mathsp{D}\big) \mathsp{W}^\dagger\,,
\end{equation}
where $\mathsp{W}$ and $\mathsp{V}$ are both unitary and symplectic, and $\mathsp{D}$ is of the form in Eq.~(\rf{eq:diagonalize_H_k}).  The decomposition~(\rf{eq:bolch_messiah}) is called the Bloch-Messiah reduction~\cite{bloch_canonical_1962, braunstein_squeezing_2005}.\si{Bloch-Messiah reduction}  It states that a multi-mode Bogoliubov transformation (Gaussian unitary) can be decomposed into a multiport beamsplitter, followed by a set of single-mode squeezers on each mode, followed by yet another multiport beamsplitter.

Consider the time evolution~(\rf{eq:heisenberg_evolutions_time}) for the symplectic matrix $\mathsp{S}$,
\begin{equation}
 \dt{\mathsp{S}}(t)= -i \mathsp{Z} \mathsp{H}(t)\mathsp{S}(t)
\end{equation}
where $\mathsp{H}(t)$ is a time-dependent Hamiltonian matrix.  Noticing that
\begin{equation}
 \mathsp{K}^2(t) = \mathsp{S}(t) \mathsp{S}^\dagger(t)
\end{equation}
we have
\begin{align}
\frac{\dif \mathsp{K}^2}{\dif t}&=\dt{\mathsp{S}} \mathsp{S}^\dagger +\mathsp{S} \dt{\mathsp{S}}^\dagger= -i \mathsp{Z} \mathsp{H} \mathsp{K}^2 + i \mathsp{K}^2 \mathsp{H} \mathsp{Z}\,.
\end{align}
Thus $\mathsp{K}^2$ satisfies a closed equation and can be determined without knowing the unitary part $\mathsp{U}$. This is particularly useful for quantifying the squeezing power of $\mathsp{S}$, which only depends on $\mathsp{K}$.

\section{Examples and Applications}

\subsection{Single-Mode Case}
\lb{subsec:single_mode_case}

In the single-mode case, any Hamiltonian matrix can be written as
\begin{equation}\lb{eq:hamiltonian_matrix_single_mode}
\mathsp{H}=\omega_0\ssp \mathsp{I}+\omega_1\mathsp{X}+\omega_2\mathsp{Y}\;,
\end{equation}
where $\omega_0$, $\omega_1$, $\omega_2$ are real parameters, $\mathsp{I}$ is the identity matrix, and $\mathsp{X}$ and $\mathsp{Y}=i\mathsp{X}\mathsp{Z}$ are Pauli matrices.\footnote{Because $\mathsp{Z}^* \neq \mathsp{X}\mathsp{Z}\mathsp{X}$, the Pauli-$\mathsp{Z}$ is not included.} From Eq.~(\rf{eq:symplectic_invariants_b}), we have the symplectic invariant
\begin{subequations}
\begin{align}
\tr(\mathsp{Z}\mathsp{H}\mathsp{Z}\mathsp{H})&=\big(\omega_0\ssp \mathsp{I}-\omega_1\mathsp{X}-\omega_2\mathsp{Y}\big)\big(\omega_0\ssp \mathsp{I}+\omega_1\mathsp{X}+\omega_2\mathsp{Y}\big)\\
&=\omega_0^2 - \omega_1^2 - \omega_2^2\\
&=\sgn(\omega_0^2 - \omega_1^2 - \omega_2^2)\, \omega^2\,,
\end{align}
\end{subequations}
where $\omega=\sqrt{\norm{\omega_0^2 - \omega_1^2 - \omega_2^2}}$.  Depending on the sign of this invariant, the canonical form of $\mathsp{H}$ bifurcates: the Hamiltonian $\sH$ is equivalent to $\omega\ssp \a^\dagger \a$ for $\omega_0^2 - \omega_1^2 - \omega_2^2 >0$, or to $i \omega(\a^2-\a^{\dagger\, 2})/2$ for $\omega_0^2 - \omega_1^2 - \omega_2^2 < 0$.

To take the Hamiltonian matrix $\mathsp{H}$ into canonical form, we introduce the symplectic matrix (Bogoliubov transformation),
\begin{equation}\lb{eq:symplectic_2D_a}
\mathsp{B}=e^{i\mathspst{Z}\mathspst{M}}\;.
\end{equation}
Here the matrix $\mathsp{M}$ in the exponential shares the form of Eq.~(\rf{eq:hamiltonian_matrix_single_mode}), and the whole generator thus takes the form
\begin{equation}\lb{eq:single_mode_generator}
i\mathsp{Z}\mathsp{M}=\gamma \big(\kappa_1 \mathsp{X} + \kappa_2\mathsp{Y} - i \kappa_3\mathsp{Z}\big)\;,
\end{equation}
where $\gamma > 0$, and $\kappa_1$, $\kappa_2$, and $\kappa_3$ are real numbers satisfying $\norm{\kappa_1^2+\kappa_2^2-\kappa_3^2}=1$.  Note that the following identity holds,
\begin{align}
\big(\kappa_1\mathsp{X} + \kappa_2\mathsp{Y} - i \kappa_3\mathsp{Z}\big)^2
&=(\kappa_1^2+\kappa_2^2-\kappa_3^2)\,\mathsp{I}=\pm\ssp \mathsp{I}\;.
\end{align}
For $\kappa_1^2+\kappa_2^2-\kappa_3^2=1$, we have
\begin{subequations}
\begin{align}
\mathsp{B}
&=\sum_{j=0}^{\infty} \frac{\gamma^j}{j!}\, \big(\kappa_1\mathsp{X} + \kappa_2\mathsp{Y} -i \kappa_3\mathsp{Z}\big)^j\\
&=\bigg(\sum_{j=0}^{\infty}\frac{ \gamma^{2j}}{( 2j) !}\bigg)\,\mathsp{I}+
\bigg(\sum_{k=0}^{\infty}\frac{\gamma^{2k+1}}{(2k+1)!}\bigg)\,
\big(\kappa_1\mathsp{X} + \kappa_2\mathsp{Y} -i \kappa_3\mathsp{Z}\big)\\[4pt]
&=\mathsp{I} \cosh \gamma+\big(\kappa_1\mathsp{X} + \kappa_2\mathsp{Y} -i \kappa_3
\mathsp{Z}\big)\sinh \gamma\;. \lb{symplectic_2D_b}
\end{align}
\end{subequations}
Similarly, for $\kappa_1^2+\kappa_2^2-\kappa_3^2=-1$, we have
\begin{equation}\lb{symplectic_2D_c}
\mathsp{B}
=\mathsp{I} \cos \gamma+\big(\kappa_1\mathsp{X} + \kappa_2\mathsp{Y} -i \kappa_3
\mathsp{Z}\big)\sin \gamma\;.
\end{equation}
We divide the symplectic transformation into two steps,
\begin{equation}
 \mathsp{B}= \mathsp{B}_{\nsp \mathrm{sqz}}(\vartheta) \mathsp{B}_{\nsp \mathrm{rot}}(\theta)\,,
\end{equation}
where $\mathsp{B}_{\nsp \mathrm{rot}}$ is a rotation and $\mathsp{B}_{\nsp \mathrm{sqz}}$ a squeezer.  The rotation part takes the form
\begin{equation}
\mathsp{B}_{\nsp \mathrm{rot}}(\theta)= e^{i\theta\mathspst{Z}/2}\;,
\end{equation}
which is both symplectic and unitary; it brings $\mathsp{H}$ into the form
\begin{align}
\mathsp{H}'&=\mathsp{B}_{\nsp \mathrm{rot}}^\dagger\nsp (\theta)\,  \mathsp{H} \, \mathsp{B}_{\nsp \mathrm{rot}}(\theta)=\omega_0\ssp \mathsp{I}+\sqrt{\omega_1^2+\omega_2^2}\; \mathsp{X}\;,
\end{align}
where $\tan\theta= - \omega_2/ \omega_1$.  The squeezer, which is both symplectic and Hermitian, takes the form
\begin{equation}
 \mathsp{B}_{\nsp \mathrm{sqz}}(\vartheta)=e^{i\vartheta \mathspst{Z}\mathspst{Y}/2}=e^{\vartheta\mathspst{X}/2}\,.
\end{equation}
Applying $\mathsp{B}_{\nsp \mathrm{sqz}}$ to $\mathsp{H}'$, we have
\begin{align}
\mathsp{H}'_\vartheta&=\mathsp{B}^\dagger_{\nsp \mathrm{sqz}}(\vartheta)\, \mathsp{H}'\, \mathsp{B}_{\nsp \mathrm{sqz}}(\vartheta)=\xi(\vartheta)\ssp \mathsp{I}+ \eta(\vartheta)\mathsp{X}\;,
\end{align}
where $\xi(0)=\omega_0$ and $\eta(0)=\sqrt{\omega_1^2+\omega_2^2}$.
To determine the functions $\xi(\vartheta)$ and $\eta(\vartheta)$, we notice that $\mathsp{H}'_\vartheta$ satisfies the differential equation
\begin{equation}
\di{\mathsp{H}'_\vartheta}{\vartheta}=\half\,\big(\mathsp{X}\mathsp{H}'_\vartheta+\mathsp{H}'_\vartheta\mathsp{X}\big)\;,
\end{equation}
which leads to the equations
\begin{align}\lb{eq:squeeze_evolution}
\dt \xi(\vartheta) =\eta (\vartheta)\,,\qquad \dt \eta(\vartheta) =\xi (\vartheta)\,.
\end{align}
The solutions to Eqs.~(\rf{eq:squeeze_evolution}) depend on the initial conditions, and we discuss them separately.  For $\omega_0^2-\omega_1^2-\omega_2^2 > 0$, i.e., $\xi(0) >\eta(0)$, we have
\begin{align}
\xi(\vartheta)=\omega\ssp \cosh(\phi+\vartheta)\,,\qquad
\eta(\vartheta)=\omega\ssp \sinh(\phi+\vartheta)\,,
\end{align}
where $\tanh\phi =\sqrt{\omega_1^2+\omega_2^2}\,\big/\omega_0$.  For  $\omega_0^2-\omega_1^2-\omega_2^2< 0$, i.e., $\xi(0) <\eta(0)$, we have
\begin{align}
\xi(\vartheta)=\omega\ssp \sinh(\phi+\vartheta)\,,\qquad
\eta(\vartheta)=\omega\ssp \cosh(\phi+\vartheta)\,,
\end{align}
where $\tanh\phi =\omega_0\big/\sqrt{\omega_1^2+\omega_2^2}\,$.  By choosing the squeeze parameter $\vartheta=-\phi$, we have the canonical form
\begin{equation}\lb{canonical_form_2D}
\mathsp{H}_{\nsp\mathrm{can}}=\left\{
\begin{aligned}
&\omega\,\mathsp{I} &\text{for}\,\:  \omega_0^2 - \omega_1^2 - \omega_2^2 >  0&\;,\\
&\omega \mathsp{X}   &\text{for}\,\:  \omega_0^2 - \omega_1^2 - \omega_2^2 <     0& \;.
\end{aligned}\right.
\end{equation}
The symplectic eigenvalues of $\mathsp{H}$ are real in the first case, and $\mathsp{H}$ is symplectic equivalent to the rotation Hamiltonian (phase shift); the symplectic eigenvalues are pure imaginary in the second case, and $\mathsp{H}$ is symplectic equivalent to the squeezing Hamiltonian.  When $\omega_0^2 - \omega_1^2 - \omega_2^2 = 0$, it could be the trivial case $\mathsp{H}=0$, or
\begin{equation}
\mathsp{H}_{\nsp\mathrm{psq}}= \mathsp{I}-\mathsp{X}=
\begin{pmatrix}
1   &  -1\,\\
-1  &  \,1\,
\end{pmatrix}\,,
\end{equation}
where the subscript means $\hatp^2$ for the reason we will see soon.  Note that $\mathsp{H}_{\nsp\mathrm{psq}}$ is nilpotent, i.e., $(\mathsp{Z}\mathsp{H}_{\nsp\mathrm{psq}})^2=0$, which means that it only has one eigenstate, $(1,\,1)^T$, whose eigenvalue is zero; therefore, it is impossible to bring this Hamiltonian matrix into one of the canonical forms~(\rf{eq:canonical_form_matrix}).  The exponentiation of $\mathsp{H}_{\nsp\mathrm{psq}}$, however, is quite easy because of the nilpotent condition,
\begin{equation}
 \exp\big(\mathord{-}it\mathsp{Z}\mathsp{H}_{\nsp\mathrm{psq}}\big)=\mathsp{I}-it\mathsp{Z}\mathsp{H}_{\nsp\mathrm{psq}}\,.
\end{equation}
The evolution is linear in time instead of oscillatory or exponential, because the Hamiltonian describes a free particle,
\begin{align}
 \sH_{\nsp\mathrm{psq}} &= \half\,\big(\a^\dagger\nsp\a+ \a\ssp \a^\dagger-\a^2-\a^{\dag\, 2}\big)= \hatp^2\;.
\end{align}

\subsection{Decomposing a Squeeze Operation into a Product of Rotations and the \texorpdfstring{$\boldsymbol{e^{-it\hatp^2}}$}{exp(-\textit{itp}\hspace{-1pt}\^{}2)} Operation}\si{Squeeze operator}

Sometimes, the squeeze operation $\sS(\gamma) = e^{\gamma (\a^2 - \a^{\dagger\, 2})/2}$ is not readily available for certain physical systems, but the rotation $\sR(\theta)=e^{-i\theta(\a^\dagger\nsp \a+\a\ssp \a^\dagger)/2}$ and the $e^{-it\hatp^2}$ operation are.  In this subsection, I show how to decompose the squeeze operation into a product of the other two.

The rotation, squeeze, and $\hatp^2$ Hamiltonians take the forms
\begin{subequations}
\begin{align}
&\sH_{\mathrm{rot}}=\half\big(\a^\dagger\nsp \a + \a\a^\dagger \big)\,,\\[3pt]
&\sH_{\mathrm{sqz}}=\frac{i}{2}\big(\a^2-\a^{\dag\, 2}\big)\,,\\[3pt]
&\sH_{\mathrm{psq}}=\half \big(\a^\dagger\nsp \a+\a\ssp\a^\dagger-\a^2-\a^{\dagger\, 2}\big)=\hatp^2\,,
\end{align}
\end{subequations}
where we always use symmetric ordering of the annihilation and creation operators.  The corresponding Hamiltonian matrices are
\begin{equation}\hspace{-1em}
\mathsp{H}_{\nsp \mathrm{rot}}=
\begin{pmatrix}
\,1\,  &  \,0\,\\
\,0\,  &  \,1\,
\end{pmatrix}\;,\qquad
\mathsp{H}_{\nsp \mathrm{sqz}}=
\begin{pmatrix}
\,0\,  &  \,-i\,\\
\,i\,  &  \,0\,
\end{pmatrix}\;,\qquad
\mathsp{H}_{\nsp \mathrm{psq}}=
\begin{pmatrix}
1   &  -1\,\\
-1  &  \,1\,
\end{pmatrix}\;.
\end{equation}
The evolution matrices, exponential in the Hamiltonian matrices, take the forms
\begin{subequations}
\begin{align}
&\mathsp{S}_{\nsp \mathrm{rot}}(\theta) = e^{-i\theta\mathspst{Z}\mathspst{H}_{\nsp \mathrm{rot}}} = \mathsp{I} \cos \theta-i \mathsp{Z} \sin \theta\,,\\
&\mathsp{S}_{\nsp \mathrm{sqz}}(\gamma) = e^{-i\gamma \mathspst{Z}\mathspst{H}_{\nsp \mathrm{sqz}} } = \mathsp{I} \cosh \gamma-\mathsp{X} \sinh \gamma\,,\\
&\mathsp{S}_{\nsp \mathrm{psq}}(t) = e^{-it\mathspst{Z}\mathspst{H}_{\nsp \mathrm{psq}} }=\mathsp{I}-it\mathsp{Z}\mathsp{H}_{\nsp \mathrm{psq}}=\mathsp{I}-i t \mathsp{Z}-t \mathsp{Y}\;.
\end{align}
\end{subequations}
Multiplying $\mathsp{S}_{\nsp \mathrm{rot}}$ and $\mathsp{S}_{\nsp \mathrm{psq}}$ together, we have
\begin{subequations}
\begin{align}
\mathsp{S}_{\nsp \mathrm{rot}}(\theta)\, \mathsp{S}_{\nsp \mathrm{psq}}(t)&\nonumber\\
&\hspace{-5em}=(\mathsp{I} \cos \theta-i \mathsp{Z} \sin \theta) (\mathsp{I}-i t \mathsp{Z}-t \mathsp{Y})\\
&\hspace{-5em}=\mathsp{I} (\cos \theta-t \sin \theta) - i \mathsp{Z} (t \cos \theta+\sin \theta) -t \mathsp{Y} \cos \theta + t \mathsp{X} \sin \theta \;.
\end{align}
\end{subequations}
By choosing $t=-\tan \theta$, $\theta\in (-\pi/2,\pi/2)$, we have
\begin{equation}
\mathsp{S}_{\nsp \mathrm{rot}}(\theta)\, \mathsp{S}_{\nsp \mathrm{psq}}(-\tan \theta)
=\mathsp{I} \sec \theta + \big(\mathsp{Y} \cos \theta - \mathsp{X} \sin \theta\big)\tan \theta \;.
\end{equation}
The terms with $\mathsp{X}$ and $\mathsp{Y}$ can be combined by inserting two rotations of opposite angles,
\begin{subequations}
\begin{align}
&\hspace{-1.em}\mathsp{S}_{\nsp \mathrm{rot}}(-\theta/2+\pi/4)\, \mathsp{S}_{\nsp \mathrm{rot}}(\theta)\, \mathsp{S}_{\nsp \mathrm{psq}}(-\tan \theta)\, \mathsp{S}_{\nsp \mathrm{rot}}(\theta/2-\pi/4)\nonumber\\
&=e^{i(\theta/2-\pi/4)\mathspst{Z}}\,\Big(\mathsp{I} \sec \theta + \big(\mathsp{Y} \cos \theta - \mathsp{X} \sin \theta\big)\tan \theta\Big)\, e^{-i(\theta/2-\pi/4)\mathspst{Z}}\\
&=\mathsp{I} \sec \theta- \mathsp{X} \tan \theta\;.
\end{align}
\end{subequations}
By choosing $\tan \theta= \sinh \gamma$, we have
\begin{subequations}
\begin{align}
\mathsp{S}_{\nsp \mathrm{sqz}}(\gamma)
&=\mathsp{I} \cosh \gamma-\mathsp{X} \sinh \gamma\\
&=\mathsp{S}_{\nsp \mathrm{rot}}(\theta/2+\pi/4)\, \mathsp{S}_{\nsp \mathrm{psq}}(-\tan \theta)\, \mathsp{S}_{\nsp \mathrm{rot}}(\theta/2-\pi/4)\;,
 \end{align}
\end{subequations}
Thus, we have decomposed the squeeze operator into three terms, each of which is either a rotation or an $e^{-it\hatp^2}$ operation.

\subsection{Decoupling the Two-Mode Squeeze Hamiltonian}\si{Two-mode squeeze}

Here, I show how to use the symplectic transformation to decouple the following two-mode squeeze Hamiltonian into a sum of two single-mode squeeze Hamiltonians,
\begin{equation}\lb{eq:two_mode_squeeze}
\sH_{\mathrm{tms}}=\omega (\a^\dagger \a+\b^\dagger \b+1)
 +i\gamma(\a\b - \a^\dagger\b^\dagger)\,.
\end{equation}
In matrix form, we have
\begin{equation}
\sH_{\mathrm{tms}}=\half\,
\begin{pmatrix}
\a^\dagger & \b^\dagger & \a & \b
\end{pmatrix}
\begin{pmatrix}
\omega     & 0            & 0            & -i \gamma\\
0          & \omega       & -i \gamma    & 0\\
0          & i \gamma     & \omega       & 0\\
i \gamma   & 0            & 0            & \omega
\end{pmatrix}
\begin{pmatrix}
\a\\ \b\\ \a^\dagger\\ \b^\dagger
\end{pmatrix}\;,
\end{equation}
and the corresponding Hamiltonian matrix is
\begin{equation}
\mathsp{H}_{\nsp \mathrm{tms}}=
\scalebox{.95}{\mbox{$
\begin{pmatrix}
\omega I    & -i \gamma X\\[4pt]
i \gamma X  &  \omega I
\end{pmatrix}$}}\;,
\end{equation}
where $I$ and $X$ are the $2\times 2$ identity and Pauli-$X$ matrices.  Consider the following symplectic transformation of a beamsplitter,
\begin{equation}
\mathsp{B}_{\mathrm{bsp}}
=\exp\left[-\frac{i}{4}\pi\mathsp{Z}\begin{pmatrix}X & 0 \\[1pt] 0 & X\end{pmatrix}\right]
=\frac{1}{\sqrt{2}}\,
\begin{pmatrix}
I-i X &  0\\[4pt]
0     &  I+i X
\end{pmatrix}\;,
\end{equation}
which brings the Hamiltonian matrix to
\begin{subequations}
\begin{align}
\hspace{-2em}\mathsp{B}_{\mathrm{bsp}}^\dagger\,  \mathsp{H}_{\nsp \mathrm{tms}}\,  \mathsp{B}_{\mathrm{bsp}} &=\frac{1}{2}\,
\scalebox{.94}{\mbox{$
\begin{pmatrix}
I+i X &  0\\
0     &  I-i X
\end{pmatrix}
\begin{pmatrix}
\omega I     & -i \gamma X\\
i \gamma X   & \omega I
\end{pmatrix}
\begin{pmatrix}
I-i X  &  0\\
0      &  I+i X
\end{pmatrix}$}}\\[4pt]
&=\scalebox{.94}{\mbox{$
\begin{pmatrix}
\omega I   &  \gamma I\\
\gamma I   &  \omega I
\end{pmatrix}$}}\;. \lb{eq:decomposed_two_mode_squeeze}
\end{align}
\end{subequations}
The Hamiltonian matrix~(\rf{eq:decomposed_two_mode_squeeze}) represents two identical independent single-mode squeezers.  Although this problem can be done without any knowledge of symplectic methods, when the Hamiltonian matrix becomes more complicated, e.g., the frequencies $\omega$ for the two modes are different, it is quite difficult to simplify just by looking at the Hamiltonian, and the symplectic methods come to be a useful tool.

\subsection{Symplectic Eigenvalues for Two Modes}
\lb{subsec:solve_two_mode}

In this subsection, I discuss how to solve the symplectic eigenvalue problem of the two-mode case ($\nrank=2$) without solving the symplectic eigenvalue equation~(\rf{eq:characteristic_equation}).

It turns out that the two symplectic invariants, Eqs.~(\rf{eq:symplectic_invariants_b}) and (\rf{eq:symplectic_invariants_a}), are particularly useful,
\begin{equation}\lb{eq:invariants_two_mode}
 \chi_{\mathrm d}=\det(\mathsp{H})\,, \qquad \chi_{\mathrm t}=\half\,\tr(\mathsp{Z}\mathsp{H}\mathsp{Z}\mathsp{H})\,.
\end{equation}
Denoting the four eigenvalues of $\mathsp{Z}\mathsp{H}$ by $\{\pm \lambda_1,\, \pm \lambda_2\}$, we have
\begin{equation}\lb{eq:invariants_eigenvalues}
\chi_{\mathrm d},\,\chi_{\mathrm t}=\left\{
\begin{aligned}
\norm{\lambda_1}^2\ssp \norm{\lambda_2}^2\,,&\quad &&\norm{\lambda_1}^2+\norm{\lambda_2}^2\,,\quad &&\mbox{$\lambda_1$ and $\lambda_2$ are both real,}\\
\mathord{-}\norm{\lambda_1}^2\ssp \norm{\lambda_2}^2\,,&\quad &&\norm{\lambda_1}^2-\norm{\lambda_2}^2\,,\quad &&\mbox{$\lambda_1$ is real and $\lambda_2$ pure imaginary,}\\
\norm{\lambda_1}^2\ssp \norm{\lambda_2}^2\,,&\quad &&\!\!\!\!\mathord{-}(\norm{\lambda_1}^2+\norm{\lambda_2}^2)\,,\quad &&\mbox{$\lambda_1$ and $\lambda_2$ are both pure imaginary,}\\
\norm{\lambda}^4\,,\quad\;\;  &&&\hspace{-2em}2\ssp (\norm{\re\lambda}^2-\norm{\im\lambda}^2)\,,\quad &&\mbox{$\lambda_1=\lambda_2^*=\lambda$ are complex.}
\end{aligned}\right.
\end{equation}
Note that we leave out the case in which $\lambda_1$ is pure imaginary and $\lambda_2$ real, because can be obtained from the second case by swapping $\lambda_1$ and $\lambda_2$.  Assuming all the eigenvalues are nonzero, we can tell the type of the eigenvalues by the following procedure: (i)~if $\chi_{\mathrm d} < 0$, $\mathsp{H}$ has a pair of real and a pair of pure imaginary eigenvalues; (ii)~if $\chi_{\mathrm d} > 0$ and $\chi_{\mathrm t}^2 < 4 \chi_{\mathrm d}$, $\mathsp{H}$ has a quadruplet of complex eigenvalues; (iii)~if $\chi_{\mathrm d} > 0$ and $\chi_{\mathrm t}^2 > 4 \chi_{\mathrm d}$ and $\chi_{\mathrm t} > 0$, $\mathsp{H}$ has two pairs of real eigenvalues; (iv)~if $\chi_{\mathrm d} > 0$ and $\chi_{\mathrm t}^2 > 4 \chi_{\mathrm d}$ and $\chi_{\mathrm t} < 0$, $\mathsp{H}$ has two pairs of pure imaginary eigenvalues.

After determining the classes of the eigenvalues, we can solve for them by using Eq.~(\rf{eq:invariants_eigenvalues}).  Knowing the symplectic eigenvalues, however, is not equivalent to knowing the canonical forms~(\rf{eq:canonical_form_matrix}); for example, both $\pm \mathsp{H}_{\nsp\mathrm{real}}$ have the same set of symplectic eigenvalues.

\subsection{Positiveness of the Hamiltonian Matrix}

In some cases, one wants to know if a quadratic Hamiltonian is stable (has a well defined ground state).  For that purpose, one need only check whether the Hamiltonian matrix is positive definite [see Eq.~(\rf{eq:positive_stable})].  One way to do that is by the following condition for matrices of the spinor form~(\rf{eq:Hamiltonian_matrix}):
\begin{equation}
\mbox{\parbox{0.76\textwidth}{$\mathsp{H}$ is nonnegative if and only if $H_1\geq 0$ and ${\displaystyle H_2=\sqrt{H_1}\, C\, \big(\sqrt{H_1}\ssp\big)^*}$
for some contraction $\,C$.\footnotemark}}
\end{equation}\footnotetext{A contraction is an operator whose operator norm is less than one.}This theorem is pretty handy, because it reduces the positivity problem of a $2\nrank\times 2\nrank$ matrix to that of an $\nrank\times \nrank$ matrix.

\subsection{Fluctuations Allowed by Quantum Mechanics}\lb{subsec:fluctuation_quantum_mechanics}\si{Quantum fluctuation}

For applications like quantum metrology or quantum computation, we may want to suppress quantum fluctuations to prevent noise.  Zero quantum fluctuation is not possible because of the Heisenberg uncertainty principle (the covariance matrix cannot vanish in quantum mechanics).  A natural question to ask is the following: what are the constrains on a given positive definite matrix $\mathsp{R}$, so that there exist a quantum state whose covariance matrix is $\mathsp{R}$?  This is an easy question for the single-mode case.  The determinant of the given positive definite matrix must be greater than half of the Planck constant $\hbar$; for the multi-mode case, however, the answer is not so obvious.

In the basis of the quadratures, the covariance matrix of a quantum state is a $2\nrank\times 2\nrank$ real matrix,\si{Covariance matrix}
\begin{equation}\lb{eq:covariance_matrix_a}
\mathsp{R}
=\begin{pmatrix}
    R_1   &  R_2^\transp\\
    R_2   &  R_3
   \end{pmatrix}\;,
\end{equation}
where $R_{1,\,jk}=2\, \av{\Delta\hatx_j \Delta\hatx_k}$, $R_{2,\,jk}=\av{\Delta\hatx_k \Delta\hatp_j+ \Delta\hatp_j \Delta\hatx_k}$, and $R_{3,\,jk}=2\,\av{\Delta\hatp_j \Delta\hatp_k}$ with $\Delta\hatx_j= \hatx_j-\av{\hatx_j}$ and $\Delta\hatp_j= \hatp_j-\av{\hatp_j}$.
In the basis of the annihilation and creation operators, we have
\begin{equation}
\Delta\a_j=\frac{1}{\sqrt 2}\, (\Delta\hatx_j+i\Delta\hatp_j)\,,\qquad \Delta\a_j^\dagger=\frac{1}{\sqrt 2}\, (\Delta\hatx_j-i\Delta\hatp_j)\,,
\end{equation}
and the covariance matrix takes the new form
\begin{equation}\lb{eq:covariance_matrix_b}
 \mathsp{C}=
 \begin{pmatrix}
    C_1 & C_2^*\\
    C_2 & C_1^*
   \end{pmatrix}
  =\half
   \begin{pmatrix}
    \identity & \;\,i\identity\\
    \identity & \mathord{-}i\identity
   \end{pmatrix}
   \begin{pmatrix}
    R_1   & R_2^\transp\\
    R_2   & R_3
   \end{pmatrix}
   \begin{pmatrix}
    \;\identity            & \identity\\
    \mathord{-}i\identity  & i\identity
   \end{pmatrix}\;,
\end{equation}
where $C_{1,\,jk}=\av{\Delta\a_j \Delta\a_k^\dagger+\Delta\a_k^\dagger \Delta\a_j}$, $C_{2,\,jk}=2\,\av{\Delta\a_j^\dagger \Delta\a_k^\dagger}$.  Note that $C_1^\dagger=C_1$ and $C_2^T=C_2$, and the covariance matrix $\mathsp{C}$ has the same symmetry as the Hamiltonian matrix $\mathsp{H}$.  The good news is that the covariance matrix $\mathsp{C}$ is always positive definite, therefore, all of its eigenvalues are real.  Thus, it can be brought into the following Williamson normal form by a symplectic transformation,
\begin{equation}\lb{eq:variance_matrix_canonical_form}
\mathsp{C}_{\nsp\mathrm{can}}=\mathsp{B}^\dagger \mathsp{C} \mathsp{B}=
\begin{pmatrix}
 \diag(\lambda_1,\ldots, \lambda_\nrank)  & \nullmatrix\\
 \nullmatrix                        &  \diag(\lambda_1,\ldots, \lambda_\nrank)
\end{pmatrix}\;.
\end{equation}
Quantum mechanics constrains $\mathsp{C}_{\nsp\mathrm{can}}$ to satisfy
\begin{equation}\lb{eq:quantum_fluctuation}
\lambda_j=\avb{\Delta\b^\dagger_j \Delta\b_j + \Delta\b_j \Delta\b^\dagger_j\,}\geq 1\,,\quad\mbox{for $j=1,2,\ldots,\nrank$}\;,
\end{equation}
where $\Delta\b_j$ and $\Delta\b_j^\dagger$ are the annihilation and creation operators in the new basis, with their mean removed.  On the other hand, we can always find a quantum state whose variance matrix is Eq.~(\rf{eq:variance_matrix_canonical_form}) if all its diagonal elements are greater than $1$.  Thus, given a $2\nrank\times 2\nrank$ matrix $\mathsp{C}$ which satisfies $\mathsp{C}^\dagger= \mathsp{C}$ and $\mathsp{C}^*=\mathsp{X}\mathsp{C}\mathsp{X}$, we have\\[-8pt]
\begin{equation}\lb{eq:covariance_matrix_physical}
 \mbox{\parbox{0.76\textwidth}{$\mathsp{C}$ is the covariance matrix for some quantum state if and only if its canonical form $\mathsp{C}_{\nsp\mathrm{can}}$ is greater or to than the identity $\mathsp{I}$.}}
\vspace{3pt}
\end{equation}\\[-8pt]
This condition, which requires the canonical form, is not convenient to use; an equivalent one is
\begin{equation}\lb{eq:C_>_Z_a}
 \mathsp{C}_{\nsp\mathrm{can}}\geq \mathsp{Z}\,.
\end{equation}
Because $\mathsp{Z}$ is symplectically invariant, we have
\begin{equation}\lb{eq:C_>_Z_b}
 \mathsp{C}\geq \mathsp{Z}\;.
\end{equation}
Note that this condition Eq.~(\rf{eq:C_>_Z_b}) does not require the canonical form, and it automatically excludes the pure-imaginary-eigenvalue and complex-eigenvalue cases.

For pure Gaussian states, the equality in Eq.~(\rf{eq:quantum_fluctuation}) is satisfied, i.e.,
\begin{equation}
 \mathsp{C}_\mathrm{can}= \mathsp{B}\mathsp{C}\mathsp{B}^\dagger=\mathsp{I}\;\;\Longleftrightarrow\;\;
 \mathsp{C}= \mathsp{Z}\mathsp{B}^\dagger\mathsp{B}\mathsp{Z}
\end{equation}
Consequently, we have
\begin{equation}
 \mathsp{C}^{-1} = \mathsp{B}^\dagger\mathsp{B} = \mathsp{Z}\mathsp{C}\mathsp{Z}\;,
\end{equation}
which implies the condition~(\rf{eq:C_>_Z_b}).

For the two-mode case, if we already know $\mathsp{C}> 0$, then it is easy to restate the condition Eq.~(\rf{eq:covariance_matrix_physical}) in terms of symplectic invariants [see Subsec.~\rf{subsec:solve_two_mode} for details],
\begin{equation}
 \chi_{\mathrm d}=\det(\mathsp{C})=\lambda_1^2\ssp \lambda_2^2\,, \qquad \chi_{\mathrm t}=\half\,\tr(\mathsp{Z}\mathsp{C}\mathsp{Z}\mathsp{C})=\lambda_1^2+\lambda_2^2\,,
\end{equation}
where $\lambda_1\geq \lambda_2 > 0$ represent the positive eigenvalues. Since $\lambda_1$ and $\lambda_2$ must be greater or equal than one, we have
\begin{equation}
 \chi_{\mathrm t}-\sqrt{\chi_{\mathrm t}^2-4\chi_{\mathrm d}} = \lambda_1^2+\lambda_2^2-\normb{\lambda_1^2-\lambda_2^2}\geq 2\,.
\end{equation}
This inequality, a necessary and sufficient condition for $\mathsp{C}$ to be physical, can be simplified to
\begin{equation}\lb{eq:invariants_covariance}
\chi_{\mathrm t} \leq \chi_{\mathrm d}+1\;.
\end{equation}
For more detailed discussion on the two-mode case, see~\cite{pirandola_correlation_2009}.

\subsection{Dynamics of the Covariance Matrix}

Consider a quantum state evolving under the quadratic Hamiltonian~(\rf{eq:quadratic_Hamiltonian_c}),
\begin{equation}
\sH=\half\,
\begin{pmatrix}
\abf^\dagger & \abf
\end{pmatrix}
\begin{pmatrix}
H_1(t) & H_2^*(t) \\ H_2(t) & H_1^*(t)
\end{pmatrix}
\begin{pmatrix}
\abf \\  \; \abf^\dagger
\end{pmatrix}\;.
\end{equation}
The time evolution of the annihilation and creation operators in the Heisenberg picture is
\begin{equation}
\begin{pmatrix}
\abf(t)\\ \,\abf^\dagger\nsp (t)
\end{pmatrix}
=\mathsp{S}(t)
\begin{pmatrix}
\abf\\ \;\abf^\dagger
\end{pmatrix}\;,
\end{equation}
where the evolution matrix $\mathsp{S}$ satisfies
\begin{equation}
\dt{\mathsp{S}}(t)
=\mathord{-}i \mathsp{Z}\mathsp{H}(t)\mathsp{S}(t)\;.
\end{equation}
The covariance matrix~(\rf{eq:covariance_matrix_b}) at time $t$ thus takes the form
\begin{align}
 \mathsp{C}(t) = \mathsp{S}(t)\mathsp{C}(0)\ssp \mathsp{S}^\dagger(t)\;,
\end{align}
and the governing equation is
\begin{align}
 \dt{\mathsp{C}}(t) = \mathord{-}i \mathsp{Z}\mathsp{H}(t) \mathsp{C}(t)+i\mathsp{C}(t)\mathsp{H}(t)\mathsp{Z}\;.
\end{align}

\renewcommand{\lb}[1]{\label{gp_stable:#1}}
\renewcommand{\rf}[1]{\ref{gp_stable:#1}}

\chapter{The Gross-Pitaevskii Ground State Is Stable}

\begin{quote}
A method is more important than a discovery, since the right method will lead to new and even more important discoveries.\\[4pt]
-- Lev Landau\ai{Landau, Lev}
\end{quote}

\noindent In this Appendix, we investigate whether the ground state of the Gross-Pitaevskii Equation (GPE) is stable\si{GP ground state} and show that it is because all the Bogoliubov excitations increase the energy of the condensate.  The proof is constructed by noticing that both the perturbations of the condensate wavefunction and the Bogoliubov excitations are described by the Bogoliubov-de Gennes (BdG) equations~\cite{bogoliubov_new_method_1958, de_gennes_superconductivity_1966}.\ai{de Gennes, Pierre-Gilles}\ai{Bogoliubov, Nikolay}\si{Bogoliubov-de Gennes equations}

Consider a Bose-Einstein Condensate (BEC) govern by the Hamiltonian
\begin{equation}
\sH=\int\uppsi^\dagger (\xbf)  \Big(\mathord{-}\frac{\hbar^2}{2m}\boldsymbol\nabla^2 +V(\xbf)\Big) \uppsi(\xbf)+\frac{g}{2}\, \big[\uppsi^\dagger (\xbf)\big]^2\,\uppsi^2(\xbf)\,\dif \xbf\;.
\end{equation}
The mean-field energy for the product state $\ket{\psi(\xbf)}^{\otimes N}$ is
\begin{equation}
E_\mathrm{mf}=\int N\, \psi^*(\xbf) \Big(\mathord{-}\frac{\hbar^2}{2m}\boldsymbol\nabla^2 +V(\xbf)\Big) \psi(\xbf)+\frac{g}{2}\, N(N-1) [\psi^*(\xbf)]^2\,\psi^2(\xbf)\,\dif \xbf\;.
\end{equation}
The Gross-Pitaevskii (GP) ground state $\psinot(\xbf)$ is the state that minimize the mean-field energy $E_\mathrm{mf}$, and a perturbation around $\psinot(\xbf)$ takes the form
\begin{equation}\lb{eq:perturbation}
\psinot(\xbf) \rightarrow \sqrt{1-\epsilon^2}\, \psinot(\xbf)+\epsilon\, \psinot_\perp (\xbf)\;,
\end{equation}
where $\epsilon$ is a small parameter, and $\psinot_\perp (\xbf)$ is some normalized state orthogonal to $\psinot(\xbf)$.  We neglect the change of an overall phase in Eq.~(\rf{eq:perturbation}), because it does not affect the mean-field energy.  The first order perturbation of the mean-field energy for the GP ground state is zero, i.e., $\delta E_\mathrm{mf}^{(1)}=0$, which gives rise to the GPE
\begin{equation}\lb{eq:gp_equation}
\Big(\mathord{-}\frac{\hbar^2}{2m}\boldsymbol\nabla^2 +V(\xbf)+g (N-1)
\norm{\psinot(\xbf)}^2-\mu\Big)\,\psinot(\xbf)=0\;,
\end{equation}
where $\mu$ is the chemical potential, and hereafter we neglect the difference between $N-1$ and $N$.  In the second order perturbation, the value of $E_\mathrm{mf}$ cannot decrease for the GP ground state; i.e., the following quantity is nonnegative,
\begin{equation}\lb{eq:2nd_order_perturbation}\hspace{-2em}
\begin{split}
\frac{\delta E_\mathrm{mf}^{(2)}}{\epsilon^2 N}&=
\int \psinot^*_\perp (\xbf) \Big(\mathord{-}\frac{\hbar^2}{2m}\boldsymbol\nabla^2 +V(\xbf)\Big) \psinot_\perp (\xbf) - \psinot^*(\xbf)\Big(\mathord{-}\frac{\hbar^2}{2m}\boldsymbol\nabla^2 +V(\xbf)\Big) \psinot (\xbf)\\[3pt]
&\eqindent +g N\bigg( 2\,\normb{\psinot(\xbf)}^2\, \normb{\psinot_\perp (\xbf)}^2+\half\,\Big([\psinot^*(\xbf)]^2\, \psinot_\perp^2 (\xbf)+ \mathrm{c.c.}\Big)-\normb{\psinot(\xbf)}^4\bigg)\,\dif \xbf\;.
\end{split}
\end{equation}
With the GPE~(\rf{eq:gp_equation}), we notice that
\begin{align}\lb{eq:chemecal_potential_condition}\hspace{-2em}
\mu=\mu\int\psinot^*_\perp (\xbf) \psinot_\perp (\xbf)\,\dif \xbf=\int\psinot^*(\xbf) \Big(\mathord{-}\frac{\hbar^2}{2m}\boldsymbol\nabla^2 +V(\xbf)+g N
\norm{\psinot(\xbf)}^2\,\Big) \psinot(\xbf)\,\dif \xbf\;.
\end{align}
Putting Eq.~(\rf{eq:chemecal_potential_condition}) into Eq.~(\rf{eq:2nd_order_perturbation}), we have
\begin{equation}\hspace{-1em}
\begin{split}
\frac{\delta E_\mathrm{mf}^{(2)}}{\epsilon^2 N}&=
\int \psinot^*_\perp (\xbf) \Big(\mathord{-}\frac{\hbar^2}{2m}\boldsymbol\nabla^2 +V(\xbf)-\mu\Big) \psinot_\perp (\xbf)+2g N\,\normb{\psinot(\xbf)}^2\, \normb{\psinot_\perp (\xbf)}^2\\[3pt]
&\eqindent +\frac{g N}{2}\,\Big([\psinot^*(\xbf)]^2\, \psinot_\perp^2 (\xbf)+ \mathrm{c.c.}\Big)\,\dif \xbf\;.
\end{split}
\end{equation}
In matrix form, we have
\begin{equation}\lb{eq:matrix_form}\hspace{-2em}
\frac{\delta E_\mathrm{mf}^{(2)}}{\epsilon^2 N}
=\half\ssp
\begin{pmatrix}
\,\bra{ \psinot_\perp }& \bra{\psinot^*_\perp}\,\,
\end{pmatrix}\!
\begin{pmatrix}
H_\mathrm{gp}+g  N  Q \norm{\psinot}^2 Q  & g
N Q\ssp \psinot^2 Q^*\\[6pt] g
N Q^* (\psinot^*)^2 Q &H_\mathrm{gp}+g  N Q^* \norm{\psinot}^2 Q^*
\end{pmatrix}\!
\begin{pmatrix}
\ket{\psinot_\perp }\\[7pt] \ket{\psinot^*_\perp }
\end{pmatrix}\,,
\end{equation}
where $Q=\identity-\proj{\psinot}$, and
\begin{equation}
H_\mathrm{gp}= \mathord{-}\frac{\hbar^2}{2m}\boldsymbol\nabla^2 +V+g N
\norm{\psinot}^2-\mu\;.
\end{equation}
The GP ground state being stable is equivalent to the following matrix being positive definite in the subspace orthogonal to the condensate wavefunction $\ket{\psinot}$,
\begin{align}
 \mathsp{H}_\mathrm{ncb}  =
\begin{pmatrix}
H_\mathrm{gp}+g  N  Q \norm{\psinot}^2 Q  & g
N Q\ssp \psinot^2 Q^*\\[6pt] g
N Q^* (\psinot^*)^2 Q &H_\mathrm{gp}+g  N Q^* \norm{\psinot}^2 Q^*
\end{pmatrix}                                                                                         \end{align}

Because the vector $\big(\;\ket{\psinot_\perp }\quad \ket{\psinot^*_\perp }\,\big)^T$ in Eq.~(\rf{eq:matrix_form}) is of a restricted form, an immediate conclusion that $\mathsp{H}_\mathrm{ncb}$ is positive definite is not available.  We notice, however, the following symmetry,
\begin{equation}
\mathsp{H}_\mathrm{ncb}^*=\mathsp{X} \mathsp{H}_\mathrm{ncb}\ssp \mathsp{X}\,,\qquad \mbox{where}\:\,
\mathsp{X}=
\begin{pmatrix}
\,\nullmatrix\, &\, \identity\; \\ \,\identity\, &\, \nullmatrix\;
\end{pmatrix}\,.
\end{equation}
Suppose $\ket{\ffq}$ is an eigenvector of the Hermitian matrix $\mathsp{H}_\mathrm{ncb}$, we have
\begin{subequations}
\begin{align}
\mathsp{H}_\mathrm{ncb}\, \ket{\ffq}=\lambda\, \ket{\ffq}
&\;\Longrightarrow\;  \mathsp{X} \mathsp{H}_\mathrm{ncb}\ssp \mathsp{X}\mathsp{X} \ket{\ffq}=\lambda\, \mathsp{X} \ket{\ffq}\\
&\;\Longrightarrow\; \mathsp{H}_\mathrm{ncb}^*\, \mathsp{X}\ket{\ffq}=\lambda\, \mathsp{X}\ket{\ffq}\\
&\;\Longrightarrow\;  \mathsp{H}_\mathrm{ncb}\, \mathsp{X} \ket{\ffq}^*=\lambda\, \mathsp{X} \ket{\ffq}^*\,.
\end{align}
\end{subequations}
Thus, both $\ket{\ffq}$ and $\mathsp{X} \ket{\ffq}^*$ are eigenvectors of $\mathsp{H}_\mathrm{ncb}$.  More generally, any linear combination of them is also an eigenvector of $\mathsp{H}_\mathrm{ncb}$ for the same eigenvalue $\lambda$.  Particularly, we introduce
\begin{gather}
\ket{\ffq_+}=\ket{\ffq}+\mathsp{X} \ket{\ffq}^*\quad\mbox{and}\;\;\ket{\ffq_-}=i\ssp \big(\ket{\ffq}-\mathsp{X} \ket{\ffq}^*\big)\;,
\end{gather}
which satisfy the condition
\begin{equation}\lb{eq:X*_condition}
\ket{\ffq_{\pm}}=\mathsp{X} \ket{\ffq_{\pm}}^*\;.
\end{equation}
Using this construction, we can always find a complete set of eigenvectors of $\mathsp{H}_\mathrm{ncb}$ satisfying  Eq.~(\rf{eq:X*_condition}), and this is the same restriction on the vector $\big(\;\ket{\psinot_\perp }\quad \ket{\psinot^*_\perp }\,\big)^T$. Thus, Eq.~(\rf{eq:matrix_form}) says that the minimum eigenvalue of $\mathsp{H}_\mathrm{ncb}$ is nonnegative; i.e., $\mathsp{H}_\mathrm{ncb}$ is positive definite.

We can ask another question: Is the GP ground state the only GP eigenstate that is stable under Bogoliubov excitations?  The answer is NO, as one simple example shows.  Consider a BEC confined in a one-dimensional interval, $x\in [0,L]$, with periodic boundary condition. Any momentum eigenstate, i.e., $\psi_k(x) = e^{ikx}/\sqrt{L}$, is also an eigenstate of the GP equation,
\begin{equation}
 \Big(\mathord{-}\frac{\hbar^2}{2m}\frac{\partial^2}{\partial x^2} +g N
\norm{\psi_k}^2-\mu_k\Big)\psi_k = 0\,, \quad \mbox{where}\;\; \mu_k=\frac{\hbar^2 k^2}{2m}+g\ssp n\;,
\end{equation}
with $n=N/L$.  Stability requires that 
\begin{equation}
\mathsp{H}_\mathrm{ncb} =
\begin{pmatrix}
H_\mathrm{gp}+g \ssp n\,  Q_k  & g
\ssp n\, Q_k\ssp e^{2ikx}\ssp Q^*_k\\[6pt] g
\ssp n\, Q^*_k\ssp e^{-2ikx}\ssp Q_k & H_\mathrm{gp}+g  \ssp n\, Q^*_k
\end{pmatrix}\,,
\end{equation}
be nonnegative, where $Q_k$ ($Q_k^*$) is the projector onto the orthogonal space of $\ket{\psi_k}$ (\,$\ket{\psi_{-k}}$).  Because the matrix $\mathsp{H}_\mathrm{ncb}$ only couples the vector $\big(\;\ket{\psi_{k+p}}\quad \ket{\psi^*_{k+p} }\,\big)^T/\sqrt{2}\,$ to $\big(\;\ket{\psi_{k-p}}\quad \ket{\psi^*_{k-p} }\,\big)^T/\sqrt{2}$ for $p\neq 0$, we have the following in the subspace spanned by these two vectors,
\begin{equation}
\mathsp{H}_\mathrm{ncb}^{(p)} =
\begin{pmatrix}
{\displaystyle\frac{\hbar^2 (p^2+2kp)}{2m}}  +g\ssp n & g\ssp n
\\[6pt] g\ssp n  & {\displaystyle\frac{\hbar^2 (p^2-2kp)}{2m}}  +g\ssp n
\end{pmatrix}\,,
\end{equation}
which is positive definite if and only if
\begin{align}\hspace{-1.5em}
 \hbar^2 p^2/2m+g\ssp n \geq \sqrt{ (\hbar^2 kp/m)^2 +g^2\ssp n^2}\;\; \Longleftrightarrow\;\; \hbar^2 k^2/m \leq \hbar^2 p^2/4m + g\ssp n\;.
\end{align}
Stability requires that this condition holds for all $p$, and the strongest of which is when $L\rightarrow \infty$,
\begin{align}\lb{eq:condition_of_stability}
 \norm{k} \leq \sqrt{g\ssp n\ssp m}/\hbar\;.
\end{align}
Since the right hand side of Eq.~(\rf{eq:condition_of_stability}) is of finite value, a BEC can be stable even if it is not in the GP ground state.

\renewcommand{\lb}[1]{\label{at_factorization:#1}}
\renewcommand{\rf}[1]{\ref{at_factorization:#1}}

\chapter{The Autonne-Takagi Factorization}
\label{ch:at_factorization}

\begin{quote}
The mathematical facts worthy of being studied are those which, by their analogy with other facts, are capable of leading us to the knowledge of a physical law.
\\[4pt]
-- Henri Poincare\ai{Poincare, Henri}
\end{quote}

\noindent In this Appendix, I discuss the Autonne-Takagi factorization\si{Autonne-Takagi factorization} theorem (see Corollary 4.4.4 (c) in~\cite{horn_matrix_2013}) in a way that physicists might found more approachable.  The Autonne-Takagi factorization is useful for writing down the Schmidt decomposition form of an arbitrary two-boson wavefunction~\cite{paskauskas_quantum_2001}.  Also, it is the tool to decomposing a multimode squeeze operator into single-mode squeezers (see Sec.~\chref{symplectic:sec:the_polar_decomposition}).

The Autonne-Takagi factorization theorem says that any complex symmetric matrix $\Phi = \Phi^T$\si{Symmetric matrix} can be factorized into the following form:
\begin{equation}\lb{eq:at_factorization}
U^{T}\nsp \Phi\ssp\ssp U=\Lambda\;,
\end{equation}
where $U$ is a unitary matrix and $\Lambda=\diag(\lambda_1,\lambda_2,\ldots,\lambda_\nrank)$.  Note that the phases of the $\lambda_j$s are arbitrary; one choice is to make the $\lambda_j$s nonnegative by absorbing half of these phases into the column vectors of $U$ and half to the row vectors of $U^T$.  Consider the following time-reversal eigenvalue equation,
\begin{align}\lb{eq:eigenvalue}
\Phi\,\ket{u_j}=\lambda_j\ssp \ket{u_j}^*\;,
\end{align}
where the column vector $\ket{u_j}$ is normalized, and $\ket{u_j}^*\equiv\ket{u_j^*}$ is the complex conjugate of $\ket{u_j}$ in the computational basis $\ket{e_j}$.  The eigenvalue $\lambda_j=\norm{\lambda_j}\ssp e^{i\theta_j}$ is generally complex, but the phase $e^{i\theta_j}$ can always be absorbed into the eigenvector,
\begin{equation}
\Phi\ssp \big(e^{-i\ssp \theta_j/2}\,\ket{u_j}\big)=\norm{\lambda_j}\,\big(e^{-i\ssp \theta_j/2}\,\ket{u_j}\big)^*\;;
\end{equation}
without loss of generality, we assume $\lambda_j\geq 0$ for $j=1,2,\ldots,\nrank$.  The transpose of the eigenvalue equation~(\rf{eq:eigenvalue}) reads
\begin{equation}
\bra{u_k^*} \Phi=\lambda_k\ssp \bra{u_k}\;,
\end{equation}
and we have the following condition for the eigenvectors $\ket{u_j}$ and $\ket{u_k}$,
\begin{equation}
\bra{u_k^*} \Phi\,\ket{u_j}=\lambda_j\braket{u_k^*}{u_j^*}=\lambda_k\braket{u_k}{u_j}\;.
\end{equation}
For $\lambda_j\neq \lambda_k$ the above condition is equivalent to
\begin{equation}
\braket{u_k}{u_j}=0\;,
\end{equation}
and thus $\{\,\ket{u_j}\;\vert\;j=1,2,\ldots\,\nrank\}$ form an orthonormal basis when there are no degenerate eigenvalues.\footnote{The generalization to the degenerate case can be done by linearly combining the eigenvectors of the same eigenvalue.}  With the eigenvectors, the matrix $\Phi$ takes the form
\begin{equation}
\Phi=\sum_j \lambda_j\ssp \ket{u_j^*}\bra{u_j}\;,
\end{equation}
which is manifestly symmetric.  Defining the unitary matrix $U = \sum_j \ket{u_j}\bra{e_j}$, we have
\begin{equation}
 U^T \Phi\ssp\ssp U = \sum_j \lambda_j\ssp \ket{e_j^*} \bra{e_j} = \sum_j \lambda_j\ssp \ket{e_j}\bra{e_j} = \Lambda\;,
\end{equation}
where the fact $\ket{e_j^*}=\ket{e_j}$ is used.

One interesting question is what equation determines the eigenvalues $\lambda_j$. By using Eq.~(\rf{eq:eigenvalue}) twice, we have
\begin{equation}
\Phi^\dagger \Phi\ssp\ket{u_j}=\Phi^*\lambda_j\ssp \ket{u_j^*}=\norm{\lambda_j}^2\ssp \ket{u_j}\;.
\end{equation}
Thus, $\ket{u_j}$ is an eigenvector of the nonnegative matrix $\Phi^\dagger \Phi$, and we have the following equation that determines its eigenvalues,
\begin{equation}
\det\ssp(\ssp \Phi^{\dag} \Phi-\norm{\lambda}^2\ssp \identity \ssp)=0\;,
\end{equation}
where $\identity$ is the identity matrix.

To end this Appendix, I discuss how to validate the eigenvalue equation~(\rf{eq:eigenvalue}) from the conventional eigenvalue equation
\begin{equation}\lb{eq:eigenvalue_double_a}
\Phi^\dagger \Phi\ssp\ket{u}=\Phi^* \Phi\ssp\ket{u}=\norm{\lambda}^2\ssp \ket{u}\;.
\end{equation}
By introduce the vector $\ket{v} = \Phi^*\ssp \ket{u^*}$, Eq.~(\rf{eq:eigenvalue_double_a}) takes the form
\begin{equation}\lb{eq:eigenvalue_double_b}
\Phi\ssp \ket{u}=\ssp\ket{v^*}\;,\quad \mbox{and}\;\;\Phi^*\ssp \ket{v^*}=\norm{\lambda}^2 \ssp\ket{u}\;.
\end{equation}
Note that $\ket{v}$ is also an eigenvector of $\Phi^\dagger \Phi$ with the same eigenvalue $\norm{\lambda}^2$,
\begin{equation}
\Phi^\dagger \Phi\ssp\ket{v}=\Phi^* \Phi\ssp\ket{v}=\norm{\lambda}^2\ssp\Phi^*\ket{u^*}=\norm{\lambda}^2\ssp\ket{v}\;.
\end{equation}
As a consequence, $\ket{v}$ is a multiple of $\ket{u}$ if the matrix $\Phi^\dagger \Phi$ is nondegenerate.  Note also that we have the following normalization relation for the two vectors,
\begin{align}
 \braket{v}{v}=\braket{v^*}{v^*} = \bra{u} \Phi^\dagger\Phi\ssp \ket{u} = \norm{\lambda}^2 \braket{u}{u}\;.
\end{align}
Thus we can always choose the phase of $\lambda$ such that $\ket{v}=\lambda\ssp \ket{u}$, and Eq.~(\rf{eq:eigenvalue_double_b}) is then equivalent to Eq.~(\rf{eq:eigenvalue}).

\backmatter

\newcommand{\etalchar}[1]{$^{#1}$}
\def\alphabibitem#1{%
\expandafter\ifx\csname firstalphabibitem\endcsname
\relax\def\firstalphabibitem{}
\else\expandafter
\ifx\csname alphabibitemskip\endcsname
\relax\bigskip
\else\alphabibitemskip
\fi
\fi
\penalty-200\expandafter
\ifx\csname alphabibitemlabel\endcsname
\relax\expandafter
\ifx\csname alphabibitemcontent\endcsname\relax
\item[\LARGE\bfseries\MakeUppercase{#1}]
\hfill\vskip 1ex\hrule\medskip
\else\item[\Large\bfseries\MakeUppercase{#1}]
\alphabibitemcontent{\MakeUppercase{#1}}%
\fi
\else\expandafter
\ifx\csname alphabibitemcontent\endcsname\relax
\item[\alphabibitemlabel{\MakeUppercase{#1}}]
\vskip 1ex\hrule\medskip
\else\item[\alphabibitemlabel{\MakeUppercase{#1}}]
\alphabibitemcontent{\MakeUppercase{#1}}%
\fi
\fi}

\cleardoublepage
\thispagestyle{plain}
\phantomsection
\printindex{ai}{Author Index}
\chaptermark{Author Index}
\thispagestyle{plain}
\printindex{si}{Subject Index}
\chaptermark{Subject Index}

\end{document}